\documentclass[a4paper,11pt]{article}
\pdfoutput=1
\usepackage{jheppub}
\usepackage{amsmath,amssymb,bm,graphicx,epsf,colordvi}
\usepackage{slashed}
\def\bea#1\eea{\begin{align}#1\end{align}} 
\usepackage{diagbox}
\usepackage{mathtools}
\usepackage{lipsum}
\usepackage{soul}
\usepackage{enumitem}
\usepackage{braket}
\usepackage{bm}
\usepackage{algorithm}
\usepackage[noend]{algpseudocode}

\usepackage{tikz}
\usepackage{tikzit}
\usetikzlibrary{snakes}
\usetikzlibrary{decorations.markings}
\tikzset{snake it/.style={decorate, decoration=snake}}

\allowdisplaybreaks 
\usepackage{color}
\usepackage[dvipsnames]{xcolor}

\usepackage{mathrsfs}
\usepackage{appendix}
\usepackage{bm}
\usepackage{lipsum}

\newcommand{\bef}{\begin{figure}[hbt]\centering}
\newcommand{\eef}{\end{figure}}
\definecolor{or}{rgb}{0.6,0.,1.}
\definecolor{forest}{RGB}{34,139,34}
\usepackage{mathtools}
\usepackage{subfigure}
\usepackage{booktabs}
\usepackage{graphicx}
\graphicspath{ {./figure/} }
\usepackage{caption}
\usepackage{lipsum}

\newcommand{\ENC}{E$^{N}$C }

\newcommand{\beq}{\begin{equation}}
\newcommand{\eeq}{\end{equation}}
\def\bea#1\eea{\begin{align}#1\end{align}}

\def \be  {\begin{equation}}
\def \ee  {\end{equation}}
\def \ba  {\begin{eqnarray}}
\def \ea  {\end{eqnarray}}
\newcommand{\df}{\mathrm{d}}

\newcommand{\nn}{\nonumber}

\allowdisplaybreaks

\usepackage{graphicx}

\newcommand{\w}{\omega}
\newcommand{\e}{\epsilon}
\newcommand{\de}{\delta}
\newcommand{\half}{\frac{1}{2}}
\newcommand{\tl}[1]{\widetilde{#1}}

\def\Fig#1{Fig.~{\ref{#1}}}
\DeclareRobustCommand{\Sec}[1]{Sec.~\ref{#1}}
\DeclareRobustCommand{\Eq}[1]{Eq.~(\ref{#1})}

\newcommand{\fd}[2]{\parbox{#1}{\includegraphics[width=#1]{#2}}}

\usepackage{stackengine}

\title{Dissecting Parton Showers with Multi-Point Energy Correlators}

\author[1]{Mark Gonzalez,}

\author[2]{Philip Harris,}

\author[3]{Kyle Lee,}

\author[1]{Ian Moult,}

\author[2]{Simon Rothman}

\affiliation[1]{Department of Physics, Yale University, New Haven, CT 06511}
\affiliation[2]{Massachusetts Institute of Technology, Cambridge, MA 02139}
\affiliation[3]{High Energy Physics Division, Argonne National Laboratory, Lemont, IL, USA}
\affiliation[1]{Department of Physics, Yale University, New Haven, CT 06511}

\abstract{The last several years have seem tremendous progress in the ability to both compute and measure multi-point correlations in energy flux. 
The highly differential nature of energy correlators makes them ideal probes of multi-collinear factorization and azimuthal structure within jets.
In this paper, we explore the phenomenology of four-point correlators in jet substructure. 
We identify experimentally realizable projections that probe different factorization channels onto splitting tensors and splitting functions.
We perform a detailed phenomenological study using both Herwig and Pythia. By comparing parton shower results with analytic calculations in kinematic limits, we are able to disentangle intrinsic spin correlations from kinematic azimuthal correlations. 
In experimentally accessible kinematic regions, we find the  spin correlations are subdominant, strongly motivating a complete calculation of the four-point correlator in QCD to provide a test of the parton shower results.
We also present parameterizations and analysis algorithms that can be used experimentally. 
Our work sets the stage for the experimental measurement of these observables at the LHC, and their use as probes of the next generation of parton showers.
}

\begin{document}

\maketitle

\section{Introduction}

Colliders provide one of the most powerful means of studying a variety of systems in high energy and nuclear physics. In the last decade, there has been tremendous progress in our ability to extract more detailed information from collider experiments, using jet substructure. For reviews, see \cite{Larkoski:2017jix,Asquith:2018igt}. The study of the detailed internal structure of jets has provided new ways to search for physics beyond the Standard Model, perform precision measurements of Standard Model parameters, and probe hot and cold nuclear matter. Continued progress relies on the exploration of new jet substructure observables, which ultimately lead to new tools for phenomenological applications.

Energy correlators are a specific class of jet substructure observables where there has been tremendous progress in the last few years. They are defined as correlation functions of energy flow operators
\cite{Sveshnikov:1995vi,Tkachov:1995kk,Korchemsky:1999kt,Bauer:2008dt,Hofman:2008ar,Belitsky:2013xxa,Belitsky:2013bja,Kravchuk:2018htv}
\begin{align}\label{eq:ANEC_op}
\mathcal{E}(\vec n) = \lim_{r\to \infty}  \int\limits_0^\infty \df t~ r^2 n^i T_{0i}(t,r \vec n)\,.
\end{align}
These observables have a long history, originally being proposed for $e^+e^-$ collisions  \cite{Basham:1979gh,Basham:1978zq,Basham:1978bw,Basham:1977iq}. The two-point correlator (also referred to as the Energy Energy Correlator, or EEC), was measured at SLD and LEP  \cite{SLD:1994idb,L3:1992btq,OPAL:1991uui,TOPAZ:1989yod,TASSO:1987mcs,JADE:1984taa,Fernandez:1984db,Wood:1987uf,CELLO:1982rca,PLUTO:1985yzc}. Recently they have been proposed as phenomenological jet substructure observables at the LHC \cite{Dixon:2019uzg,Chen:2020vvp,Lee:2022uwt}. This has led to their measurement in a wide variety of systems, yielding the most precise extraction of the strong coupling constant from jet substructure \cite{CMS:2024mlf}, new ways to probe the quark-gluon plasma \cite{CMS:2024ovv}, and new probes of heavy-quark dynamics~\cite{ALICE:2025igw}, amongst many other applications. For a review of recent applications of energy correlators, see \cite{Moult:2025nhu}.

As compared to the original incarnation of energy correlators in the LEP era, one of the new aspects in the study of energy correlators is the possibility to both calculate and measure higher-point correlators, analogous to non-gaussianities in the study of the cosmic microwave background (CMB). On the experimental side, this has been made possible by the combination of high-resolution detectors with the extremely high energies of the LHC, which enables multi-point correlators to be measured in a regime where they manifest the perturbative dynamics of QCD. On the theoretical side, it has been driven by remarkable advances in perturbative quantum field theory, which have made the analytic calculation of multi-point correlators possible. The advent of precision calculations on tracks \cite{Chang:2013iba,Chang:2013rca,Chen:2022muj,Jaarsma:2023ell,Chen:2022pdu,Li:2021zcf,Jaarsma:2022kdd,Lee:2023npz,Lee:2023tkr,Lee:2023xzv,Jaarsma:2025tck,Lee:2026hub} has made it possible to interface theory and experiment for these complicated observables.

The three-point energy correlator was first computed in the collinear limit in \cite{Chen:2019bpb}. It was experimentally measured inside high energy jets at the LHC using CMS Open Data in \cite{Chen:2022swd,Komiske:2022enw}. Since then, there has been tremendous progress in the theoretical calculation of multi-point correlators in both $\mathcal{N}=4$ super Yang-Mills, and QCD. This includes the calculation of the four-point collinear limit of the energy correlator in $\mathcal{N}=4$ super Yang-Mills \cite{Chicherin:2024ifn}, the calculation of the full three-point correlators in $\mathcal{N}=4$ super Yang-Mills \cite{Yan:2022cye} and QCD \cite{Yang:2022tgm,Yang:2024gcn}, and progress towards simplifying and extending these calculations to higher-point correlators \cite{He:2025zbz,Ma:2025qtx,Volovich:2026pup}. In turn, this has motivated more formal studies, leading in particular, to the determination of the integrand for the energy correlator in the collinear limit of $\mathcal{N}=4$ super Yang-Mills up to 12 points \cite{He:2024hbb}. The availability of these analytic results for multi-point correlators opens up many opportunities for improving our understanding of the substructure of jets at hadron colliders.

\begin{figure}[h]
    \centering
        \includegraphics[width=0.3\textwidth]{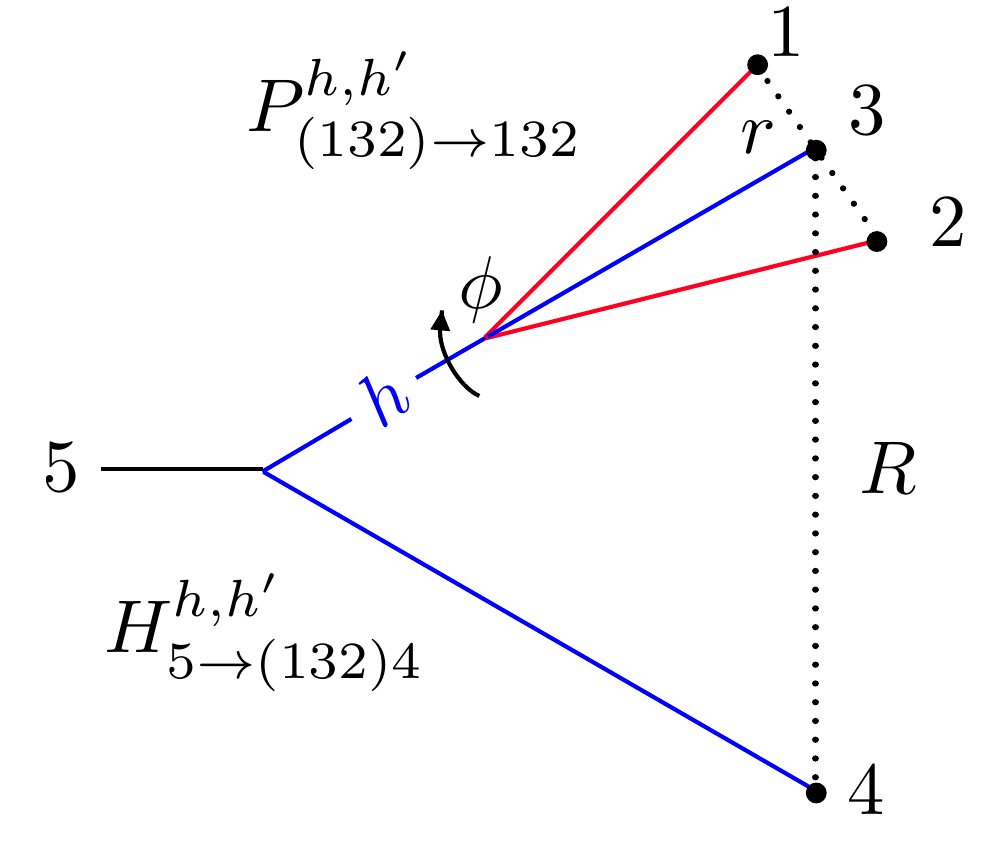} \hspace{1.5cm}
         \raisebox{0.5cm}{\includegraphics[width=0.4\textwidth]{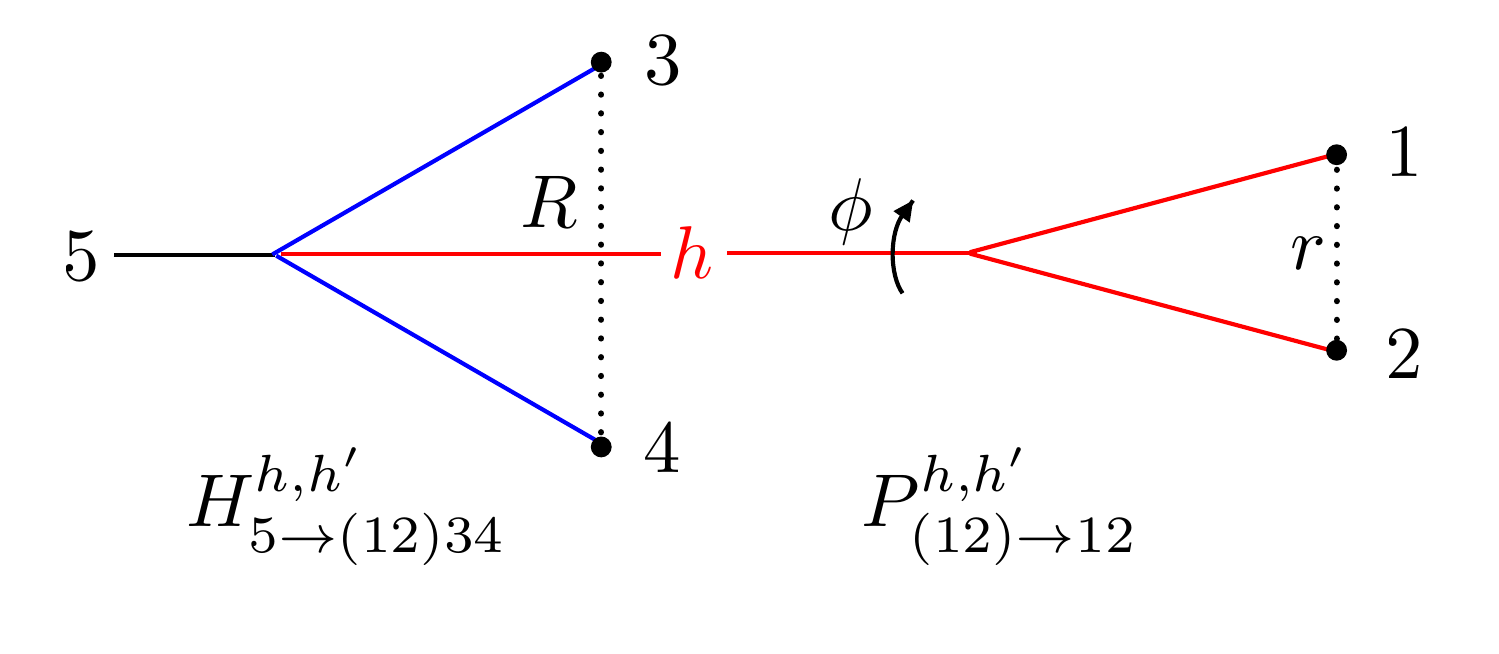} }
    \includegraphics[width=0.45\textwidth]{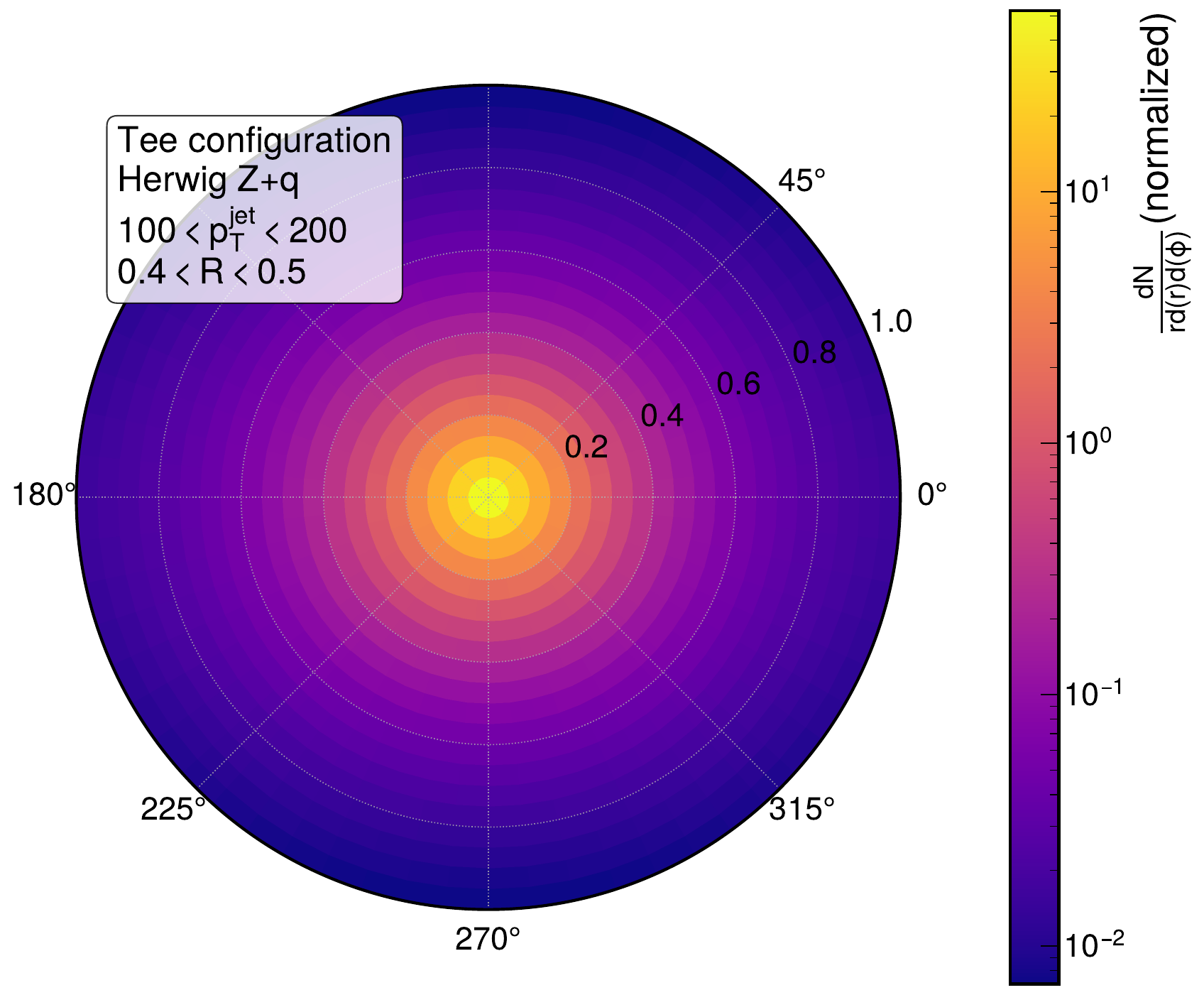} 
        \includegraphics[width=0.45\textwidth]{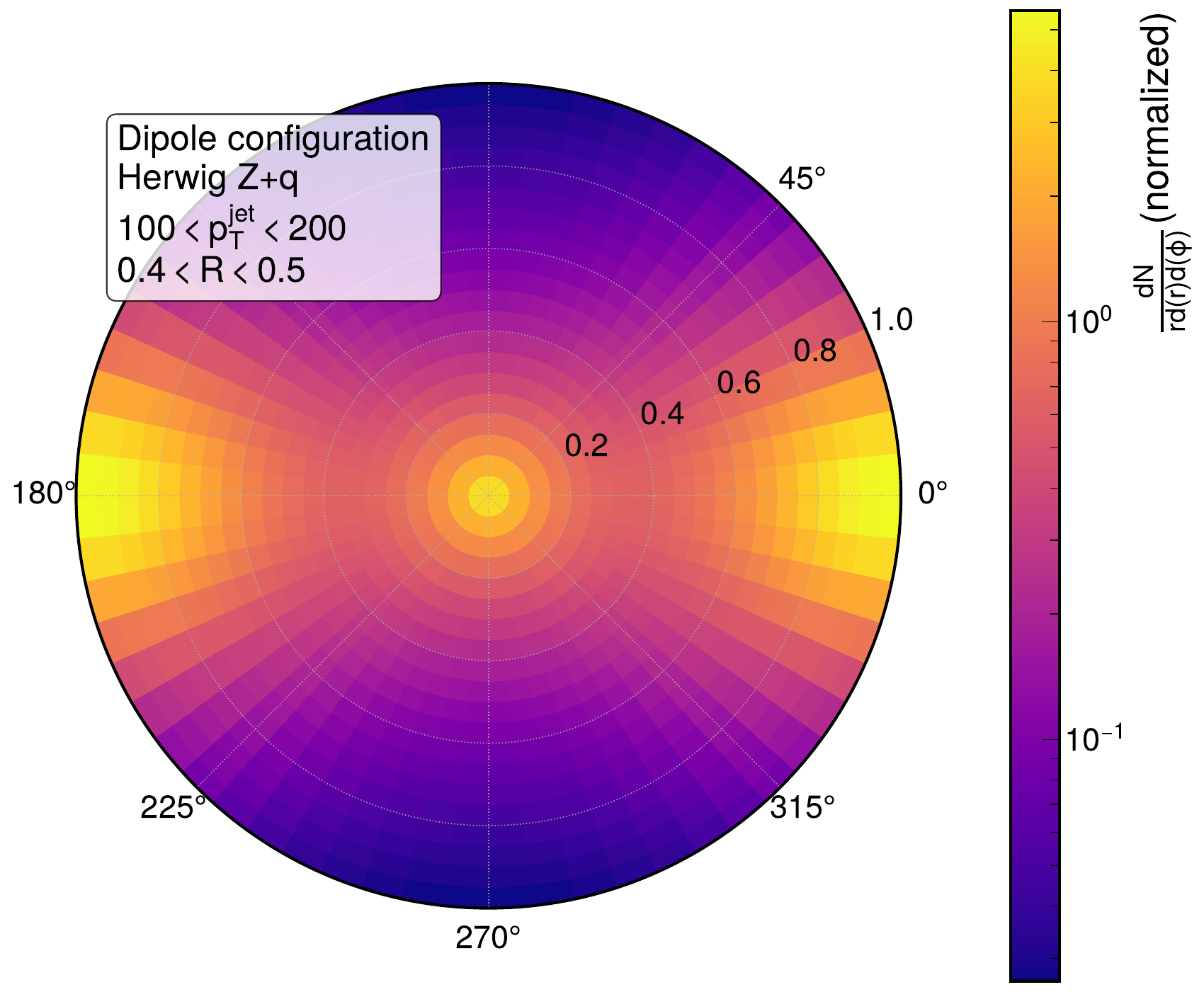} 
    \caption{Projections of the four-point correlator, which we call ``the tee" and ``the dipole", are designed to probe different factorization channels of the four-point splitting function, providing valuable information for the development of parton shower simulations.}
    \label{fig:intro_plot}
\end{figure}

Parton shower algorithms are based on the idea of factorization. The simplest algorithms capture only the leading logarithmic approximation, in which (loosely speaking) the shower is built from iterated $1\to 2$ splittings. In the last decade, driven by ever increasing experimental demands, there has been a push to improve the theoretical description of parton showers by incorporating the physics that this approximation leaves out: spin correlations \cite{Knowles:1988hu,Knowles:1987cu,Knowles:1988vs,Bahr:2008pv,Bellm:2015jjp,Bellm:2019zci,Richardson:2018pvo,Karlberg:2021kwr,Hamilton:2021dyz}, subleading color effects, higher order DGLAP evolution, and higher-point splitting functions \cite{Hoche:2017iem}. For a collection of references on additional progress in parton showers, see \cite{Dasgupta:2020fwr,Hamilton:2020rcu,Hoche:2017hno,Hoche:2017iem,vanBeekveld:2025lpz,vanBeekveld:2024qxs,FerrarioRavasio:2023kyg,Hamilton:2023dwb,vanBeekveld:2022ukn,vanBeekveld:2022zhl,Hoche:2024dee}. Each of these ingredients modifies the radiation pattern in a characteristic way. As this program progresses, it is therefore important to have not only complete theoretical calculations to compare against, but also observables that are specifically sensitive to the new ingredients, in particular to the iterated structure of higher-point splittings and to their spin correlations.

A particularly appealing aspect of the energy correlators in this regard is that they are fully differential in angle. This makes them ideal probes of multi-collinear factorization, since clusters of correlators can be taken collinear, exactly mapping them onto collinear factorization of the underlying amplitude (squared). Additionally, they retain dependence on azimuthal correlations, enabling one to probe spin correlations. While jet substructure at the LHC is mostly studied in an unpolarized setting, by rotating clusters of correlators \emph{with respect to each other}, one can study spin correlations inside jets. 

This was studied in detail for the three-point correlator in \cite{Chen:2021gdk,Chen:2020adz}. The three-point correlator has only a single factorization channel: when two of the correlators become collinear, it factorizes, directly probing the factorization of the underlying three-point splitting function into a $1\to 2$ splitting tensor and a $1\to 2$ splitting function. Analytic predictions of spin correlations in this limit \cite{Chen:2020adz} were used as a validation of their implementation in parton showers  \cite{Karlberg:2021kwr}.

While the three-particle splitting function has been known for a long time \cite{Catani:1998nv,Campbell:1997hg}, there has been recent progress to simplify it for implementation in parton showers, including incorporating masses \cite{Braun-White:2022rtg,Craft:2023aew,Hoche:2025vto,Dhani:2023uxu,Campbell:2025lrs}. The four-particle splitting functions have also been computed  \cite{DelDuca:2020vst,DelDuca:2019ggv}. We are therefore led to explore analytically calculable observables that are capable of probing the spin structure of $1\to 3$ splitting functions and splitting tensors, and that can be measured in data. These will provide stringent tests of the next generation of parton shower generators, as well as motivate new QCD calculations.

Beyond the case of massless QCD, many applications of the three-point correlator have been proposed. These include distinguishing different physical effects of nuclear modification in heavy ion collisions \cite{Bossi:2024qho,Barata:2025fzd}, and measurements of the top quark mass \cite{Holguin:2023bjf,Holguin:2022epo,Holguin:2024tkz}. It would also be interesting to incorporate mass effects from charm and beauty quarks~\cite{Barata:2025uxp,Craft:2022kdo,Andres:2023ymw,Gao:2026xuq}. New related observables will surely also lead to phenomenological advances in these areas.

In this paper, we perform a systematic exploration of the four-point energy correlator as a phenomenological jet substructure observable, focusing on dissecting the structure of higher order parton showers. While the full four-point correlator is a multi-variable object which is difficult to visualize, we introduce several simple projections, which are both experimentally feasible and probe specific factorization channels of the four-point splitting amplitudes. Two of these configurations, along with their simulated results, are shown in \Fig{fig:intro_plot}. In the first configuration, which we refer to as ``the tee", a cluster of three correlators is taken to be collinear, probing the factorization of the $1\to 4$ splitting function into a $1\to 2$ splitting tensor and a $1\to 3$ splitting function. In the second configuration, which we call ``the dipole", one pair of correlators is taken collinear and rotated with respect to the other pair, probing the complementary factorization into a $1\to 3$ splitting tensor and a $1\to 2$ splitting function. We also identify a third projection, which we refer to as ``the tripole". As compared to the three-point correlator, one of the reasons the four-point correlator is particularly interesting is the presence of these multiple factorization channels, and we are able to cleanly access them all. We believe that observables probing these configurations will be invaluable for the development of the next generation of parton showers.

In both cases, our projections provide access to the spin correlations in the different factorized limits, which can be studied as a function of the ratio of the sizes of the two clusters. This separation of variables allows us to disentangle perturbative power corrections to the factorization from genuine spin correlations. While we will ultimately find that genuine spin correlations are small for phenomenological purposes, we believe that they will be important for validation of parton shower simulations, as was the case for the three-point correlator \cite{Chen:2020adz}.

The goal of this paper is to lay the groundwork for an experimental measurement of the four-point correlator, and future comparisons with parton showers. We begin by presenting explicit parameterizations for different configurations of interest of the four-point correlator. Although the four-point correlator has not yet been analytically computed in QCD, we use the known splitting functions and splitting tensors to compute it in iterated limits. This provides a baseline for the leading power factorized results in these limits, as well as a calculation of the intrinsic spin correlations. We then perform a detailed parton shower study, using both Pythia and Herwig, of the four-point correlator in our proposed kinematic limits. In particular, we study the effect of soft and collinear spin correlations, as implemented in these showers, on our observables. We find that these are small, consistent with our analytic calculations, compared to relatively large azimuthal modulations observed in our distributions, which we therefore interpret as kinematic power corrections associated with our factorizations. This strongly motivates the complete calculation of the four-point energy correlator in QCD. 

To facilitate future experimental measurements of the four-point correlator at the LHC, we also provide explicit algorithms for computing the correlators, as well as a discussion of the treatment of statistical uncertainties for multi-point correlators. We believe these will have applications in other experimental studies of energy correlators. Our work lays the foundations for the experimental study of the four-point correlator, as well as for future phenomenological applications.

An outline of this paper is as follows. In \Sec{sec:4ptProj} we discuss different configurations of the four-point correlator that are interesting from a physics perspective, and show how they can be parameterized. We also present plots for these configurations using the known results in $\mathcal{N}=4$ super Yang-Mills.  In \Sec{sec:proj} we discuss in detail the structure of multi-point energy correlators in QCD, their calculation from splitting functions, and their factorization in iterated limits. Although we are not currently able to compute the full four-point correlator in QCD, we compute it in certain kinematic limits using iterated triple collinear splitting functions, focusing in particular on the magnitude of spin interference effects. In \Sec{sec:pheno} we perform a detailed phenomenological study, using the parton shower generators Pythia and Herwig. We present practical algorithms and parameterizations which can be used for experimental analysis of the four-point correlator.   We conclude in \Sec{sec:conc}. We provide a number of appendices describing in detail the experimental practicalities of identifying the tee, dipole and tripole configurations, with the hope that this will aid their experimental implementation.

\section{Projections of Four-Point Correlators}
\label{sec:4ptProj}

\begin{figure}
\begin{center}
\includegraphics[scale=0.2]{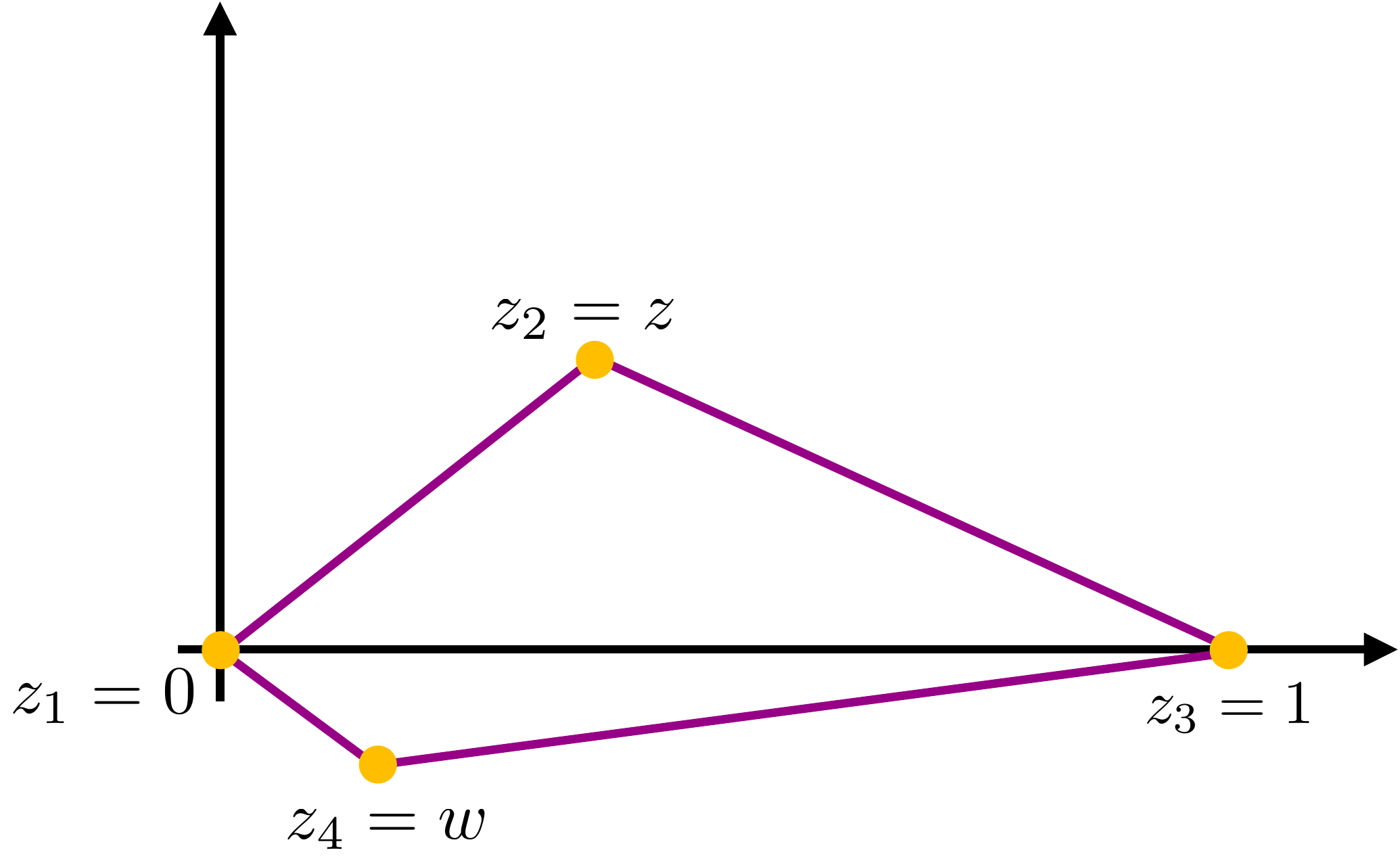} 
\end{center}
\caption{Parameterization of a generic labeled four-detector configuration in the collinear plane. After using translations, rotations, and an overall rescaling to place a chosen reference pair at $z_1=0$ and $z_3=1$, the remaining shape degrees of freedom are encoded by the two complex variables $z_2=z$ and $z_4=w$. }
\label{fig:param}
\end{figure}

In this section, we introduce several projections of the four-point correlator, which we will focus on in our phenomenological studies. For applications to the LHC, we are interested in the study of the four point correlator in the collinear limit, namely inside a single high energy jet.  The collinear limit of the four-point correlator depends on a scaling variable, $R_L$, which sets its overall size, and two complex variables, $z$, $w$, which parameterize its shape \cite{Chicherin:2024ifn}. The $z$ and $w$ variables can be drawn in the complex plane to visualize the shape of the four-point correlator, and are illustrated in \Fig{fig:param}

While these variables are theoretically convenient, and indeed the analytic results for the correlator in $\mathcal{N}=4$ sYM are expressed in terms of these variables, experimentally they are challenging, since they define a four dimensional space that is difficult to bin.  For experimental studies, we would like to identify specific projections of the full four-point correlator that contain interesting physics effects and provide explicit parameterizations of these projections, allowing them to be directly studied in data, including understanding binning uncertainties.

For the three-point correlator, this was achieved in \cite{Komiske:2022enw}. A modification of this parameterization that is convenient in the case of more uniform distributions, as present in heavy ion collisions, was given in \cite{Bossi:2024qho}. Another approach to parameterizations of multi-point correlators was presented in \cite{Alipour-fard:2024szj}.

In this section, we present several projections of the four-point correlator that we believe are of interest experimentally and show how they can be parameterized. We refer to these configurations as ``the dipole", ``the tee", and the ``the tripole". As we will explain,  these projections are explicitly designed to probe different factorization channels of the $1\to 4$ splitting function. In particular, ``the dipole" configuration is designed to probe the factorization $1\to 4$ splitting function into a $1\to 3$ splitting tensor and a $1\to 2$ splitting function, while ``the tee" is designed to probe the factorization into a $1\to 2$ splitting tensor and a $1\to 3$ splitting function. We believe that observables that probe these configurations will be invaluable for the next generation of parton showers. 
We hope that these inspire the experimental measurements of the four-point correlator in real LHC data. 

One of our primary interests in designing the projections is to expose angular correlations within jets, which can arise from spin correlations from perturbative gluons.  There has been significant recent work incorporating spin correlations into improved parton showers \cite{Hamilton:2021dyz,Karlberg:2021kwr}. The effect of spin correlations was studied for the three-point correlator in \cite{Chen:2020adz,Chen:2021gdk}. However, it was found that spin correlations in the three-point correlator were tiny due to cancellations between quarks and gluons. Although we will find that spin correlations are small also in the four-point correlator, we still believe that the observables that we introduce will be important for improving our understanding of parton showers.

As we introduce our different parameterizations, we also show plots of these configurations using the complete analytic results for the four-point correlator in $\mathcal{N}=4$ sYM \cite{Chicherin:2024ifn}. We emphasize that an important difference between QCD and sYM, is that supersymmetry suppresses spin correlations in sYM. In \Sec{sec:proj}, we will perform analytic calculations to identify the magnitude of the spin correlations in QCD.

\subsection{The Dipole Configuration}\label{sec:proj_a}

\begin{figure}
\begin{center}
\raisebox{0.0cm}{\includegraphics[scale=0.30]{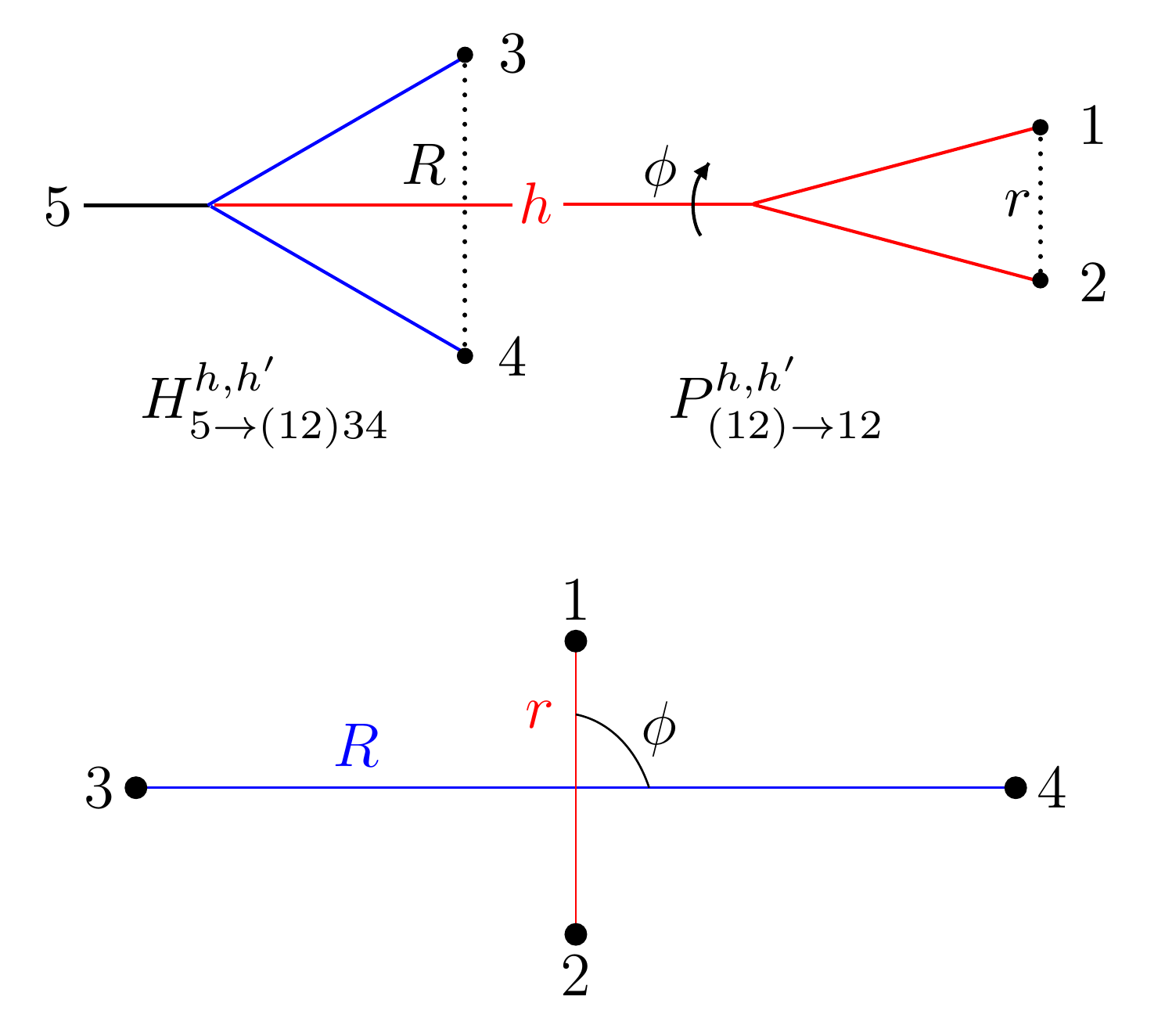}}
\includegraphics[scale=0.30]{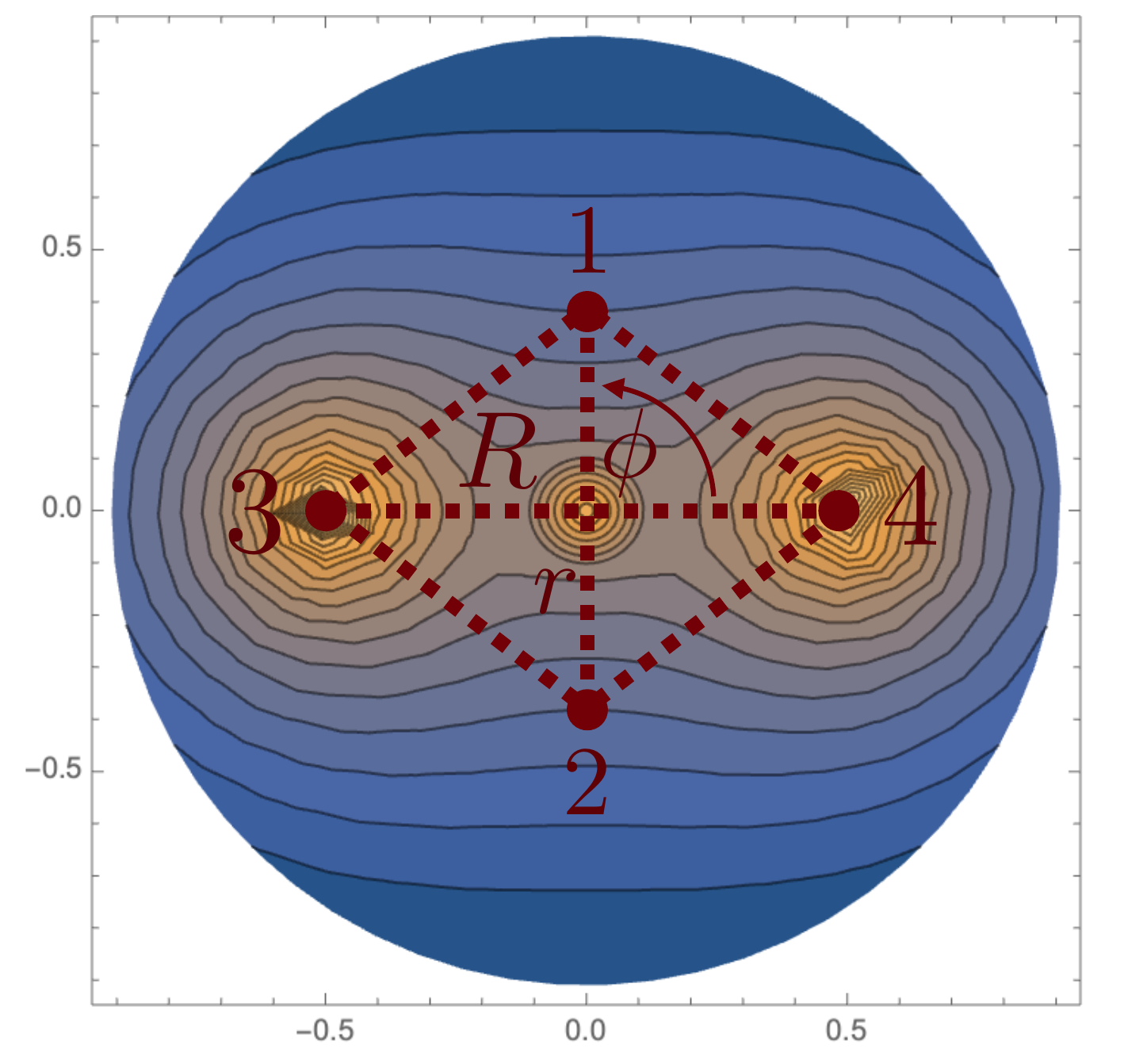} 
\end{center}
\caption{The ``dipole" configuration: On the left we show the kinematic configuration of the four detectors. Two dipoles of size $r$ and $R$ are rotated with respect to each other by an angle $\phi$. On the right we show a plot of this configuration in $\mathcal{N}=4$ sYM. The location of the detectors in the larger dipole remain fix, while the smaller dipole is rotated. 
}
\label{fig:dipole}
\end{figure}

The first configuration we introduce is referred to as ``the dipole". In this configuration, illustrated in \Fig{fig:dipole}, we consider the rotation of two dipoles, one of size $r$ and one of size $R$, with respect to each other by an angle, $\phi$. As we will describe, this configuration exhibits a simple factorization in the limit $r\ll R$, into the polarized $1\to 3$ splitting tensor \cite{DelDuca:2019ggv,DelDuca:2020vst} and a $1\to 2$ splitting function, thus providing clean access to these ingredients. In particular, through the variable $\phi$, it provides access to potential spin corelations between the two-dipoles (i.e. in the $1\to 3$ splitting tensor).  One motivation for studying this configuration is that by controlling the ratio between $r$ and $R$, one can control the length that the intermediate parton sourcing the smaller dipole propagates, providing clean studies of this factorization. Additionally,  this configuration could be quite interesting in the case of the QGP, as it would allow one to separately control the length of the medium probed by the intermediate parton. This would decouple it from the overall size of the QGP, allowing for more control. This is motivated by similar configurations studied for cosmological correlators \cite{Arkani-Hamed:2015bza,Arkani-Hamed:2018kmz}.

In computing this projection, rather than using the generic parametrization shown in Fig.~\ref{fig:param}, we choose our coordinates such that the origin lies in the center of the two dipoles. This is simply for computational convenience, as it avoids subleading terms in the ratio of length scales $\rho \equiv r/R$ which arise for a general choice of the origin.\footnote{One finds the same leading behavior in $\rho$ in both cases, hence our simpler choice.} This configuration is realized by the following values of $z_i$
\begin{align}
z_1 &=\frac{\rho}{2}\left(\cos \phi + i \sin \phi\right) \,, \qquad
z_2 =-\frac{\rho}{2}\left(\cos \phi + i \sin \phi\right)\,, \nn \\
z_3 &=\, -\frac{1}{2}\,, \qquad 
\hspace{2.3cm}z_4 =\, \frac{1}{2}\,.
\end{align}
Relative to the generic coordinates discussed above, these are realized via a permutation of the point labels combined with a shift by $-1/2$.

\subsection{The Tee Configuration}\label{sec:proj_b}

\begin{figure}
\begin{center}
\raisebox{0.0cm}{\includegraphics[scale=0.30]{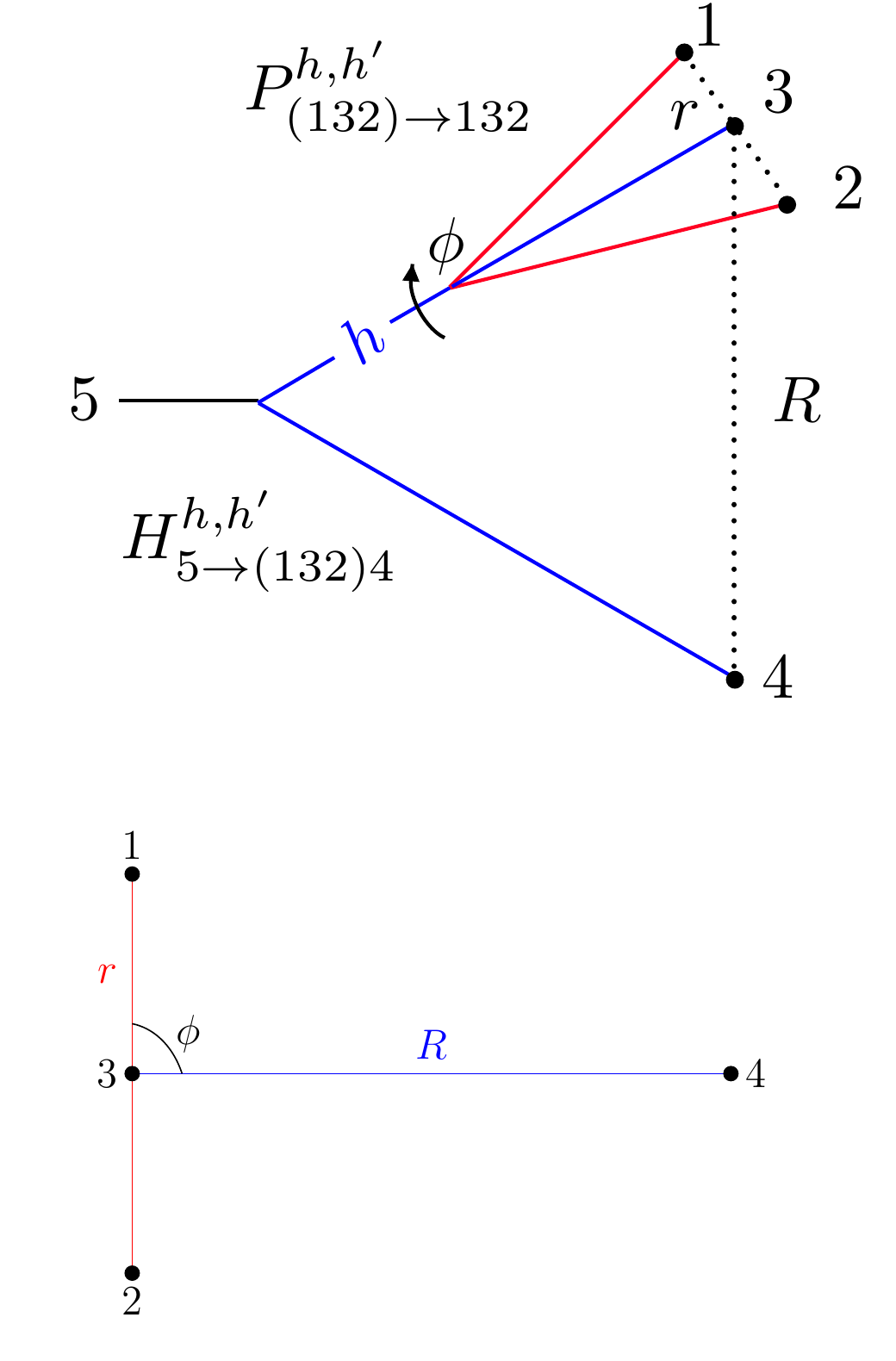}} \qquad
\includegraphics[scale=0.30]{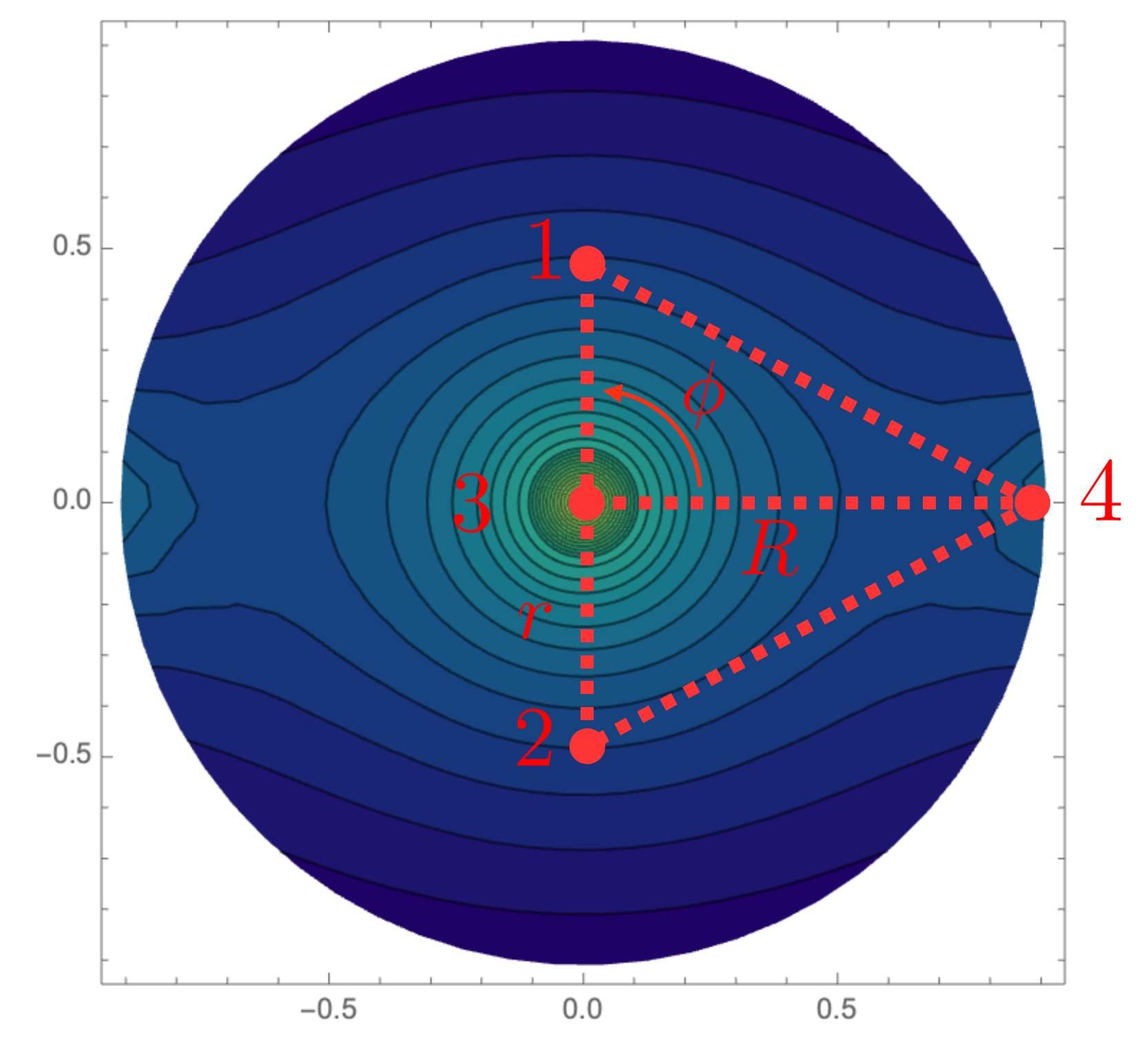} 
\end{center}
\caption{The ``tee configuration":  On the left we show the kinematic configuration of the four detectors. A set of three detectors is rotated with respect to a dipole of size $R$ by an angle $\phi$. On the right we show a plot of this configuration in $\mathcal{N}=4$ sYM. The location of the detectors in the larger dipole remain fix, while the set of three-detectors are rotated.  }
\label{fig:tee}
\end{figure} 

In the second configuration, which we refer to as ``the tee" configuration we rotate three detectors in a line configuration, each with separation $r$, with respect to a larger dipole of size $R$.  This configuration is a generalization of the squeezed limit of the three-point correlator studied in \cite{Chen:2022jhb,Chen:2021gdk,Chen:2020adz}. In the limit $r\ll R$, this configuration exhibits a clean factorization into a polarized $1\to 2$ splitting tensor, multiplying a polarized $1\to 3$ splitting function. This provides an exact compliment to the ``dipole" configuration, providing access to the other factorization channel. We believe that by having access to all factorization channels, we will be able to provide stringent tests on parton showers. As with the case of the dipole, the tee configuration also provides access to spin effects through the angle $\phi$. Although we will find that these effects are small, we believe that they are important for improving our description of parton showers.

Using the generic coordinates of \Fig{fig:param}, we can represent this configuration as
\bea
z_1 =&\, 0\,, \qquad
z_2 =\, z = \frac{\rho}{2}(\cos \phi, \sin \phi)\,, \qquad 
z_3 =\, 1\,,\qquad z_4 =\,w = -\frac{\rho}{2}(\cos \phi, \sin \phi)\,.
\eea
In practice, we parametrize this as
\bea
z_1= \frac{\rho}{2}\,, \qquad z_2 = -\frac{\rho}{2}\,, \qquad z_3 = 0\,, \qquad z_4 = \cos \phi + i \sin \phi\,,
\eea
through a permutation of the point labels and $\phi \rightarrow - \phi$.

\subsection{The Tripole Configuration}\label{sec:proj_c}

\begin{figure}
\begin{center}
\includegraphics[scale=0.35]{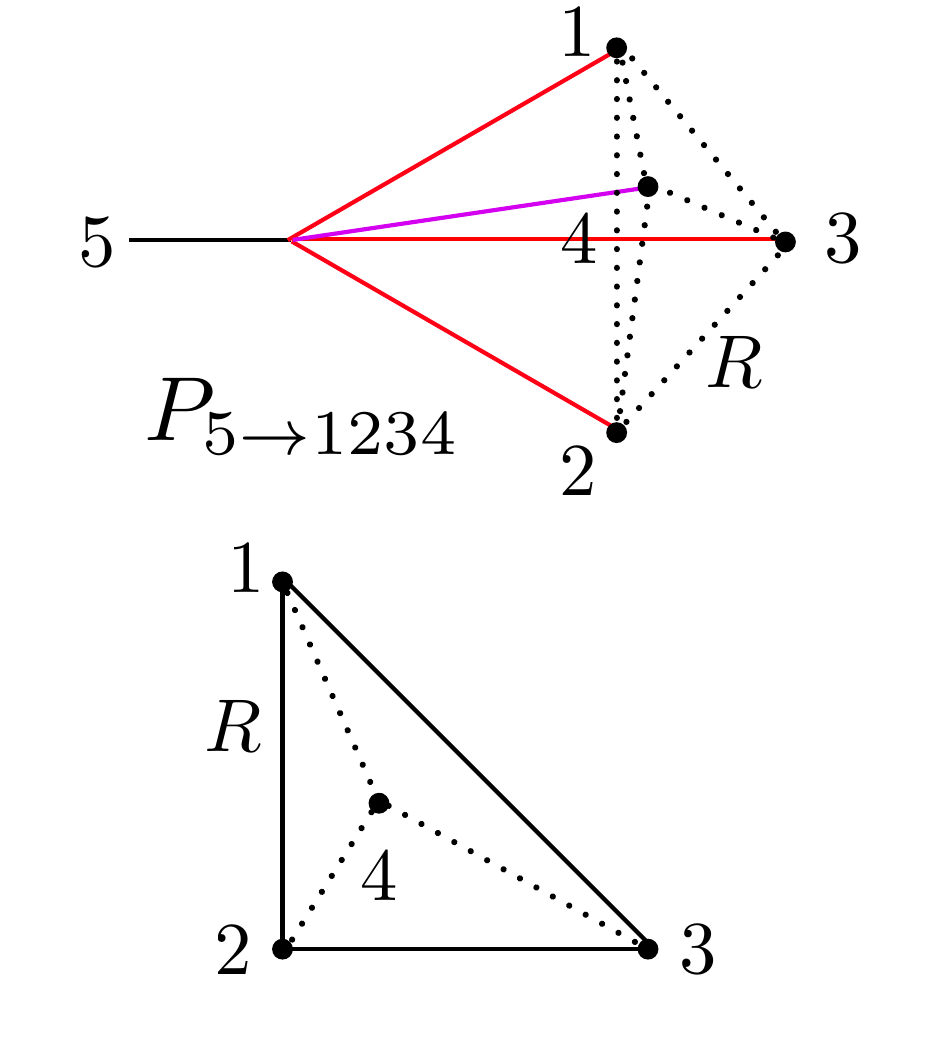} 
\includegraphics[scale=0.30]{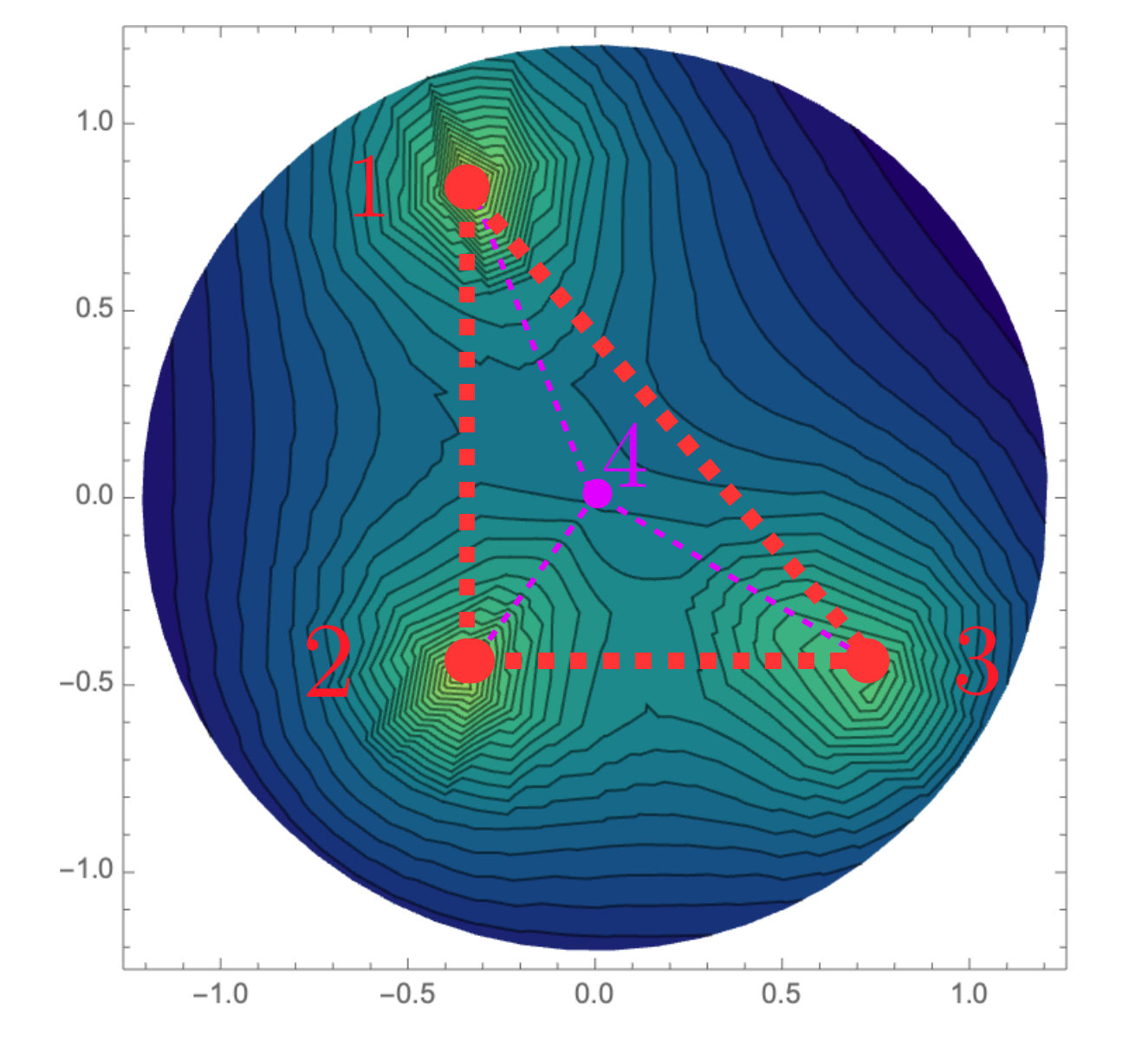} 
\end{center}
\caption{The ``tripole configuration": Three-detectors (shown in red) are fixed in a particular triangle configuration, in our case a ``3-4-5" triangle, and the fourth detector (shown in purple) is moved with respect to these three detectors. On the right we show a plot of this distribution in $\mathcal{N}=4$ sYM.
}
\label{fig:equilat_n4}
\end{figure}

In the third configuration, we fix three of the detectors to be in a particular triangular configuration, and use the fourth detector to map out the energy distribution around these three detectors.  This is illustrated in \Fig{fig:equilat_n4}. This configuration is designed to probe the hard splitting structure within a jet, and we believe that it will be ideal for probing the structure of the full four-point splitting function, away from factorized limits.  This configuration can of course be generalized to any other fixed shape of the frozen triangle. We will refer to this configuration as ``The Tripole".

To break the symmetry of the configuration, in our phenomenological studies, we take the triangle to be a 3-4-5 right triangle configuration. We parametrize this configuration by
\bea
z_1 = 0\,, \qquad z_2 =&\, z = \left(\frac{9}{25}, \frac{12}{25}\right)\,, \qquad 
z_3 = 1 \,, \qquad
z_4 =\, w= \rho \left(\cos \phi, \sin \phi\right)\,. \qquad 
\eea

\section{Theoretical Formalism for Multi-Point Correlators in Jets}\label{sec:proj}

In this section, we develop the framework used to calculate multi-point correlators inside a collimated jet. While our main phenomenological interest is in the four-point correlator, we keep the discussion general, working with $N$-point correlators where possible. In \Sec{sec:factorization}, we first review the factorization of the correlator in the collinear limit and set up our parameterization of the angular variables. In \Sec{sec:jetfunctions}, we introduce the \ENC jet functions and their leading-order computation in terms of $1\to N$ splitting functions. In \Sec{sec:iterated}, we derive the factorization of the correlator in iterated collinear limits, where a subset of the detectors becomes collinear at a parametrically smaller angle. Finally, in \Sec{sec:calc}, we apply this formalism to the four-point correlator in QCD, and compute the intrinsic spin correlations that our projections are designed to expose, presenting explicit results for the tee configuration.

\subsection{Factorization in the Collinear Limit and Parameterization}
\label{sec:factorization}
The object of interest is the general $N$-point correlation of energy flow operators, measured inside a collimated jet defining the state $|\Psi\rangle=|\Psi(p_T,\eta)\rangle$, which we often denote as \ENC for short: $\langle\mathcal{E}(\vec{n}_1)\cdots \mathcal{E}(\vec{n}_N)\rangle$ $\equiv \langle\Psi|\mathcal{E}(\vec{n}_1)\cdots \mathcal{E}(\vec{n}_N)|\Psi\rangle$. In terms of the particles observed inside a jet, the correlator is measured as
\begin{align}
\label{eq:EECinjet}
\langle\mathcal{E}(\vec{n}_1)\mathcal{E}(\vec{n}_2)\cdots\mathcal{E}(\vec{n}_N)\rangle=\sum_J\int \frac{d\sigma_J}{dp_{T} d\eta} \sum_{i_1,i_2,\cdots,i_N \in J}\;\prod_{a=1}^{N}\frac{p_{T,i_a}}{p_T}\,\delta^{2}\big(\vec{n}_a - \vec{n}_{i_a}\big)\,,
\end{align}
where the first sum runs over the jets in the sample, the second sum runs over all $N$-tuples of particles inside the jet, and $\vec{n}_{i_a}$ denotes the angular direction of particle $i_a$.

The information contained in the $N$ detector directions can be equivalently expressed in terms of the pairwise angles $\theta_{ij}$ between the detectors, together with the overall orientation of the configuration. For applications inside high-energy jets, we are interested in correlations at angles much smaller than the jet radius $R$, i.e. in the collinear limit of the correlator. In this limit, all $N$ detectors lie within a small angular patch of the celestial sphere, which can be approximated by a plane. The configuration of $N$ points in a plane, modulo overall translations and rotations, is specified by $2N-3$ real parameters, up to a discrete reflection ambiguity to which we return below. Consequently, although there are $\binom{N}{2}$ pairwise angles between the detectors, only $2N-3$ of them are independent. To construct a convenient set of independent angles, we first note that an on-shell momentum can be parameterized in terms of a complex coordinate $z_i$ on the celestial sphere,
\begin{align}
\label{eq:celestial}
p_i^\mu = p_{T,i}(1+ |z_i|^2,z_i + \bar{z}_i,-i(z_i-\bar{z}_i),1-|z_i|^2)\,,
\end{align}
where $p_{T,i}$ denotes the transverse momentum of particle $i$.
A convenient set of $2N-3$ independent angles can then be constructed by the following procedure:
\begin{enumerate}
    \item Label the $N$ directions using the complex coordinates $z_1, z_2, \dots, z_N$ of Eq.~\eqref{eq:celestial}, as illustrated in Fig.~\ref{fig:param} for the case $N=4$. (Note that Fig.~\ref{fig:param} technically depicts a rescaled version of these coordinates, which we will discuss in detail below.) In the collinear limit, the angle $\theta_{ij}$ between two directions is related to the Euclidean distance between the corresponding points on the complex plane by
\[
|z_{ij}|^2 \equiv |z_i - z_j|^2 = \frac{\theta_{ij}^2}{4}\,,
\]
which follows directly from Eq.~\eqref{eq:celestial} in the collinear limit.

    \item Measure the $N-1$ angles between consecutive directions:
     \[
     |z_{12}|^2 = \frac{\theta_{12}^2}{4} = x_1,\,|z_{23}|^2=\frac{\theta_{23}^2}{4} = x_2,\,\cdots, |z_{N-1,N}|^2=\frac{\theta_{N-1,N}^2}{4} = x_{N-1}\,.
     \]
    \item Next, choose a reference direction (say, $z_1$) and measure the remaining $N-2$ angles between $z_1$ and the other directions, excluding $z_2$ (since $\theta_{12}$ is already measured). That is, measure
    \[
    |z_{13}|^2=\frac{\theta_{13}^2}{4} = x_{N},\, |z_{14}|^2=\frac{\theta_{14}^2}{4} = x_{N+1},\, \dots,\, |z_{1N}|^2=\frac{\theta_{1N}^2}{4} = x_{2N-3}.
    \]
\end{enumerate}
These $2N-3$ independent angles fix the configuration only up to discrete reflections: identical values of the measured angles can correspond to genuinely different shapes.\footnote{As an example, take the three points $r_1=(0,0)$, $r_2=(2,0)$, $r_3=(1,1)$ in the plane. If the fourth point is taken to be $r_4=(1,2)$, or its reflection across the line through $r_1$ and $r_3$, $r_4'=(2,1)$, all of the pairwise distances constructed in the steps above are identical, while the unmeasured distance between $r_2$ and the fourth point differs: the two configurations are genuinely different shapes.} The $2^{N-2}$ configurations consistent with the measured angles are, however, not all of distinct shape: a configuration and its global mirror image share all pairwise distances, so the configurations form $2^{N-3}$ mirror pairs, corresponding to $2^{N-3}$ genuinely distinct shapes ($1$ for $N=3$ and $2$ for $N=4$). Since QCD is parity invariant, the two members of a mirror pair contribute identically to the correlator. In this paper, as discussed in \Sec{sec:4ptProj}, we are always interested in a specific shape of the four-point correlator, so no discrete ambiguity arises; if one instead places no restriction on the shape, the measured distribution sums over the $2^{N-3}$ distinct shapes. On the other hand, we can also parameterize the shape by $(N-2)$ complex variables and these complex variables uniquely selects one of the $2^{N-3}$ distinct shapes. We find the discussion in relative angles useful to match the experimental algorithm in measuring these shapes.

If we are only interested in the independent pairwise angles of a specific shape, we can therefore equivalently represent the correlator as
\begin{align}
\label{eq:xdist}
\frac{d^{2N-3}\Sigma}{dx_1 \cdots dx_{2N-3}} \equiv&\, \sum_J\int \frac{d\sigma_J}{dp_{T} d\eta} \sum_{i_1,\cdots,i_N \in J}\; \prod_{a=1}^{N}\frac{p_{T,i_a}}{p_T}\; \prod_{k=1}^{2N-3}\delta\left(x_k - \frac{\theta_{i_{a_k} i_{b_k}}^2}{4}\right) \nn\\
=&\, \mathcal{N}\,\langle\mathcal{E}(\vec{n}_1)\mathcal{E}(\vec{n}_2)\cdots \mathcal{E}(\vec{n}_N)\rangle\,,
\end{align}
where $(a_k,b_k)$ denote the measured pairs of the procedure above; for $N=4$, the five measured angles are $\theta_{i_1i_2}^2/4$, $\theta_{i_2i_3}^2/4$, $\theta_{i_3i_4}^2/4$, $\theta_{i_1i_3}^2/4$, and $\theta_{i_1i_4}^2/4$. Here and below, both sides are understood to be evaluated on the specified shape: the measurement on the left-hand side is restricted to particle configurations of that shape, i.e., to the corresponding mirror pair. The accompanying normalization factor $\mathcal{N}$ is purely geometric, depending only on the measured angles:
\begin{align}
\label{eq:Nfactor}
\mathcal{N} = \frac{2\pi}{8^{N-2}\,\prod_{k=3}^{N}A_k}\,,
\end{align}
where the factor of $2$ counts the two mirror-image configurations of the specified shape, which contribute equally, and $A_k$ is the area of the triangle formed by the detectors $(\vec{z}_1, \vec{z}_{k-1}, \vec{z}_k)$, determined directly from the measured angles by Heron's formula,
\begin{align}
\label{eq:heron}
16 A_k^2 = 2\left(x_a x_b + x_b x_c + x_c x_a\right) - x_a^2 - x_b^2 - x_c^2\,,
\end{align}
with $(x_a, x_b, x_c) = \big(|z_{1,k-1}|^2,\, |z_{k-1,k}|^2,\, |z_{1,k}|^2\big)$. 

In the collinear limit, the angular distribution of the \ENC measured on an inclusive jet sample factorizes as~\cite{Dixon:2019uzg,Lee:2022uwt,Lee:2024icn,Lee:2025okn}
\begin{align}
\label{eq:JetFactorization}
\frac{d^{2N-3}\Sigma}{dx_1 \cdots dx_{2N-3}}=&\sum_{i,j} \mathcal{H}_i\left(p_T / \xi, \eta, \mu\right)  \otimes \int_0^1 d y\, y^N \mathcal{J}_{i j}\left(\xi, y, \ln \frac{p_T^2 R^2}{\xi^2\,\mu^2}, \mu\right) \nonumber\\
&\quad \times J_j^{[N]}(y^2 x_1, \cdots, y^2 x_{2N-3}, p_T,  \mu)\,,\\
&\hspace{-2cm}=\sum_{i,j} \mathcal{H}_i\left(p_T / \xi, \eta, \mu\right)  \otimes \int_0^1 d y\, y^N \mathcal{J}_{i j}\left(\xi, y, \ln \frac{p_T^2 R^2}{\xi^2\,\mu^2}, \mu\right)  J_j^{[N]}(y^2x_{L},\{\tilde{x}_k\}\backslash x_L,   p_T,\mu)\,,\nonumber
\end{align}
where $\mathcal{H}_i$ is the hard function describing the inclusive production of a parton $i$ with transverse momentum $p_T/\xi$ and rapidity $\eta$, which incorporates the parton distribution functions of the colliding hadrons, and $\otimes$ denotes a convolution in the momentum fraction $\xi$ carried by the jet. The matching coefficient $\mathcal{J}_{ij}$ encodes the dependence on the algorithm used to define the jet, and describes the distribution of collinear partons $j$ carrying a momentum fraction $y$ of the jet. Finally, $J_j^{[N]}$ is the \ENC jet function, which carries the entire dependence on the measured angles; its operator definition is given in \Sec{sec:jetfunctions}. The factor $y^N$ arises because the $N$ energy weights in Eq.~\eqref{eq:xdist} are normalized to the jet $p_T$, while the jet function is defined relative to the momentum of the collinear parton $j$. Similarly, the angles enter the jet function in the combination $y^2 x_k$, since the intrinsic scale of the collinear splitting is the transverse momentum $\sim y\, p_T\, \theta$.

We see from the convolution structure of Eq.~\eqref{eq:JetFactorization} that all $2N-3$ angles enter the jet function rescaled by the momentum fraction $y$ of the parton $j$ from which the collinear splitting occurs. In the second equality of Eq.~\eqref{eq:JetFactorization}, we have therefore traded the $2N-3$ angles for the largest angle $x_L = {\rm{max}}\{x_1,\cdots,x_{2N-3}\}$, together with the rescaled angles
\begin{align}
\tilde{x}_{k}\equiv \frac{x_k}{x_L}\,,
\end{align}
which are invariant under this rescaling. In this way, the $N$-point correlator is described by the largest angle $x_L$, which parameterizes the overall size of the configuration, and $2(N-2)$ nontrivial rescaled angles, which parameterize its shape. We denote the latter schematically as $\{\tilde{x}_{k}\}\backslash x_L$, indicating that the (trivially rescaled) largest angle is excluded from the set. The shape variables can equivalently be written in terms of cross-ratios. We note that there have been many studies of the so-called projected $N$-point correlators~\cite{Chen:2020vvp,Chen:2023zlx,Lee:2024esz,Lee:2026zyl}, in which only the largest angular separation, $x_L$, is measured, while all cross-ratios are integrated out.

\subsection{Jet Functions for Multi-Point Correlators}\label{sec:jetfunctions}
As is apparent from the factorization in Eq.~\eqref{eq:JetFactorization}, the entire dependence on the $2N-3$ angles resides in the \ENC jet function. The renormalization group evolution and convolution structure of Eq.~\eqref{eq:JetFactorization} play two roles: they generate the anomalous scaling of the correlator in the overall size $x_L$ from the OPE limit of the $N$-point correlator, and they determine the fractions of collinear quarks and gluons $j$ that initiate the splitting into the $N$ detected particles. The dependence on the shape of the configuration at fixed $x_L$, on the other hand, is entirely determined by the fixed-order computation of the \ENC jet function $J_j^{[N]}$. Therefore, up to the anomalous scaling in $x_L$, we can organize our computation as
\begin{align}
\langle\mathcal{E}(\vec{n}_1)\cdots \mathcal{E}(\vec{n}_N)\rangle \simeq f_q \langle\mathcal{E}(\vec{n}_1)\cdots \mathcal{E}(\vec{n}_N)\rangle_q + f_g \langle\mathcal{E}(\vec{n}_1)\cdots \mathcal{E}(\vec{n}_N)\rangle_g\,,
\end{align}
where $f_{q/g}$ are the quark and gluon jet fractions, respectively, and $\langle\mathcal{E}(\vec{n}_1)\cdots \mathcal{E}(\vec{n}_N)\rangle_{q/g}$ are the $N$-point correlators measured on quark and gluon jets; the identical decomposition applies to the angular distribution of Eq.~\eqref{eq:xdist}. At leading order, the correlator on a quark or gluon jet is determined by the corresponding \ENC jet function through the geometric factor of Eq.~\eqref{eq:Nfactor},
\begin{align}
\langle\mathcal{E}(\vec{n}_1)\mathcal{E}(\vec{n}_2)\cdots \mathcal{E}(\vec{n}_N)\rangle_{q/g} = \frac{1}{\mathcal{N}}J_{q/g}^{[N]}\left(x_1, x_2, \cdots,x_{2N-3}, p_T, \mu\right)\,.
\end{align}
The operator definitions of the quark and gluon \ENC jet functions are~\cite{Chen:2019bpb}
\begin{align}\label{eq:jetfundef}
& J_q^{[N]}\left(x_1, x_2, \cdots,x_{2N-3}, p_T, \mu\right)= \nonumber\\
& \qquad \int \frac{d l^{+}}{2 \pi} \frac{1}{2 N_c} \operatorname{Tr} \int d^4 x\, e^{i l \cdot x}\langle 0| \frac{\slashed{\bar{n}}}{2} \chi_n(x) \widehat{\mathcal{M}}_{{\mathrm{E}}^N{\mathrm{C}}} \delta(Q+\bar{n} \cdot \mathcal{P}) \delta^2\left(\mathcal{P}_{\perp}\right) \bar{\chi}_n(0)|0\rangle\,,\nonumber\\
& J_g^{[N]}\left(x_1, x_2, \cdots,x_{2N-3}, p_T, \mu\right)= \nonumber\\
& \qquad \int \frac{d l^{+}}{2 \pi} \frac{1}{2\left(N_c^2-1\right)} \int d^4 x\, e^{i l \cdot x}\langle 0| \mathcal{B}_{n, \perp}^{a, \mu}(x) \widehat{\mathcal{M}}_{{\mathrm{E}}^N{\mathrm{C}}} \delta(Q+\bar{n} \cdot \mathcal{P}) \delta^2\left(\mathcal{P}_{\perp}\right) \mathcal{B}_{n, \perp, \mu}^{a}(0)|0\rangle\,.
\end{align}
The measurement function implements the angular measurements described by the procedure above,
\begin{align}
\label{eq:measurement}
\widehat{\mathcal{M}}_{{\mathrm{E}}^N{\mathrm{C}}}=\sum_{i_1, i_2,\cdots, i_N} \frac{p_{T,i_1} p_{T,i_2}\cdots p_{T,i_N}}{p_T^N} \delta\left(x_1-\frac{\theta_{i_1 i_2}^2}{4}\right) \delta\left(x_2-\frac{\theta_{i_2 i_3}^2}{4}\right)\cdots \delta\left(x_{2N-3}-\frac{\theta_{i_{1} i_N}^2}{4}\right)\,,
\end{align}
with the sum running over all assignments of final-state particles to the $N$ detectors. Note that within the jet function the energy weights are normalized to the momentum of the initiating parton $j$; the mismatch with the jet-$p_T$ normalization of Eq.~\eqref{eq:xdist} is precisely compensated by the factor $y^N$ in Eq.~\eqref{eq:JetFactorization}. We emphasize that the measurement function carries no additional normalization factor: it computes the angular distribution of Eq.~\eqref{eq:xdist} directly, and the geometric factor $\mathcal{N}$ of Eq.~\eqref{eq:Nfactor} relating the latter to the fully differential correlator of Eq.~\eqref{eq:EECinjet} will emerge from the transverse reduction below.
If we impose that all pairwise angles are nonzero, so that contact terms (in which two or more detectors coincide) do not contribute, the leading-order \ENC jet function is determined by the tree-level $1\to N$ collinear splitting function $P_{j\to i_1 i_2 \cdots i_N}$, and first arises at order $\alpha_s^{N-1}$,
\bea
\label{eq:jetphase}
J_j^{[N]}\left(x_1, x_2, \cdots,x_{2N-3}, p_T, \mu\right) &= \sum_{i_1,i_2,\cdots, i_N}\int d\Phi_c^{(N)}\left(\frac{2g^2}{s_{1 \cdots N}}\left(\frac{\mu^2 e^{\gamma_E}}{4\pi}\right)^\epsilon\right)^{N-1}\\
&\times P_{j\to i_1 i_2 \cdots i_N}\left(\{y_{i_a}\},\{s_{i_a i_b}\}\right)\widehat{\mathcal{M}}_{{\mathrm{E}}^N{\mathrm{C}}}\,,\nn
\eea
where $d\Phi_c^{(N)}$ is the $N$-body collinear phase space, whose exact form we give below, $s_{1\cdots N}$ is the squared invariant mass of the $N$ collinear partons, the splitting function depends on the momentum fractions $y_{i_a}$ and the pairwise Mandelstam invariants $s_{i_a i_b}$ defined below, and the sum runs over the final-state flavor channels. We parameterize the collinear momenta as
\bea
p_{i_k}^{\mu} = y_{i_k} P^{\mu} + k_{\perp,i_k}^{\mu} - \frac{k_{\perp,i_k}^2}{y_{i_k}} \frac{n^{\mu}}{2 n \cdot P}\,,
\label{eq:colP}
\eea
where $y_{i_k}$ is the momentum fraction of parton $i_k$ with respect to the initiating parton $j$, $P$ is a null vector which defines the collinear direction, and $n$ is an auxiliary null vector with $k_{\perp} \cdot n = k_{\perp} \cdot P = 0$. Note that we also take $k_{\perp} \rightarrow \lambda k_{\perp}$, where $\lambda$ serves as a power counting parameter which organizes contributions to the collinear limit ($\lambda \rightarrow 0$).

Both the phase space and splitting functions are functions of the momentum fractions, as well as the Mandelstam invariants between two partons, given as
\begin{align}
\label{eq:Mandel}
s_{ij} = 2p_i\cdot p_j = y_i y_j p_T^2 \theta_{ij}^2 = y_i y_j 4p_T^2 |z_i-z_j|^2\equiv  y_i y_j 4p_T^2 |z_{ij}|^2 \,.
\end{align}
Here, we used Eq.~\eqref{eq:celestial} with $p_{T,i}=y_i p_T$,
which makes manifest the relation $|z_{ij}|^2=\theta_{ij}^2/4$ quoted above, valid up to power corrections in the angles. These complex coordinates are shown for the case $N=4$ in Fig.~\ref{fig:param}. By rescaling with $x_L$ as discussed above, we can enforce that the longest side of the configuration has unit length, and use the freedom of translations and rotations of the plane to place one of its endpoints at $0$ and the other at $1$. After the rescaling, $2(N-2)$ degrees of freedom remain, which can be characterized by $N-2$ complex coordinates in the plane. For $N=4$, these are the two complex variables $z$ and $w$ shown in Fig.~\ref{fig:param}. This parameterization can also be thought of in the language of a conformal $(N+1)$-point correlator, built from the $N$ detectors together with the local operator sourcing the state: conformal symmetry can be used to place three of the five points (for $N=4$) at $\{0,1,\infty\}$, with the point at $\infty$ corresponding to the source of the \ENC jet function. In general, one can fix two points at $0$ and $1$ in the complex plane and parameterize the remaining $N-2$ points using complex variables.
Importantly, $N-2$ complex coordinates uniquely specify the oriented shape of a generic $N$-point configuration in the plane. A global reflection gives the opposite orientation of the same unoriented shape. Since QCD is parity invariant, the two mirror configurations contribute equally, and summing over both gives an overall factor of two as discussed in Eq.~\eqref{eq:Nfactor}.

The $N$-body collinear phase space is given by
\begin{align}
\label{eq:PhiDef}
d\Phi_c^{(N)} = \left[\prod_{i=1}^{N} \frac{d^d p_i}{(2\pi)^{d-1}}\,\delta^{+}(p_i^2)\right] 2Q\,(2\pi)^{d-1}\, \delta\Big(Q-\sum_{i}\bar{n}\cdot p_i\Big)\, \delta^{d-2}\Big(\sum_{i}\vec{p}_{\perp,i}\Big)\,,
\end{align}
where $Q \equiv \bar{n}\cdot p_j = 2p_T$ is the large light-cone momentum of the initiating parton $j$. Since the LO jet function is finite for generic angles, we may also set $\epsilon=0$ from here on. Writing $\bar{n}\cdot p_i = y_i Q$ and introducing the transverse positions
\begin{align}
\vec{z}_i \equiv \frac{\vec{p}_{\perp,i}}{\bar{n}\cdot p_i}\,,
\end{align}
which are precisely the complex coordinates $z_i$ of Eq.~\eqref{eq:celestial} viewed as real two-vectors\footnote{The corresponding measure is $d^2z_i \equiv d\,{\rm Re}\, z_i\; d\,{\rm Im}\, z_i$, related to the solid angle of the detector by $d\Omega_i = 4\, d^2 z_i/(1+|z_i|^2)^2 \simeq 4\, d^2 z_i$ in the collinear limit.}, the phase space takes the form
\begin{align}
\label{eq:measure}
d\Phi_c^{(N)}\Big|_{d=4} = \frac{2^{1-N}\, Q^{2(N-1)}}{(2\pi)^{3(N-1)}} \left[\prod_{i=1}^N y_i\, dy_i\right] \delta\Big(1-\sum_i y_i\Big) \left[\prod_{i=1}^{N} d^2 z_i \right] \delta^{2}\Big(\sum_i y_i\, \vec{z}_i\Big)\,,
\end{align}
where the two-dimensional $\delta$-function fixes the energy-weighted centroid of the configuration to the direction of the initiating parton. Note that the phase-space measure carries a single power of each momentum fraction $y_i$, while the measurement function in Eq.~\eqref{eq:measurement} supplies one additional power through the energy weight of each detected parton.

The transverse integrals in Eq.~\eqref{eq:jetphase} can now be carried out exactly against the angular $\delta$-functions of the measurement. The centroid constraint absorbs the overall translations; placing detector $2$ relative to detector $1$ uses the measured angle $x_1$ together with the overall rotation, producing a factor of $\pi$; and each subsequent detector $k=3,\dots,N$ is then fixed by the two measured angles connecting it to detectors $1$ and $k-1$, up to a two-fold reflection ambiguity, with Jacobian $|\partial(x_{k-1}, x_{N+k-3})/\partial \vec{z}_k| = 8A_k$ per solution. Restricting to the specified shape retains the two mirror-image configurations, which contribute equally for any function of the pairwise angles, giving the factor of $2$ in Eq.~\eqref{eq:Nfactor}. We therefore obtain
\begin{align}
\label{eq:reduction}
\int \left[\prod_{i=1}^{N} d^2 z_i\right] \delta^{2}\Big(\sum_i y_i \vec{z}_i\Big) \prod_{k=1}^{2N-3}\delta\big(x_k - |z_{a_k b_k}|^2\big)\, F\big(\{|z_{ij}|^2\}\big) = \mathcal{N}\, F\big(\{x_k\}\big)\,,
\end{align}
valid on the support of $\delta(1-\sum_i y_i)$ for any function $F$ of the pairwise angles, where $(a_k, b_k)$ denote the measured pairs of the parameterization procedure of \Sec{sec:factorization}, and $\mathcal{N}$ is precisely the geometric factor of Eq.~\eqref{eq:Nfactor}. On the right-hand side, $F$ is evaluated at the pairwise angles of the specified shape: the measured pairs are fixed directly to $|z_{a_k b_k}|^2 = x_k$, while the remaining $\binom{N}{2}-(2N-3)$ angles are determined functions of them.

Combining Eqs.~\eqref{eq:measure} and \eqref{eq:reduction} with the energy weights of Eq.~\eqref{eq:measurement} and the Mandelstam invariants of Eq.~\eqref{eq:Mandel}, all powers of $Q$ cancel exactly. Moreover, the geometric factor $\mathcal{N}$ produced by the transverse reduction cancels against the $1/\mathcal{N}$ relating the correlator to the jet function, and the fully differential correlator at LO takes the remarkably simple closed form
\begin{align}
\label{eq:masterLO}
\langle\mathcal{E}(\vec{n}_1)\mathcal{E}(\vec{n}_2)\cdots \mathcal{E}(\vec{n}_N)\rangle_{j} =&\, \frac{1}{\mathcal{N}}\,J_{j}^{[N]}\left(x_1, x_2, \cdots,x_{2N-3}, p_T, \mu\right) \nn\\
=&\, c_N \sum_{i_1,i_2,\cdots, i_N}\int \prod_i dy_i\, \delta\Big(1-\sum_i y_i\Big)\,(y_{i_1}y_{i_2}\cdots y_{i_N})^2\nn\\
&\times \left(\frac{2g^2}{\hat{s}_{1 \cdots N}}\right)^{N-1} P_{j\to i_1 i_2 \cdots i_N}\big(\{y_{i_a}\},\{\hat{s}_{i_a i_b}\}\big)\,,
\end{align}
where $c_N \equiv 2^{1-N}/(2\pi)^{3(N-1)}$, $\hat{s}_{i_a i_b} \equiv s_{i_a i_b}/Q^2 = y_{i_a} y_{i_b} |z_{i_a i_b}|^2$ are the Mandelstam invariants in angular units, $\hat{s}_{1\cdots N} = \sum_{a<b}\hat{s}_{i_a i_b}$, and the square of the momentum fractions combines the single power from the phase-space measure with the energy weights of the measurement function. 

\subsection{Iterated Collinear Limits of Multi-Point Correlators}\label{sec:iterated}
Now we consider the case where some subset of the collinear momenta, the sub-collinear set, becomes hierarchically more collinear, with pairwise angles parametrically smaller than the overall collinear scale $x_L$. In this iterated OPE limit, the subset is itself described by a further collinear factorization, akin to Eq.~\eqref{eq:JetFactorization}, which gives rise to an anomalous scaling in the angle parameterizing the size of the subset. At fixed order, the iterated limit simplifies the form of the \ENC jet function, factorizing both the phase-space measure given in Eq.~\eqref{eq:measure} and the $1\to N$ collinear splitting function $P_{j\to i_1 i_2 \cdots i_N}$. Throughout this and the following subsection we work in the angular units introduced in \Sec{sec:jetfunctions}, dropping the hats on the Mandelstam invariants, $s_{i_a i_b} \equiv y_{i_a} y_{i_b} |z_{i_a i_b}|^2$.

This iterated limit can be systematically defined in terms of the transverse momenta of the two sets. In general, we parameterize the momenta of the sub-collinear set as
\bea
p_{i_k}^{\mu} = y_{i_k} P'^{\mu} + \kappa_{\perp,i_k}^{\mu} - \frac{\kappa_{\perp,i_k}^2}{y_{i_k}} \frac{n'^{\mu}}{2 n' \cdot P'}\,.
\label{eq:subColP}
\eea
While we are free to choose $P \neq P'$ and $n \neq n'$ in order to describe more generic configurations, we take $P = P'$ and $n = n'$ in order to enforce that all particles are collinear to a common direction. The transverse momenta of the sub-collinear set are, however, assigned an independent scaling,
\bea
\kappa_{\perp} \rightarrow \lambda' \kappa_{\perp}\,.
\eea
The iterated collinear limit is then given by $\lambda, \lambda' \rightarrow 0$ with $\lambda \gg \lambda'$.

\begin{figure}
\centering
\resizebox{0.95\textwidth}{!}{%
\begin{tikzpicture}
    \draw [] (-7.55,1.1) -- (-7.55,-1.1);
    \draw [] (-7.55,1.1) -- (-7.45,1.1);
    \draw [] (-7.55,-1.1) -- (-7.45,-1.1);
    \draw [] (-7.4,0.9) -- (-7.4,-0.9);
    \node [left] at (-7,0) {$j$};
	\draw [] (-7,0) -- (-6,0);
    \draw [] (-6,0) -- (-4,0.22);
    \draw [] (-6,0) -- (-4,-0.22);
    \draw [] (-6,0) -- (-4,0.65);
    \draw [] (-6,0) -- (-4,-0.65);
    \node [right] at (-4,0.65) {$1$};
    \node [right] at (-4,0.22) {$2$};
    \node [right] at (-4,-0.22) {$3$};
    \node [right] at (-4,-0.65) {$4$};
    \node [left] at (-7.5,0) {$\mathscr C_{1,2}$};
    \node [right] at (-3.5,0.9) {$2$};
    \filldraw[] (-6,0) circle (1pt);
    \draw [] (-3.5,0.9) -- (-3.5,-0.9);
    \draw [] (-3,1.1) -- (-3,-1.1);
    \draw [] (-3,1.1) -- (-3.1,1.1);
    \draw [] (-3,-1.1) -- (-3.1,-1.1);
    \node [right] at (-2.7,0) {$=$};
    \node [] at (-5.5,-1.3) {$P_{j \rightarrow 1234}$};
    \draw [] (-1.9,0.9) -- (-1.9,-0.9);
    \node [left] at (-1.5,0) {$j$};
    \draw [] (-1.5,0) -- (1.5,0);
    \draw [] (-0.5,0) -- (1.5,0.65);
    \draw [] (-0.5,0) -- (1.5,-0.65);
    \filldraw[] (-0.5,0) circle (1pt);
    \node [right] at (1.5,0.65) {$h$};
    \node [right] at (1.5,0) {$3$};
    \node [right] at (1.5,-0.65) {$4$};
    \draw [] (2,0.9) -- (2,-0.9);
    \node [right] at (2,0.9) {$2$};
    \node [right] at (2.3,0) {$\times$};
    \node [] at (0,-1.3) {$H_{j \rightarrow (12)34}^{hh'}$};
    \draw [] (3.1,0.9) -- (3.1,-0.9);
    \draw [] (3.5,0) -- (4.5,0);
    \draw [] (4.5,0) -- (6.5,0.65);
    \draw [] (4.5,0) -- (6.5,-0.65);
    \filldraw[] (4.5,0) circle (1pt);
    \node [left] at (3.5,0) {$h$};
    \node [right] at (6.5,0.65) {$1$};
    \node [right] at (6.5,-0.65) {$2$};
    \draw [] (7,0.9) -- (7,-0.9);
    \node [right] at (7,0.9) {$2$};
    \node [] at (5,-1.3) {$P_{(12) \rightarrow 12}^{hh'}$};
\end{tikzpicture}%
}
\caption{The iterated collinear factorization of a spin-averaged $1\rightarrow 4$ splitting function, cf.~Eq.~\eqref{eq:splitFac}. In the limit where particles 1 and 2 become collinear, the squared amplitude factorizes into the product of two squared amplitudes which describe the splitting processes at the two distinct collinear scales, correlated through the helicity indices $h,h'$ of the intermediate particle on which we factorize. }
\label{fig:HPFac}
\end{figure}
Now, we assume that $M < N$ of the collinear partons are much more collinear with each other than they are with all the others; without loss of generality, we take these to be the first $M$. Using the parameterization discussed above, one can show that to leading order in the iterated limit, the $1\to N$ collinear splitting function factorizes as~\cite{DelDuca:2019ggv, DelDuca:2020vst}
\bea
\mathscr C_{1, \cdots, M} \left[P_{j\to i_1 i_2 \cdots i_N}\right] = \left(\frac{s_{[i_1 \cdots i_M] i_{M+1} \cdots i_N}}{s_{i_1 \cdots i_M}}\right)^{M-1} H^{h h'}_{j\to i_{[1 \cdots M]}i_{M+1} \cdots i_N} P^{h h'}_{i_{[1 \cdots M]} \rightarrow i_1 \cdots i_M}\,,
\label{eq:splitFac}
\eea
with an implicit sum over the helicity indices $h, h'$ of the intermediate parton. Here, $\mathscr C_{1, \cdots, M}$ denotes the sub-collinear limit of the first $M$ particles, and the flavor of their parent parton is labeled as $i_{[1 \cdots M]}$. The polarized \emph{splitting tensor} $H^{h h'}_{j\to i_{[1 \cdots M]}i_{M+1} \cdots i_N}$ describes the $1\to(N-M+1)$ collinear splitting of parton $j$ into the parent parton $i_{[1 \cdots M]}$ and the $(N-M)$ other partons. The \emph{splitting function} $P^{h h'}_{i_{[1 \cdots M]} \rightarrow i_1 \cdots i_M}$ describes the subsequent splitting of the parent parton into the $M$ sub-collinear particles. In this paper, we only consider the tree-level splitting objects. The ratio of Mandelstam invariants in Eq.~\eqref{eq:splitFac} accounts for the mismatch between the collinear pole factor $s_{1\cdots N}^{-(N-1)}$ extracted in Eq.~\eqref{eq:masterLO} and the pole factors $s_{[i_1\cdots i_M]i_{M+1}\cdots i_N}^{-(N-M)}$ and $s_{i_1\cdots i_M}^{-(M-1)}$ appropriate to the two factorized splittings. Note that the two splitting structures are correlated by the helicity indices $hh'$ of the intermediate parent parton. These helicity correlations give sensitivity to spin effects when the sub-collinear set is rotated with respect to the remaining partons. When traced over the helicity indices, the splitting structures reduce to the corresponding unpolarized splitting functions~\cite{Catani:1999ss,DelDuca:2019ggv,DelDuca:2020vst},
\bea
\delta^{h h'} H^{h h'}_{j\to i_{[1 \cdots M]}i_{M+1} \cdots i_N} &= P_{j\to i_{[1 \cdots M]}i_{M+1} \cdots i_N}\,,\nonumber\\
\delta^{h h'} P^{h h'}_{i_{[1 \cdots M]} \rightarrow i_1 \cdots i_M} &= N_{\rm pol}\, P_{i_{[1 \cdots M]} \rightarrow i_1 \cdots i_M}\,,
\eea
where $N_{\rm pol}$ is the number of physical polarization states of the parent of the sub-collinear set ($N_{\rm pol}=2$ for a quark and $2(1-\e)$ for a gluon). For the case when $i_{[1 \cdots M]}=q$ (or $\bar q$), helicity conservation along the quark line makes the helicity dependence trivial,
\begin{align}
H^{h h'}_{j\to q\,i_{M+1} \cdots i_N} =\frac{1}{2}\delta^{hh'}P_{j\to q\,i_{M+1} \cdots i_N}\,,
\end{align}
and thus spin correlations are absent. On the other hand, if $i_{[1 \cdots M]}=g$, the helicity dependence is non-trivial (see App.~\ref{sec:pert} for examples of gluon splitting objects), giving rise to spin correlations.

Inserting the factorized splitting function of Eq.~\eqref{eq:splitFac} into the master formula of Eq.~\eqref{eq:masterLO}, we can now take the iterated limit of the \ENC itself. In order to factorize the phase space, we introduce an additional integral over the energy fraction $\zeta$ of the parent of the sub-collinear set, $1 = \int d\zeta\, \delta(\zeta - y_{i_1} - \cdots - y_{i_M})$. This gives
\bea
\label{eq:EECSplit}
\langle\mathcal{E}(\vec{n}_1)\cdots \mathcal{E}(\vec{n}_N)\rangle_{j}
=& c_N \sum_{i_{[1\cdots M]}, i_{M+1}, \cdots, i_N} \sum_{i_1,i_2,\cdots, i_M} \int d y_{i_1} \cdots d y_{i_N} \, d \zeta \nn \\
&\times \delta \left(\zeta - y_{i_1} - \cdots - y_{i_M}\right) \delta \left(1 - \zeta - y_{i_{M+1}} - \cdots - y_{i_N}\right) \left(y_{i_1} \cdots y_{i_N}\right)^2 \nn \\
&\times \left(\frac{2 g^2}{s_{[i_1 \cdots i_M]i_{M+1}\cdots i_N}}\right)^{N-M} H^{h h'}_{j\to i_{[1 \cdots M]}i_{M+1} \cdots i_N}(y_{i_{M+1}},\cdots,y_{i_N},\zeta) \nn \\
&\times \left(\frac{2 g^2}{s_{i_1 \cdots i_M}}\right)^{M-1} P^{h h'}_{i_{[1 \cdots M]} \rightarrow i_1 \cdots i_M}(y_{i_1},\cdots, y_{i_M})\,.
\eea
The factorization breaks up the single sum over partonic channels into two distinct sums over the channels of the two splittings. Furthermore, we utilize
\bea
s_{1 \cdots N} = s_{[i_1 \cdots i_M]i_{M+1}\cdots i_N} + \mathcal{O}(\lambda')\,,
\label{eq:MandSubLead}
\eea
and neglect the subleading terms in $\lambda'$. In fact, we justify neglecting such terms by considering the leading scaling behavior in the iterated collinear limit. We expect the leading terms to go as $\lambda^{-2(N-M)} \lambda'^{-2(M-1)}$. The Mandelstams depend on the square of the transverse momenta. Thus, Mandelstams involving only particles in the sub-collinear set are $\mathcal{O}(\lambda'^2)$, whereas all others are $\mathcal{O}(\lambda^2)$. Comparing the expected leading behavior in $\lambda$ and $\lambda'$ with Eq.~\eqref{eq:EECSplit}, we see that the Mandelstams in the denominators already saturate the expected scaling. This means that we keep only the $\mathcal{O}(1)$ terms of the splitting kernels.

Continuing with the factorization, we rescale the momentum fractions of the sub-collinear set by the parent energy fraction $\zeta$,
\bea
y_{i_a} \rightarrow \zeta \w_{i_a}\,, \quad a = 1,\cdots, M\,,
\eea
so that $\w_{i_a}$ is the momentum fraction of parton $i_a$ within the sub-collinear set, normalized as $\sum_a \w_{i_a}=1$. We find
\bea
\langle\mathcal{E}(\vec{n}_1)\cdots \mathcal{E}(\vec{n}_N)\rangle_{j}
=& c_N \sum_{i_{[1\cdots M]}, i_{M+1}, \cdots, i_N} \sum_{i_1,i_2,\cdots, i_M} \int d \w_{i_1} \cdots d \w_{i_M}\, d y_{i_{M+1}} \cdots d y_{i_N} \, d \zeta \nn \\
&\hspace{-3cm}\times \delta \left(1 - \w_{i_1} - \cdots - \w_{i_M}\right) \delta \left(1 - \zeta - y_{i_{M+1}} - \cdots - y_{i_N}\right) \left(\w_{i_1} \cdots \w_{i_M}\, y_{i_{M+1}} \cdots y_{i_N}\right)^2 \zeta^{3 M - 1} \nn \\
&\times \left(\frac{2 g^2}{s_{[i_1 \cdots i_M]i_{M+1}\cdots i_N}}\right)^{N-M} H^{h h'}_{j\to i_{[1 \cdots M]}i_{M+1} \cdots i_N}(y_{i_{M+1}},\cdots,y_{i_N},\zeta) \nn \\
&\times \left(\frac{2 g^2}{s_{i_1 \cdots i_M}}\right)^{M-1} P^{h h'}_{i_{[1 \cdots M]} \rightarrow i_1 \cdots i_M}(\w_{i_1},\cdots,\w_{i_M},\zeta)\,,
\eea
where the factor $\zeta^{3M-1}$ collects $\zeta^{M}$ from the integration measure, $\zeta^{-1}$ from the $\delta$-function, and $\zeta^{2M}$ from the energy weights. The change of variables acts homogeneously on the Mandelstam invariants of the sub-collinear set,
\bea
s_{i_1 \cdots i_M} = \zeta^2 \sum_{a<b} \w_{i_a} \w_{i_b} |z_{i_a i_b}|^2 \equiv \zeta^2 \bar s_{i_1 \cdots i_M}\,,
\eea
consistently with the angular units used throughout. Finally, we can always write the splitting kernels in projective form, i.e., in terms of the ratios $\w_{i_a}/\sum_b \w_{i_b}$ and ratios of Mandelstam invariants, in which case the splitting function $P^{hh'}$ is independent of $\zeta$. Thus, the iterated limit of the correlator factorizes as \cite{He:2024hbb}
\bea
\mathscr C_{1,\cdots, M} \langle\mathcal{E}(\vec{n}_1)\cdots \mathcal{E}(\vec{n}_N)\rangle_{j} =& \sum_{i_{[1\cdots M]}, i_{M+1}, \cdots, i_N} \sum_{i_1,i_2,\cdots, i_M} \biggl[\Bigl\langle \overline{\mathcal{E}}(\vec{n}_1)\cdots \overline{\mathcal{E}}(\vec{n}_M)\Bigr\rangle_{i_{[1 \cdots M]} \rightarrow i_1 \cdots i_M}\biggr]^{h h'} \nn \\
&\times \left[\left\langle \tl{\mathcal{E}}(\vec{n}_{M+1})\cdots \tl{\mathcal{E}}(\vec{n}_N)\right\rangle_{j\to i_{[1 \cdots M]}i_{M+1} \cdots i_N}\right]^{h h'}\,,
\label{eq:factorization}
\eea
with the helicity indices $h,h'$ summed, and where
\bea
\biggl[\Bigl\langle \overline{\mathcal{E}}(\vec{n}_1)\cdots \overline{\mathcal{E}}(\vec{n}_M)&\Bigr\rangle_{i_{[1 \cdots M]} \rightarrow i_1 \cdots i_M}\biggr]^{h h'} = c_M \int d \w_{i_1} \cdots d \w_{i_M}\, \delta \left(1 - \w_{i_1} - \cdots - \w_{i_M} \right) \nn \\
&\times \left(\w_{i_1} \cdots \w_{i_M}\right)^2 \left(\frac{2 g^2}{\bar s_{i_1 \cdots i_M}}\right)^{M-1} P^{h h'}_{i_{[1 \cdots M]} \rightarrow i_1 \cdots i_M}(\w_{i_1},\cdots,\w_{i_M})\,, \label{eq:EBar}\\
\biggl[\Bigl\langle \tl{\mathcal{E}}(\vec{n}_{M+1})\cdots \tl{\mathcal{E}}(\vec{n}_N)&\Bigr\rangle_{j\to i_{[1 \cdots M]}i_{M+1} \cdots i_N}\biggr]^{h h'} = c_{N-M+1} \int d y_{i_{M+1}} \cdots d y_{i_N}\,d \zeta \nn \\
&\times \delta \left(1 - \zeta - y_{i_{M+1}} - \cdots - y_{i_N}\right) \left(y_{i_{M+1}} \cdots y_{i_N}\right)^2 \zeta^{M + 1}  \\
&\times \left(\frac{2 g^2}{s_{[i_1 \cdots i_M]i_{M+1}\cdots i_N}}\right)^{N-M} H^{h h'}_{j\to i_{[1 \cdots M]}i_{M+1} \cdots i_N}(y_{i_{M+1}},\cdots,y_{i_N},\zeta)\,. \nn 
\eea
The overall factors distribute exactly between the two objects: the constant of Eq.~\eqref{eq:masterLO} satisfies the identity $c_N = c_M\, c_{N-M+1}$, with the couplings splitting accordingly, so that Eq.~\eqref{eq:factorization} holds with no leftover normalization. The first object, $\langle \overline{\mathcal{E}}(\vec{n}_1)\cdots \overline{\mathcal{E}}(\vec{n}_M)\rangle$, describes the splitting of the sub-collinear set: it is precisely a standard $M$-point energy correlator, including its overall normalization, now computed with the polarized splitting function. The second object, $\langle \tl{\mathcal{E}}(\vec{n}_{M+1})\cdots \tl{\mathcal{E}}(\vec{n}_N)\rangle$, describes the initial splitting process which produces the parent of the sub-collinear set together with all the other particles.

\subsection{Iterated Limits of Four-Point Correlators in QCD and Spin Correlations}\label{sec:calc}

Using the formalism developed in the previous subsections, we now evaluate iterated limits of the four-point correlator in QCD, for the tee projection introduced in \Sec{sec:4ptProj}. To perform these calculations, we need to carry out the contractions of the helicity indices between the two splitting objects in Eq.~\eqref{eq:factorization}. As discussed in \Sec{sec:iterated}, these contractions are non-trivial only when the intermediate parton on which we factorize is a gluon, in which case the helicity indices become transverse Lorentz indices. The spin of the initiating parton $j$ is averaged throughout, as appropriate for an unpolarized jet function, so that the splitting tensor carries only the helicity indices of the intermediate parent (see App.~\ref{sec:pert}). The splitting tensors then have terms proportional to the gluon polarization tensor $d^{\mu\nu}$ and to the boost-invariant transverse momenta $\tl k_{i_a}$,
\begin{align}
\label{eq:sTnForm}
H^{\mu \nu}_{j \rightarrow i_{[1\cdots M]} i_{M+1}\cdots i_N} =&\, \tl A_{j \rightarrow i_{[1\cdots M]} i_{M+1}\cdots i_N}(\{y_{i_a}, s_{i_a i_b}\})\, d^{\mu \nu}(P,n) \nn \\
&+ \sum_{a,b=M+1}^N \tl B_{ab}(\{y_{i_a}, s_{i_a i_b}\}) \frac{\tl k^{\mu}_{i_a} \tl k^{\nu}_{i_b}}{s_{[i_1 \cdots i_M] i_{M+1} \cdots i_N}}\,,
\end{align}
where
\bea
d^{\mu\nu}(P,n) = -g^{\mu\nu} + \frac{P^{\mu}n^{\nu}+n^{\mu}P^{\nu}}{P\cdot n}
\eea
is the polarization sum for a gluon moving in the light-like direction $P$ with reference vector $n$ (when the reference vector is clear from context, this is abbreviated as $d^{\mu\nu}(P)$, as in App.~\ref{sec:pert}), and $\tl A$ and $\tl B_{ab}$ are channel-specific coefficients which are functions of momentum fractions and Mandelstams. The boost-invariant transverse momenta \cite{DelDuca:2019ggv,DelDuca:2020vst} can be expressed in terms of the physical transverse momenta which appear in Eq.~\eqref{eq:colP},
\bea
\tl k_{i_a}^{\mu} = k_{\perp,i_a}^{\mu} - \frac{y_{i_a}}{\sum_{b=1}^N y_{i_b}} \sum_{c=1}^N k_{\perp,i_c}^{\mu}\,,
\eea
where the sums run over the full set of $N$ partons, for which the second term is in fact trivial: $\sum_b y_{i_b} = 1$ and the total transverse momentum vanishes by Eq.~\eqref{eq:PhiDef}, so that $\tl k_{i_a}^{\mu} = k_{\perp,i_a}^{\mu}$. The same combination with the sums restricted to the sub-collinear set $\{i_1, \cdots, i_M\}$, for which $\sum_b y_{i_b} = \zeta$ and the total transverse momentum does not vanish, is non-trivial, and defines the boost-invariant transverse momenta of the sub-collinear set,
\bea
\tl \kappa_{i_a}^{\mu} =&\, \tl k_{i_a}^{\mu} - \w_{i_a} \sum_{b=1}^M \tl k_{i_b}^{\mu}  = \tl k_{i_a}^{\mu} - \w_{i_a} \tl k_{P}^{\mu}\,,
\label{eq:kappa}
\eea
where $\tl k_P \equiv \sum_{b=1}^{M} \tl k_{i_b}$ is the boost-invariant transverse momentum of the parent of the sub-collinear set\footnote{These relations use that the collinear directions of the two sets coincide, $P = P'$. The generic case where these vectors are not aligned is discussed in \cite{DelDuca:2019ggv, DelDuca:2020vst}.}. The $\tl \kappa_{i_a}$ are precisely the boost-invariant version of the transverse momenta $\kappa_{\perp,i_a}$ of Eq.~\eqref{eq:subColP}, and they are the natural variables for the second splitting, whose ingredients are all defined with respect to the parent of the sub-collinear set.

The splitting function has a very similar structure, but with the metric in place of the polarization tensor,
\bea
P^{\mu \nu}_{i_{[1\cdots M]} \rightarrow i_1 \cdots i_M} = A_{i_{[1\cdots M]} \rightarrow i_1 \cdots i_M}(\{y_{i_a}, s_{i_a i_b}\})\, g^{\mu \nu} + \sum_{a,b=1}^M B_{ab}(\{y_{i_a}, s_{i_a i_b}\}) \frac{\tl \kappa^{\mu}_{i_a} \tl \kappa^{\nu}_{i_b}}{s_{i_1 \cdots i_M}}\,,
\label{eq:sFnForm}
\eea
where $A$ and $B_{ab}$ are once again channel-specific. The decompositions of Eqs.~\eqref{eq:sTnForm} and \eqref{eq:sFnForm} are the most general forms these objects can take: they are symmetric rank-two tensors, and only their components on the $(d-2)$-dimensional transverse subspace, which is spanned by the transverse metric and products of the boost-invariant transverse momenta, contribute to the contractions below. The splitting tensor is written directly in this transverse form, with $d^{\mu\nu}(P,n)$ playing the role of the transverse metric, while the splitting function conventionally carries $g^{\mu\nu}$ \cite{Catani:1999ss}. The explicit coefficients for all channels are given in \cite{Catani:1999ss,DelDuca:2019ggv,DelDuca:2020vst}, and the ingredients needed here are collected in App.~\ref{sec:pert}.

The contractions between the splitting objects amount to constants and products of the transverse momenta,
\bea
d_{\mu \nu}(P,n)\, g^{\mu \nu} =& -2\left(1 - \e\right)\,, \nn \\
d_{\mu \nu}(P,n)\, \tl k_{i_a}^{\mu} \tl k_{i_b}^{\nu} =& -\tl k_{i_a} \cdot \tl k_{i_b}\,, \nn \\
\tl \kappa^{\mu}_{i_a} \tl \kappa^{\nu}_{i_b} g_{\mu \nu} =&\, \tl \kappa_{i_a} \cdot \tl \kappa_{i_b}\,, \nn \\
\tl \kappa^{\mu}_{i_{a}} \tl \kappa^{\nu}_{i_{b}} \tl k_{i_{c},\mu} \tl k_{i_{d},\nu} =&\, \left(\tl \kappa_{i_{a}} \cdot \tl k_{i_{c}}\right) \left(\tl \kappa_{i_{b}} \cdot \tl k_{i_{d}}\right)\,,
\label{eq:contractions}
\eea
where we used that the $\tl k$ and $\tl \kappa$ are transverse to both $P$ and $n$. The contractions of the boost-invariant transverse momenta $\tl k_i$ can then be expressed in terms of Mandelstams and momentum fractions,
\bea
\tl k_{i_{a},\,\mu} \tl k_{i_{b}}^{\mu} = \half \left[s_{i_{a}i_{b}} - \sum_{c=1}^N \left(\frac{y_{i_{a}}}{\sum_{l=1}^N y_{i_l}} s_{i_{b} i_c} + \frac{y_{i_b}}{\sum_{l=1}^N y_{i_l}} s_{i_a i_c}\right) + \frac{2 y_{i_a} y_{i_b}}{\left(\sum_{l=1}^N y_{i_l}\right)^2}s_{1\cdots N}\right]\,.
\label{eq:boostkContraction}
\eea
As seen in Eq.~\eqref{eq:MandSubLead}, the Mandelstams encode terms that are subleading in the iterated collinear limit. Those Mandelstams that only involve particles in the sub-collinear set have uniform scaling in $\lambda'$,
\bea
s_{i_a i_b} \rightarrow \lambda'^2 s_{i_a i_b}\,, \quad a,b \in \{1,\dots,M\}\,.
\eea
For those that involve particles from the two sets, we have
\bea
s_{i_a i_c} =&\, 2 p_{i_a} \cdot p_{i_c} 
=\, 2 p_{i_c} \cdot \left(\zeta\,\omega_{i_a} P + \kappa_{\perp,i_a} - \frac{\kappa_{\perp,i_a}^2}{\zeta\,\omega_{i_a}} \frac{n}{2 n \cdot P}\right) \nn \\
\rightarrow &\, 2\, \zeta\,\omega_{i_a} \left(p_{i_c} \cdot  P\right) + \mathcal{O}(\lambda') \nn \\
=&\, \omega_{i_a} s_{[i_1\cdots i_M]i_c} + \mathcal{O}(\lambda')\,, \quad a \in\{1,\dots, M\}\,, \quad c \in \{M+1,\dots, N\}\,,
\eea
where in the last step we used that the parent momentum is $p_{i_{[1\cdots M]}} = \zeta P + \mathcal{O}(\lambda')$, so that $s_{[i_1\cdots i_M]i_c} = 2 \zeta \left(P \cdot p_{i_c}\right) + \mathcal{O}(\lambda')$, and we again neglect terms of $\mathcal{O}(\lambda')$. This power counting, along with Eqs.~\eqref{eq:Mandel}, \eqref{eq:contractions}, and \eqref{eq:boostkContraction}, allows one to express the contribution to the iterated limit in terms of the momentum fractions and the physical scales of the problem.

Note that all dependence on the azimuthal angle $\phi$, which rotates the sub-collinear set with respect to the remaining detectors, arises from the mixed contraction in the last line of Eq.~\eqref{eq:contractions}, in which the transverse momenta of the two sets are dotted into one another. These are precisely the terms in which the helicity of the intermediate gluon differs between the amplitude and the conjugate amplitude, i.e., the interference of its $\pm 1$ helicity states. Since the transverse vectors of the two sets enter quadratically, the azimuthal dependence takes the form
\bea
\tl \kappa^{\mu}_{i_{a}} \tl \kappa^{\nu}_{i_{b}} \tl k_{i_{c},\mu} \tl k_{i_{d},\nu} \sim&\, f(\omega_{i_{a}},\omega_{i_{b}},y_{i_{c}},y_{i_{d}},\zeta) \cos^2 \phi \nn \\
=&\, \half f(\omega_{i_{a}},\omega_{i_{b}},y_{i_{c}},y_{i_{d}},\zeta) \left(1 + \cos(2 \phi)\right)\,,
\eea
where $f$ is some function of the momentum fractions (and of the fixed angular configuration). At leading power, the iterated limit therefore contains a constant contribution and a pure $\cos(2 \phi)$ modulation: no $\cos\phi$ term can appear, since the helicity-flip interference is by two units, and $\sin(2\phi)$ terms are forbidden by parity. For the four-point correlator, the result then takes the form
\bea
\mathscr{C}_{1,\cdots,M} \langle \mathcal{E}(\vec n_1) \cdots \mathcal{E}(\vec n_4)\rangle_{j} = c_4 \left(2g^2\right)^{3} \frac{\alpha_j + \beta_j \cos(2 \phi)}{R^{8-2M}\, r^{2M-2}}\,,
\label{eq:resultForm}
\eea
where $r$ and $R$ are the sizes of the sub-collinear set and of the full configuration, measured as distances in the celestial coordinates of the parameterizations of \Sec{sec:4ptProj} ($R$ is not to be confused with the jet radius appearing in Eq.~\eqref{eq:JetFactorization}). That is, $r^2$ and $R^2$ are squared celestial distances of the form $|z_{ij}|^2 = \theta_{ij}^2/4$: for the dipole, $r$ and $R$ are the separations of the small and large pairs, while for the tee, $r$ is the separation of the outer detectors of the sub-collinear line and $R$ the distance of the fourth detector from its center. In particular, the overall scale is $x_L = R^2$ at leading power in $r/R$. With the overall constant $c_4$ and the couplings made explicit, $\alpha_j$ and $\beta_j$ are pure numbers that depend only on the shape of the configuration. We compute these explicitly for the tee configuration below; note that we must always have $\alpha_j \geq |\beta_j|$ to ensure positivity of the observable. We emphasize that at leading power in $r/R$, the dependence on the two angular scales and on the azimuth completely factorizes: the relative modulation $\beta_j/\alpha_j$ is independent of $r$. For the QCD prediction, any $r$-dependence of the measured azimuthal modulation is therefore a direct signal of power corrections to the iterated collinear limit.

\begin{figure}
\begin{center}
\includegraphics[scale=0.46]{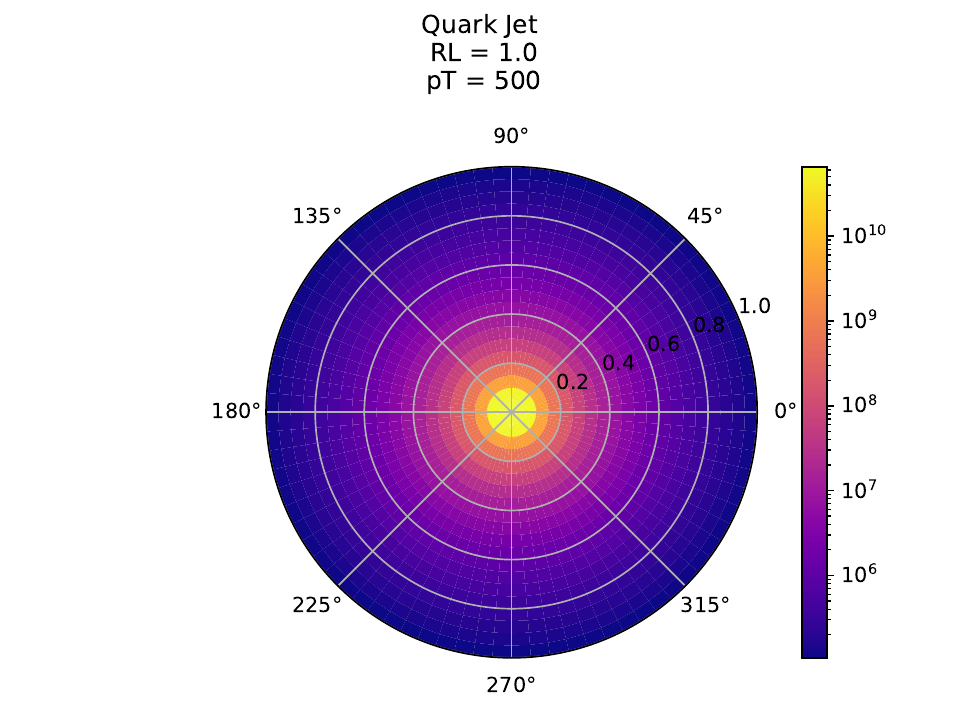} 
\includegraphics[scale=0.46]{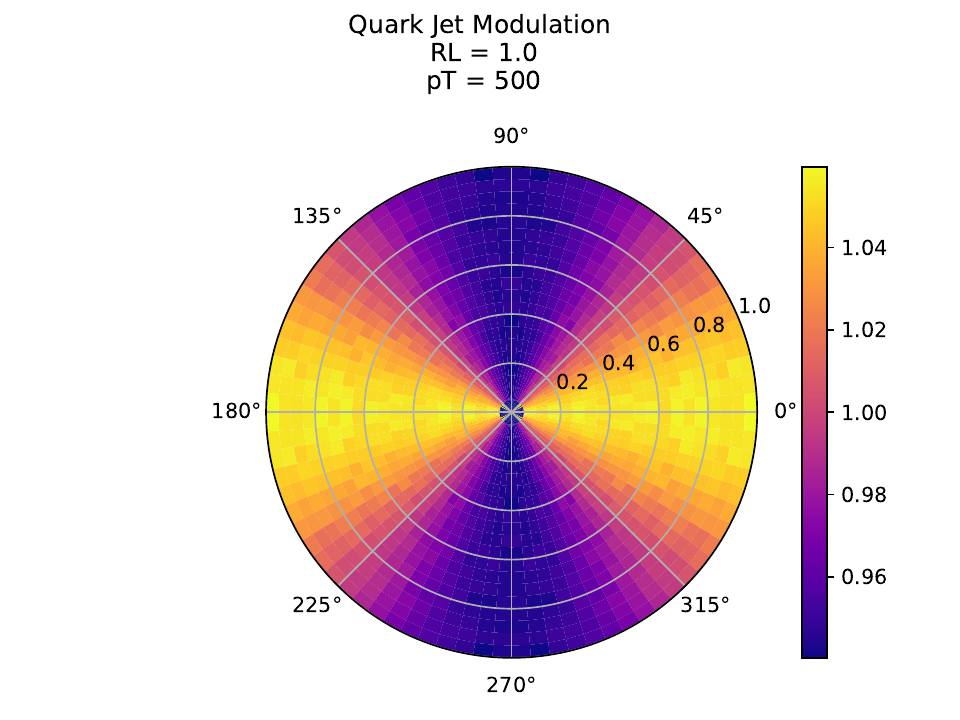} 
\end{center}
\caption{Leading-power analytic result for the tee configuration on quark jets, displayed in polar coordinates $(r,\phi)$. Left: the full result, which is dominated by the $1/r^4$ scaling of the collinear pole. Right: the same result normalized in each $r$ bin to its azimuthal average, exposing the ${\approx}5\%$ $\cos(2\phi)$ modulation, which at leading power is independent of $r$.}
\label{fig:tee_quark}
\end{figure}

\begin{figure}
\begin{center}
\includegraphics[scale=0.46]{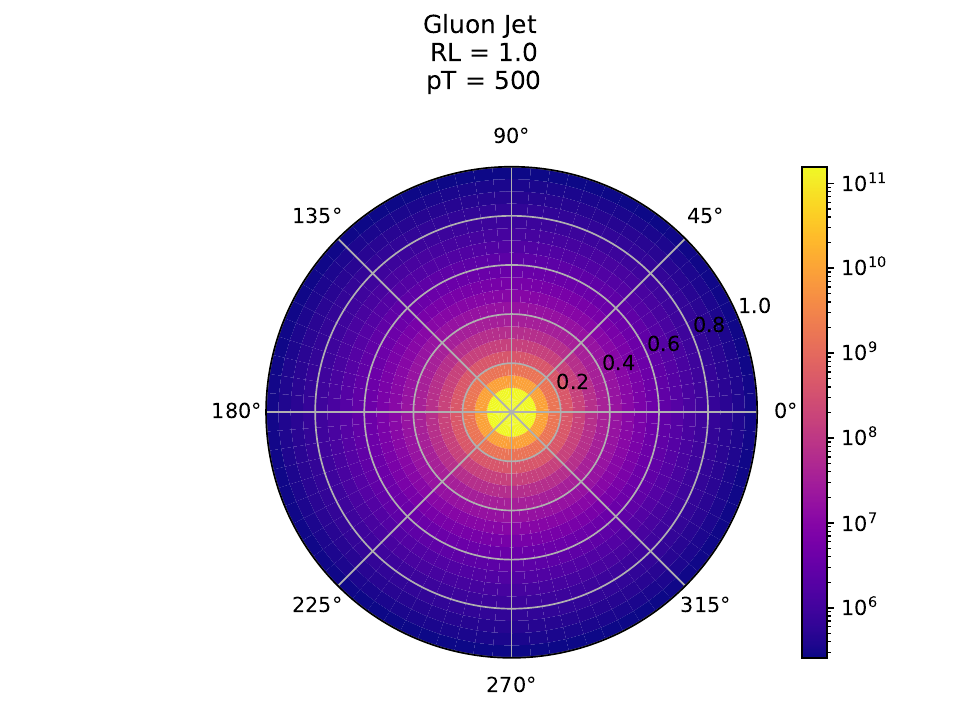}
\includegraphics[scale=0.46]{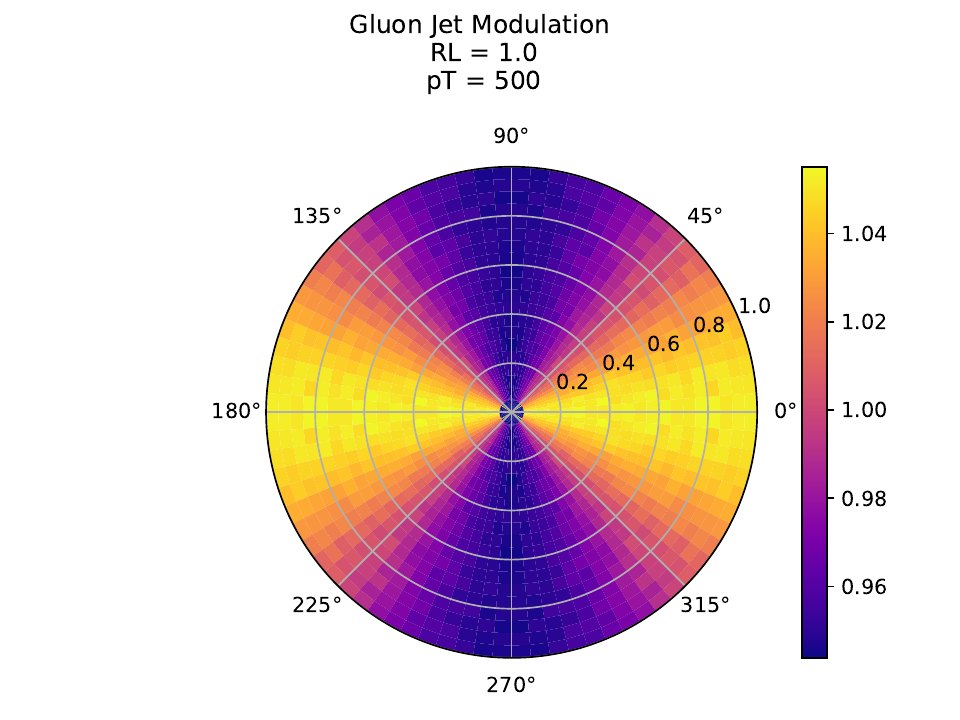}
\end{center}
\caption{Same as Fig.~\ref{fig:tee_quark}, for gluon jets. The relative modulation is again at the $5\%$ level.}
\label{fig:tee_gluon}
\end{figure}

We now present explicit results for the tee configuration, whose iterated limit $\mathscr{C}_{1,2,3}$ requires a $1 \rightarrow 2$ splitting tensor and a polarized $1 \rightarrow 3$ splitting function; both ingredients are compact, and are collected in App.~\ref{sec:pert}. (The dipole configuration instead probes the $1\to 3$ splitting tensors of \cite{DelDuca:2019ggv,DelDuca:2020vst} together with the $1\to 2$ splitting functions, and can be computed in complete analogy.) We have analytically computed the coefficients $\alpha_j$ and $\beta_j$ for both quark and gluon jets. The final results can be expressed in terms of dilogarithms evaluated at specific rational values, and are collected in App.~\ref{sec:pert}. Here we quote their numerical values, defined in the normalization of Eq.~\eqref{eq:resultForm}, in which the overall constant $c_4$ and the coupling factor $(2g^2)^3$ have been made explicit. Evaluating the results numerically gives, for the tee configuration ($M=3$),
\bea\label{eq:3point_answer}
\alpha_q + \beta_q \cos(2\phi)
=&\, 7.921\, C_F^3 + 2.920\, C_F^2 C_A + \left(0.167 + 0.847\, n_f\right) C_F^2 T_R \nn \\
&+ 0.651\, C_F C_A T_R n_f + 28.933\, C_F C_A^2 \nn \\
&+ \left[-0.471\, C_F^2 T_R n_f - 0.320\, C_F C_A T_R n_f + 2.279\, C_F C_A^2\right] \cos (2 \phi) \nn \\
=&\, 381.689+2.055\, n_f + \left(27.351 - 1.059\, n_f\right) \cos (2 \phi) \nn \\
=&\, 391.963+ 22.056\cos(2 \phi)\,, \\
\alpha_g + \beta_g \cos(2\phi)
=&\, 0.588\, C_F^2 T_R n_f + 1.010\, C_F C_A T_R n_f \nn \\
&+ \left(0.012 + 0.012\, n_f\right) C_F T_R^2 n_f + 0.759\, C_A^2 T_R n_f + 33.727\, C_A^3 \nn \\
&+ \left[-0.471\, C_F C_A T_R n_f - 0.320\, C_A^2 T_R n_f + 2.279\, C_A^3\right] \cos (2 \phi) \nn \\
=&\, 910.635+5.961\, n_f +0.004\, n_f^2 +\left(61.539 - 2.383\, n_f\right) \cos (2 \phi) \nn \\
=&\, 940.544 + 49.625 \cos(2 \phi)\,,
\eea
where we take $N_c = 3$ and $n_f = 5$ in the last step. 

Two features of these results are worth highlighting. First, the modulation coefficients of quark and gluon jets are related by an overall Casimir rescaling,
\bea
\beta_g = \frac{C_A}{C_F}\, \beta_q\,,
\eea
term by term in the color structures. This follows directly from the structure of the factorization: the modulation is generated entirely by the helicity-flip part of the $1\to 2$ splitting tensor of the intermediate gluon, which has an identical functional form for quark and gluon jets up to the color factor $C_F$ versus $C_A$, see~Eqs.~\eqref{eq:gqST} and \eqref{eq:ggST}. This exact rescaling is special to tee-like configurations, in which the sub-collinear set contains all but one of the detectors, so that the helicity flip enters through a $1\to 2$ splitting tensor. In the dipole configuration, the modulation instead probes the helicity-flip parts of the $1\to 3$ splitting tensors, which are not related between quark and gluon jets by an overall color factor. Second, the relative size of the modulation is small,
\bea
\frac{\beta_q}{\alpha_q} \simeq 5.6\%\,, \qquad \frac{\beta_g}{\alpha_g} \simeq 5.3\%\,.
\eea
We see that the intrinsic spin correlation in the tee configuration is at the level of $5\%$, similar to the magnitude of the spin effect in the squeezed limit of the three-point correlator \cite{Chen:2020adz}. The results are plotted in \Fig{fig:tee_quark} and \Fig{fig:tee_gluon}. While this effect is small, and probably challenging to observe experimentally, knowing its magnitude as computed in the leading-power factorization is crucial for the comparison with parton shower simulations, in order to disentangle genuine spin correlations from kinematic power corrections. As we will see in \Sec{sec:pheno}, in the kinematic regions accessible at the LHC, the simulated distributions from parton-shower generators exhibit azimuthal modulations of order $40\%$, whose relative size depends strongly on $r$, in sharp contrast with the small, $r$-independent modulation predicted by the leading-power factorization. This shows that the azimuthal structure of the simulated distributions is not governed by the leading-power iterated limit. Current parton showers are built from iterated $1\to 2$ splittings: they contain neither the exact $1\to 3$ splitting function that enters the leading-power iterated limit, nor the full $1\to 4$ splitting function that governs the power corrections away from it, so there is no reason to expect them to be accurate in either regime. Therefore, our leading-power predictions already provide a significant stress test of parton-shower generators, directly probing the iterated structure of the splittings and whether it is implemented correctly. Furthermore, this strongly motivates the complete calculation of the four-point correlator in QCD, which would determine the size of these power corrections from first principles.

\section{Phenomenology with Parton Shower Simulations}\label{sec:pheno}

In this section we perform a phenomenological study of our different projections of the four-point correlator using parton shower Monte Carlos. 
The goal of this phenomenological study is three-fold. First, we would like to develop algorithms for the measurement of configurations of the four-point energy correlators in experiment, as well as for the treatment of statistical uncertainties.
Second, we would like to illustrate that for experimentally reasonable values of jet radii, $R$, and $p_T$ available at the LHC, it is possible to access the four-point correlator in a perturbative regime, where we can resolve its features, and compare it with perturbative calculations.
Third, we would like to systematically explore how different parton showers, and parton shower settings modify the dependence of our projections of the four-point correlator on both the relative size of the dipoles, $r$, and on the azimuthal angle, $\phi$. To do this, we study the observables for both quark and gluon jets, in both Pythia and Herwig, and for various settings of spin correlations (the details of which we discuss in \Sec{sec:ev_gen}). We find surprising discrepancies between the different showers, which motivate the experimental measurement of these observables, as well as their complete calculation in QCD.

In \Sec{sec:ev_gen} we describe in detail the event generation, including the different configurations for spin effects. In \Sec{sec:practical} we present the algorithms for computing the different configurations of the four point correlator experimentally, and discuss the treatment of statistical uncertainties. In \Sec{sec:results} we present detailed studies of our different configurations of the four-point correlator.

\subsection{Event Generation}\label{sec:ev_gen}

We simulate $Z+g$ and $Z+q$ (5-flavor) production at tree level with Madgraph5~\cite{Alwall:2011uj}. 
These simulated events are then interfaced with either Pythia8~\cite{Sjostrand:2014zea} with the CP5 tune~\cite{CMS:2022awf} or Herwig7~\cite{Bahr:2008pv,Bellm:2015jjp} with the CH3 tune~\cite{Gieseke:2012ft}.
The same generated events are showered by both programs, and no matching is performed as there is exactly one final-state parton. 
No detector simulation is performed.

In order to estimate the sensitivity of these observables to various physical effects, we generate several variations in addition to the nominal CP5 and CH3 tunes. 
One physical effect of particular interest is azimuthal correlations throughout the parton shower due to the spin-1 nature of the gluon. 
This effect has recently been observed by the CMS collaboration through azimuthal correlations in the reconstructed shower history~\cite{CMS:2026jnt}.
From the perspective of Monte Carlo parton showers there are two effects: a ``hard spin correlations'' effect in the limit that two branchings are collinear, and a ``soft spin correlations effect'' related to the emission of very soft gluons.
Herwig implements spin correlations across the parton shower at each stage of the splitting~\cite{Richardson:2018pvo}, and exposes independent toggles for both the hard and soft spin correlations to the end user (\texttt{SpinCorrelations} and \texttt{SoftCorrelations}). Pythia applies an approximate angular rescaling to gluons in their decay frame to reflect polarization~\cite{Fischer:2017htu}, and only exposes a toggle for the hard spin correlations effect ({\tt TimeShower:phiPolAsym}). 
More details on these toggles and their impact on the parton shower kinematics can be found in Appendix \ref{appendix:sctoggles}. 
We therefore generate the following matrix of generator settings:

\begin{table}[H]
    \centering
    \begin{tabular}{c|c|c}
        Parton shower program & Hard spin correlations & Soft spin correlations \\
        \hline\hline
       Pythia  & ON & ON \\
       Pythia & OFF & ON \\
       \hline
       Herwig & ON & ON \\
       Herwig & OFF & ON \\
       Herwig & ON & OFF \\
       Herwig & OFF & OFF
    \end{tabular}
    \caption{Matrix of different parton shower settings for studying the impact of spin correlations.}
    \label{tab:ps_spin_setting}
\end{table}

We also consider the impact of hadronization on our four-point EEC observables, in order to understand the sensitivity of these observables to nonperturbative effects, and the borders between perturbative and nonperturbative regimes. 
Pythia8 uses a Lund string fragmentation model for hadronization~\cite{Andersson:1983ia,Sjostrand:1984ic}, while Herwig7 uses a cluster model~\cite{Webber:1983if,Gieseke:2012ft}. 
We are only able to disable hadronization in Pythia, which we use to study the impact of this process in isolation. We generate the following matrix of generator settings:
\begin{table}[H]
    \centering
    \begin{tabular}{c|c|c}
        Parton shower program & Spin correlations & Hadronization \\
        \hline\hline
       Pythia  & ON  & ON\\
       Pythia & ON  & OFF
    \end{tabular}
    \caption{Matrix of different parton shower settings for studying the impact of hadronization. In all cases both hard and soft spin correlations are enabled.}
    \label{tab:ps_settings}
\end{table}

We perform a realistic event selection and reconstruction procedure, modeled off what might be done in a measurement at the LHC.
We trigger on the $Z\rightarrow\mu\mu$ decay, requiring that the two leading generated muons satisfy:
\begin{itemize}
    \item $|\eta| < 2.4$
    \item leading (subleading) muon $p_T > 26 (15)$ GeV
    \item The two muons have opposite charge
    \item The dimuon invariant mass must be within 20 GeV of the nominal $Z$ mass of 91.1876 GeV \cite{ParticleDataGroup:2024cfk}. 
\end{itemize}
We then reconstruct the leading generated jet with the anti-kt algorithm \cite{Cacciari:2008gp}, with radius parameter 0.8. 
The large radius parameter is chosen to minimize edge effects from the jet clustering and give as wide a view as possible of the jet formation dynamics. 
The reconstructed jets are required to satisfy (again mimicking what might be done at the LHC):
\begin{itemize}
    \item $|\eta| < 1.7$
    \item $\Delta R(J,\mu_1), \Delta R(J, \mu_2) > 0.8$ for the two leading muons
    \item Less than 80\% of the jet energy is due to muons
    \item Less than 80\% of the jet energy is due to electrons
    \item Less than 90\% of the jet energy is due to photons
    \item Less than 90\% of the jet energy is due to neutral hadrons
    \item There is at least one jet constituent 
\end{itemize}
where $\Delta R$ is the distance in the pseudorapidity-azimuth plane ($\Delta R = \sqrt{\left(\Delta \eta\right)^2 + \left(\Delta \phi\right)^2}$). 
In practice the energy composition cuts have minimal effect.

We then compute the resolved four-point EEC observables on these jets, binning in the jet $p_T$ as well as the three EEC coordinates ($R, r, \phi$) for each of the three configurations.

\subsection{Practical Details}\label{sec:practical}

In this section we provide algorithms to measure our projections of the four-point correlator in data, as well as to treat statistical uncertainties. Several additional details are collected in \Sec{sec:exp}. Our goal is to be as explicit as possible, so that these procedures can be used in future experimental analyses. Readers not interested in these details can freely skip to the next section.

\subsubsection{Observable Definitions}

The definitions of our three configurations in Section \ref{sec:4ptProj} are of measure zero, so we require the insertion of some finite tolerance in order to come up with experimentally practical observables. We operationalize the three four-point configurations as follows:

\paragraph{The dipole}

The dipole configuration is defined as two pairs of particles whose separations in the $(\eta, \phi)$ plane have a common midpoint, within a finite tolerance. Concretely, we use the following procedure:
\begin{algorithmic}[1]
\State $t \gets$ tolerance
\For{\{i, j, k, l\} $\in$ particles}
\For{(A,B,C,D) $\in$ \{(i,j,k,l),(i,k,j,l),(i,l,j,k)\}}
\If{$\Delta R$(A,B) $<$ $\Delta R$(C,D)}
\State swap((A,B), (C,D))
\EndIf
\State $m_1\gets$ midpoint $\overline{AB}$
\State $m_2\gets$ midpoint $\overline{CD}$ 
\If{$\Delta R(m_1, m_2) < t$}
\State (A, B,C, D) is a \textbf{dipole}
\State $R \gets \Delta R(A,B)$
\State $r \gets \Delta R(C,D)/\Delta R(A,B)$
\State $\phi \gets $ the angle between $\overline{AB}$ and$\overline{CD}$
\EndIf
\EndFor
\EndFor
\end{algorithmic}
A proof that this procedure does not introduce any bias is given in Appendix \ref{sec:algoproof}.

\paragraph{The tee}

The tee configuration is defined as two pairs of particles whose separations in the $(\eta, \phi)$ plane have the property that the midpoint between one pair lines within a finite tolerance of one member of the other pair. Concretely, we use the following procedure:
\begin{algorithmic}[1]
\State $t \gets$ tolerance
\For{\{i, j, k, l\} $\in$ particles}
\For{(A,B,C,D) $\in$ \{(i,j,k,l),(i,k,j,l),(i,l,j,k)\}}
\If{$\Delta R(A,B) < \Delta R(C,D)$}
\State swap((A,B), (C,D))
\EndIf
\State $m_1\gets$ midpoint $\overline{AB}$
\State $m_2\gets$ midpoint $\overline{CD}$ 
\If{$\Delta R(m_2, A) < t$ OR $\Delta R(m_2,B) <t$}
\State (A,B,C,D) is a \textbf{tee}
\State $R \gets d(A,B)$
\State $r \gets \Delta R(C,D)/\Delta R(A,B)$
\State $\phi \gets $ the angle between $\overline{AB}, \overline{CD}$
\EndIf
\EndFor
\EndFor
\end{algorithmic}
Again, a proof that this this procedure does not introduce any bias is given in Appendix \ref{sec:algoproof}.

\paragraph{The triangle}

The triangle configuration is the simplest to define experimentally. We simply look for a set of three particles that are within some finite tolerance of being arranged in the appropriate triangle in the $(\eta, \phi)$ plane, and then allow the fourth point to be any other member of the jet. As no selection is being made on the fourth particle, it is impossible for any procedure to result in a systematic bias. We use the following procedure:
\begin{algorithmic}[1]
\State $t \gets$ tolerance
\State $LoM \gets 5/4$ 
\State $LoS \gets 5/3$
\For{\{i, j, k, l\} $\in$ particles}
\For{(A,B,C,D) $\in$ \{(i,j,k,l),(j,k,l,i),(k,l,i,j),)(l,i,j,k)\}}
\State sort (A, B, C) such that $\Delta R(A,B) > \Delta R(A,C) > \Delta R(B, C)$
\If{$|\Delta R(A, B)/\Delta R(A,C) - LoM|< t$ AND $|\Delta R(A, B)/\Delta R(B,C) - LoS|< t$}
\State (A,B,C,D) is a \textbf{triangle}
\State $R \gets \Delta R(A,B)$
\State $r \gets \Delta R(C,D)/\Delta R(A,B)$
\State $\phi \gets $ the angle between $\overline{CD}$ and $\overline{AC}$
\EndIf
\EndFor
\EndFor
\end{algorithmic}
Note that we loop over all four possible triangles, so it is possible for a given set of four points to contribute more than one triangle (eg if $(i,j,k)$ and $(j,k,l)$ are both $(3,4,5)$ triangles). 

In this paper we take the tolerance $t=0.05$ for all three configurations.

\subsubsection{Uncertainties}

An important property of energy-energy correlator observables is that each event contributes to combinatorially many bins of the observable histogram. The standard Poisson prescription for the covariance between two histogram bins $x^\mu$ and $x^\nu$ would be
\begin{equation}
    \text{Cov}\left[x^\mu,x^\nu\right] = \delta_{\mu\nu}\sum_i  \left(w_i^{(\mu)}\right)^2 
    \label{eqn:wrong_poisson}
\end{equation}
where the sum is over all of the weights contributing to histogram bin $\mu$. 

This prescription is not correct, as it treats the individual EEC entries as the statistical degrees of freedom, rather than entire events. Instead, we must construct the full covariance matrix by filling a per-event histogram $\vec{x}^{(i_{evt})}$ and taking the outer product:
\begin{equation}
    C = \sum_{i_{evt}} \vec{x}^{i_{evt}} \left(\vec{x}^{i_{evt}}\right)^T
\end{equation}
This gives a covariance matrix which typically has large off-diagonal contributions.

\begin{figure}
    \centering
    \includegraphics[width=0.55\textwidth]{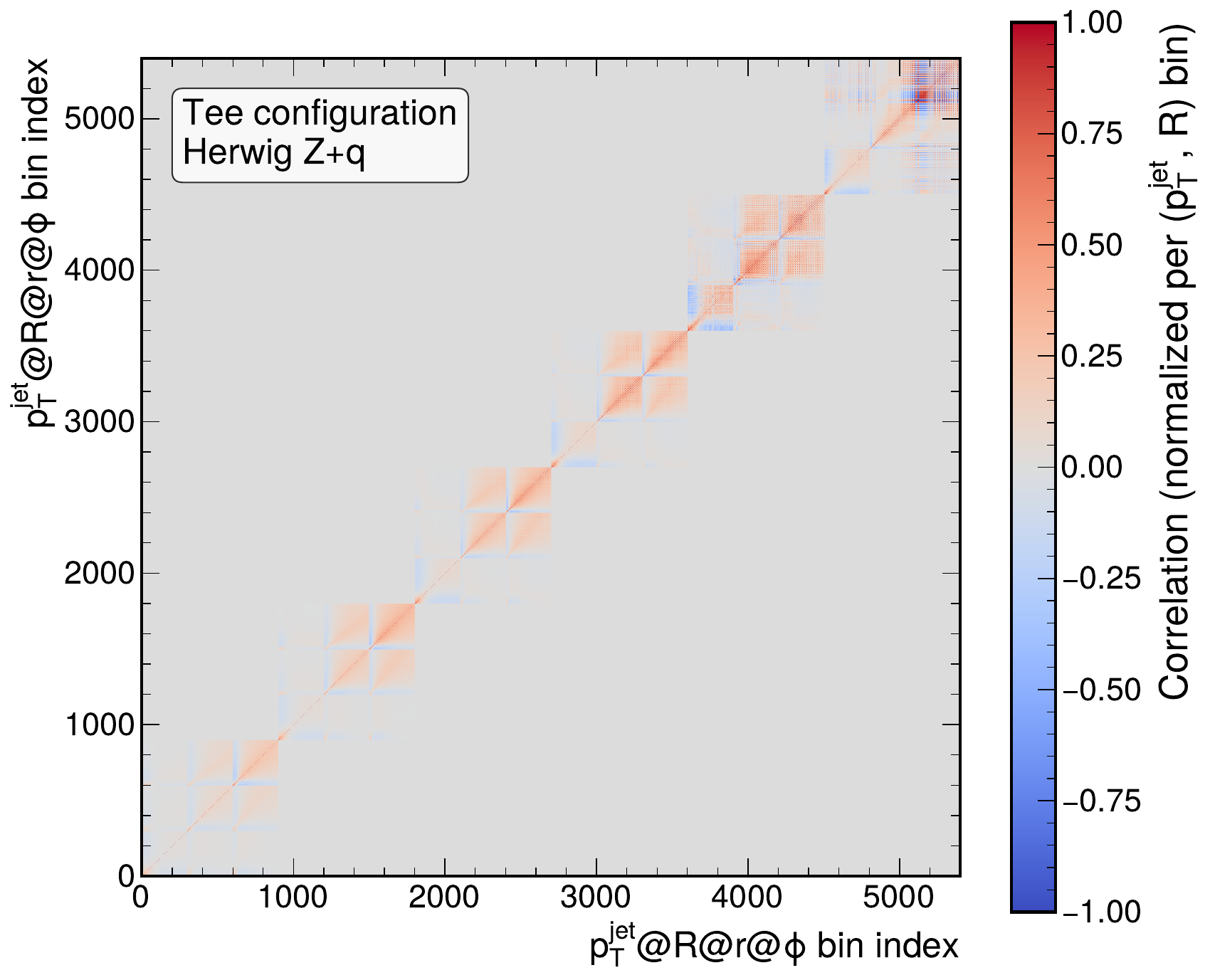}
    \caption{Example of a covariance matrix. The off-diagonal contributions are substantial, and must be taken into account when performing any statistical analysis of the data.}
    \label{fig:covariance-example}
\end{figure}

Figure~\ref{fig:covariance-example} shows an example of a covariance matrix for the dipole configuration. The off-diagonal contributions are substantial, and must be taken into account when performing any statistical analysis of the data.

\subsection{Phenomenological Results}\label{sec:results}

In this section we present detailed phenomenological studies in Pythia and Herwig of the different configurations of the four-point correlator.

\subsubsection{The Dipole Configuration}\label{sec:proj_a_pheno}

We begin by considering the dipole configuration. For convenience, we recall that the dipole configuration probes the factorization of the four-point splitting function into a $1\to 3$ splitting tensor, and a $1\to 2$ splitting function
\begin{align}
\fd{8cm}{figures/dipole_solo}\,.
\end{align}
For fixed $R$, it is defined in terms of the variable $r$, which controls the relative size of the dipoles, and hence the factorization, and the variable $\phi$, which allows the study of potential spin correlations. We will study discrepancies between the $r$ and $\phi$ dependence of the different parton showers.

\begin{figure}[h]
    \centering
    \includegraphics[width=0.45\textwidth]{figures/MCstudy/dipole/herwig_q/histogram_shapes_radial2D_pt1_R1.pdf} 
    \includegraphics[width=0.45\textwidth]{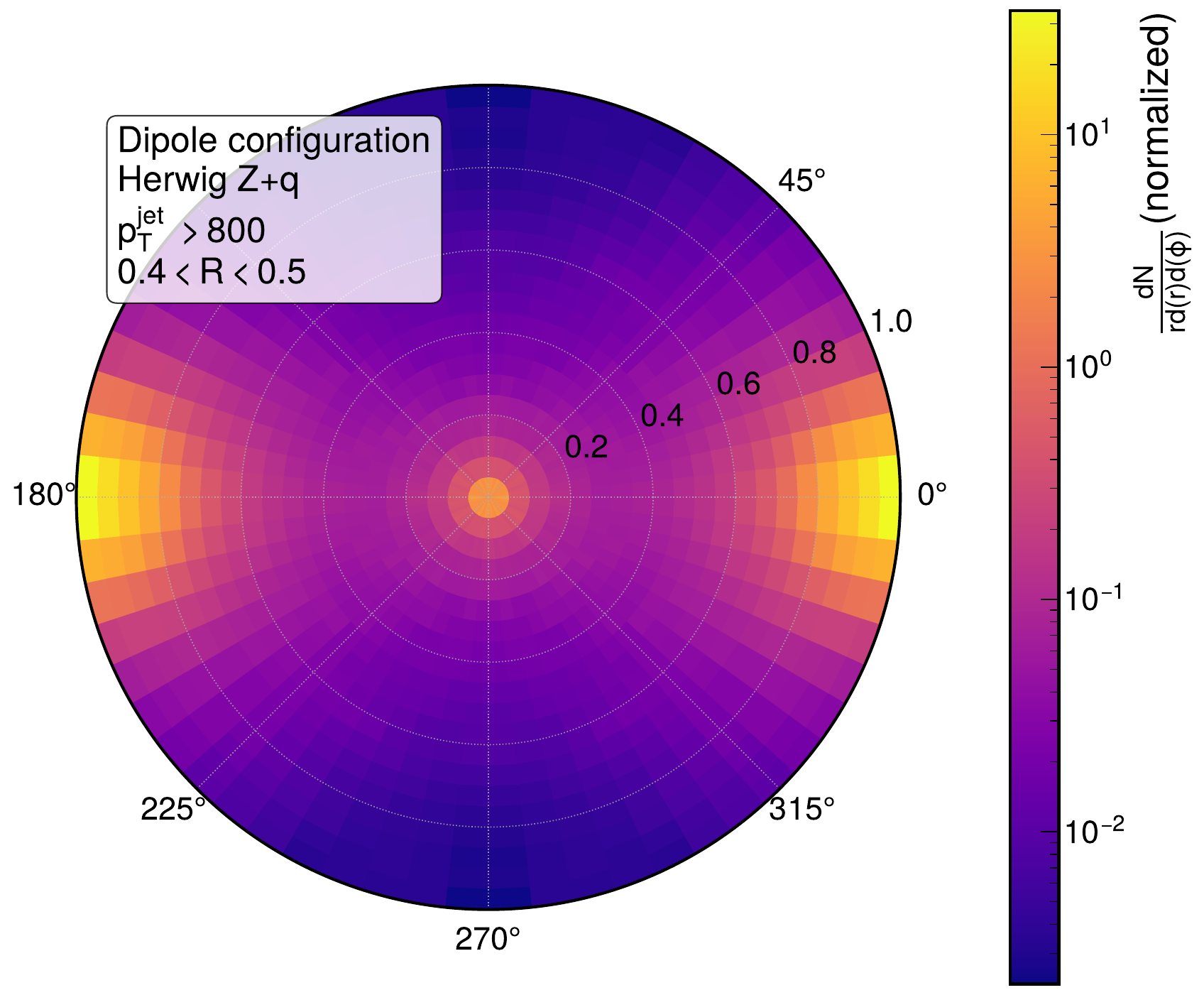} 
    \caption{Simulated results for the dipole configuration in Herwig $Z+q$ events, for $100$ GeV $\leq p_T \leq 200$ GeV, and $p_T \geq 800$ GeV.}
    \label{fig:dipole_herwig_q}
\end{figure}

We first begin with a baseline plot to illustrate the general structure of the dipole configuration of the four-point correlator. 
In Figure~\ref{fig:dipole_herwig_q} we show results for the dipole configuration in $Z+q$ events in Herwig. 
The result is shown for two different $p_T$ bins, namely $100$ GeV $\leq p_T \leq 200$ GeV, and $p_T \geq 800$ GeV. 
As expected, the distribution is dominated by the collinear poles at $r\rightarrow 0$ and $r\rightarrow 1, \phi \rightarrow 0,\pi$, which are regulated by confinement. 
As the $p_T$ is increased, the collinear poles become significantly enhanced, since they remain in a perturbative regime for a wider range of angles.

\begin{figure}[h]
    \centering
    \includegraphics[width=0.45\textwidth]{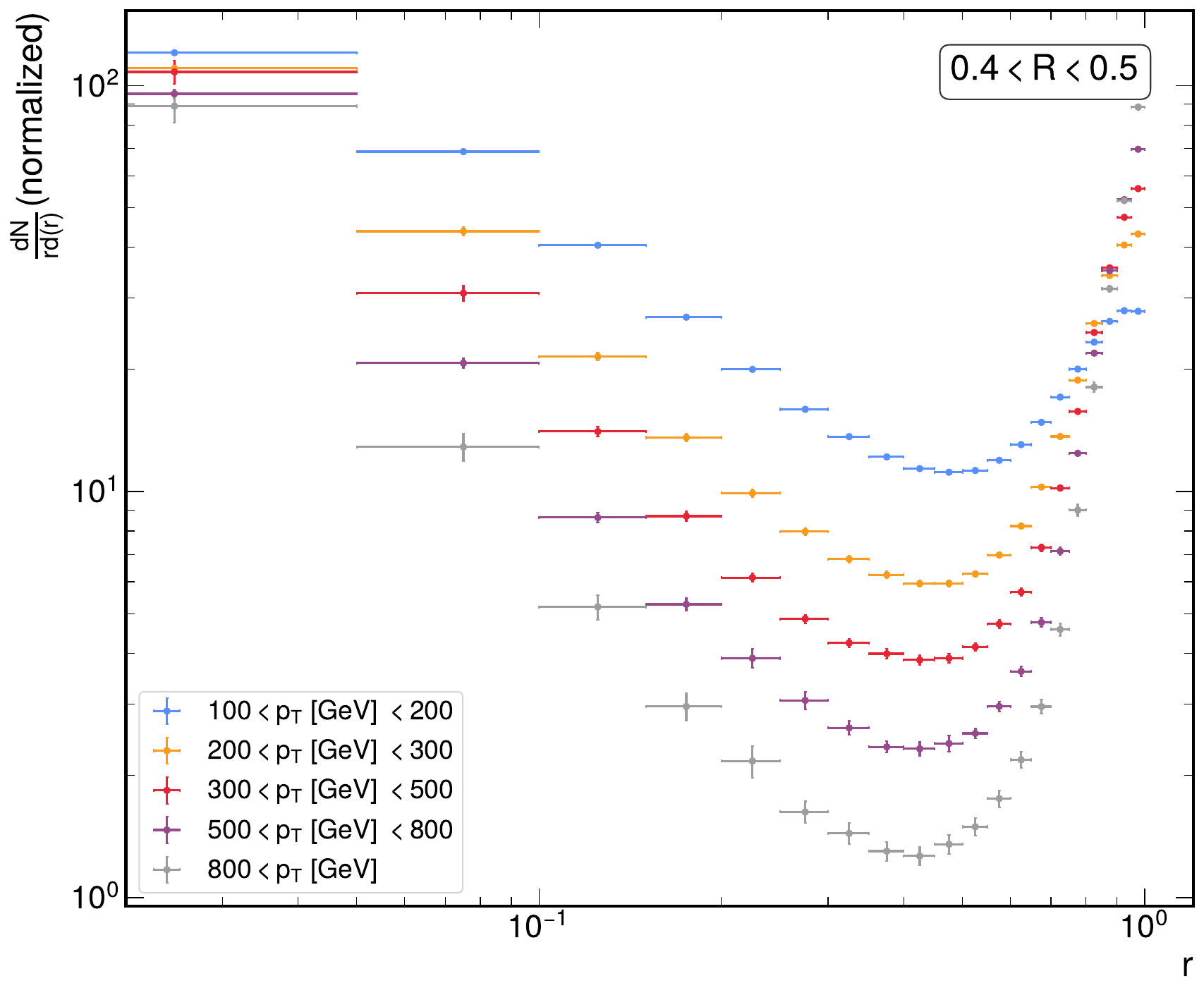} 
    \includegraphics[width=0.45\textwidth]{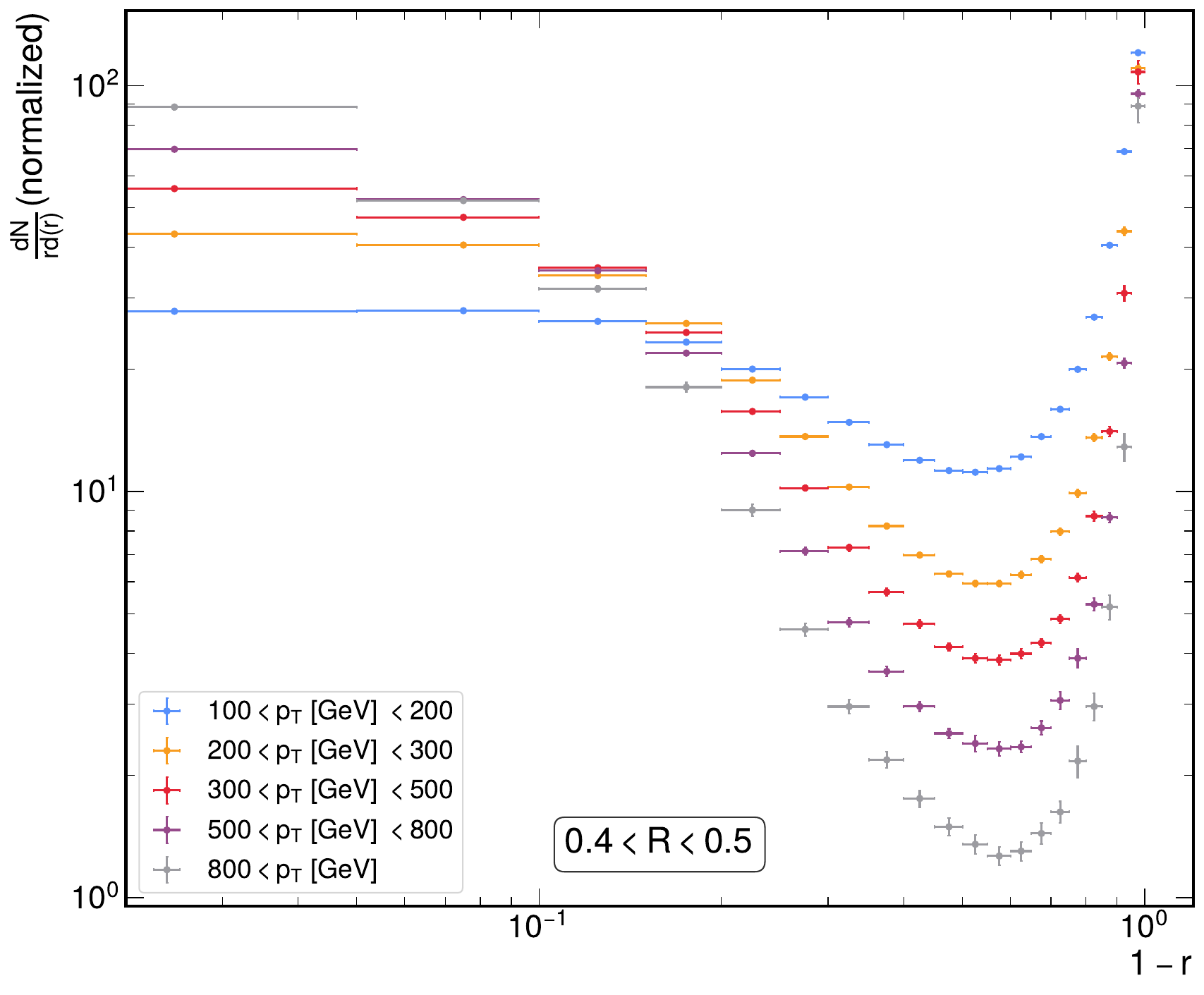} 
    \caption{Simulated results for the dipole configuration in Herwig $Z+q$ events, integrated over $\phi$.
    Results are shown in terms of both $r$ (left) and $1-r$ (right), to illustrate the different behavior of the two collinear poles.}
\label{fig:dipole_herwig_q_radial}
\end{figure}

To highlight the structure of the collinear poles, we can  fix $R$ and study the scaling with the variable $r$. In the small $r\ll 1$ region, the iterated limits are expected to give the scaling predicted by Eq.~\eqref{eq:resultForm}. Fig.~\ref{fig:dipole_herwig_q_radial} shows that small-$r$ exhibits the scaling
\begin{align}
\frac{d\Sigma}{r\,dr} \sim r^{-2+\delta}\,,
\end{align}
where $0<\delta<1$ arises from anomalous scaling. The classical scaling exponent agrees exactly with that expected for the $M=2$ case in Eq.~\eqref{eq:resultForm}. 

The opposite endpoint, $1-r\ll 1$, is not described by the particular iterated limit considered in Eq.~\eqref{eq:resultForm}. Instead, it corresponds to a limit in which the two branches produced by the primary $1\to 2$ splitting each undergo a further $1\to 2$ splitting. Figure~\ref{fig:dipole_herwig_q_radial} also displays the associated collinear enhancement as a function of $1-r$. It is amusing that these radial slices behave almost like the standard two-point energy correlator, but now, within a four-point correlator.

\paragraph{Quark vs gluon jets:}

\begin{figure}[h]
    \centering
    \includegraphics[width=0.4\textwidth]{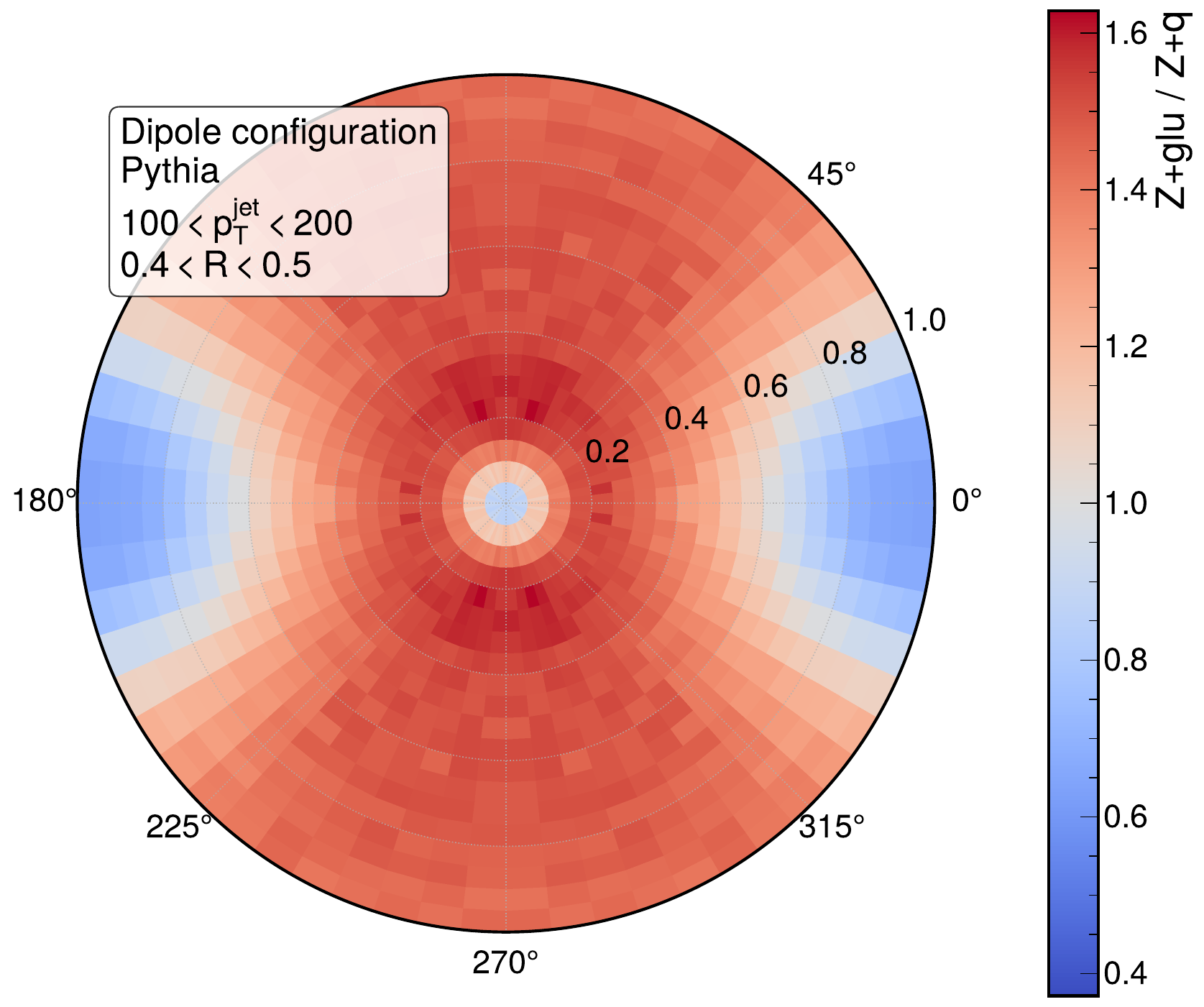} 
    \includegraphics[width=0.4\textwidth]{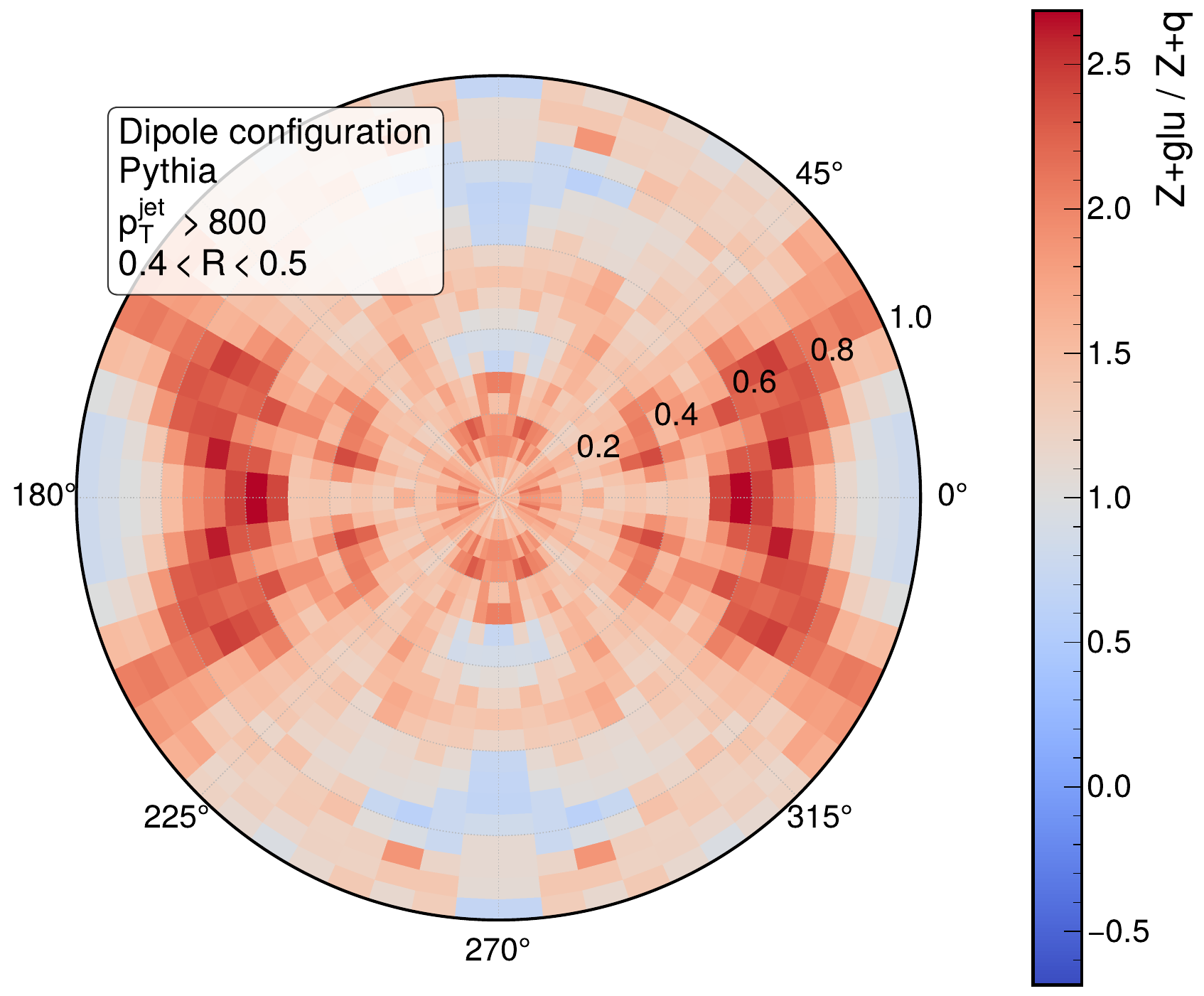} 
        \includegraphics[width=0.4\textwidth]{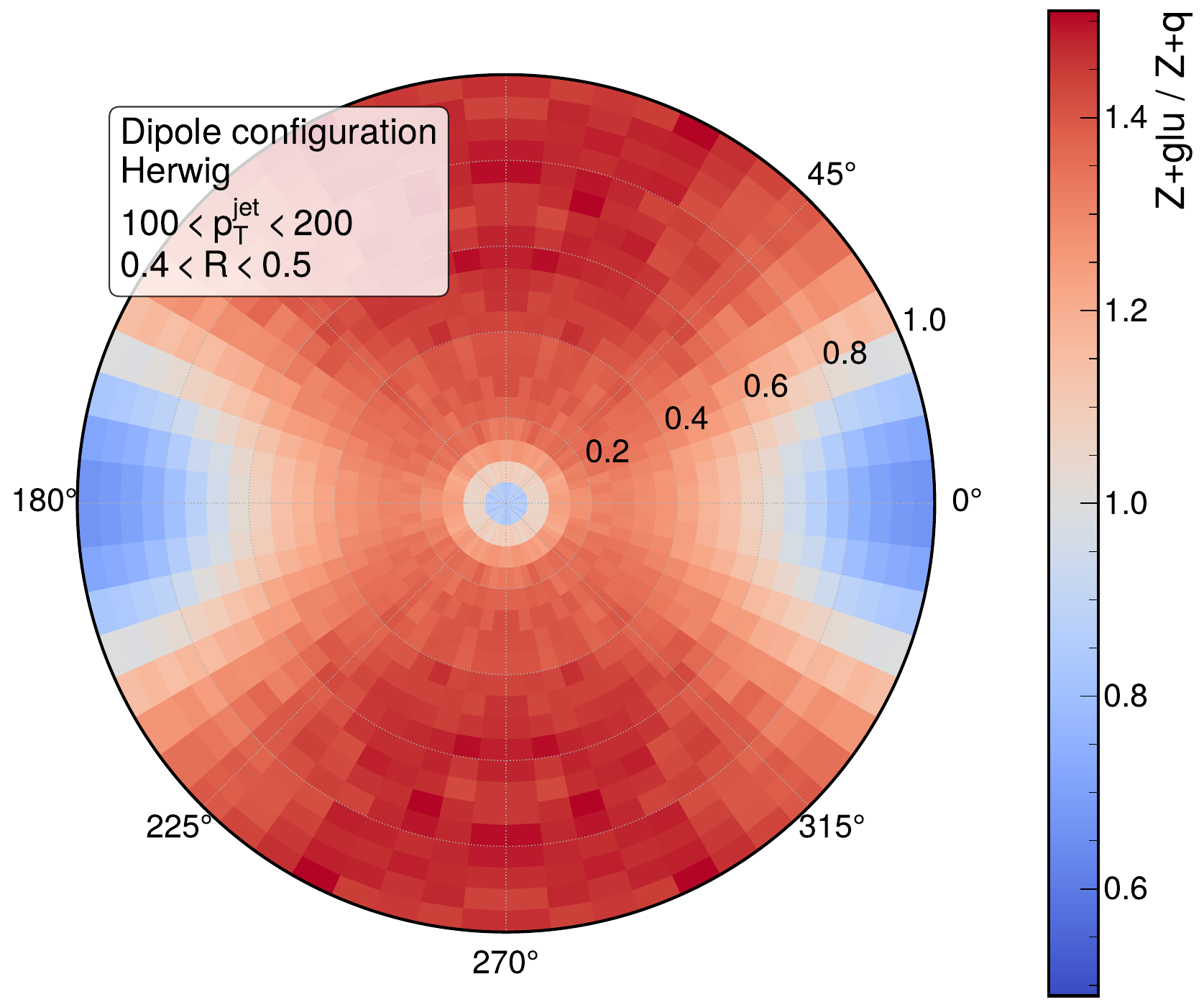} 
    \includegraphics[width=0.4\textwidth]{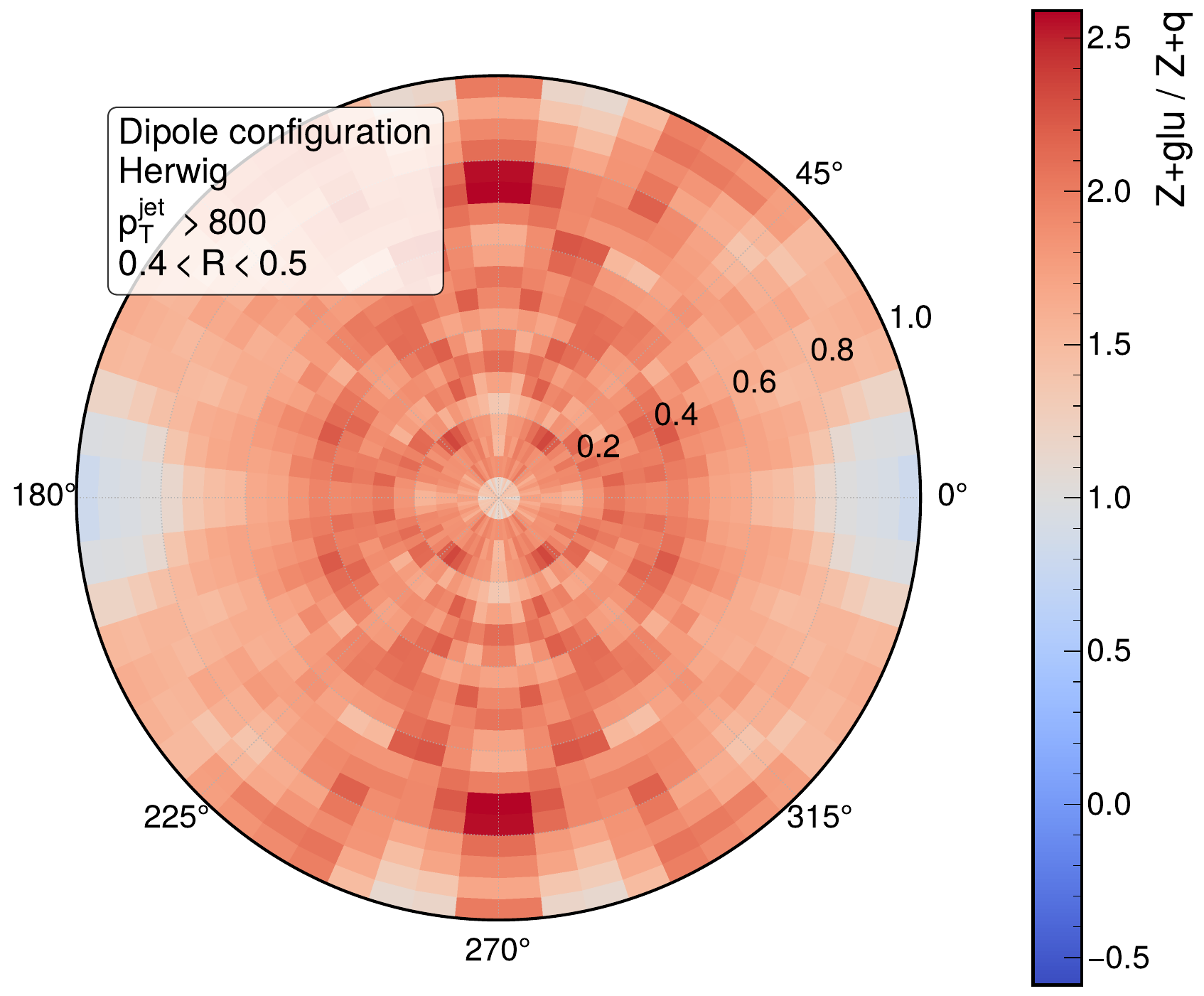} 
    \caption{Ratio between energy correlator distributions in ``the dipole" configuration for gluon-initiated and quark-initiated jets in Pythia (top row) and Herwig (bottom row) simulations. Results are shown for for $100$ GeV $\leq p_T \leq 200$ GeV (left), and $p_T \geq 800$ GeV (right).}
    \label{fig:dipole_pythia_glu_vs_pythia_q}
\end{figure}

\begin{figure}[h]
    \centering
    \includegraphics[width=0.4\textwidth]{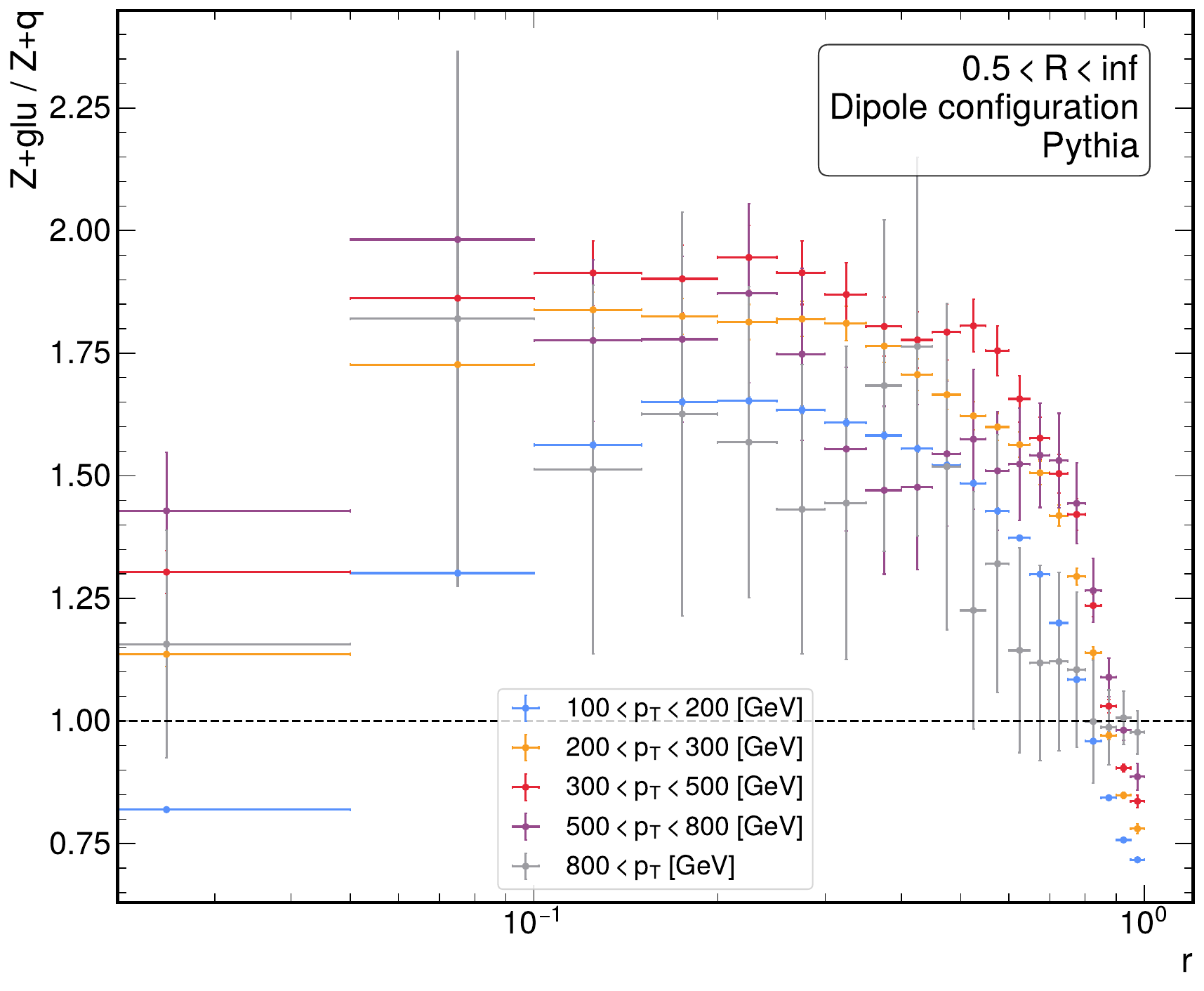} 
    \includegraphics[width=0.4\textwidth]{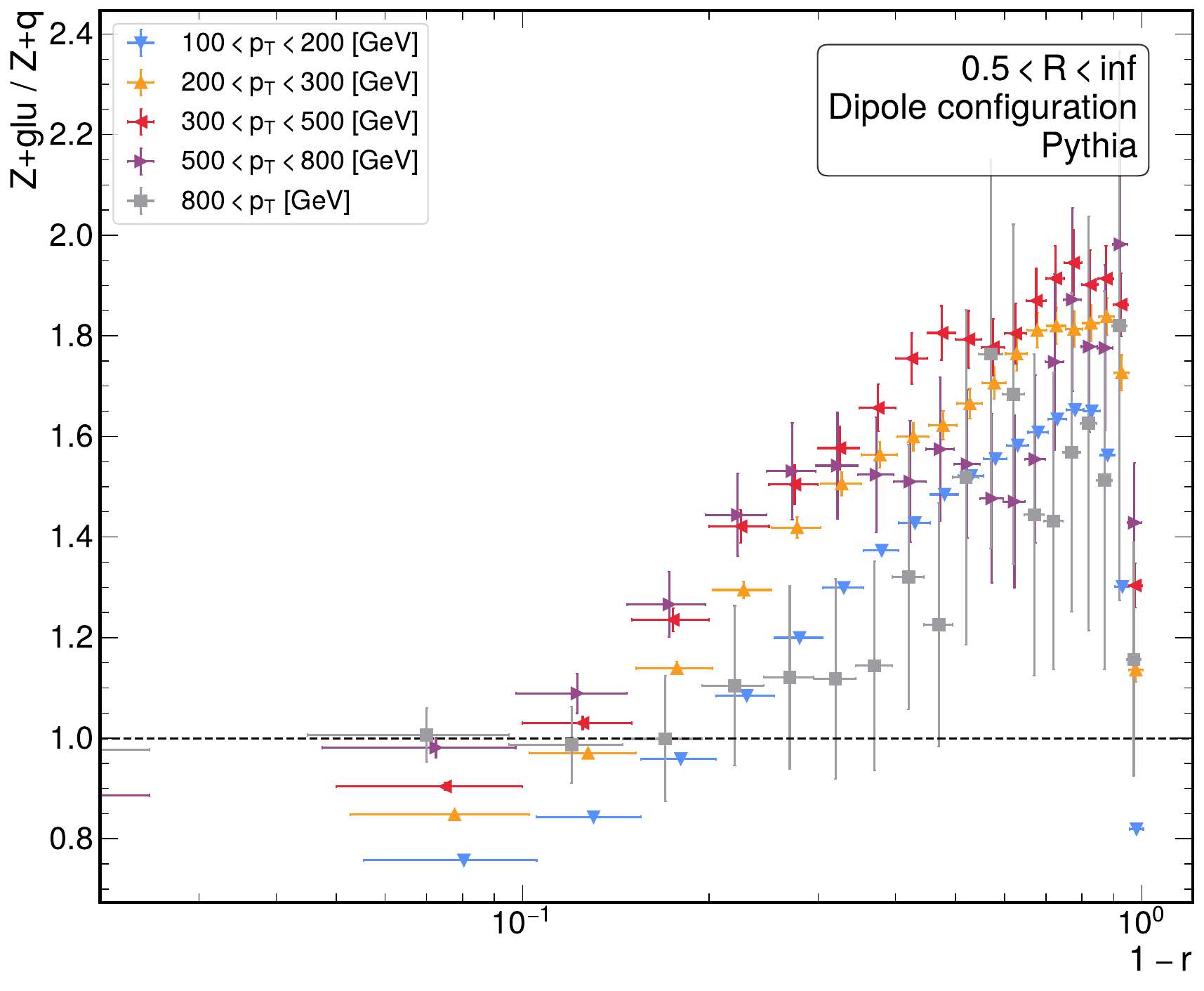} 
    \includegraphics[width=0.4\textwidth]{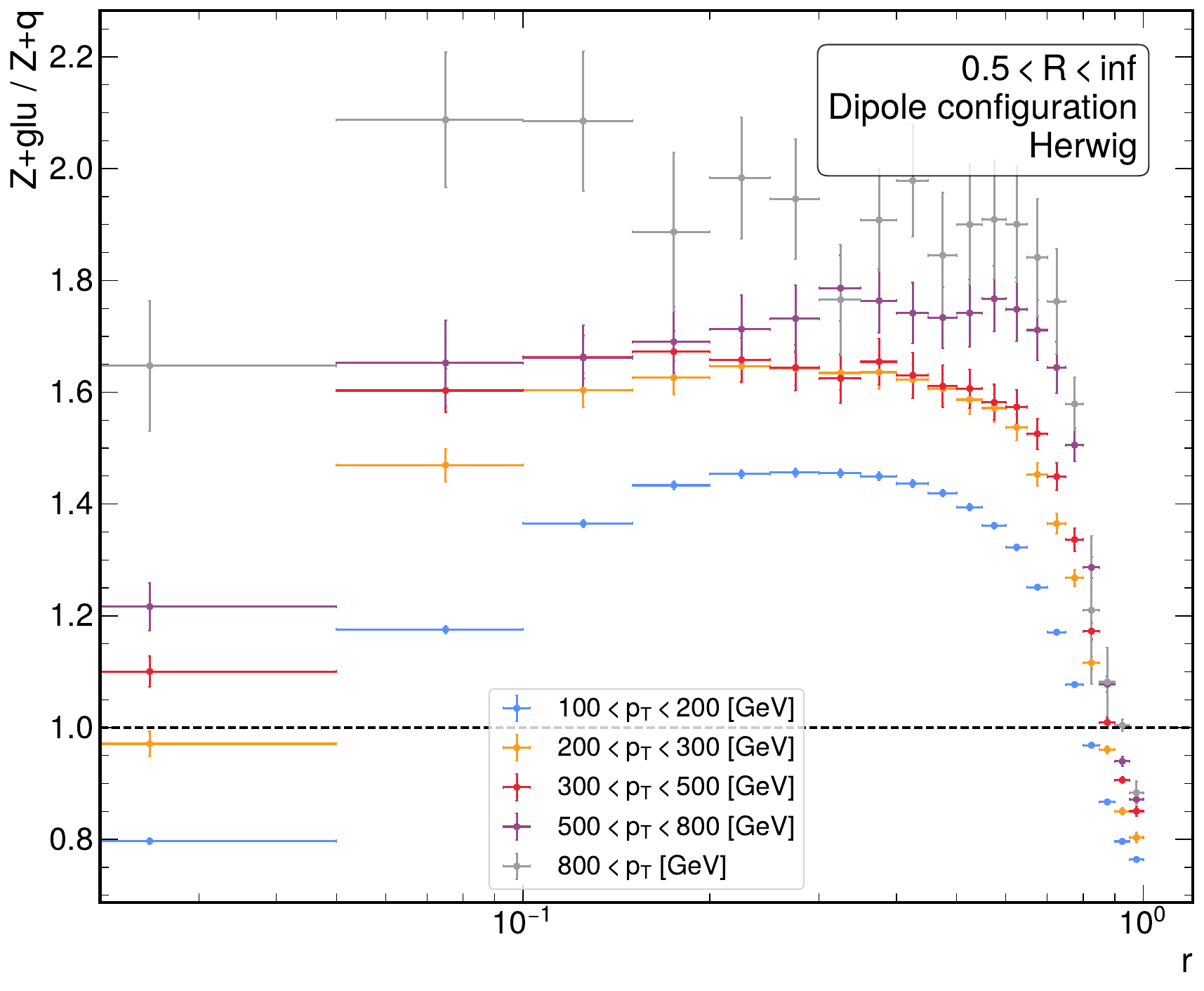} 
    \includegraphics[width=0.4\textwidth]{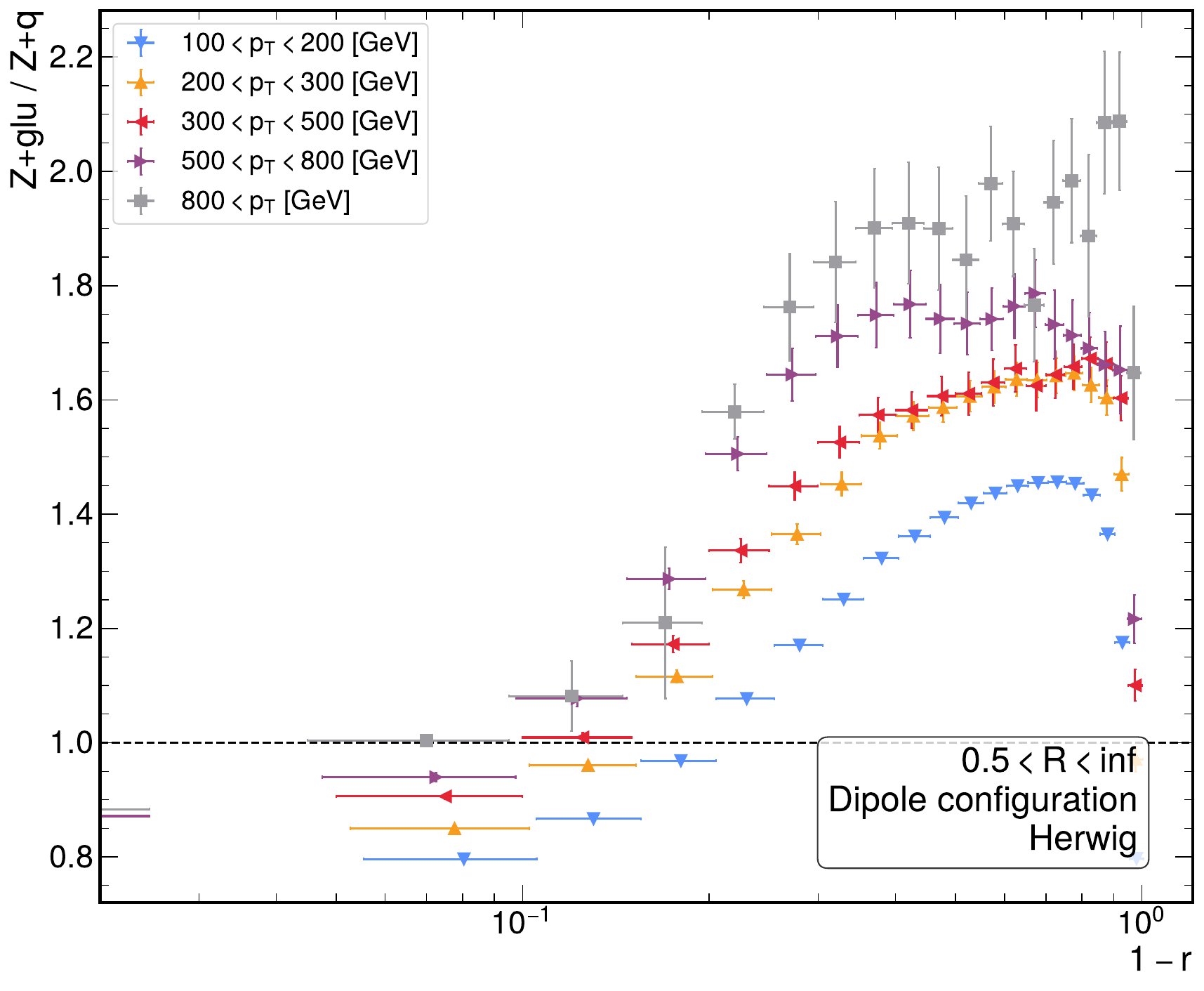} 
    \caption{Ratio between energy correlator distributions in ``the dipole" configuration for gluon-initiated and quark-initiated jets in Pythia (top row) and Herwig (bottom row) simulations, integrated over $\phi$.
    Results are show in terms of both $r$ (left) and $1-r$ (right), to illustrate the different behavior of the two collinear poles.}
\label{fig:dipole_glu_vs_q_radial}
\end{figure}

Quark-initiated and gluon-initiated jets obey different QCD anomalous dimensions, and should have correspondingly different four-point correlator distributions.
Figure~\ref{fig:dipole_pythia_glu_vs_pythia_q} shows the ratio between the simulated distributions for quark-initiated and gluon-initiated jets in Pythia (top row) and Herwig (bottom row).
The main effect is a change in the slope of the collinear poles. This can be studied in more detail by focusing on radial projections. In \Fig{fig:dipole_glu_vs_q_radial}, we show the ratio of angular projections between quark and gluon jets, as simulated in both Pythia and Herwig.

\paragraph{Impact of the parton shower model}

\begin{figure}[h]
    \centering
    \includegraphics[width=0.4\textwidth]{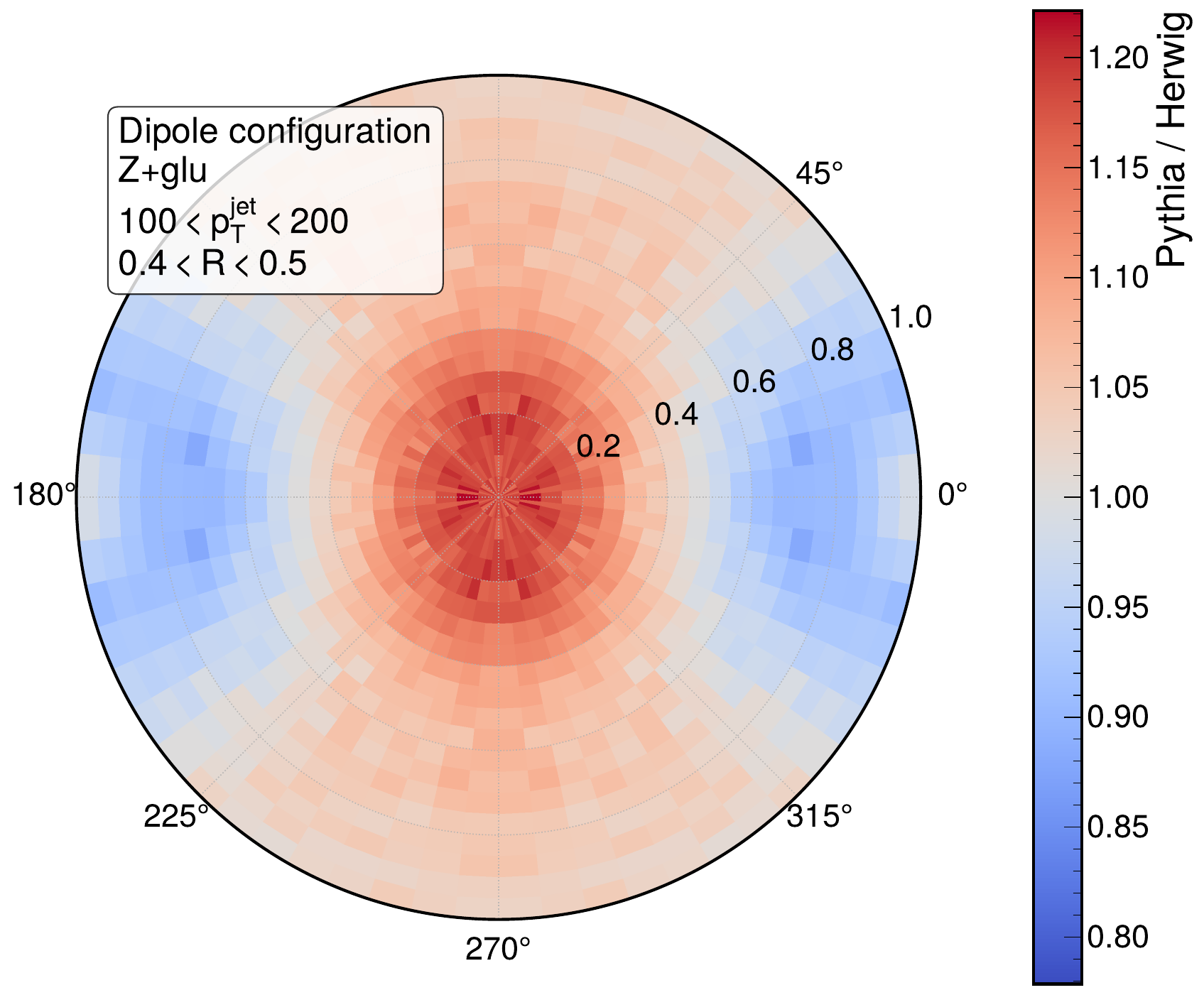} 
    \includegraphics[width=0.4\textwidth]{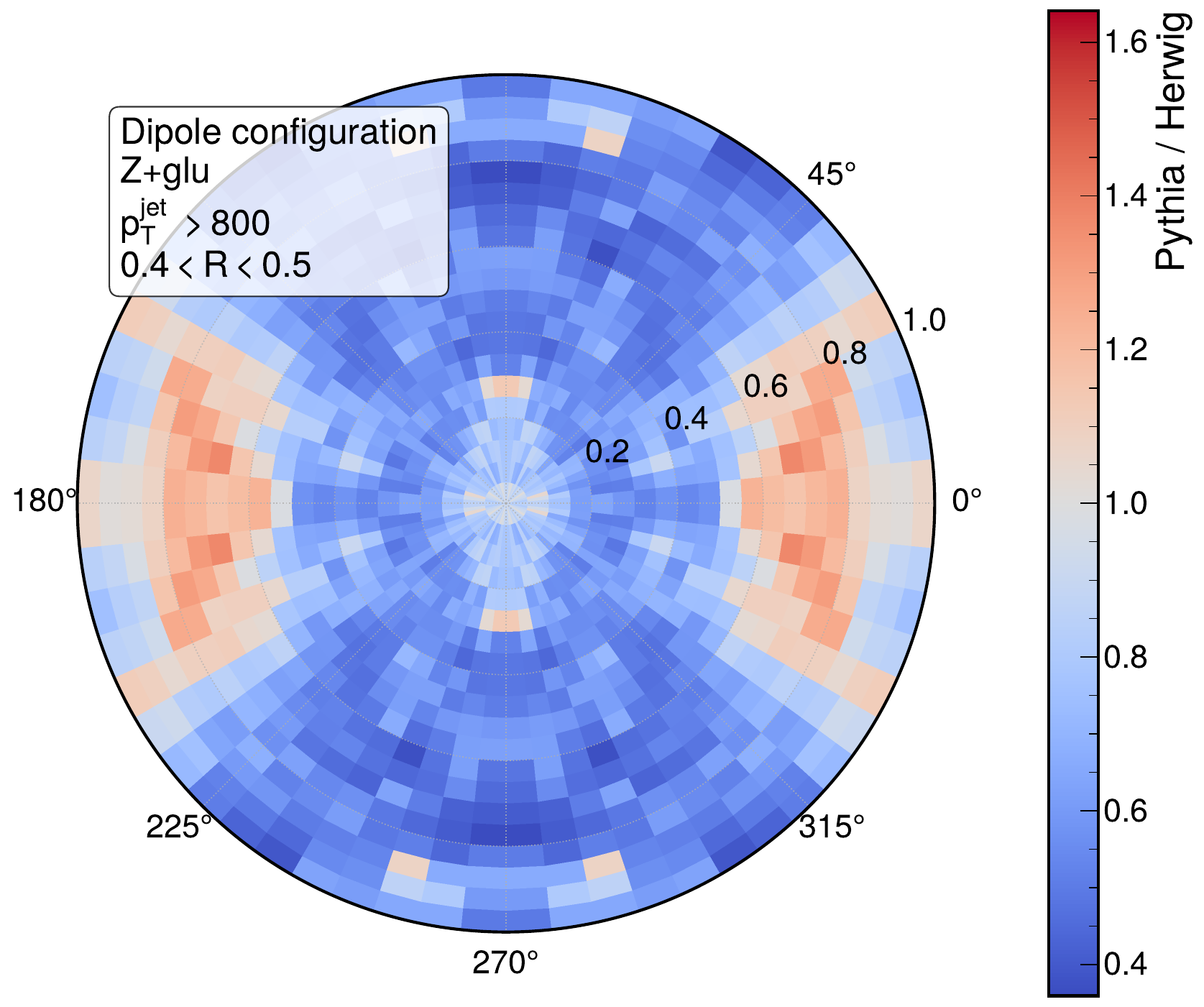} 
        \includegraphics[width=0.4\textwidth]{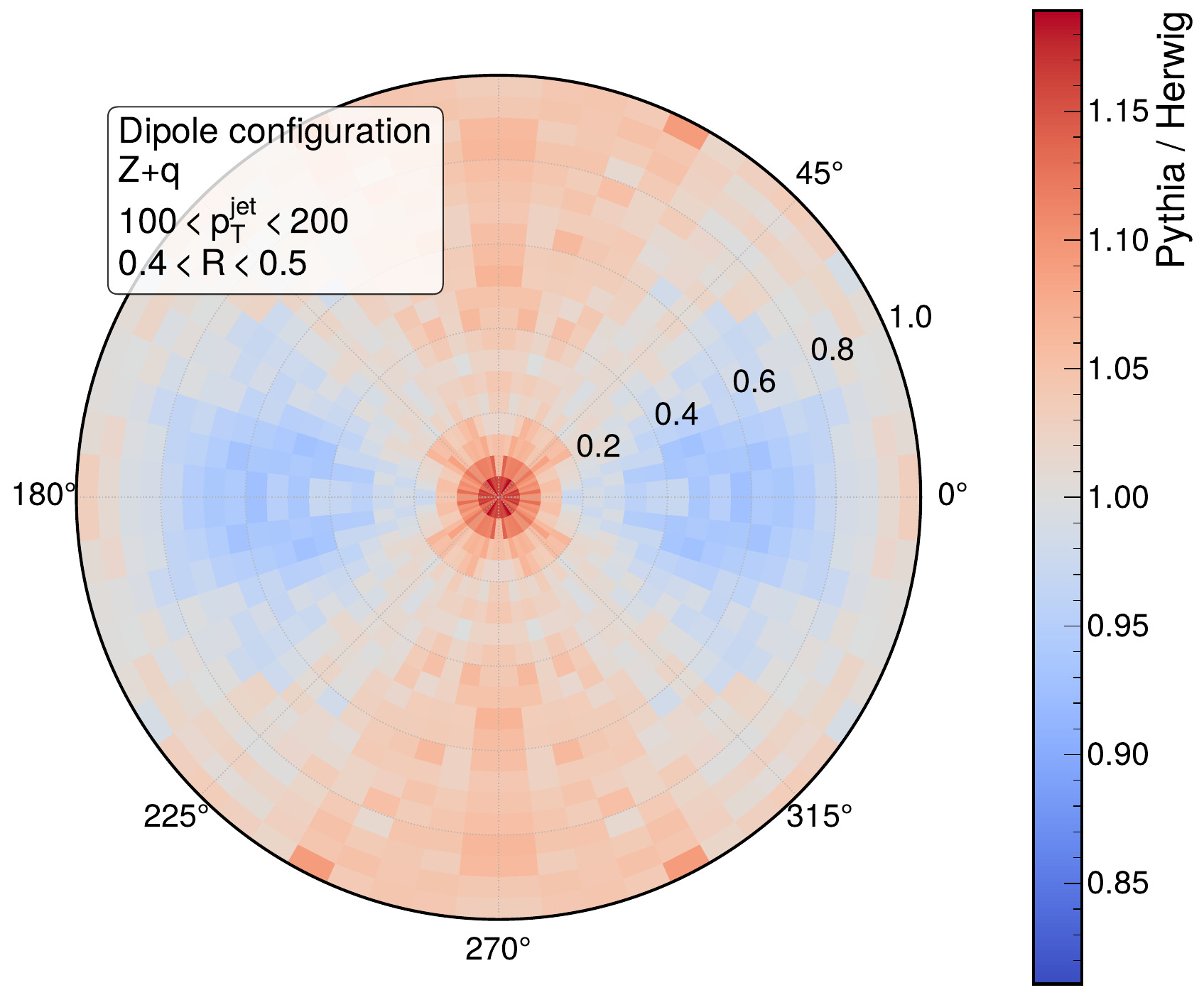} 
    \includegraphics[width=0.4\textwidth]{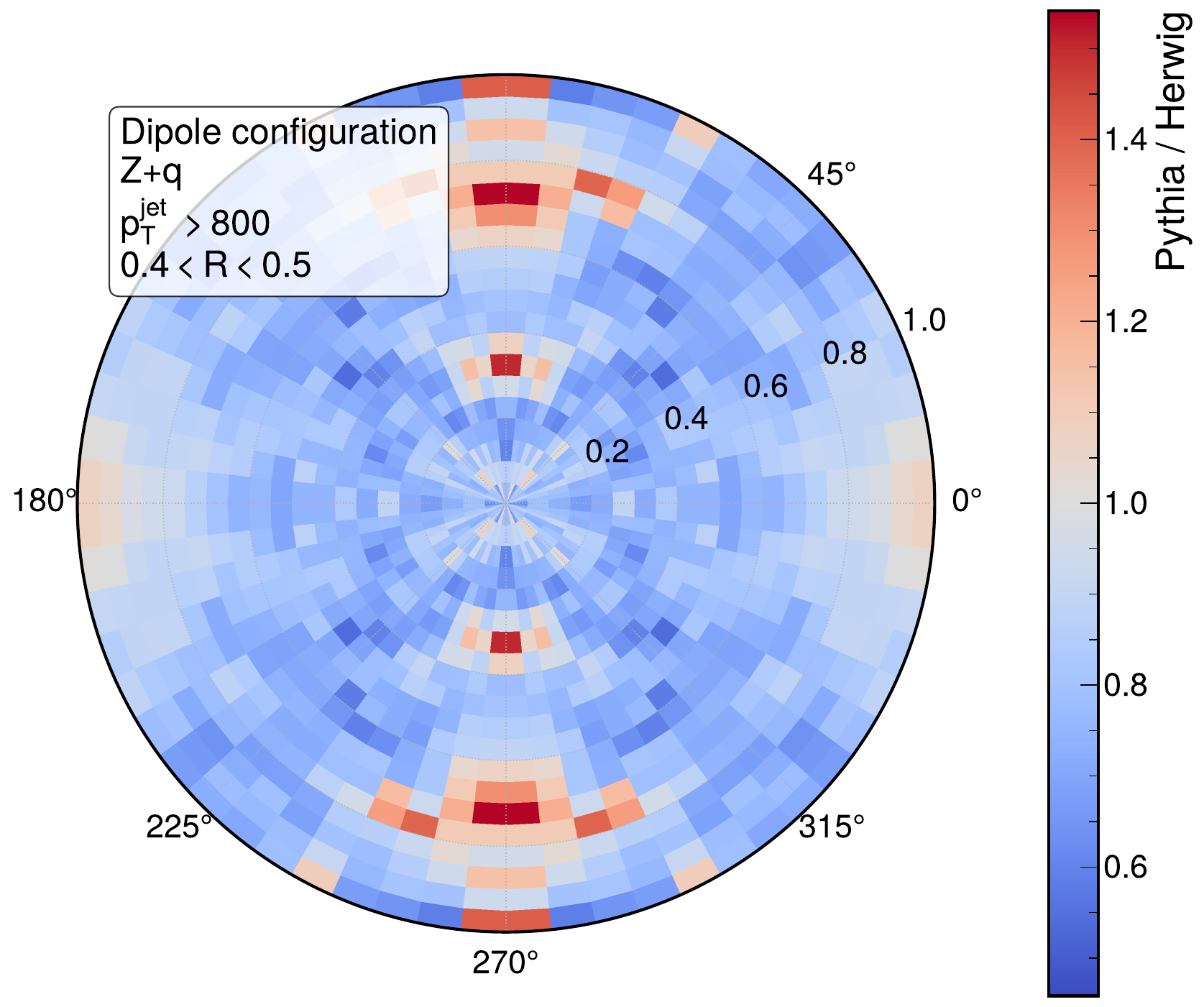} 
    \caption{Ratio between Pythia and Herwig results for  energy correlator distributions in ``the dipole" configuration for $Z+g$ (top) and $Z+q$ (bottom) events. Results are shown for $100$ GeV $\leq p_T \leq 200$ GeV (left), and $p_T \geq 800$ GeV (right).}
    \label{fig:dipole_pythia_glu_vs_herwig_glu}
\end{figure}

\begin{figure}[h]
    \centering
    \includegraphics[width=0.4\textwidth]{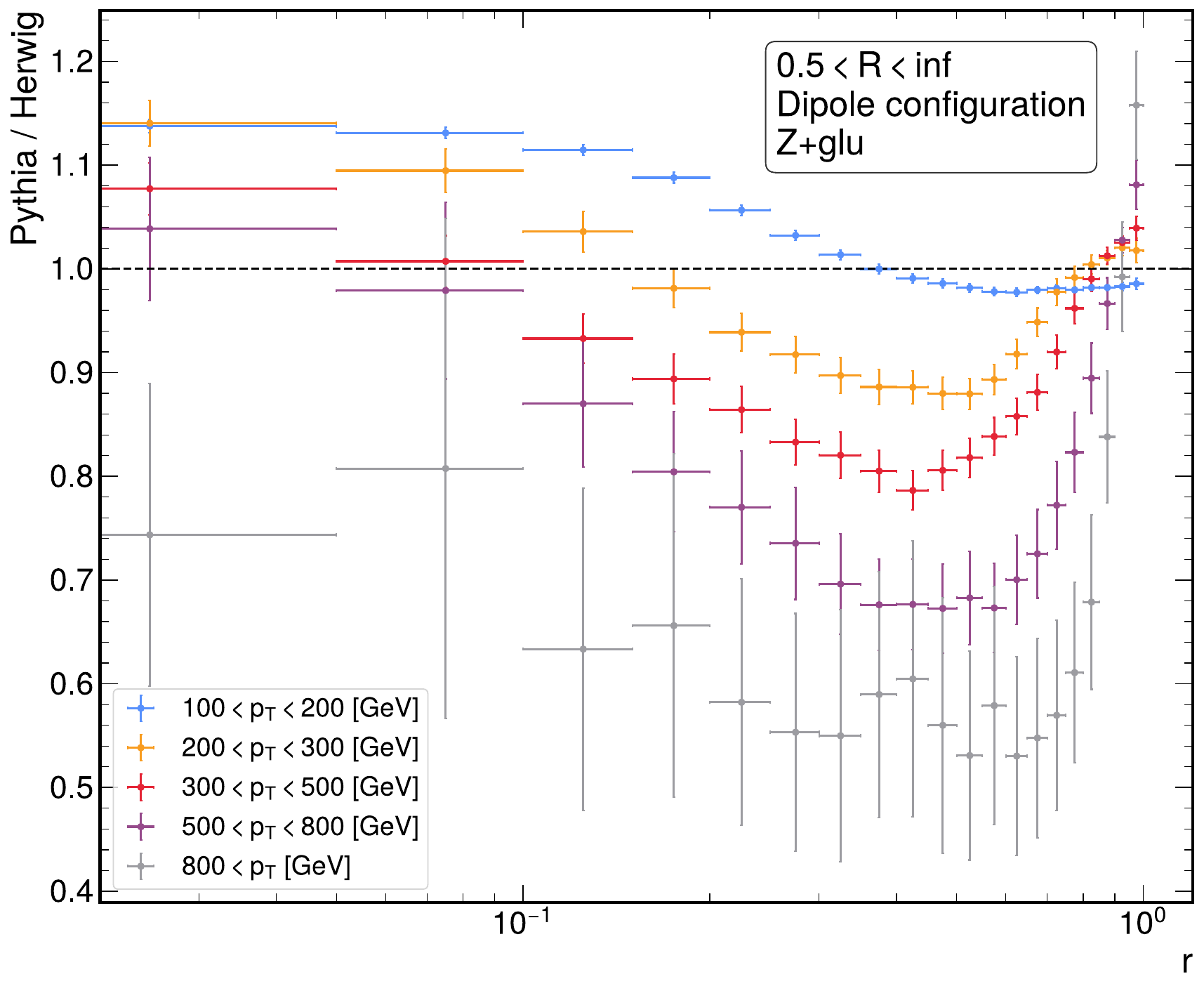} 
    \includegraphics[width=0.4\textwidth]{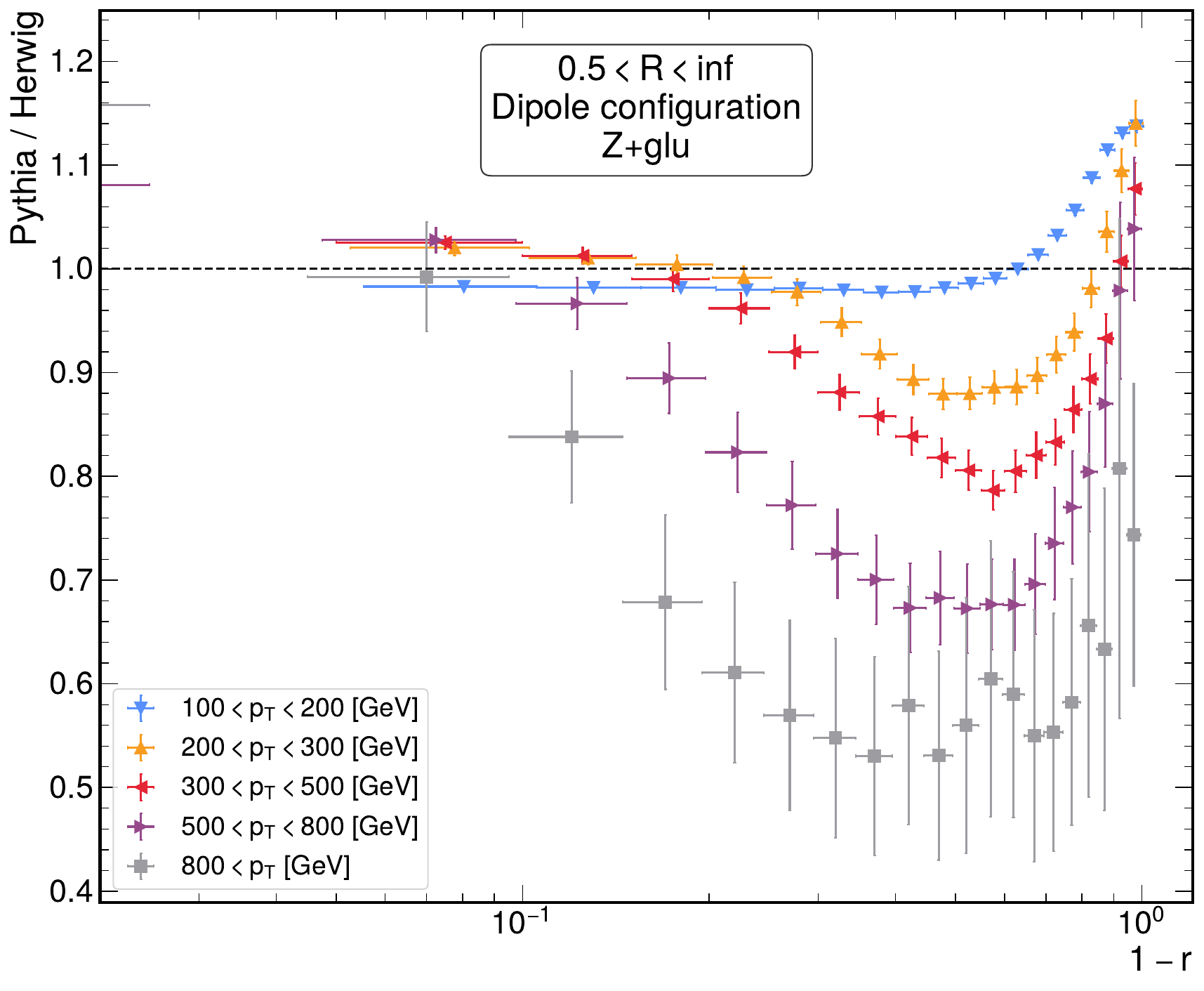} 
    \includegraphics[width=0.4\textwidth]{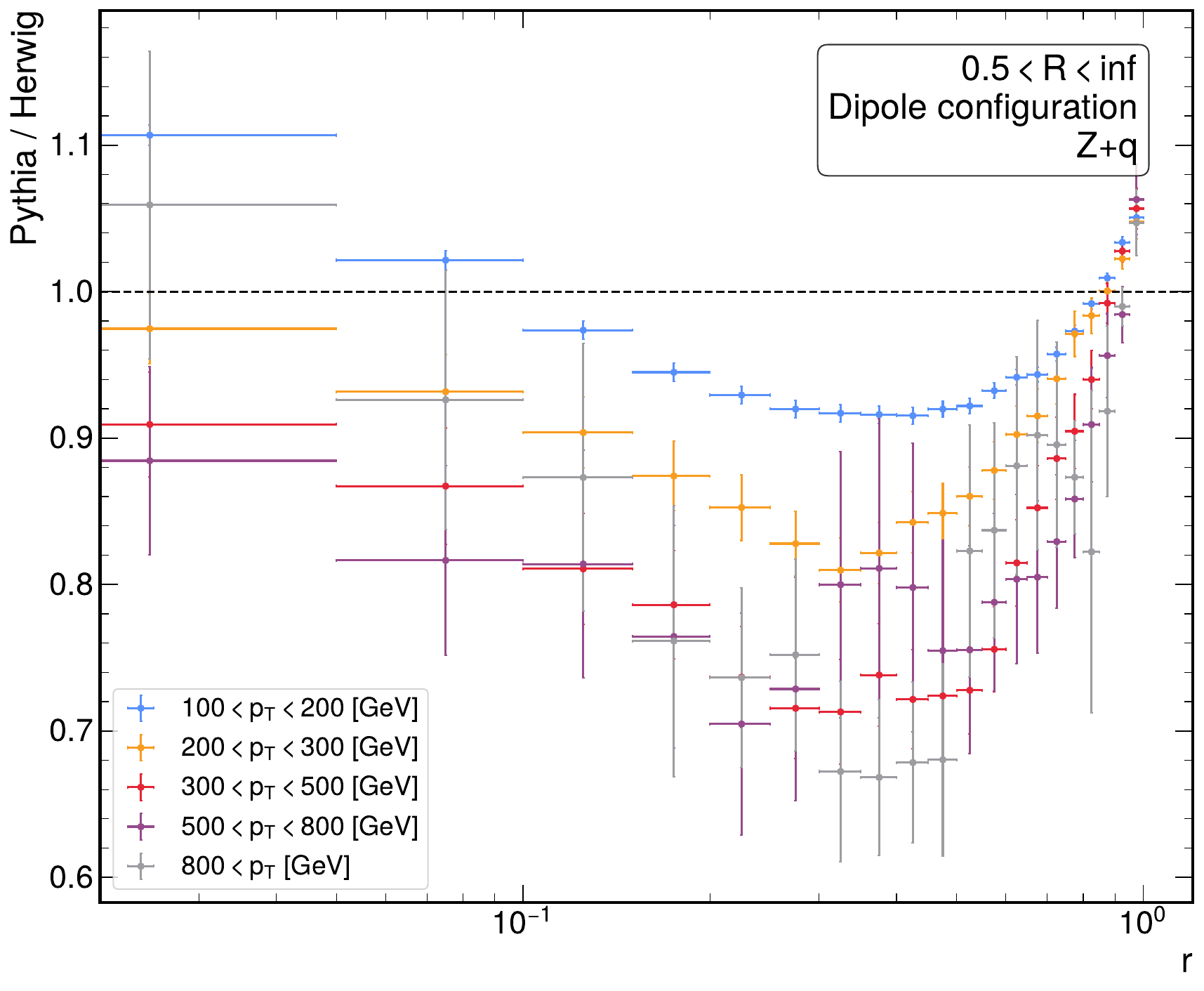} 
    \includegraphics[width=0.4\textwidth]{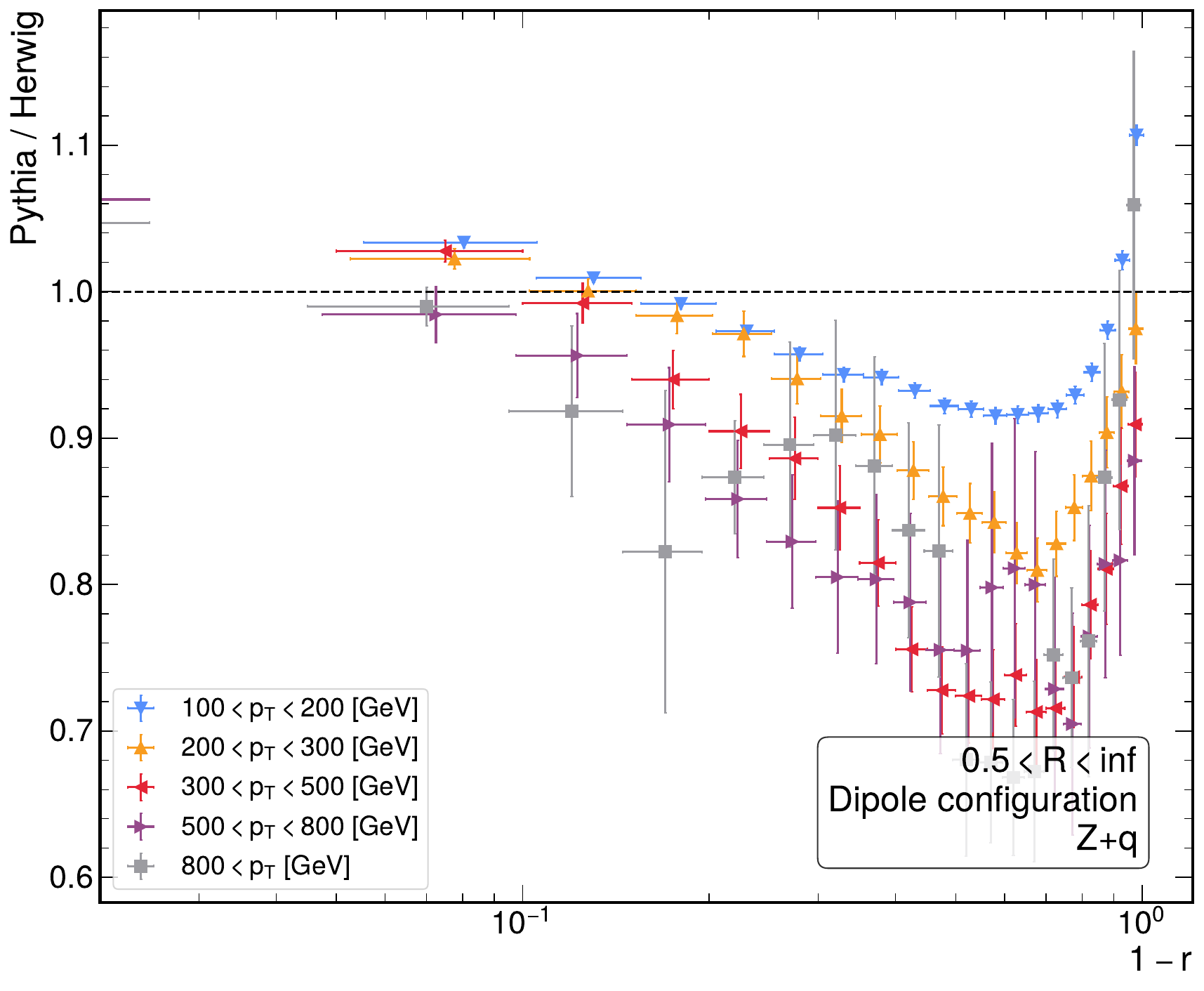} 
    \caption{Ratio between Pythia and Herwig results for  energy correlator distributions in ``the dipole" configuration for $Z+g$ (top) and $Z+q$ (bottom) events, integrated over $\phi$.
    Results are show in terms of both $r$ (left) and $1-r$ (right), to illustrate the different behavior of the two collinear poles.}
    \label{fig:dipole_pythia_vs_herwig_radial}
\end{figure}

Pythia and Herwig use very different parton shower models, and it is therefore interesting to compare the results of the two programs.
Figure~\ref{fig:dipole_pythia_glu_vs_herwig_glu} shows the ratio between the simulated distributions for Pythia and Herwig, for quark-initiated and gluon-initiated jets, respectively.
As can be seen in Figure~\ref{fig:dipole_pythia_vs_herwig_radial}, the two programs mostly agree in the nonperturbative regime (small $r$), and give different slopes in the perturbative regime (large $r$).

\paragraph{Impact of hadronization}

\begin{figure}[h]
    \centering
    \includegraphics[width=0.4\textwidth]{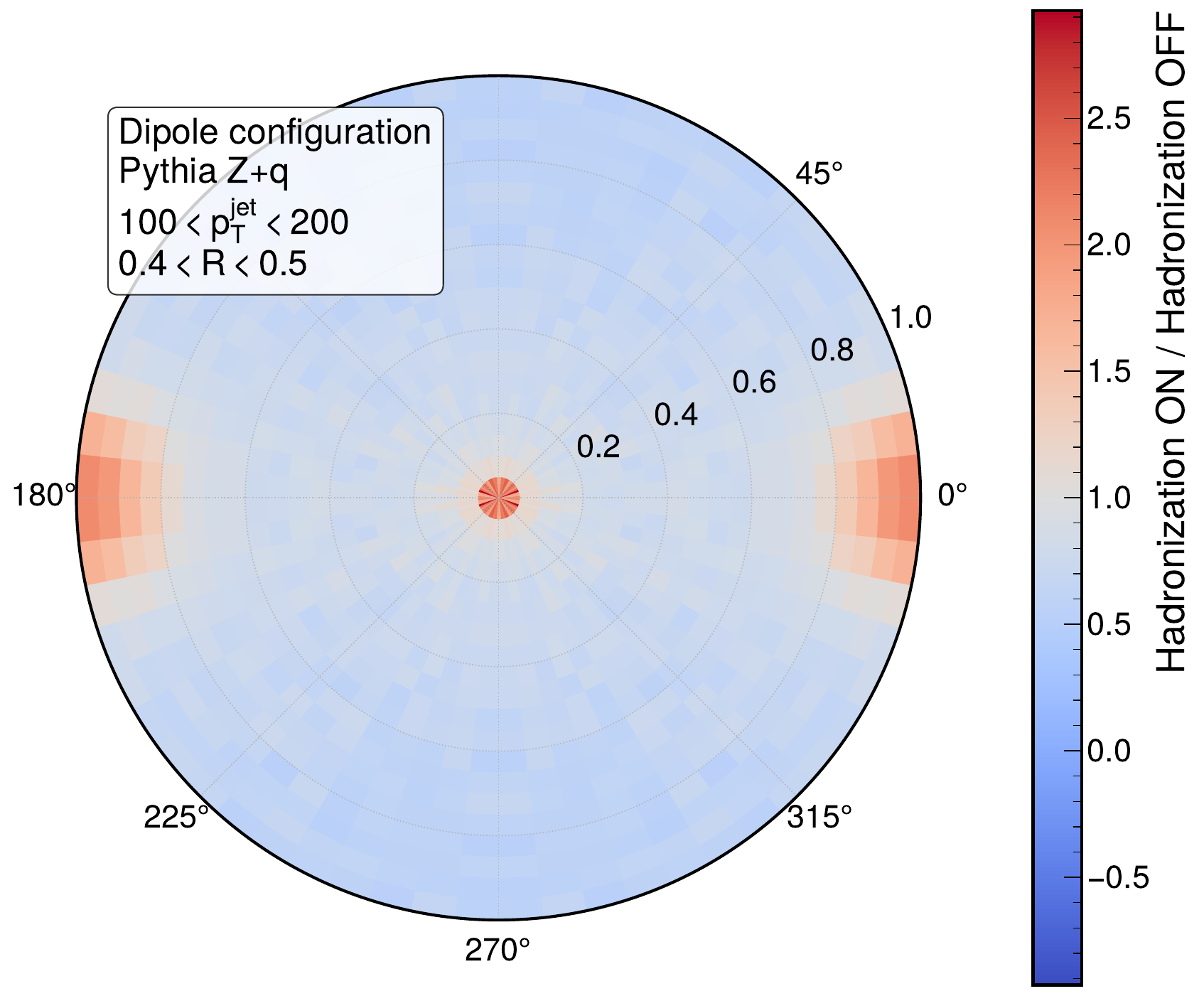} 
    \includegraphics[width=0.4\textwidth]{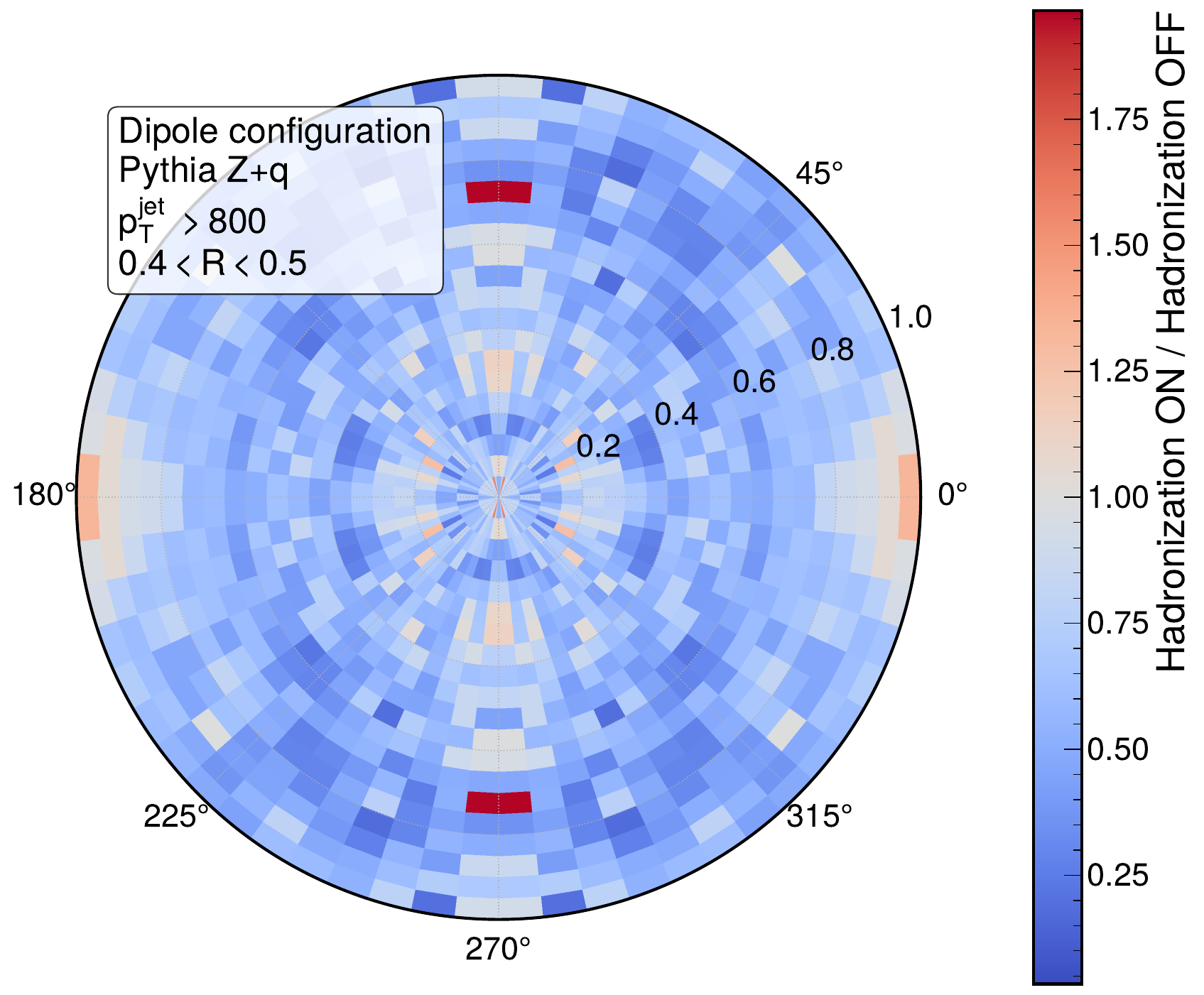} 
    \caption{Ratio between hadronized and parton-level energy correlator distributions in ``the dipole" configuration for Pythia $Z+q$ events. Results are shown for $100$ GeV $\leq p_T \leq 200$ GeV (left), and $p_T \geq 800$ GeV (right).}
    \label{fig:dipole_pythia_q_nom_vs_nohad}
\end{figure}

The Pythia parton shower allows us to turn off hadronization and compute the four-point correlators at the parton level. Figure~\ref{fig:dipole_pythia_q_nom_vs_nohad} shows the ratio between the hadronized and parton-level distributions for the dipole configuration in Pythia $Z+q$ events.
The ratio is approximately flat far from the collinear poles, while hadronization has a large effect near $r\rightarrow 0$ and $r\rightarrow 1, \phi \rightarrow 0,\pi$, as expected. Overall, the effects are quite mild, illustrating that for kinematic configurations accessible at the LHC, measurements will be able to probe the perturbative structure of the four-point correlator.

\subsubsection{The Tee Configuration} \label{sec:proj_b_pheno}

Next, we consider the tee configuration. For convenience, we recall that the tee configuration probes the factorization of the four-point splitting function into a $1\to 2$ splitting tensor, and a $1\to 3$ splitting function
\begin{align}
\fd{4cm}{figures/tee_solo}\,.
\end{align}
For fixed $R$, it is defined in terms of the variable $r$, which controls the relative size of the dipoles, and hence the factorization, and the variable $\phi$, which allows the study of potential spin correlations. We once again study discrepancies between the $r$ and $\phi$ dependence of the different parton showers.

\begin{figure}[h]
    \centering
    \includegraphics[width=0.4\textwidth]{figures/MCstudy/tee/herwig_q/histogram_shapes_radial2D_pt1_R1.pdf} 
    \includegraphics[width=0.4\textwidth]{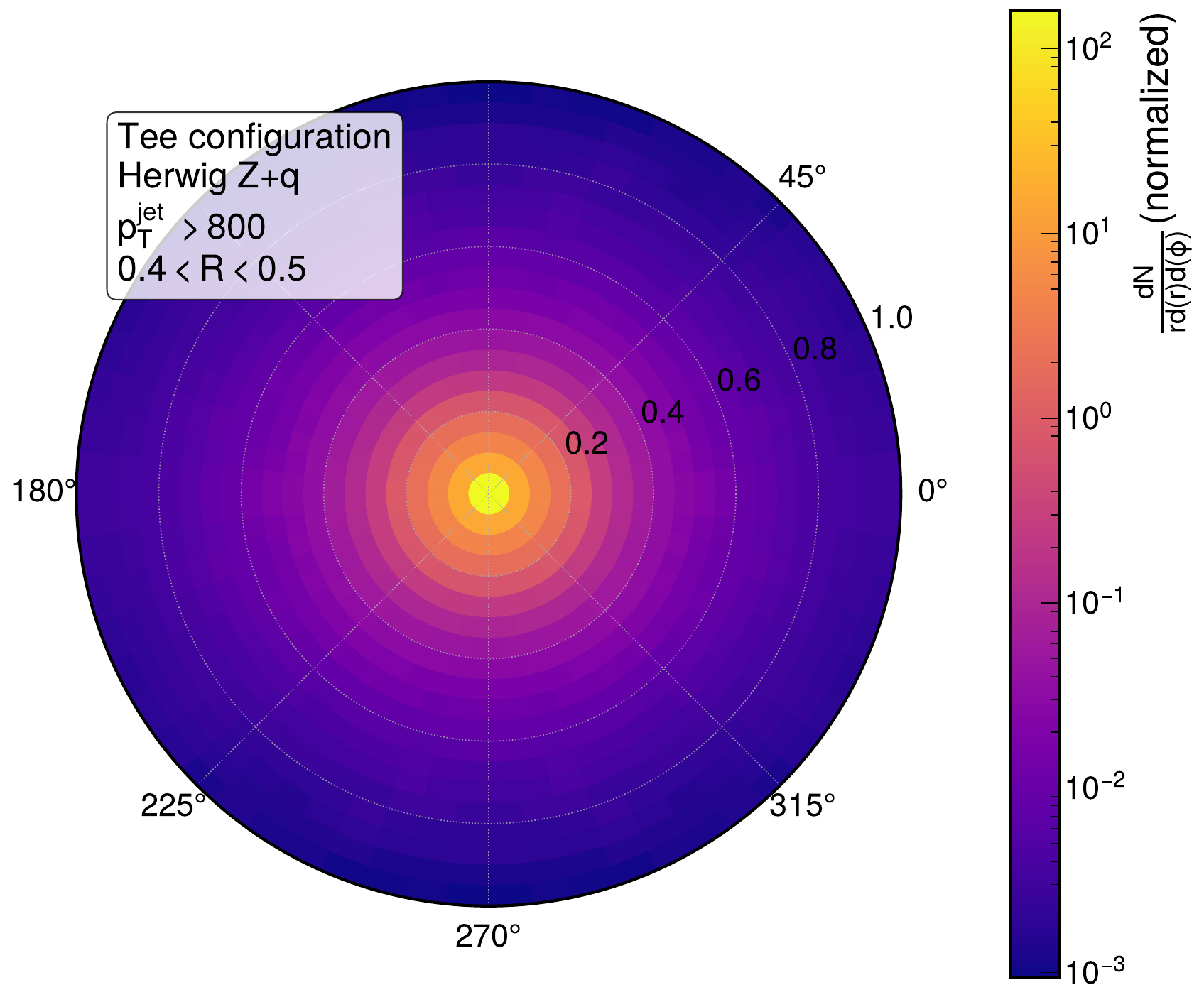} 
    \caption{Simulated results for ``the tee" configuration of the four-point correlator in Herwig $Z+q$ events. Results are shown for $100$ GeV $\leq p_T \leq 200$ GeV (left), and $p_T \geq 800$ GeV (right).}
    \label{fig:tee_herwig_q}
\end{figure}

We first begin with a baseline plot to illustrate the general structure of the tee configuration of the four-point correlator.
Figure~\ref{fig:tee_herwig_q} shows the simulated results for the tee configuration in Herwig $Z+q$ events, for $100$ GeV $\leq p_T \leq 200$ GeV (left), and $p_T \geq 800$ GeV (right). The distribution is highly dominated by the collinear pole at $r\rightarrow 0$, which becomes narrower with increasing jet $p_T$, as the distribution becomes increasingly perturbative.

\begin{figure}[h]
    \centering
    \includegraphics[width=0.45\textwidth]{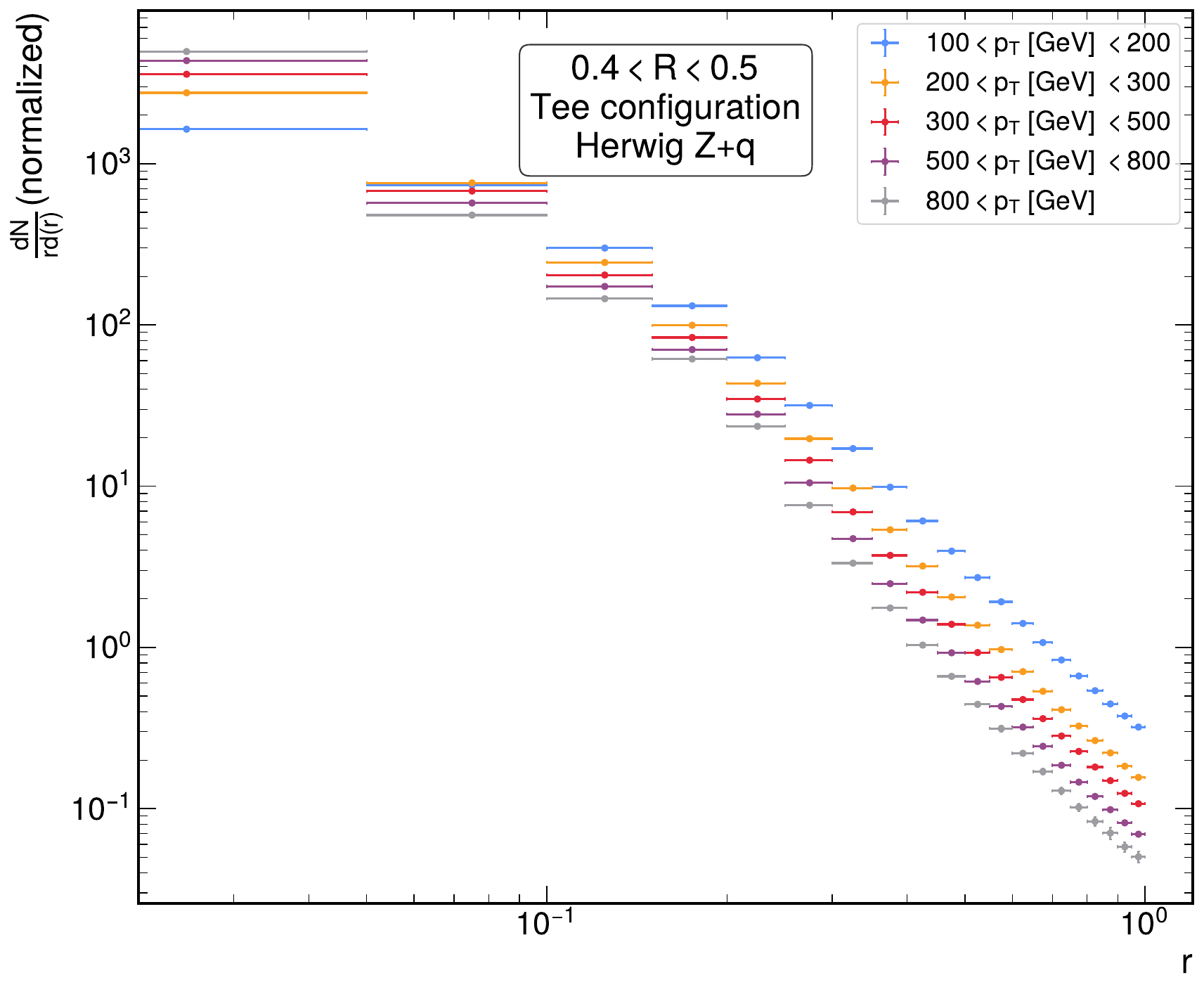} 
    \caption{Radial profile of the tee configuration in Herwig $Z+q$ events, measured by integrating over $\phi$. 
    The slope of the distribution as a function of $r$ is a power law in the perturbative regime, with a slope that is sensitive to the running of the strong coupling.
    At small $r$ the power law scaling breaks due to non-perturbative confinement effects. }
    \label{fig:tee_radial_herwig_q}
\end{figure}

An interesting feature of the tee configuration is that we can directly focus on the scaling in the $r$ variable for fixed $R$. The scaling of the three-point correlator was first studied in \cite{Komiske:2022enw}, and is predicted to be a power law. Factorization predicts that this scaling is universal, and for small $r$ should be the same \emph{within} the four-point correlator, as for the three-point correlator itself. This will be an important test of implementations of three-point splitting functions in parton showers.  This scaling is illustrated in Figure \ref{fig:tee_radial_herwig_q}, which shows the radial profile of the tee distribution, integrated over $\phi$.
At large $r$ the distribution is dominated by a power law scaling, as predicted by the perturbative calculations of Section~\ref{sec:proj}.
This scaling breaks down at small $r$ due to non-perturbative confinement effects.  We see that for LHC accessible energies, there is a reasonably wide window where perturbative scaling should be observable. In the perturbative scaling region, we observe
\begin{align}
\frac{d\Sigma}{r\,dr} \sim r^{-4+\delta}\,,
\end{align}
where $0<\delta<1$ arises from anomalous scaling. The classical scaling exponent agrees exactly with that of the $M=3$ case in Eq.~\eqref{eq:resultForm}.

\begin{figure}[h]
    \centering
    \includegraphics[width=0.4\textwidth]{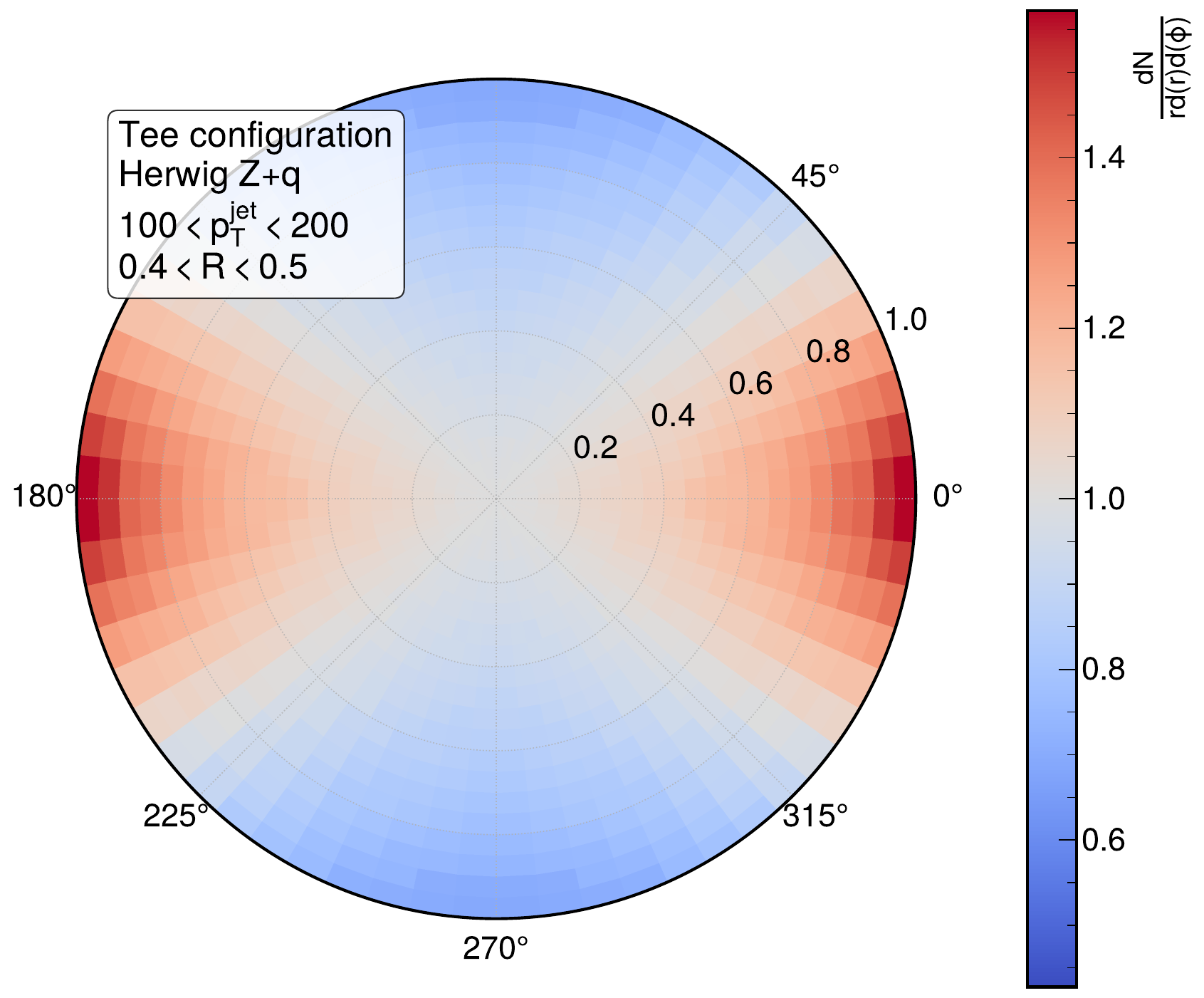} 
    \includegraphics[width=0.4\textwidth]{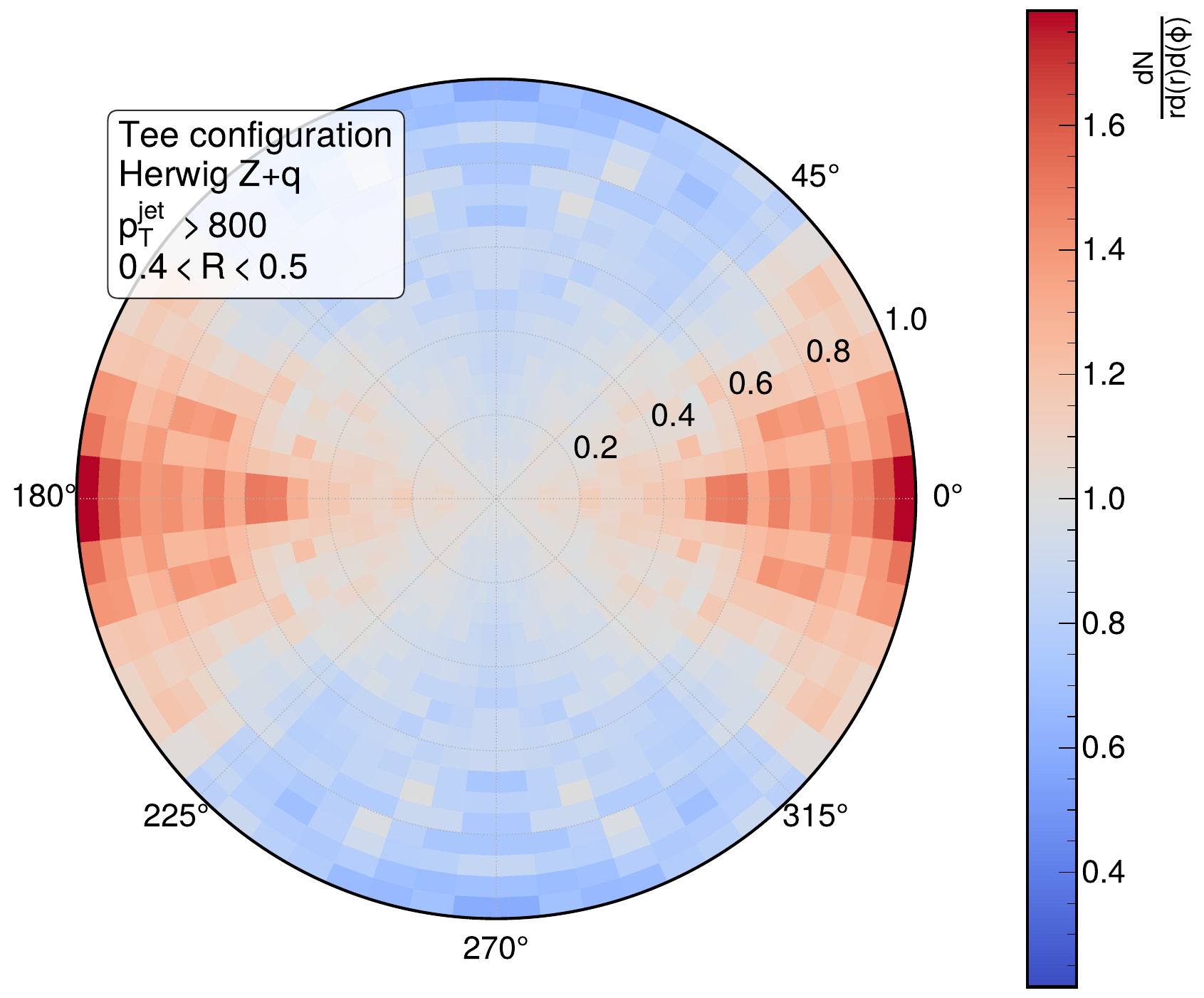} 
    \caption{The simulated results for the tee configuration in Herwig $Z+q$ events from Figure~\ref{fig:tee_herwig_q}, with the $r$-dependence factorized out to reveal any angular modulations. Results are shown for $100$ GeV $\leq p_T \leq 200$ GeV (left), and $p_T \geq 800$ GeV (right).}
    \label{fig:tee_angular2D_herwig_q}
\end{figure}

\begin{figure}[h]
    \centering
    \includegraphics[width=0.4\textwidth]{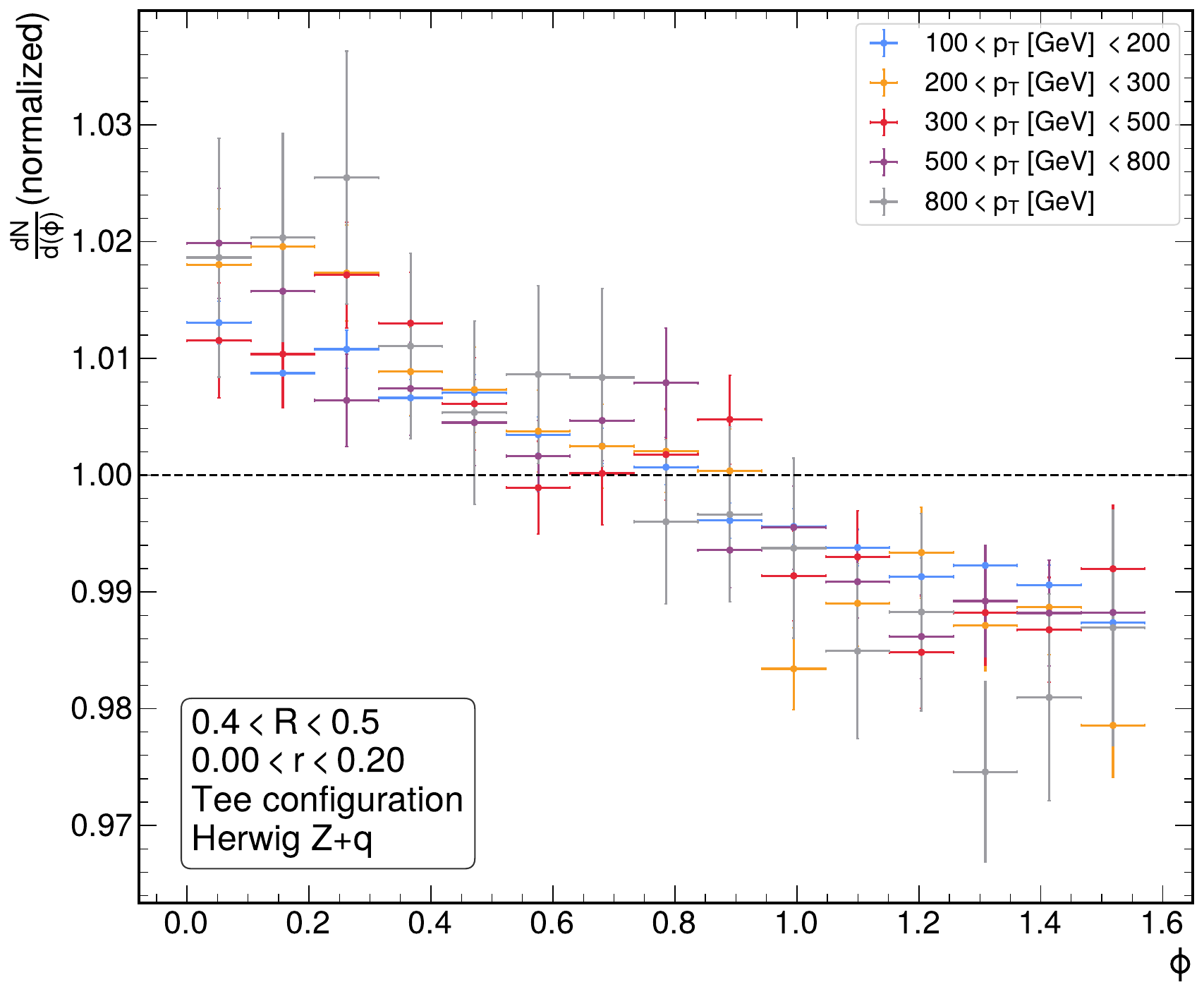} 
    \includegraphics[width=0.4\textwidth]{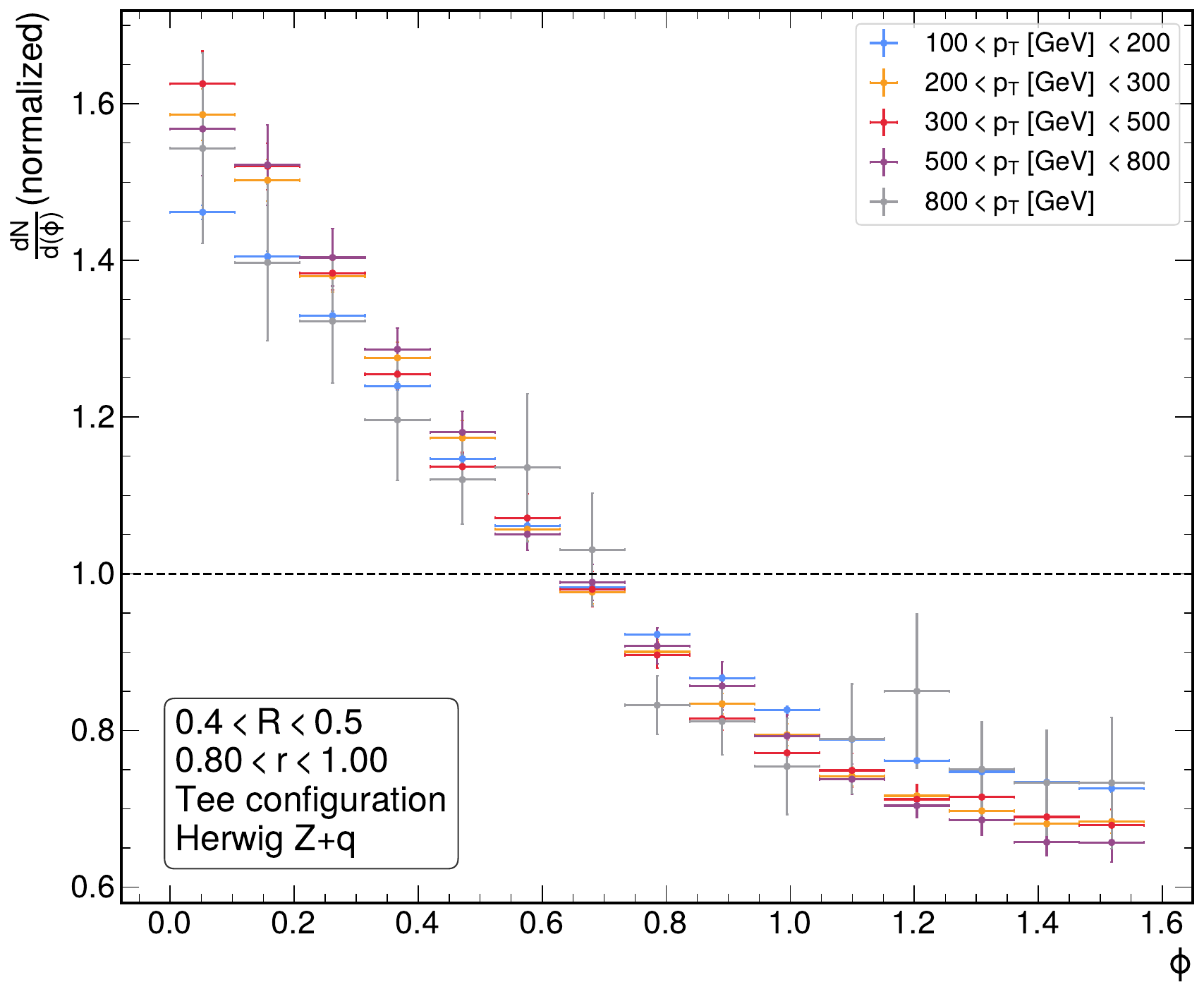} 
    \caption{Angular modulation of the tee configuration in Herwig $Z+q$ events, for two slices of $r$: $0\leq r \leq 0.2$ and $0.8\leq r \leq 1$.}
    \label{fig:tee_angular1D_herwig_q}
\end{figure}

The collinear pole at $r\rightarrow 0$ dominates the distribution, obscuring any azimuthal dependence that may be present. 
We isolate the azimuthal dependence by factoring out the collinear pole, dividing each bin by the integral over $\phi$ at that $r$.
The resulting angular dependence is shown in Figure~\ref{fig:tee_angular2D_herwig_q}. 
The simulated distributions are schematically similar to the analytic calculations of Section~\ref{sec:proj}, with an approximate $\cos(2\phi)$ modulation.
However, the simulated distributions are different from the perturbative calculation in two important ways which suggest that these effects are kinematic power corrections and not spin correlations. 
First, the modulation in the simulation is much larger ($\sim 40\%$) than in the calculation ($\sim 4\%$).
Second, the calculation predicts that the magnitude of the modulation should be independent of $r$, while the simulation shows a strong $r$-dependence, with the modulation being largest at large $r$. 
This can be seen most clearly by comparing the angular distributions in different $r$ bins, as shown in Figure~\ref{fig:tee_angular1D_herwig_q}.

This strongly suggests that for the region of $r$ that is perturbative, the modulation in $\phi$ is dominated by kinematic power corrections. We will provide  further evidence for this shortly, when we study the dependence on the inclusion of spin correlations in the shower. Parton showers do not currently capture the full $1\to 4$ splitting function, and therefore most likely do not reliably compute this distribution. Its large magnitude suggests that until it is properly under control, it will be difficult to make any statement about spin correlations in this configuration. This strongly motivates the complete calculation of the four-point correlator in QCD, which will resolve this issue and provide an excellent benchmark for parton showers.

\paragraph{Quark vs gluon jets}

\begin{figure}[h]
    \centering
    \includegraphics[width=0.4\textwidth]{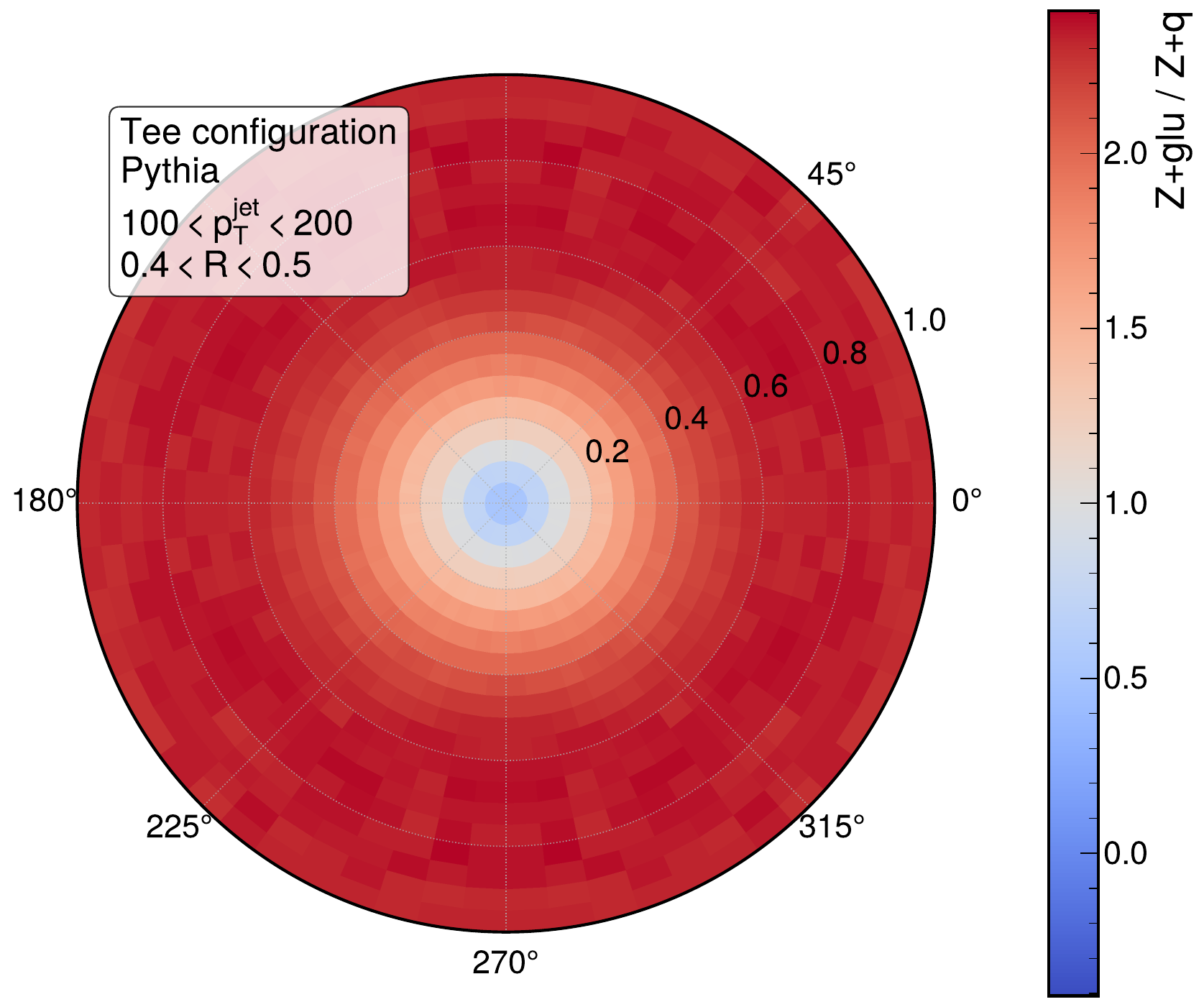} 
    \includegraphics[width=0.4\textwidth]{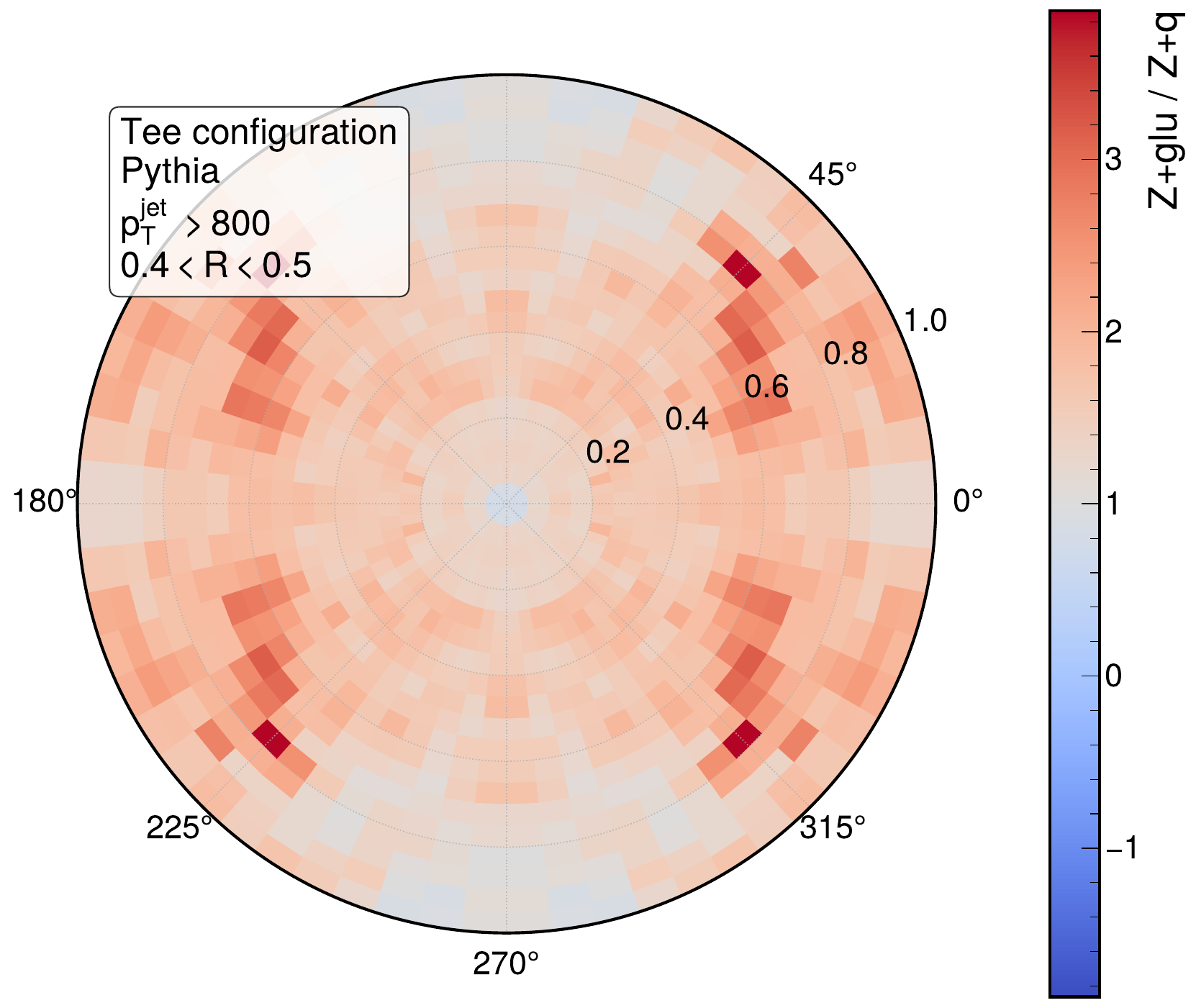} 
    \includegraphics[width=0.4\textwidth]{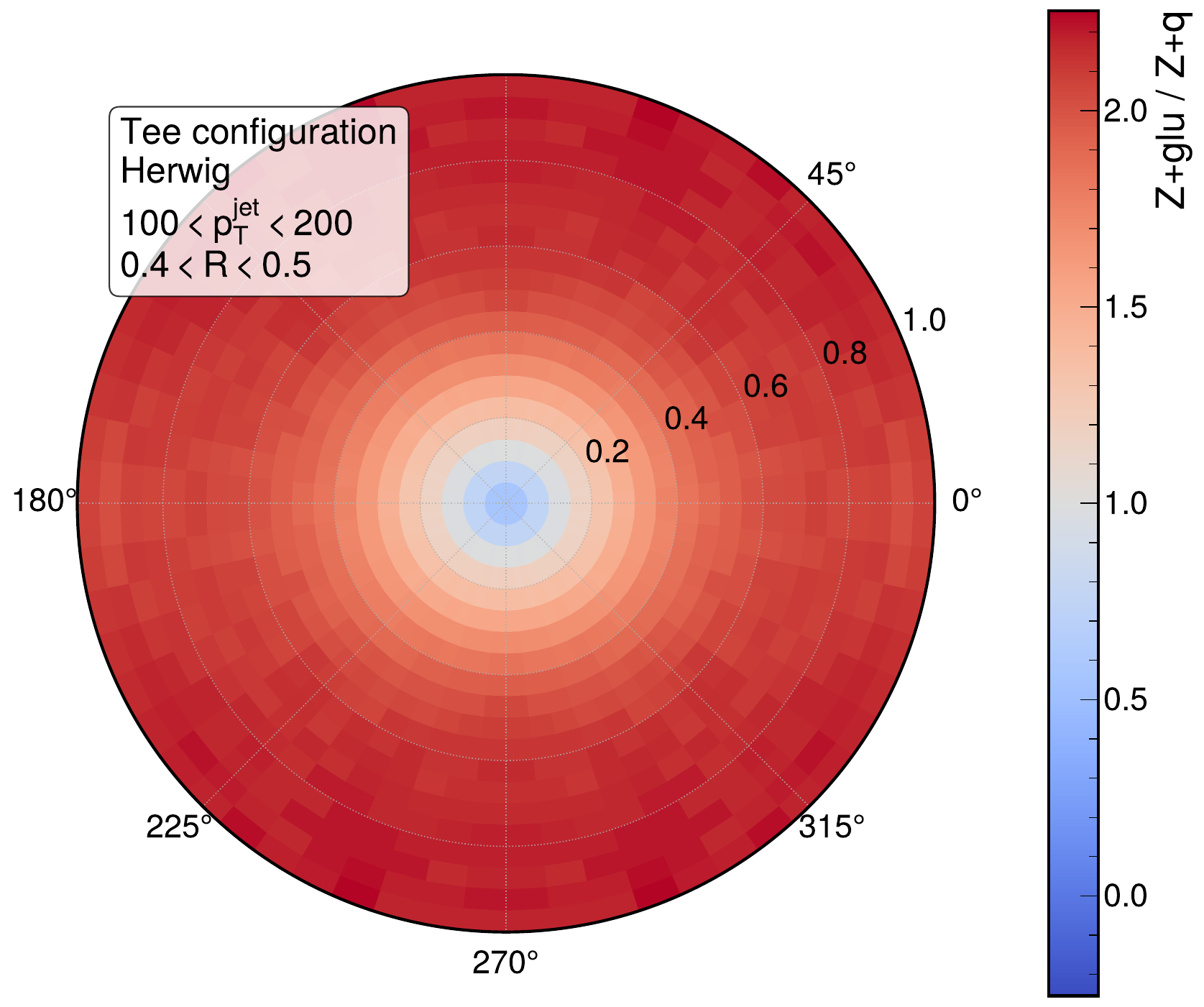} 
    \includegraphics[width=0.4\textwidth]{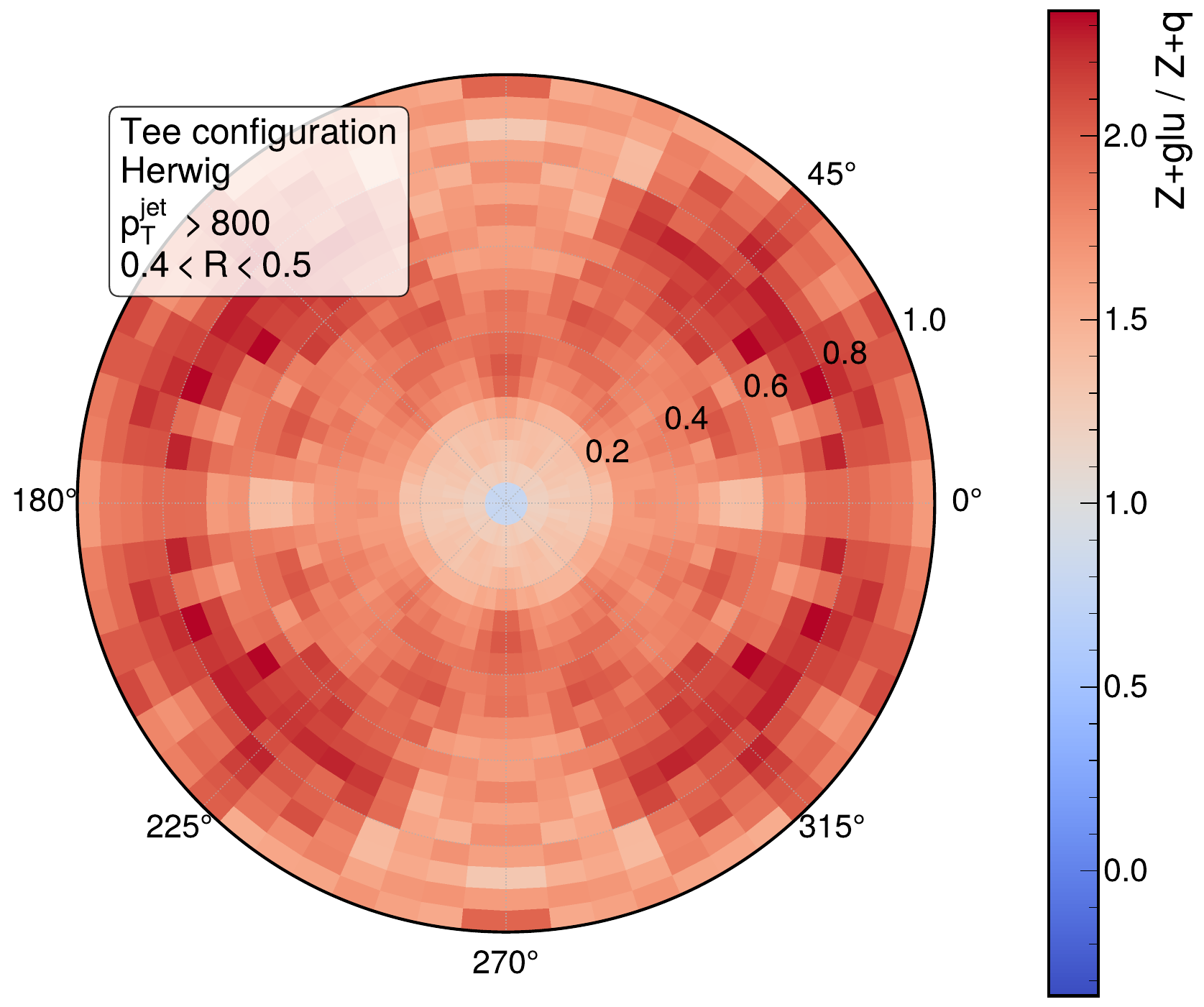} 
    \caption{Ratio between E4C tee distributions for gluon-initiated and quark-initiated jets in Pythia (top) and Herwig (bottom) simulations. Results are shown for $100$ GeV $\leq p_T \leq 200$ GeV (left), and $p_T \geq 800$ GeV (right).}
    \label{fig:tee_pythia_glu_vs_pythia_q}
\end{figure}

Quarks and gluons have distinct splitting functions, and should therefore generate qualitatively different four-point correlator distributions. In Figure~\ref{fig:tee_pythia_glu_vs_pythia_q} we show the ratio between the simulated distributions for quark-initiated and gluon-initiated jets in Pythia and Herwig, respectively. Interestingly, there is not a large angular dependence, rather, the difference is dominated by the $r$ dependence. This is easily understood: the scaling in $r$ is a power-law with an anomalous dimension which depends on the quark or gluon nature of the parton initiating the splitting into the three-partons which are taken to be collinear.   This is more pronounced at lower values of $p_T$ where $\alpha_s$ is larger.  

\begin{figure}[h]
    \centering
    \includegraphics[width=0.4\textwidth]{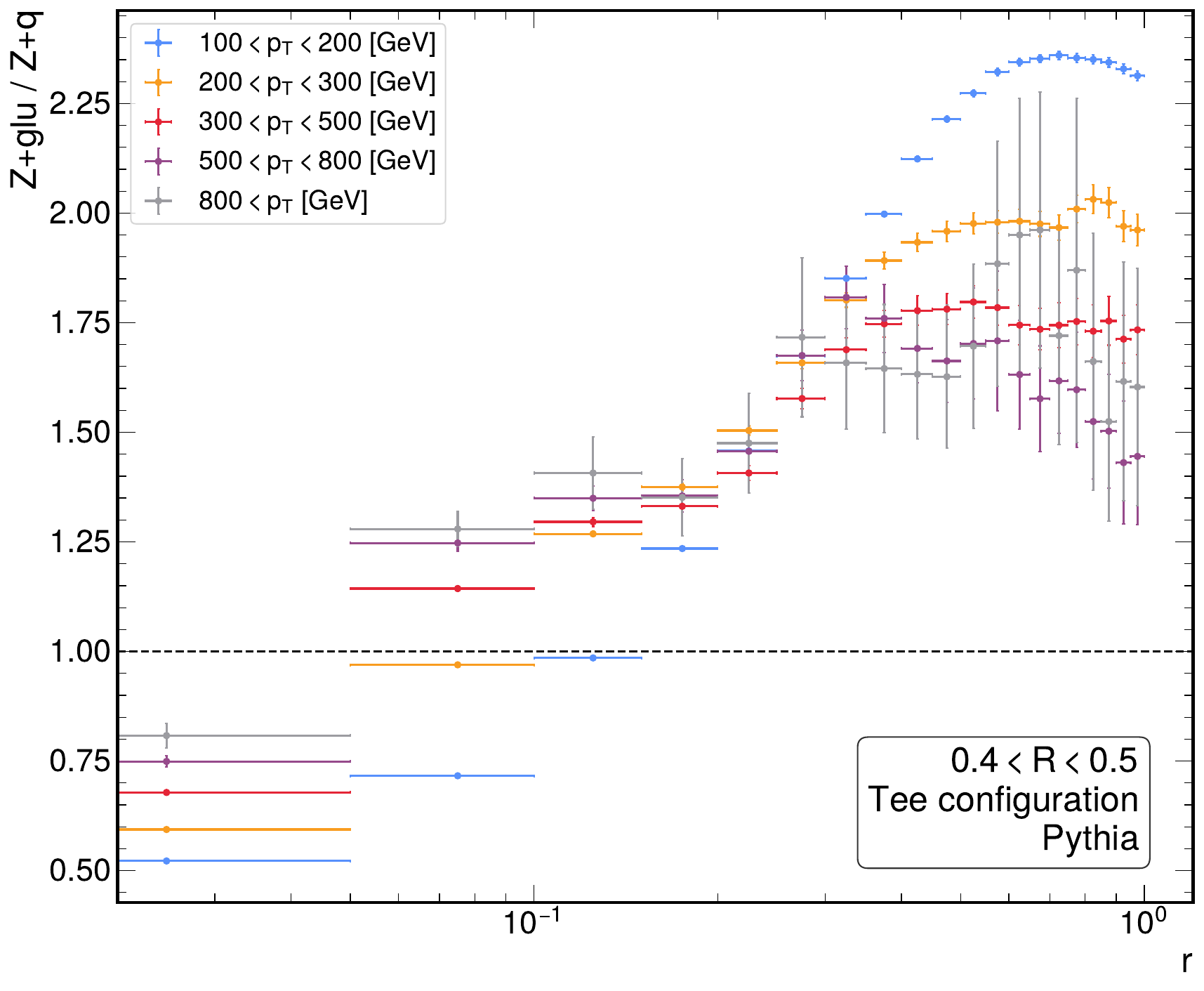} 
    \includegraphics[width=0.4\textwidth]{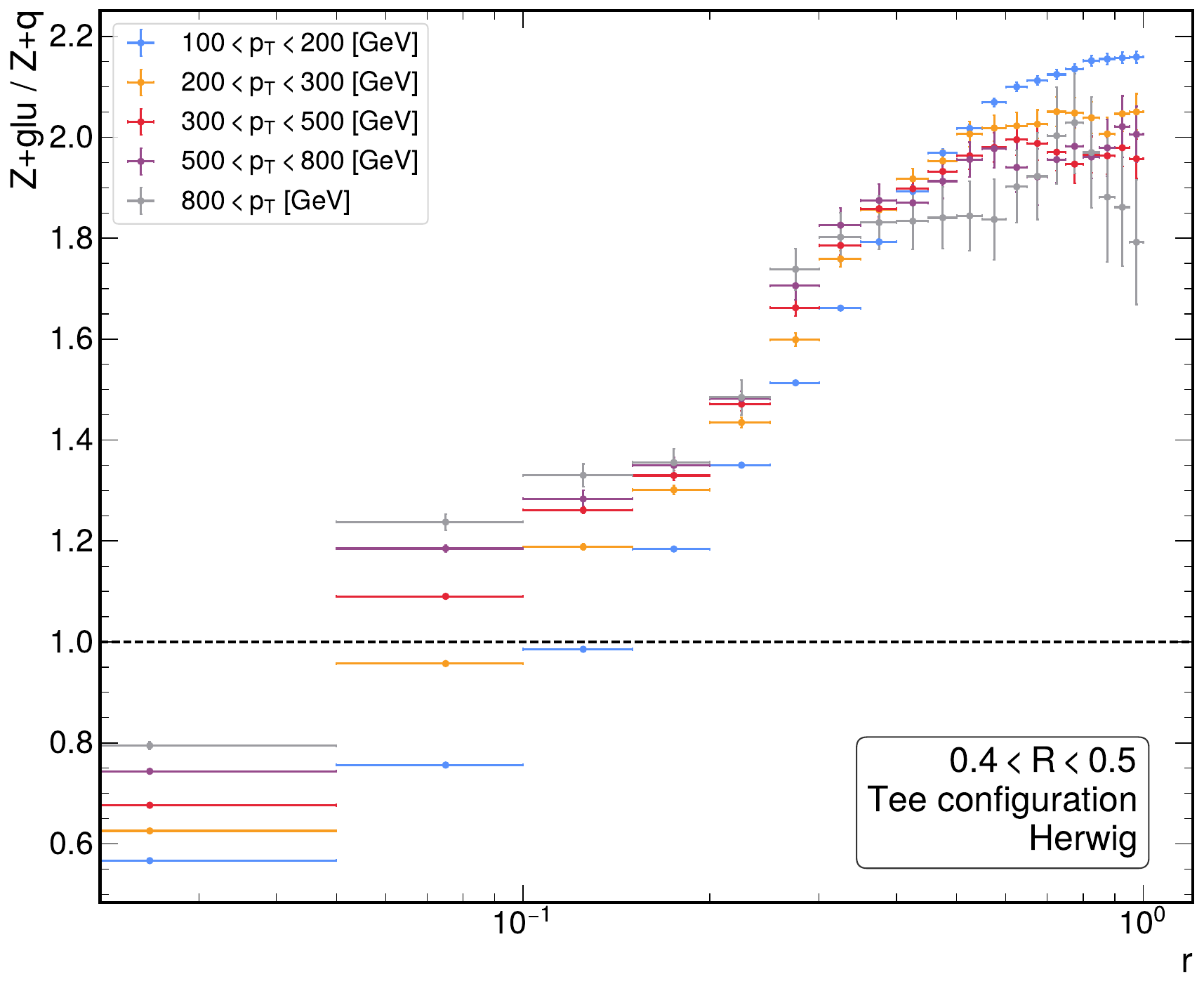} 
    \caption{Ratio between E4C tee distributions for gluon-initiated and quark-initiated jets. \textbf{Left:} Pythia; \textbf{Right:} Herwig.}
    \label{fig:tee_radial_glu_vs_q}
\end{figure}

To further investigate this effect, in Figure~\ref{fig:tee_radial_glu_vs_q} we show the radial profiles of the distribution (integrating with respect to $\phi$). Interestingly, the difference between quark and gluon jets scales more strongly with the jet $p_T$ in Pythia than in Herwig. It would be interesting to investigate this in more detail, and compare it with predictions for the scaling behavior of the three-point correlator itself (not inside the four-point function).

\paragraph{Impact of the parton shower model}

\begin{figure}[h]
    \centering
    \includegraphics[width=0.4\textwidth]{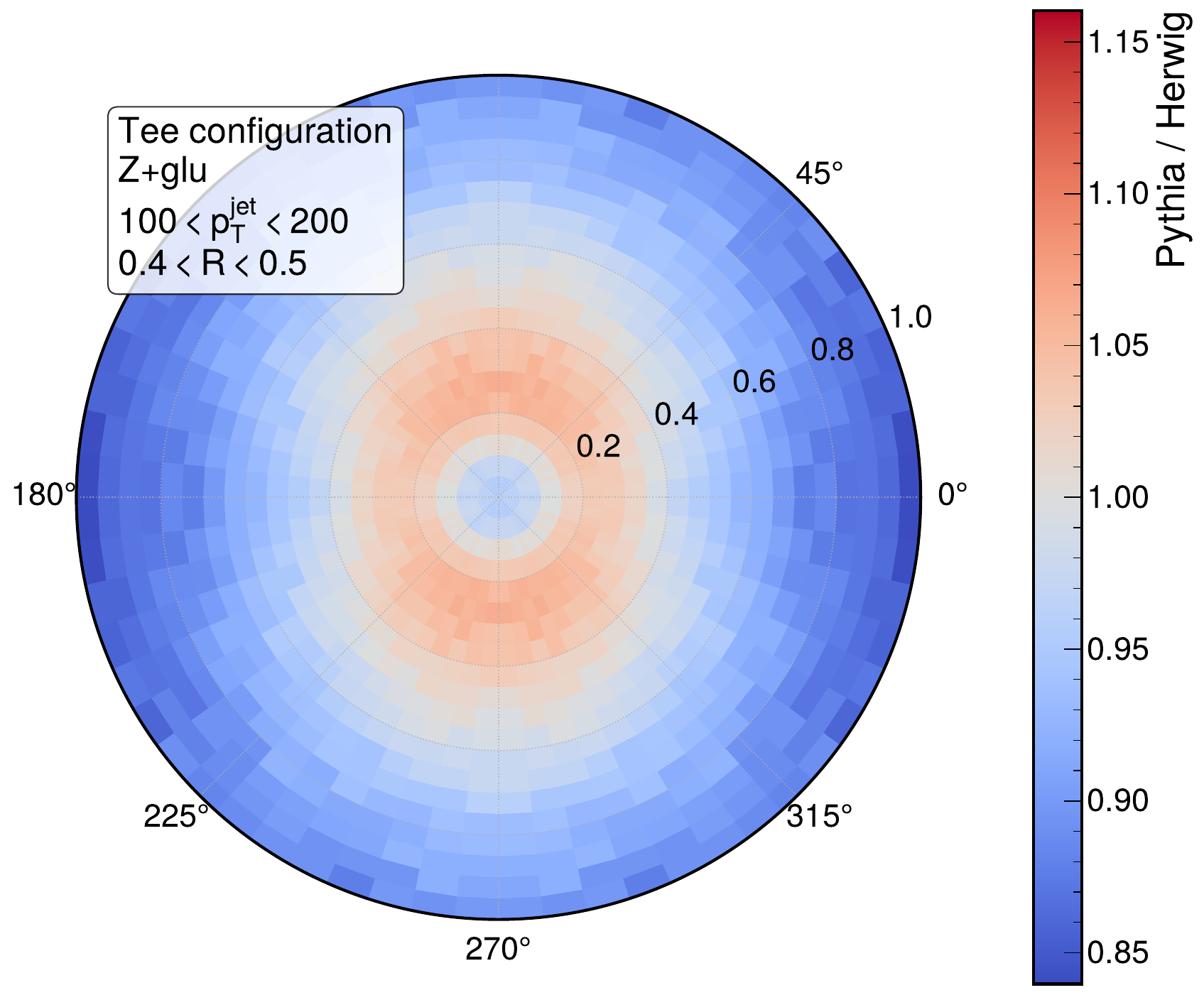} 
    \includegraphics[width=0.4\textwidth]{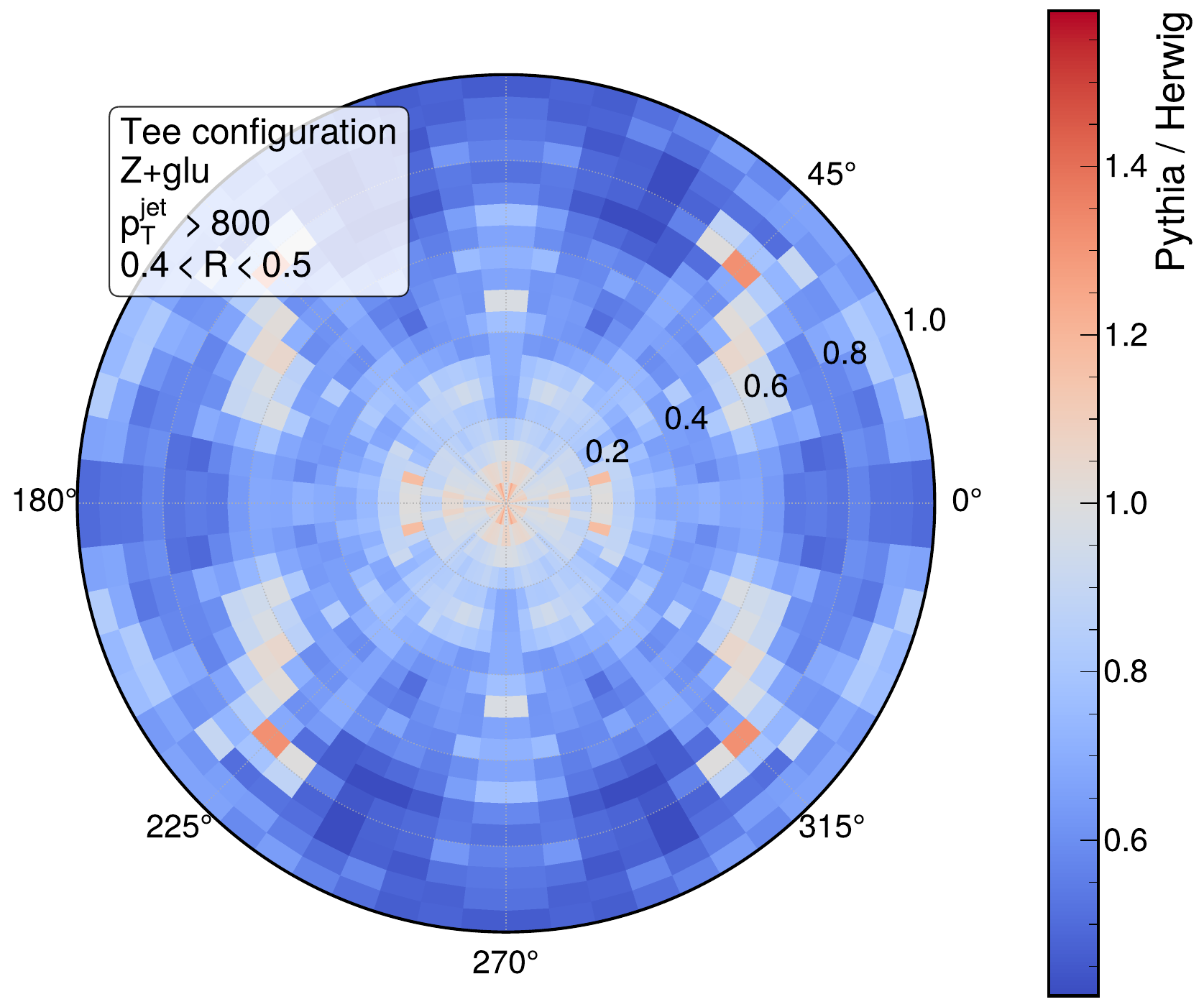} 
    \includegraphics[width=0.4\textwidth]{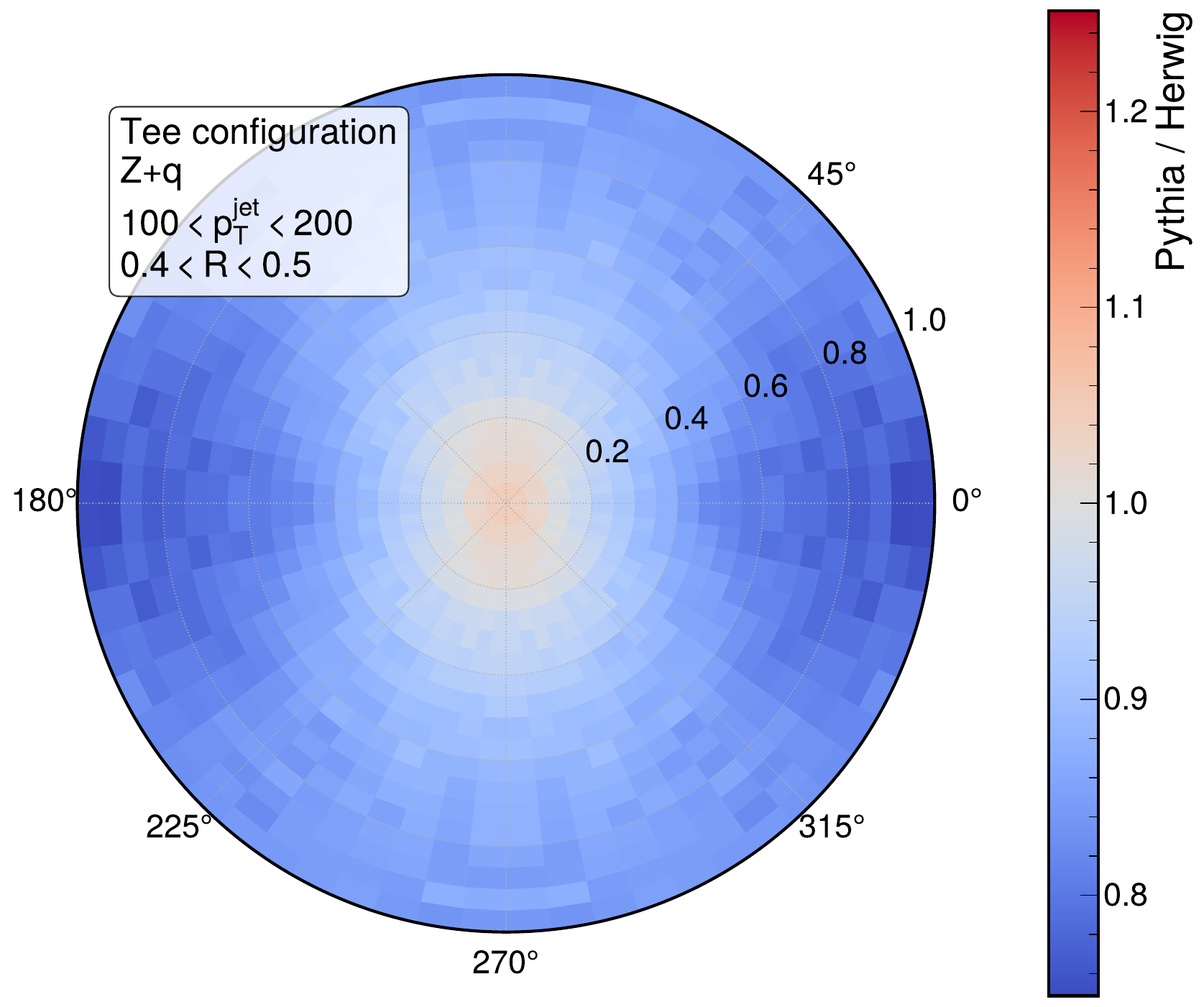} 
    \includegraphics[width=0.4\textwidth]{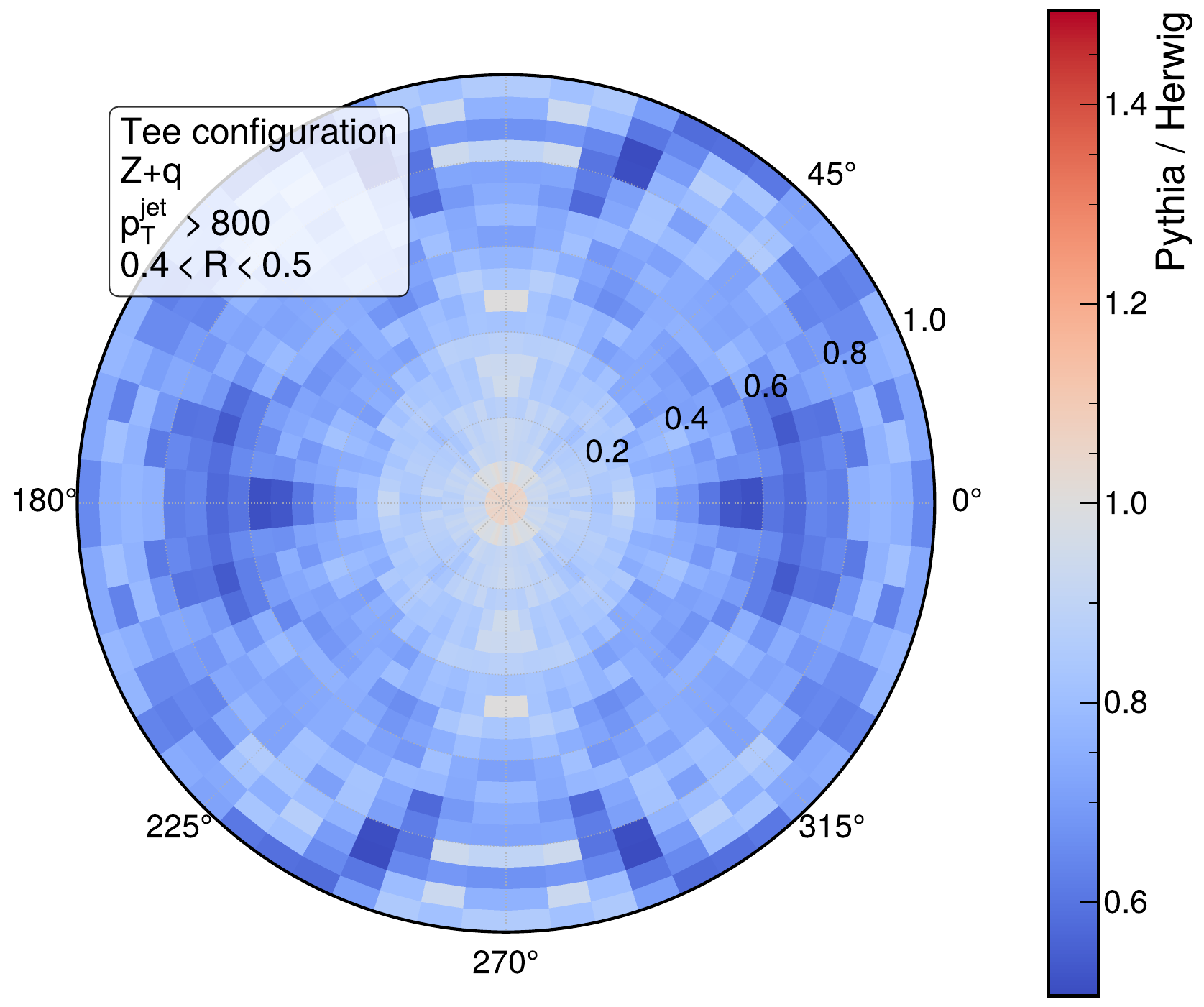} 
    \caption{Ratio between E4C tee distributions produced by Pythia and Herwig in $Z+g$ (top) and $Z+q$ (bottom) events. Results are shown for $100$ GeV $\leq p_T \leq 200$ GeV (left), and $p_T \geq 800$ GeV (right).}
    \label{fig:tee_pythia_glu_vs_herwig_glu}
\end{figure}

\begin{figure}[h]
    \centering
    \includegraphics[width=0.4\textwidth]{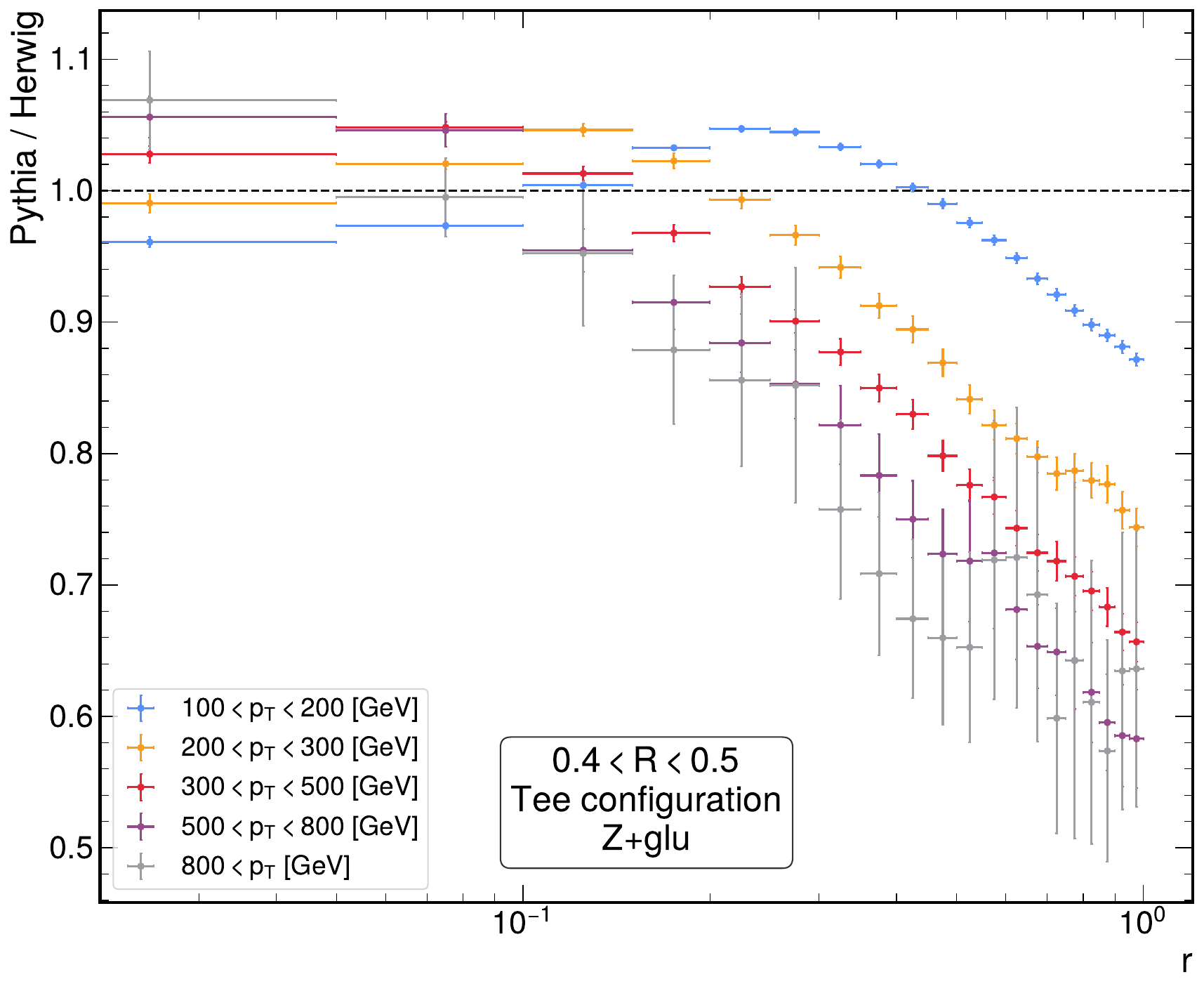} 
    \includegraphics[width=0.4\textwidth]{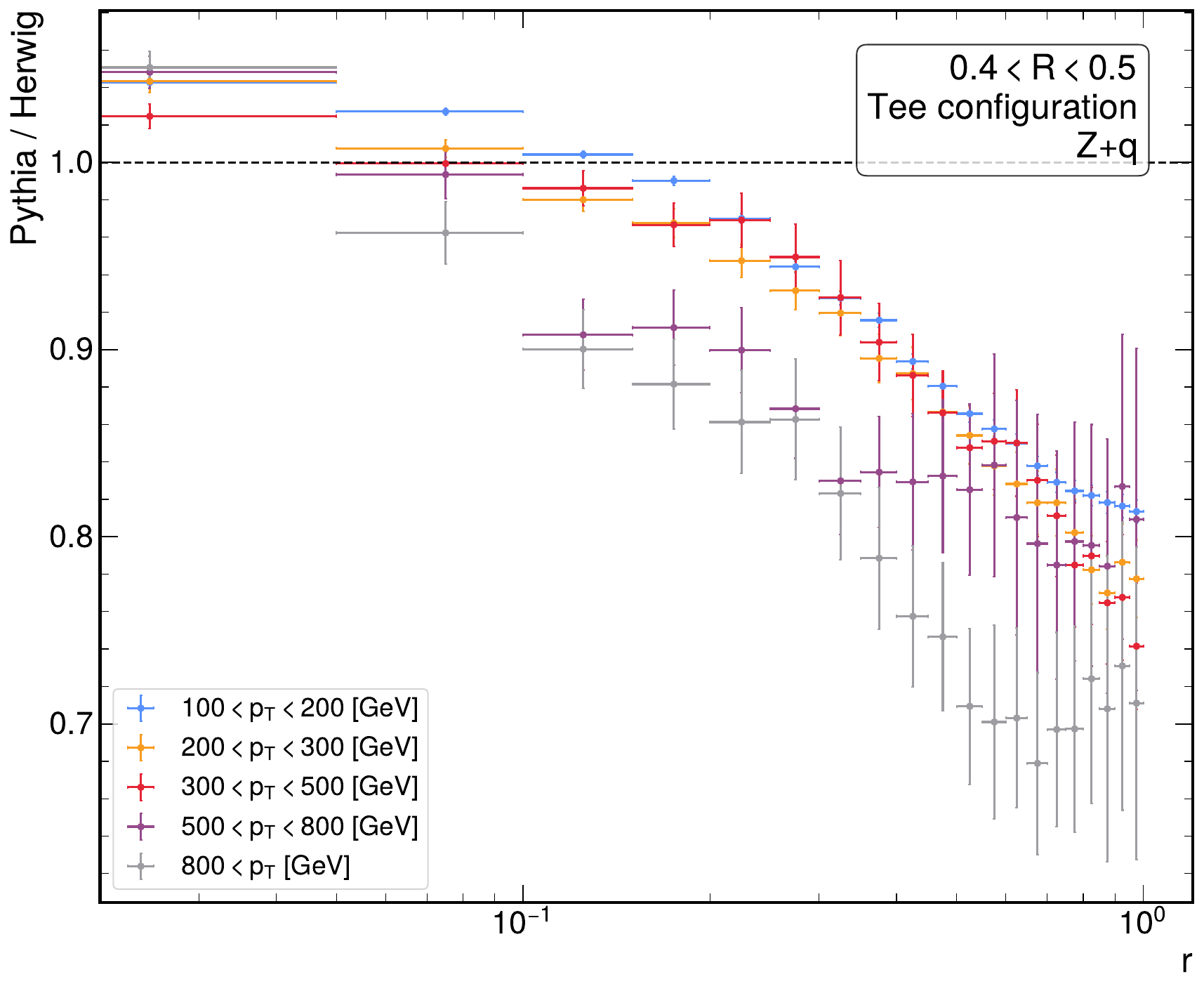} 
    \caption{Ratio between E4C tee distributions produced by Pythia and Herwig. \textbf{Left:} $Z+g$; \textbf{Right:} $Z+q$.}
    \label{fig:tee_radial_pythia_vs_herwig}
\end{figure}

Pythia and Herwig use very different parton shower models, and it is therefore interesting to compare the results of the two programs.
Figure~\ref{fig:tee_pythia_glu_vs_herwig_glu} shows the ratio between the simulated distributions for Pythia and Herwig, for quark-initiated and gluon-initiated jets, respectively.
As can be seen in Figure~\ref{fig:tee_radial_pythia_vs_herwig}, the two programs mostly agree in the non-perturbative regime (small $r$), and give different slopes in the perturbative regime (large $r$).

\paragraph{Impact of hadronization}

\begin{figure}[h]
    \centering
    \includegraphics[width=0.4\textwidth]{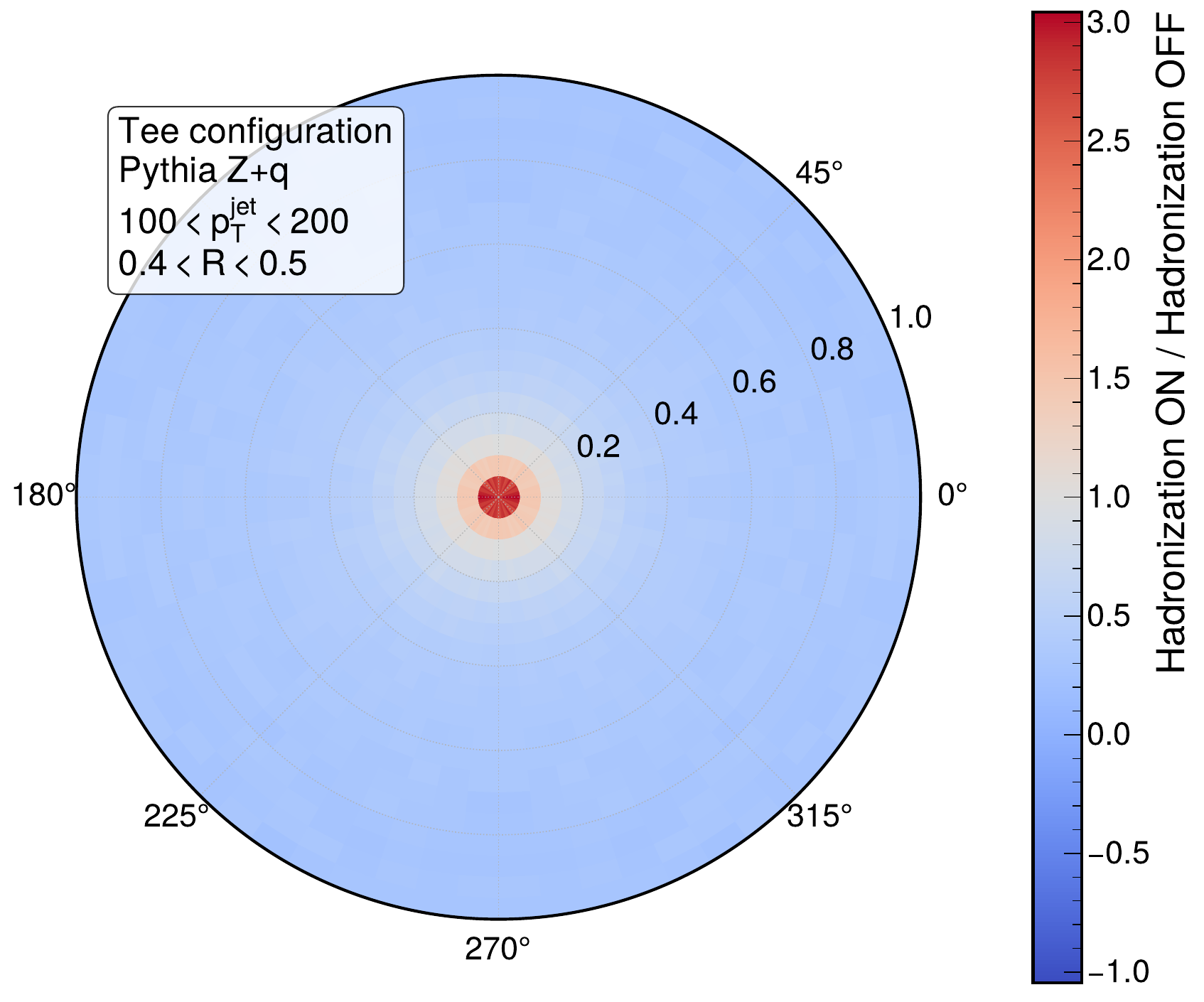} 
    \includegraphics[width=0.4\textwidth]{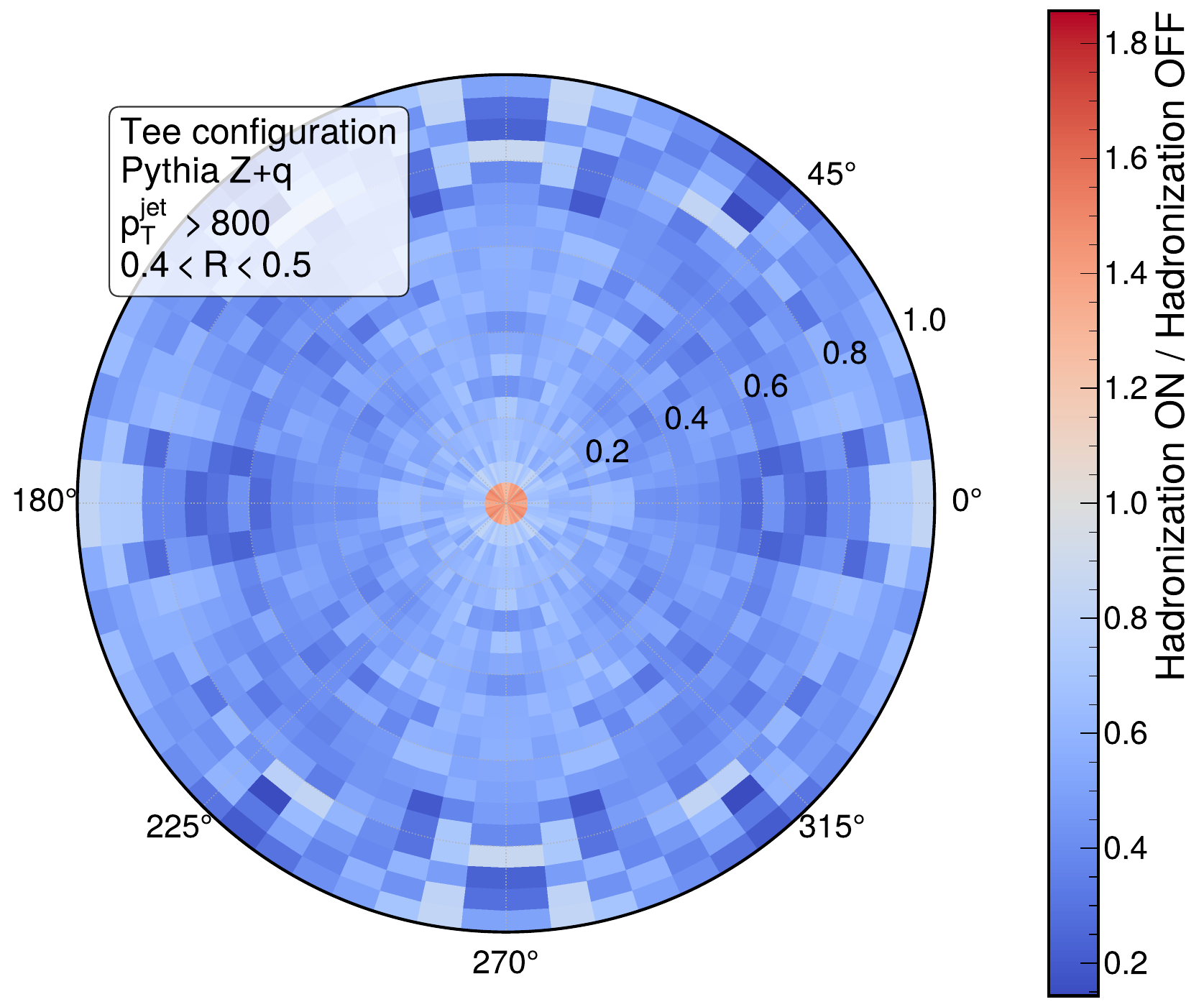} 
    \caption{Ratio between the hadronized and parton-level distributions for the tee configuration in Pythia $Z+q$ events. Results are shown for $100$ GeV $\leq p_T \leq 200$ GeV (left), and $p_T \geq 800$ GeV (right).}
    \label{fig:tee_had_vs_nohad_2D}
\end{figure}

\begin{figure}[h]
    \centering
    \includegraphics[width=0.4\textwidth]{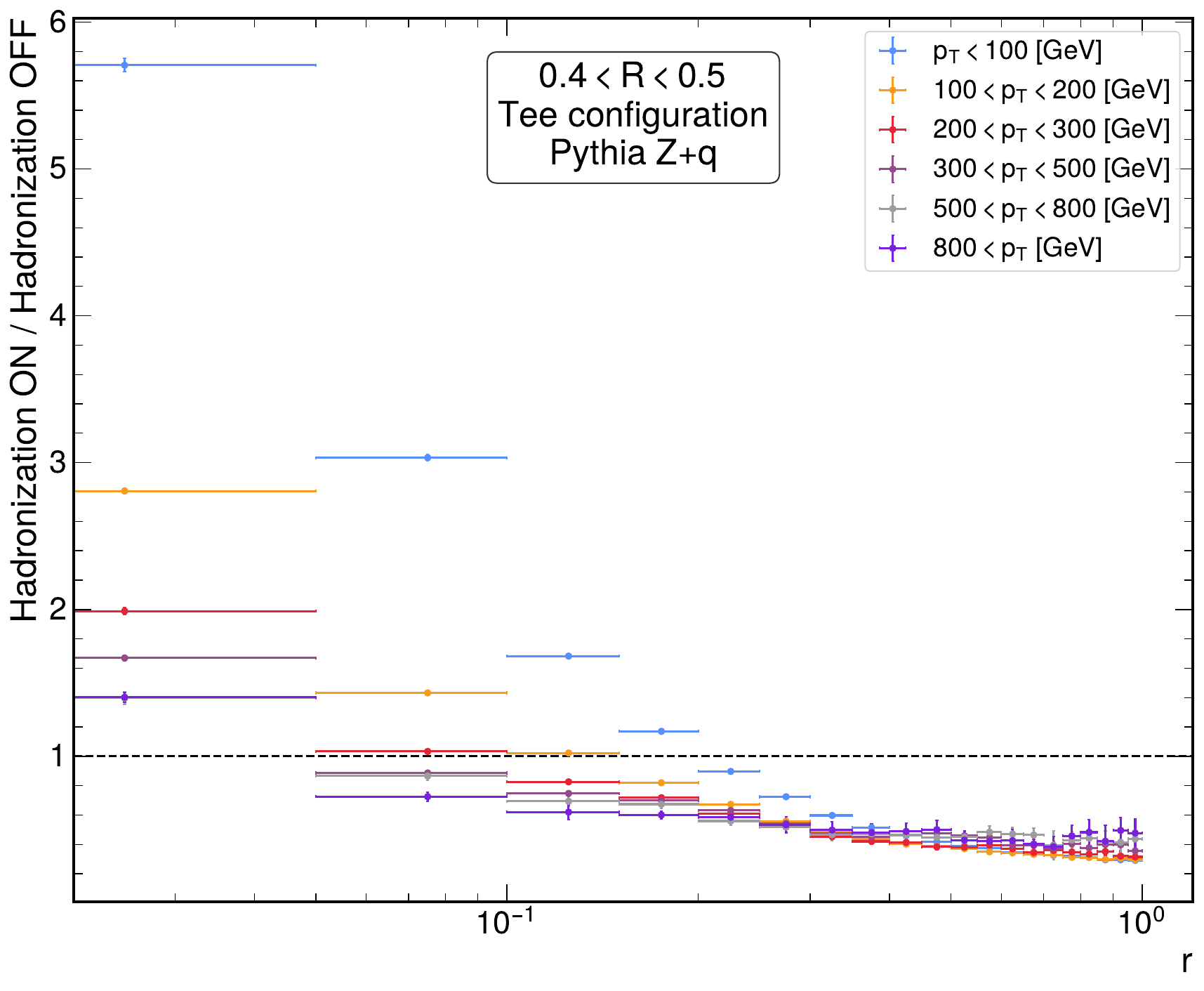} 
    \caption{Ratio between the hadronized and parton-level distributions for the tee configuration in Pythia $Z+q$ events, integrated over $\phi$.}
    \label{fig:tee_had_vs_nohad_1D}
\end{figure}

The Pythia parton shower allows us to turn off hadronization and compute the four-point correlators at the parton level. Figure~\ref{fig:tee_had_vs_nohad_2D} shows the ratio between the hadronized and parton-level distributions for the tee configuration in Pythia $Z+q$ events.
The ratio is approximately flat at large $r$, where we expect the perturbative calculation to be valid, and has a strong $r$-dependence at small $r$, where we expect confinement to dominate.
This is made explicit in Figure~\ref{fig:tee_had_vs_nohad_1D}, which shows the ratio of the radial profiles for the hadronized and parton-level distributions. 
It is clear that hadronization dominates at small $r$, and that the threshold between the perturbative and non-perturbative parts of the distribution is a function of the jet momentum.

\paragraph{Impact of spin correlations}

Here we provide further evidence for the azimuthal modulations we observe being dominated by kinematic power corrections by studying the effects of spin correlations in the parton shower. 

Both Pythia and Herwig allow the user to disable hard spin correlations. Additionally, Herwig allows the user to disable \emph{soft} spin correlations. In Appendix \ref{appendix:sctoggles} we provide a comprehensive study of all possible combinations. Here we present only the results in Herwig where we have disabled both the collinear and soft spin correlations. We explicitly verify in Appendix \ref{appendix:sctoggles} that in this limit the gluon behaves like a scalar. 

\begin{figure}[h]
    \centering
     \includegraphics[width=0.4\textwidth]{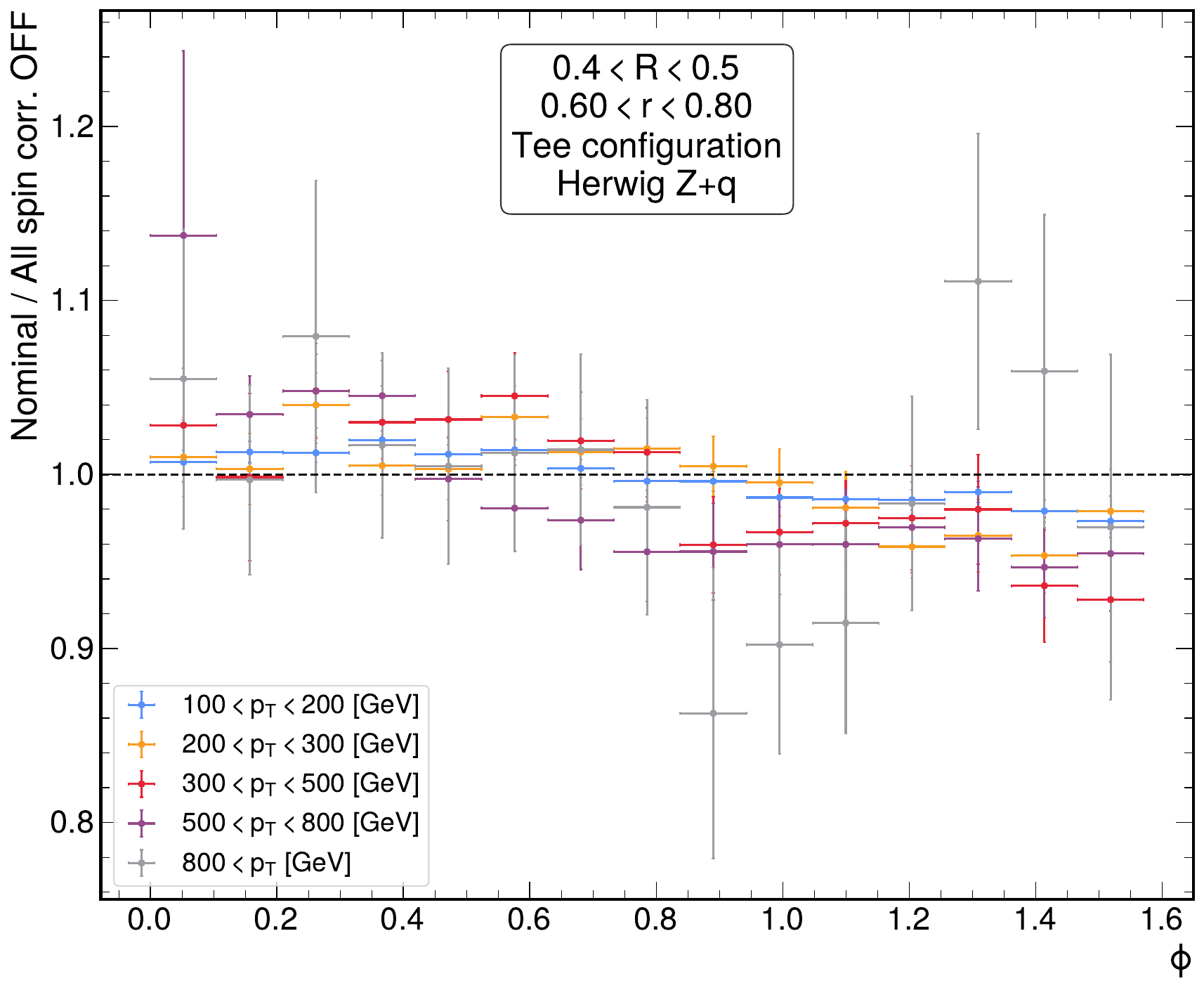} 
    \includegraphics[width=0.4\textwidth]{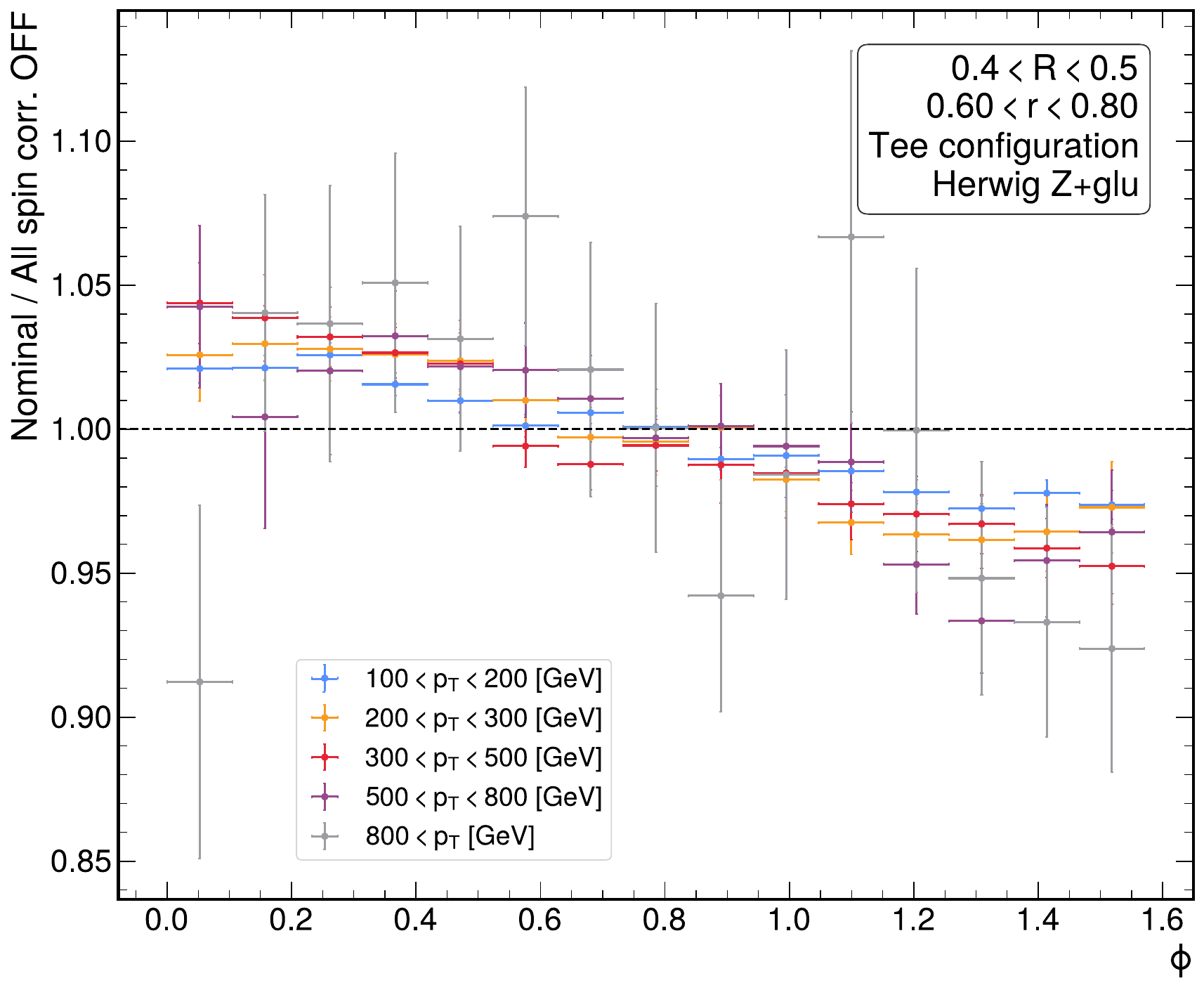} 
    \caption{Ratio between the tee distributions in Herwig $Z+q$ (left) and $Z+g$ (right) events with and without both hard and soft spin correlations (i.e.~comparing gluons behaving as spin-1 vs spin-0).}
    \label{fig:tee_herwigG_spin_intext}
\end{figure}

In Figure~\ref{fig:tee_herwigG_spin_intext} we show the ratio between the tee distributions in Herwig $Z+q$ (left) and $Z+g$ (right) events with and without both hard and soft spin correlations (i.e. comparing gluons behaving as spin-1 vs spin-0). The statistical uncertainty in many of the bins makes it difficult to draw firm conclusions, however, it is clear that this effect is much smaller than the overall azimuthal dependence, confirming that it is a kinematic effect. In both cases, but particularly for the gluon, we see a slight $\cos(2\phi)$ modulation of order $5\%$, similar in magnitude to what is predicted by the theoretical calculations. 

We believe that our study emphasizes an important subtlety about observing the effects of spin correlations experimentally using multi-point correlators (or similar observables) in jet substructure. For most kinematics that can be accessed in jet substructure, there will be two contributing factors to azimuthal asymmetries: kinematics and intrinsic spin correlations. In many setups, such as the one studied here, the azimuthal asymmetries generated by kinematic correlations are much larger than those generated by spin correlations. This is concerning, since parton showers based on $1\to 2$ splittings will certainly not generate the correct kinematic correlations in multi-point correlation functions. Unambiguously observing spin correlation effects in data will require much better control of the modulation in $\phi$ arising from kinematic effects. It seems that to properly identify the effects of spin correlation, will require the complete QCD calculation with and without spin-correlations. For the four-point energy correlator this is achievable, and could be directly compared to data, much in analogy with the original discovery of the gluon spin using event shape observables. More generally, we emphasize that the complexity of jet substructure observables that are being measured in data now surpass what can be achieved in analytic or parton shower simulations, emphasizing the importance of further pushing theory calculations of higher-point correlators.

\subsubsection{The Tripole Configuration}\label{sec:proj_c_pheno}

Finally, we consider the tripole configuration. We take as a specific example a configuration where three of the detectors form a 3-4-5 triangle configuration.

\begin{figure}[h]
    \centering
    \includegraphics[width=0.4\textwidth]{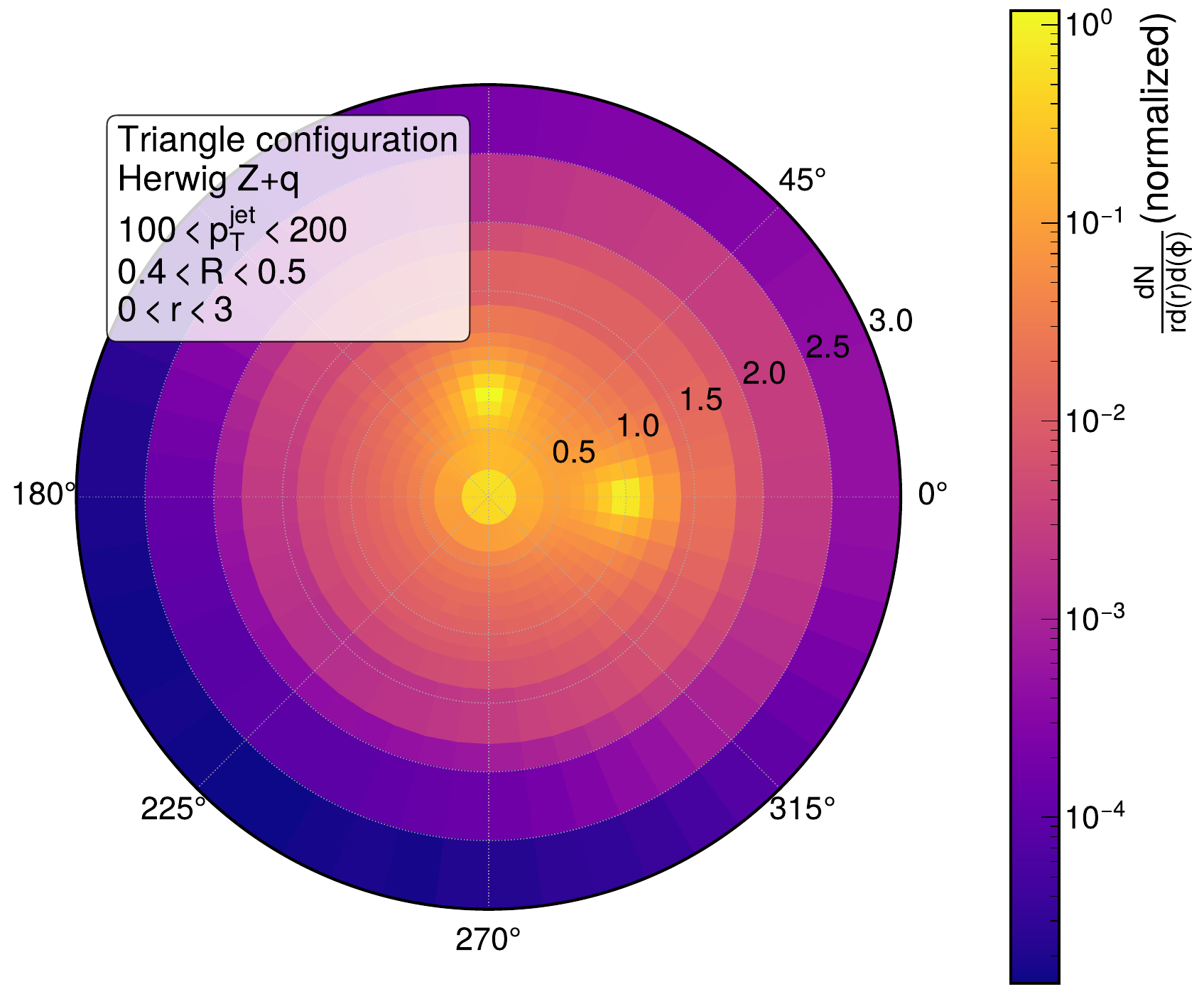} 
    \includegraphics[width=0.4\textwidth]{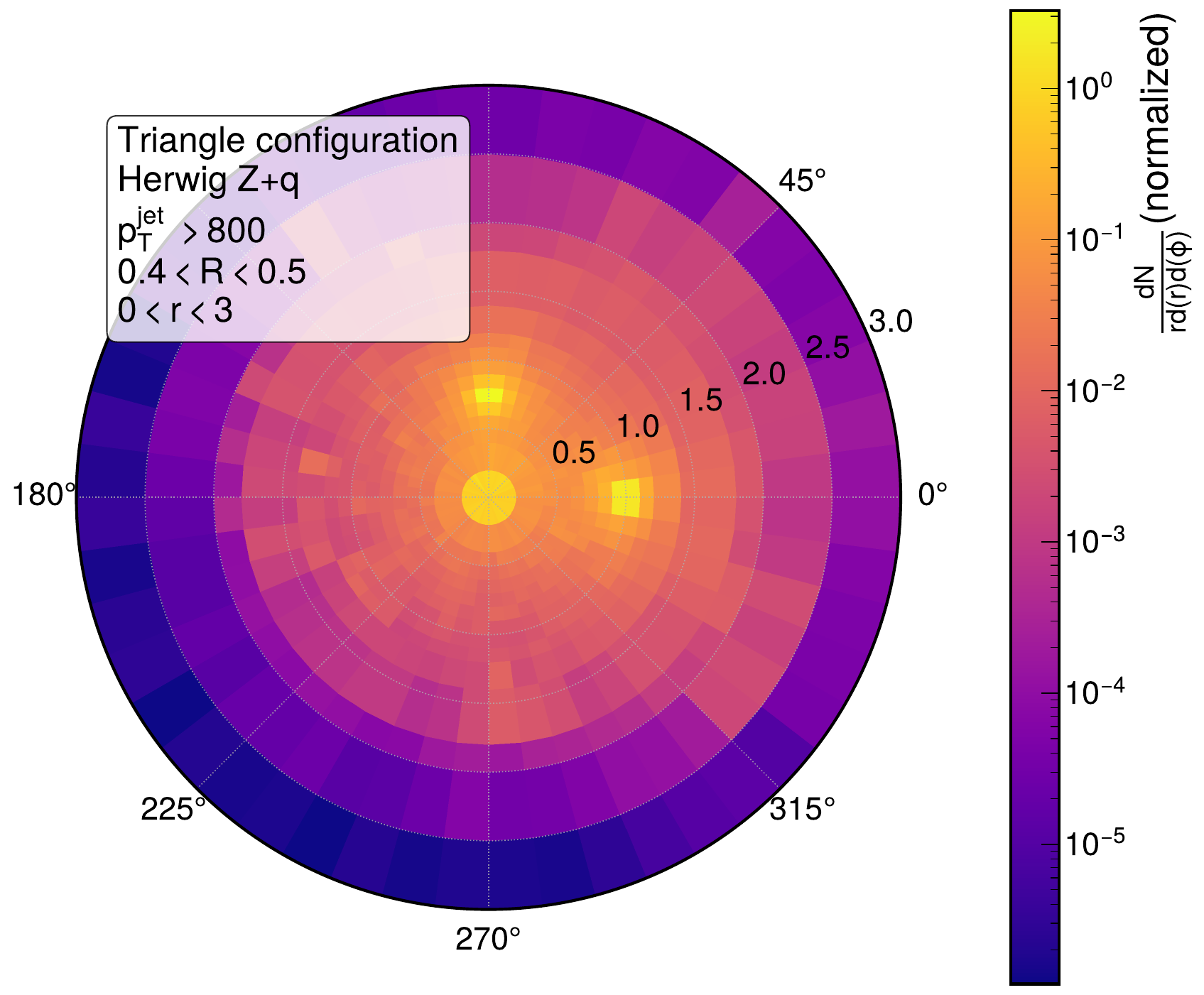} 
    \caption{Simulated results for the triangle configuration in Herwig $Z+q$ events. Results are shown for $100$ GeV $\leq p_T \leq 200$ GeV (left), and $p_T \geq 800$ GeV (right).}
    \label{fig:triangle_herwig_q}
\end{figure}

Figure~\ref{fig:triangle_herwig_q} shows the simulated results for the triangle configuration in Herwig $Z+q$ events, for $100$ GeV $\leq p_T \leq 200$ GeV (left), and $p_T \geq 800$ GeV (right). The triangle configuration is sharply visible, and becomes increasingly sharp at higher $p_T$ as the distribution becomes perturbative. We find this distribution quite striking, and hope that it can be measured experimentally!

\section{Conclusions}\label{sec:conc}

In the last several years there has been tremendous progress in the understanding of the internal substructure of jets, both from improvements in parton shower simulations, and analytic calculations. One of the advances has been the analytic calculation of multi-point energy correlators, in particular, the three and four-point correlators in the collinear limit. Combined with experimental data, this provides an opportunity to significantly advance the theoretical description of jet structure.

In this paper we performed a first phenomenological exploration of the four-point correlator. We introduced three different projections of the four-point correlator: ``the dipole", ``the tee", and the ``the tripole". These projections were explicitly designed to probe different factorization channels of the $1\to 4$ splitting function, allowing these factorization channels to be studied in both parton showers and, ultimately, data. We introduced ``the dipole" configuration to probe the factorization of a $1\to 4$ splitting function into a $1\to 3$ splitting tensor and a $1\to 2$ splitting function, and ``the tee" to probe the factorization into a $1\to 2$ splitting tensor and a $1\to 3$ splitting function. In both cases, the fully differential nature of the energy correlators enables a measure of the azimuthal structure. We believe that observables that probe these configurations will be invaluable for the next generation of parton showers. 

We performed a detailed phenomenological analysis of the four-point correlator using both the Herwig and Pythia parton showers, as well as analytic calculations in kinematic limits. We considered a number of different parton shower configurations, studying the effect of spin correlations in both soft and collinear emissions. In experimentally accessible kinematic regions, we find the  spin correlations are strongly subdominant to kinematic azimuthal correlations, strongly motivating a complete calculation of the four-point correlator in QCD to compare to the parton shower results. We believe that it will only be possible to fully interpret the data in terms of spin correlations if the full QCD calculation is performed. 

Additionally, we presented algorithms for constructing the four point energy correlators in experiment and properly treating uncertainties. This paper lays the foundations for experimental analyses and complete calculations of four-point correlators in QCD. We hope to see the four-point correlator measured in data in the near future. This will represent a frontier measurement in jet substructure. The complexity of observables that are being measured in data has caught up and surpassed what can be achieved in theory, emphasizing the importance of further pushing theory calculations of higher point correlators in QCD.

\acknowledgments

We thank Matt Leblanc, Jennifer Roloff, Kai Yan and Vittorio Del Duca for useful discussions. K.L. is supported by the U.S. Department of Energy under contract DE-AC02-06CH11357. M.G. and I.M. are supported by the DOE Early Career Award DE-SC0025581, and the Sloan Foundation. P.H. and S.R are supported by DOE Early Career Award DE-SC0021943
and DOE Award DE-SC0011939.

\appendix

\section{Perturbative Ingredients}\label{sec:pert}
In this Appendix, we collect the perturbative ingredients used for the calculations in Sec.~\ref{sec:calc}.
\vspace{0.5cm}

\noindent {\bf{$1 \rightarrow 2$ Splitting Functions:}}
We use the notation $P_{ij}(x)$ where $x$ is the momentum fraction of parton $i$. The relevant splitting functions are \cite{Catani:1999ss}
\bea
P^{ss'}_{qg}(x) =&\, \delta^{s s'} C_F \left[\frac{1+x^2}{1-x} - \e(1-x)\right]\,, \\
P^{\mu \nu}_{q \bar q}(x) =&\, T_R \left[-g^{\mu \nu} + 4 x(1-x)\frac{k_{\perp}^{\mu} k_{\perp}^{\nu}}{k_{\perp}^2}\right]\,, \\
P^{\mu \nu}_{gg}(x) =&\, 2 C_A \left[-g^{\mu \nu} \left(\frac{x}{1-x} + \frac{1-x}{x}\right) - 2 (1-\e)x(1-x)\frac{k_{\perp}^{\mu} k_{\perp}^{\nu}}{k_{\perp}^2}\right]\,.
\eea
\vspace{0.5cm}

\noindent {\bf{$1 \rightarrow 3$ Splitting Functions:}}
The tree level $1\to 3$ splitting functions were first computed in \cite{Campbell:1997hg,Catani:1998nv}. Here, we use the notation $P_{i_1 i_2 i_3}(x_1,x_2,x_3)$, where $x_j$ is the momentum fraction of parton $i_j$
\bea
P_{\bar q'_1 q'_2 q_3}^{s s'}(x_1,x_2,x_3) =&\, \de^{s s'} C_F T_R \frac {s_{123}} { 2 s _ { 12 } } \Bigg[ - \frac { \left(x_1 \left( s _ { 12 } + 2 s _ { 23 } \right) - x_2 \left( s _ { 12 } + 2 s _ { 13 } \right) \right) ^ { 2 } } { \left( x_1 + x_2 \right) ^ { 2 } s _ { 12 } s _ { 123 } } \nn \\
&+ \frac { 4 x_3 + \left( x_1 - x_2 \right) ^ { 2 } } { x_1 + x_2 } + ( 1 - 2 \epsilon ) \left(x_1 + x_2 - \frac {s _ { 12 } } { s _ { 123 } } \right) \Bigg]\,, \\
P_{\bar q_1 q_2 q_3}^{(id)\,s s'}(x_1,x_2,x_3) =&\, \de^{s s'} C_F \left(C_F - \frac{1}{2} C_A\right) \Bigg\{ (1-\epsilon) \left(\frac{2 s_{23}}{s_{12}} - \epsilon\right) \nn \\
&+ \frac{s_{123}}{s_{12}} \left[\frac{1+x_1^2}{1-x_2} - \frac{2 x_2}{1-x_3} - \epsilon \left(\frac{(1-x_3)^2}{1-x_2} + 1 + x_1 - \frac{2 x_2}{1-x_3}\right) - \epsilon^2(1-x_3)\right] \nn \\
&- \frac{x_1 s_{123}^2}{2 s_{12}s_{13}} \left[\frac{1+x_1^2}{(1-x_2)(1-x_3)} - \epsilon \left(1 + 2 \frac{1-x_2}{1-x_3}\right) - \epsilon^2\right]\Bigg\} + \left(2 \leftrightarrow 3\right)\,, \\
P_{g_1 g_2 q_3}^{s s'}(x_1,x_2,x_3) =&\, \de^{s s'} C _ { F } ^ { 2 } \left\{ \frac { s _ { 123 } ^ { 2 } } { 2 s _ { 13 } s _ { 23 } } x_3 \left[ \frac { 1 + x_3 ^ { 2 } } { x_1 x_2 } - \epsilon \frac { x_1 ^ { 2 } + x_2 ^ { 2 } } { x_1 x_2 } - \epsilon ( 1 + \epsilon ) \right] + ( 1 - \epsilon ) \left[ \epsilon - ( 1 - \epsilon ) \frac { s _ { 23 } } { s _ { 13 } } \right] \right. \nn \\
&\left.+ \frac { s _ { 123 } } { s _ { 13 } } \left[ \frac { x_3 \left( 1 - x_1 \right) + \left( 1 - x_2 \right) ^ { 3 } } { x_1 x_2 } - \epsilon \left( x_1 ^ { 2 } + x_1 x_2 + x_2 ^ { 2 } \right) \frac { 1 - x_2 } { x_1 x_2 } + \epsilon ^ { 2 } \left( 1 + x_3 \right) \right] \right\} \nn \\
& + C _ { F } C _ { A } \left\{ ( 1 - \epsilon ) \left( \frac { \left( x_1 \left( s _ { 12 } + 2 s _ { 23 } \right) - x_2 \left( s _ { 12 } + 2 s _ { 13 } \right) \right) ^ { 2 } } { 4 \left( x_1 + x_2 \right) ^ { 2 } s _ { 12 } ^ { 2 } } + \frac { 1 } { 4 } - \frac { \epsilon } { 2 } \right) \right. \nn \\
& + \frac { s _ { 123 } ^ { 2 } } { 2 s _ { 12 } s _ { 13 } } \left[ \frac { 2 x_3 + ( 1 - \epsilon ) \left( 1 - x_3 \right) ^ { 2 } } { x_2 } + \frac { 2 \left( 1 - x_2 \right) + ( 1 - \epsilon ) x_2 ^ { 2 } } { 1 - x_3 } \right] \nn \\
& - \frac { s _ { 123 } ^ { 2 } } { 4 s _ { 13 } s _ { 23 } } x_3 \left[ \frac { 2 x_3 + ( 1 - \epsilon ) \left( 1 - x_3 \right) ^ { 2 } } { x_1 x_2 } + \epsilon ( 1 - \epsilon ) \right] \nn \\
& + \frac { s _ { 123 } } { 2 s _ { 12 } } \left[ ( 1 - \epsilon ) \frac { x_1 \left( 2 - 2 x_1 + x_1 ^ { 2 } \right) - x_2 \left( 6 - 6 x_2 + x_2 ^ { 2 } \right) } { x_2 \left( 1 - x_3 \right) } + 2 \epsilon \frac { x_3 \left( x_1 - 2 x_2 \right) - x_2 } { x_2 \left( 1 - x_3 \right) } \right] \nn \\
&+ \frac { s _ { 123 } } { 2 s _ { 13 } } \left[ ( 1 - \epsilon ) \frac { \left( 1 - x_2 \right) ^ { 3 } + x_3 ^ { 2 } - x_2 } { x_2 \left( 1 - x_3 \right) } - \epsilon \left( \frac { 2 \left( 1 - x_2 \right) \left( x_2 - x_3 \right) } { x_2 \left( 1 - x_3 \right) } - x_1 + x_2 \right) \right. \nn \\
&\left.- \frac { x_3 \left( 1 - x_1 \right) + \left( 1 - x_2 \right) ^ { 3 } } { x_1 x_2 } + \epsilon \left( 1 - x_2 \right) \left.\left( \frac { x_1 ^ { 2 } + x_2 ^ { 2 } } { x_1 x_2 } - \epsilon \right) \right] \right\} + ( 1 \leftrightarrow 2 )\,, \\
P_{g_1 q_2 \bar q_3}^{\mu \nu}(x_1,x_2,x_3) =&\, C_F T_R \frac{1}{2}\Bigg\{-g^{\mu \nu} \left[-2 + \frac{2 s_{123} s_{23} + (1 - \e)(s_{123} - s_{23})^2}{s_{12} s_{13}}\right] \nn \\
&+ \frac{4 s_{123}}{s_{12} s_{13}} \left(\tl k_3^{\mu} \tl k_2^{\nu} + \tl k_2^{\mu} \tl k_3^{\nu} - (1-\e)\tl k_1^{\mu} \tl k_1^{\nu}\right) \Bigg\} \nn \\
&+ C_A T_R \frac{1}{4}\Bigg\{\frac{s_{123}}{s_{23}^2} \left[g^{\mu \nu} \frac{t_{23,1}^2}{s_{123}} - 16 \frac{x_2^2 x_3^2}{x_1(1-x_1)} \left(\frac{\tl k_2}{x_2} - \frac{\tl k_3}{x_3}\right)^{\mu} \left(\frac{\tl k_2}{x_2} - \frac{\tl k_3}{x_3}\right)^{\nu}\right] \nn \\
&+ \frac{s_{123}}{s_{12} s_{13}} \left[2 s_{123} g^{\mu \nu} - 4 \left(\tl k_2^{\mu} \tl k_3^{\nu} + \tl k_3^{\mu} \tl k_2^{\nu}- (1-\e)\tl k_1^{\mu} \tl k_1^{\nu}\right)\right] \nn \\
&- g^{\mu \nu} \left[-(1-2\e) + 2 \frac{s_{123}}{s_{12}}\frac{1-x_3}{x_1(1-x_1)} + 2 \frac{s_{123}}{s_{23}} \frac{1-x_1+2 x_1^2}{x_1(1-x_1)}\right] \nn \\
&+ \frac{s_{123}}{s_{12} s_{23}} \left[-2 s_{123} g^{\mu \nu} \frac{x_2(1 - 2 x_1)}{x_1(1 - x_1)} - 16 \tl k_3^{\mu} \tl k_3^{\nu} \frac{x_2^2}{x_1(1 - x_1)} + 8(1-\e) \tl k_2^{\mu} \tl k_2^{\nu} \right. \nn \\
&+ \left. 4 \left(\tl k_2^{\mu} \tl k_3^{\nu} + \tl k_3^{\mu} \tl k_2^{\nu}\right) \left(\frac{2 x_2(x_3 - x_1)}{x_1(1 - x_1)} + (1 - \e)\right)\right]\Bigg\} + (2 \leftrightarrow 3)\,, \\
P_{g_1 g_2 g_3}^{\mu \nu}(x_1,x_2,x_3) =&\, C_A^2 \Bigg\{\frac{(1-\e)}{4 s_{12}^2} \left[-g^{\mu \nu} t_{12,3}^2 + 16 s_{123} \frac{x_1^2 x_2^2}{x_3(1-x_3)} \left(\frac{\tl k_2}{x_2} - \frac{\tl k_1}{x_1}\right)^{\mu} \left(\frac{\tl k_2}{x_2} - \frac{\tl k_1}{x_1}\right)^{\nu}\right] \nn \\
&- \frac{3}{4}(1-\e)g^{\mu \nu} + \frac{s_{123}}{x_3 s_{12}} \left[\frac{2(1-x_3)+4 x_3^2}{1-x_3} - \frac{1 - 2 x_3(1 - x_3)}{x_1(1-x_1)}\right] g^{\mu \nu} \nn \\
&+ \frac{s_{123}(1-\e)}{s_{12}s_{13}} \left[2 x_1 \left(\tl k_2^{\mu} \tl k_2^{\nu} \frac{1 - 2 x_3}{x_3(1-x_3)} + \tl k_3^{\mu} \tl k_3^{\nu} \frac{1 - 2 x_2}{x_2(1-x_2)}\right) \right. \nn \\
&+ \left. \frac{s_{123}}{2(1-\e)}g^{\mu \nu} \left(\frac{4 x_2 x_3 + 2 x_1 (1-x_1) - 1}{(1-x_2)(1-x_3)} - \frac{1 - 2 x_1(1-x_1)}{x_2 x_3}\right) \right. \nn \\
&+ \left. (\tl k_2^{\mu} \tl k_3^{\nu} + \tl k_3^{\mu} \tl k_2^{\nu}) \left(\frac{2 x_2(1-x_2)}{x_3(1-x_3)} - 3\right)\right]\Bigg\} + (\text{5 permutations})\,,
\eea
where
\bea
t_{ij,k} = 2 \frac{x_i s_{jk} - x_j s_{ik}}{x_i + x_j} + \frac{x_i - x_j}{x_i + x_j} s_{ij}\,.
\eea
\vspace{0.5cm}

\noindent{\bf{$1 \rightarrow 2$ Splitting Tensors:}}
In all splitting tensors, we average over the initial spin such that the source is unpolarized \cite{DelDuca:2019ggv,DelDuca:2020vst}. This leaves only the helicity indices for the intermediate particle we are factorizing on. It also allows us to express a number of the splitting tensors in terms of spin-averaged splitting functions\footnote{This is equivalent to averaging over the initial spins.}, up to some trivial helicity indices. We employ the notation $H_{i_{[1 \cdots M]} i_{N}}(x)$, where $i_{[1 \cdots M]}$ denotes the parent parton of the sub-collinear set (the parton whose helicity indices the splitting tensor carries) with momentum fraction $x$.
\bea
H^{hh'}_{q g}(x) =&\, \half \de^{h h'} P_{q g}(x)\,, \\
H^{hh'}_{q \bar q}(x) =&\, \half \de^{h h'} P_{q \bar q}(x)\,, \\
H^{\mu \nu}_{gq}(x) =&\, C_F \left[\half (1-x) d^{\mu \nu}(P) - \frac{2 x}{1-x}\frac{k_{\perp}^{\mu} k_{\perp}^{\nu}}{k_{\perp}^2}\right]\,, \label{eq:gqST} \\
H^{\mu \nu}_{gg}(x) =&\, C_A \left[\left(\frac{1-x}{x} + x(1-x)\right) d^{\mu \nu}(P) - \frac{2 x}{1-x}\frac{k_{\perp}^{\mu} k_{\perp}^{\nu}}{k_{\perp}^2}\right]\,. \label{eq:ggST}
\eea
The $1 \rightarrow 3$ splitting tensors can be found in \cite{DelDuca:2019ggv,DelDuca:2020vst}.

\noindent{\bf{Analytic Integrated Results:}} Additionally, we collect the analytic results for the tee configuration. The expressions for $\alpha$ and $\beta$, as defined in \Eq{eq:3point_answer} are given by
\bea
\alpha_q =& \left[-1800\, \text{Li}_2\left(-\frac{1}{3}\right)-\frac{880 \pi ^2}{3}+\frac{1436542}{2079}-900 \log ^2(3)+\frac{2741440 \log (2)}{693}\right] C_F^3 \nn \\
&+ \left[1900\, \text{Li}_2\left(-\frac{1}{3}\right)+\frac{940 \pi ^2}{3}-\frac{3128329}{4158}+950 \log ^2(3)-\frac{2896016 \log (2)}{693}\right] C_F^2 C_A \nn \\
&+ \left[-800\, \text{Li}_2\left(-\frac{1}{3}\right)-\frac{400 \pi ^2}{3}+\frac{2737759}{8316}-400 \log ^2(3)+\frac{1222192 \log (2)}{693} \right.\nn \\
&+ \left.\left(-800\, \text{Li}_2\left(-\frac{1}{3}\right)-\frac{400 \pi ^2}{3}+\frac{1368713}{4158}-400 \log ^2(3)+\frac{1222912 \log (2)}{693}\right)n_f \right] C_F^2 T_R \nn \\
&+ \left[-6200\, \text{Li}_2\left(-\frac{1}{3}\right)-\frac{3100 \pi ^2}{3}+\frac{10212463}{3960}-3100 \log ^2(3)+\frac{449708 \log (2)}{33}\right] C_F C_A T_R n_f \nn \\
&+ \left[-5100\, \text{Li}_2\left(-\frac{1}{3}\right)-840 \pi ^2+\frac{1428220}{693}-2550 \log ^2(3)+\frac{2586160 \log (2)}{231}\right] C_F C_A^2\,, \nn \\
\beta_q =& \left[-\frac{2188}{5775}-\frac{256 \log (2)}{1925}\right] C_F^2 T_R n_f + \left[-6912\, \text{Li}_2\left(-\frac{1}{3}\right)-1152 \pi ^2+\frac{249317443}{86625} \right. \nn \\
&\left. -3456 \log ^2(3)+\frac{87702368 \log (2)}{5775}\right] C_F C_A T_R n_f + \left[\frac{1992}{1925}+\frac{3456 \log (2)}{1925}\right] C_F C_A^2\,,
\eea
\bea
\alpha_g =& \left[-\frac{936\, \text{Li}_2\left(-\frac{1}{3}\right)}{7}-\frac{2288 \pi ^2}{105}+\frac{18675046}{363825}-\frac{468 \log ^2(3)}{7} +\frac{7127744 \log (2)}{24255}\right] C_F^2 T_R n_f \nn \\
&+ \left[\frac{988\, \text{Li}_2\left(-\frac{1}{3}\right)}{7}+\frac{2444 \pi ^2}{105}-\frac{40702243}{727650}+\frac{494 \log ^2(3)}{7}-\frac{37501328 \log (2)}{121275}\right] C_F C_A T_R n_f \nn \\
&+ \left[-\frac{416\, \text{Li}_2\left(-\frac{1}{3}\right)}{7}-\frac{208 \pi ^2}{21}+\frac{35590867}{1455300}-\frac{208 \log ^2(3)}{7} +\frac{15888496 \log (2)}{121275} \right. \nn \\
&+ \left. \left(-\frac{416\, \text{Li}_2\left(-\frac{1}{3}\right)}{7}-\frac{208 \pi ^2}{21}+\frac{35590867}{1455300}-\frac{208 \log ^2(3)}{7} +\frac{15888496 \log (2)}{121275}\right)n_f \right] C_F T_R^2 n_f \nn \\
&+ \left[-\frac{50592\, \text{Li}_2\left(-\frac{1}{3}\right)}{7}-\frac{8432 \pi ^2}{7}+\frac{173611871}{57750}-\frac{25296 \log ^2(3)}{7} +\frac{4368592 \log (2)}{275}\right] C_A^2 T_R n_f \nn \\
&+ \left[-\frac{41616\, \text{Li}_2\left(-\frac{1}{3}\right)}{7}-\frac{4896 \pi ^2}{5}+\frac{19423792}{8085}-\frac{20808 \log ^2(3)}{7} +\frac{35171776 \log (2)}{2695}\right] C_A^3 \nn \\
\beta_g =& \left[-\frac{2188}{5775}-\frac{256 \log (2)}{1925}\right] C_F C_A T_R n_f + \left[-6912\, \text{Li}_2\left(-\frac{1}{3}\right)-1152 \pi ^2+\frac{249317443}{86625} \right. \nn \\
&\left.-3456 \log ^2(3)+\frac{87702368 \log (2)}{5775}\right] C_A^2 T_R n_f + \left[\frac{1992}{1925}+\frac{3456 \log (2)}{1925}\right] C_A^3\,.
\eea

\section{Experimental Practicalities}\label{sec:exp}

The slices of configuration space which exactly correspond to the configurations defined in Section \ref{sec:proj} are of measure zero, making them impossible to observe in an actual measurement. Instead one must define some finite-width intervals over which to consider an observed configuration of particles to be ``close enough'' to the ideal configuration. It is in principle possible for the construction of these intervals to impose some bias on the final results, and so the definition must be chosen carefully. 

\subsection{Identifying Tees and Dipoles}

We use the following algorithm (written in pseducode) to identify tee and dipole configurations. 
    
\begin{algorithmic}[1]
\State $t \gets$ tolerance
\For{\{i, j, k, l\} $\in$ particles}
\For{(A,B,C,D) $\in$ \{(i,j,k,l),(i,k,j,l),(i,l,j,k)\}}
\If{d(A,B) $<$ d(C,D)}
\State swap((A,B), (C,D))
\EndIf
\State $m1\gets$ midpoint $\overline{AB}$
\State $m2\gets$ midpoint $\overline{CD}$ 
\If{d(m2, A) $<$ t OR d(m2,B) $<$ t}
\State (A,B,C,D) is a \textbf{tee}
\State $R_L \gets d(A,B)$
\State $r \gets d(C,D)/d(A,B)$
\State $\phi \gets $ the angle between $\overline{AB}, \overline{CD}$
\EndIf
\If{d(m1, m2) $<$ t}
\State (A, B,C, D) is a \textbf{dipole}
\State $R_L \gets d(A,B)$
\State $r \gets d(C,D)/d(A,B)$
\State $\phi \gets $ the angle between $\overline{AB}$ and$\overline{CD}$
\EndIf
\EndFor
\EndFor
\end{algorithmic}

Note that we loop over all possible pairings of the four points, and also that the identifications of ``tee'' and ``dipole'' are \textbf{not} mutually exclusive. This is necessary to avoid biasing the resulting angular distributions.

The binning variables live on the domains
\begin{align}
    0 \leq &R_L \leq \inf \\
    0 \leq &r \leq 1 \\
    0 \leq &\phi \leq \pi/2
\end{align}

A full implementation of this algorithm in \texttt{c++} is available \href{}{here}.

\subsubsection{Proof our Algorithm is Unbiased}
\label{sec:algoproof}
It is conceivable that selecting configurations in this matter could impose some angular bias on the final distributions. Thankfully it is easy to prove rigorously that our algorithm has no such issues. 

To see this, we consider the following question: say we have already sampled three points (A, B, C) in the $(\eta,\phi)$ plane from some random process (eg QCD jet formation). If we then sample a fourth point D, what is the probability that the resulting arrangement (A, B, C, D) will be identified as either a dipole or tee configuration? This probability will be
\begin{equation}
    P = \iint \rho(D | A,B,C) d\mu
\end{equation}
where the area integral is over all points $D$ such that the condition will be satisfied. Obviously we cannot compute this quantity without knowing the distribution $\rho$. However, for very small thresholds $t$ we can approximate that the probability is constant over the area, and the integral reduces to
\begin{equation}
    P = \rho \iint d\mu
\end{equation}
Thus we see that the probability (which is what we actually measure) is related to the underlying distribution, scaled by the geometric area available for the point $D$. How large is this area? 

\paragraph{The dipole configuration}

In order to simplify the algebra we assume that the underlying distribution is translation and rotation invariant, although the same argument will work regardless. Under these assumptions we can without loss of generality explicitly write the coordinates of all four points as
\begin{align}
    &A = (-R/2, 0)\\
    &B = (+R/2, 0)\\
    &C = (r \cos\phi, r\sin\phi)\\
    &D = (x, y)
\end{align}
The midpoints $m1$ and $m2$ are
\begin{align}
    &m1 = (0,0)\\
    &m2 = (\frac{x - r\cos\phi}{2}, \frac{y - r\sin\phi}{2})
\end{align}
We are interested in the distance between these points
\begin{equation}
    d(m1, m2)^2 = |m2|^2 = \left(\frac{x-r\cos\phi}{2}\right)^2 + \left(\frac{y-r\sin\phi}{2}\right)^2
\end{equation}

And our test is whether this distance is less than the threshold
\begin{equation}
    \left(\frac{x-r\cos\phi}{2}\right)^2 + \left(\frac{y-r\sin\phi}{2}\right)^2 < t^2
\end{equation}
We can reduce this inequality and solve it for $x$ and $y$ to obtain
\begin{align}
    &-2t + r \cos \phi < x < 2t + r \cos\phi \\
    &-\left[4t^2 - x^2 + 2 r x \cos\phi - r^2\cos^2\phi\right]^{1/2} + r\sin\phi < y\\
    &y < \left[4t^2 - x^2 + 2 r x \cos\phi - r^2\cos^2\phi\right]^{1/2} + r\sin\phi
\end{align}
We define new coordinates
\begin{align}
    &\chi = x + r\cos\phi\\
    &\gamma = y + r\sin\phi
\end{align}
This yields the inequalities
\begin{align}
    -2t < &\chi < 2t \\
    -\left[4t^2 - \chi^2\right]^{1/2}  < &\gamma < \left[4t^2 - \chi^2 \right]^{1/2}
\end{align}
Our geometric area integral then looks like
\begin{align}
    I&=\int_{-2t}^{+2t} d\chi \int_{-\left[4t^2 - \chi^2 \right]^{1/2}} ^{+\left[4t^2 - \chi^2 \right]^{1/2}} d\gamma \\
    &= 2 \int_{-2t}^{+2t} d\chi  \left[4t^2 - \chi^2 \right]^{1/2}
\end{align}
which evaluates to
\begin{align}
I = 4\pi t^2
\end{align}
Importantly, this quantity is \textbf{constant} with respect to the positions of the first three points. We can therefore conclude that there is no bias in the selection of dipole configurations.

\paragraph{The tee configuration}

Again we assume transnational and rotational invariance to simplify the algebra, although the same argument will apply in any case. Note that for the tee configuration the two tests we perform (on the two endpoints of the line $\overline{AB}$ are completely symmetric. We therefore need only consider one of the endpoints. 

We can without loss of generality write the coordinates of the four points as
\begin{align}
    &A = (0, 0)\\
    &B = (R, 0)\\
    &C = (r \cos\phi, r\sin\phi)\\
    &D = (x, y)
\end{align}
The midpoint of interest is
\begin{align}
    m2 = (\frac{x - r\cos\phi}{2}, \frac{y - r\sin\phi}{2})
\end{align}
And so if we perform the test w.r.t. the A endpoint we are interested in the distance
\begin{align}
    d(A, m2)^2 = |m2|^2 = \left(\frac{x-r\cos\phi}{2}\right)^2 + \left(\frac{y-r\sin\phi}{2}\right)^2
\end{align}
This is exactly the same distance as we considered above for the dipole. We can therefore conclude that there is similarly no bias in the selection of tee configurations.

\subsection{Identifying Triangles}

Because the ``triangle'' configuration places requirements on only three of the four particles it is not possible for the finite tolerances to bias the resulting angular distributions as could happen for the tees and dipoles. In the language of the proof above, this can be seen from the fact that the geometric area available for the fourth point $D$ is by construction independent of the other three points $A, B, C$, regardless of how they were identified.

For the figures shown in this paper, we use the following algorithm:
    
\begin{algorithmic}[1]
\State $t \gets$ tolerance
\State $LoM \gets 5/4$ 
\State $LoS \gets 5/3$
\For{\{i, j, k, l\} $\in$ particles}
\For{(A,B,C,D) $\in$ \{(i,j,k,l),(j,k,l,i),(k,l,i,j),)(l,i,j,k)\}}
\State sort (A, B, C) such that $d(A,B) > d(A,C) > d(B, C)$
\If{$|$d(A, B)/d(A,C) - LoM$|<$ t AND $|$d(A, B)/d(B,C) - LoS$|<$ t}
\State (A,B,C,D) is a \textbf{triangle}
\State $R_L \gets d(A,B)$
\State $r \gets d(C,D)/d(A,B)$
\State $\phi \gets $ the angle between $\overline{CD}$ and $\overline{AC}$
\EndIf
\EndFor
\EndFor
\end{algorithmic}
Note that we loop over all four possible triangles, so it is possible for a given set of four points to contribute more than one triangle (eg if $(i,j,k)$ and $(j,k,l)$ are both $(3,4,5)$ triangles). 

This configuration is a lot less symmetric than the tee and dipole configurations, so the domains for the binning variables are much larger:
\begin{align}
    0 \leq &R_L \leq \inf \\
    0 \leq &r \leq \inf \\
    0 \leq &\phi \leq 2\pi
\end{align}

The algorithm described here can of course be easily generalized to other triangle shapes by replacing the values of $LoM$ and $LoS$ with the appropriate side-length ratios. Note however that configurations with $LoM = 1$ or $LoS = 1$ have an ambiguity in the definition of $r$ and $\phi$ which may cause problems. This approach is therefore restricted to scalene triangle configurations.

\section{Monte Carlo Simulation Details}

The phenomenological study presented in this paper is designed to mimic as closely as possible a realistic analysis that could be performed at the LHC.

\subsection{Hard Scattering}

The hard scattering process is generated with Madgraph separately from the parton shower generation in Pythia and Herwig. For this generation, we perform a single Z+jet production with either a gluon or quark in the final state (5-flavors). 
The generation is done at leading order with only the tree level Z+jet diagrams included. Additionally, no matching is performed, given that only a single jet is produced in the final state.  The same events are used for both Pythia and Herwig showering. The resulting restrictions ensure that all of the parton shower variations that we consider have the same kinematics from the Madgraph matrix elements. The plots are presented separately for the two different processes: gluon jets correspond to the leading jet in  a Z boson in association with a gluon-initiated jet, and quark jets correspond to the leading production of a Z boson in association with a quark-initiated jet. In both cases the Z boson is required to have a $p_T$ greater than 150 GeV and to decay into two muons. 

\subsection{Parton Shower Settings}

In order to understand the dependence of our observables on the details on the QCD shower we then perform the parton shower with various settings of both Pythia and Herwig. Herwig is particularly useful, as it is possible to control exactly which splittings are available in the parton shower. Herwig also has a robust implementation of spin correlations in its parton shower code, which can be toggled on and off to observe the influence of gluon spin effects. 

For the default showering of Herwig, we use tune CH3\cite{CMS:2020dqt}. The spin correlations, and  are toggled  off through\\
{\tt set /Herwig/Shower/ShowerHandler:SpinCorrelations No } \\
\noindent
Additionally, the gluon splitting and quark splitting are turned off through \\
{\tt \scriptsize 
do /Herwig/Shower/SplittingGenerator:DeleteFinalSplitting g->g,g; /Herwig/Shower/GtoGGSudakov\\
do /Herwig/Shower/SplittingGenerator:DeleteInitialSplitting g->g,g; /Herwig/Shower/GtoGGSudakov\\
do /Herwig/Shower/SplittingGenerator:DeleteFinalSplitting g->u,ubar; /Herwig/Shower/GtoQQbarSudakov\\
do /Herwig/Shower/SplittingGenerator:DeleteFinalSplitting g->s,sbar; /Herwig/Shower/GtoQQbarSudakov \\
do /Herwig/Shower/SplittingGenerator:DeleteFinalSplitting g->c,cbar; /Herwig/Shower/GtoccbarSudakov \\
do /Herwig/Shower/SplittingGenerator:DeleteFinalSplitting g->b,bbar; /Herwig/Shower/GtobbbarSudakov\\
do /Herwig/Shower/SplittingGenerator:DeleteInitialSplitting g->u,ubar; /Herwig/Shower/GtoQQbarSudakov'\\
do /Herwig/Shower/SplittingGenerator:DeleteInitialSplitting g->s,sbar; /Herwig/Shower/GtoQQbarSudakov\\
do /Herwig/Shower/SplittingGenerator:DeleteInitialSplitting g->c,cbar; /Herwig/Shower/GtoccbarSudakov\\
do /Herwig/Shower/SplittingGenerator:DeleteInitialSplitting g->b,bbar; /Herwig/Shower/GtobbbarSudakov\\
}
\noindent
For Pythia, tune CP5 is utilized\cite{CMS:2022awf} as the default parton shower. Spin polarizations are turned on and off through the use of: \\
{\tt 
 TimeShower:phiPolAsym = off \\
 TimeShower:phiPolAsymHard = off \\
}
\noindent
Additionally gluon to quark decays are turned off through the parameter: \\
{\tt 
TimeShower:nGluonToQuark  = 0
}

\subsection{Event Selection and Jet Reconstruction}

We select MC events to mimic the trigger and event selections that might be made in an LHC Z+Jets analysis. In particular we select events with pairs of muons satisfying:
\begin{itemize}
    \item $|\eta| < 2.4$
    \item $p_T > 26\,(15)$ GeV for the leading (subleading) muon
    \item $|m_{\mu\mu} - M_Z| < 20$ GeV
\end{itemize}

We then construct jets with the anti-kt algorithm, with radius parameter R=0.8. These jets are further selected for EEC computation if they satisfy the following cuts:
\begin{itemize}
    \item $|\eta| < 1.7$
    \item $p_T > 30$ GeV
    \item $\Delta R({\rm jet},\,{\rm muon}) > 0.8$ for both ``triggering'' muons
    \item Number of constituent particles $\geq 2$ 
    \item Muon energy fraction $<0.8$
    \item Electron energy fraction $<0.8$
\end{itemize}

All of the jets that pass these cuts are then passed along to the EEC computation and used to fill the resulting histograms. 

\section{Exploration of Parton Shower Spin Effects}
\label{appendix:sctoggles}

In this Appendix, we validate our treatment of spin correlations in the Pythia and Herwig parton showers. First, using an observable from \cite{Hamilton:2021dyz}, we illustrate that by turning off both soft and collinear spin correlations, we are able to successfully reproduce the behavior of a ``scalar" gluon, justifying the use of this configuration for studying spin correlations in the four-point correlator, as was used in the text. 

After this justification, we also present additional plots showing the separate effects of collinear and soft spin correlations on the four-point correlator.

We begin by validating our understanding of soft and collinear spin correlations in both Pythia and Herwig.  To do so, we study the first two splittings in the parton shower history provided by the parton shower program (i.e. Pythia or Herwig):

\begin{enumerate}
    \item Reconstruct the leading anti-kt jet with radius 0.8, as described in Section \ref{sec:pheno}. 
    \item Identify the parton that initiated the jet as the highest-momentum parton in the parton shower history which is within $\Delta R < 0.4$ of the reconstructed jet. 
    \item Use the parton shower history to identify the first splitting undergone by this initial parton, skipping $1\rightarrow 1$ processes in the history. If there is no such splitting, or it is not a $1\rightarrow 2$ process, skip this event.
    \item Follow the softer of the two daughters, and identify the first splitting undergone by this daughter, again skipping any $1\rightarrow 1$ processes. Again, if there is no such splitting, or it is not a $1\rightarrow 2$ process, skip this event.
\end{enumerate}

This identifies an initial $A\rightarrow B (C\rightarrow D E)$ process in the parton shower. We can then study the properties of the two identified splittings:
\begin{align}
    z &= \frac{p_T^{\text{softer}}}{p_T^{\text{softer}} + p_T^{\text{harder}}}\\
    k_T &= p_T^{\text{softer}} \cdot \Delta R\left(\text{softer}, \text{harder}\right)
\end{align}
where ``harder'' and ``softer'' refer to the harder and softer of the two splitting daughters. Note that in this scheme $z$ is restricted to the domain $0<z<0.5$. 

The influence of spin correlations will appear in the relationship between the two splittings. In particular, we compute the angle between the splitting planes defined by $A\rightarrow BC$ and $C\rightarrow DE$ as follows:
\begin{enumerate}
    \item Characterize the splitting plane of $A\rightarrow BC$ with the normal vector
    \begin{equation}
        \hat{n}_1 = \frac{\vec{p}_B \times \vec{p}_C}{|\vec{p}_B \times \vec{p}_C|}
    \end{equation}
    \item Characterize the splitting plane of $C\rightarrow DE$ with the normal vector
    \begin{equation}
        \hat{n}_2 = \frac{\vec{p}_D \times \vec{p}_E}{|\vec{p}_D \times \vec{p}_E|}
    \end{equation}
    \item Computed the angle $\Delta\Psi$ according to 
    \begin{equation}
        \Delta\Psi = \cos^{-1}\left[\hat{n}_1 \cdot \hat{n}_2\right]
    \end{equation}
\end{enumerate}
where $\vec{p}_i$ denotes the three-momentum of parton $i$. 

If there were no spin correlations (i.e. if the gluon where spin-0), the distribution of $\Delta \Psi$ would be flat, as there would be no preference for any given direction in the splitting. 
If, however, there are spin correlations, these should manifest as modulations in the $\Delta \Psi$ distribution. 

\begin{figure}[h]
    \centering
    \includegraphics[width=0.49\textwidth]{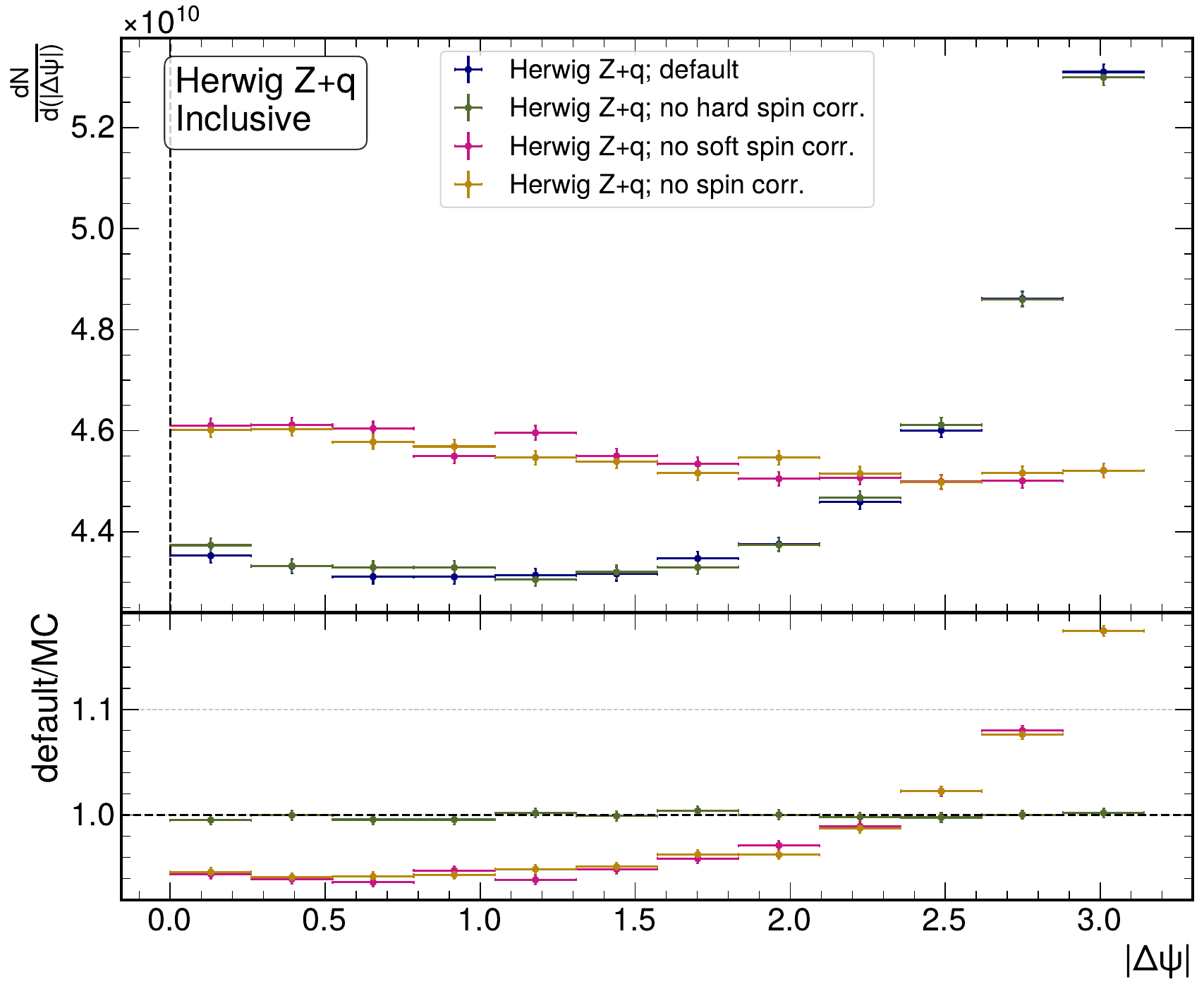}
    \includegraphics[width=0.49\textwidth]{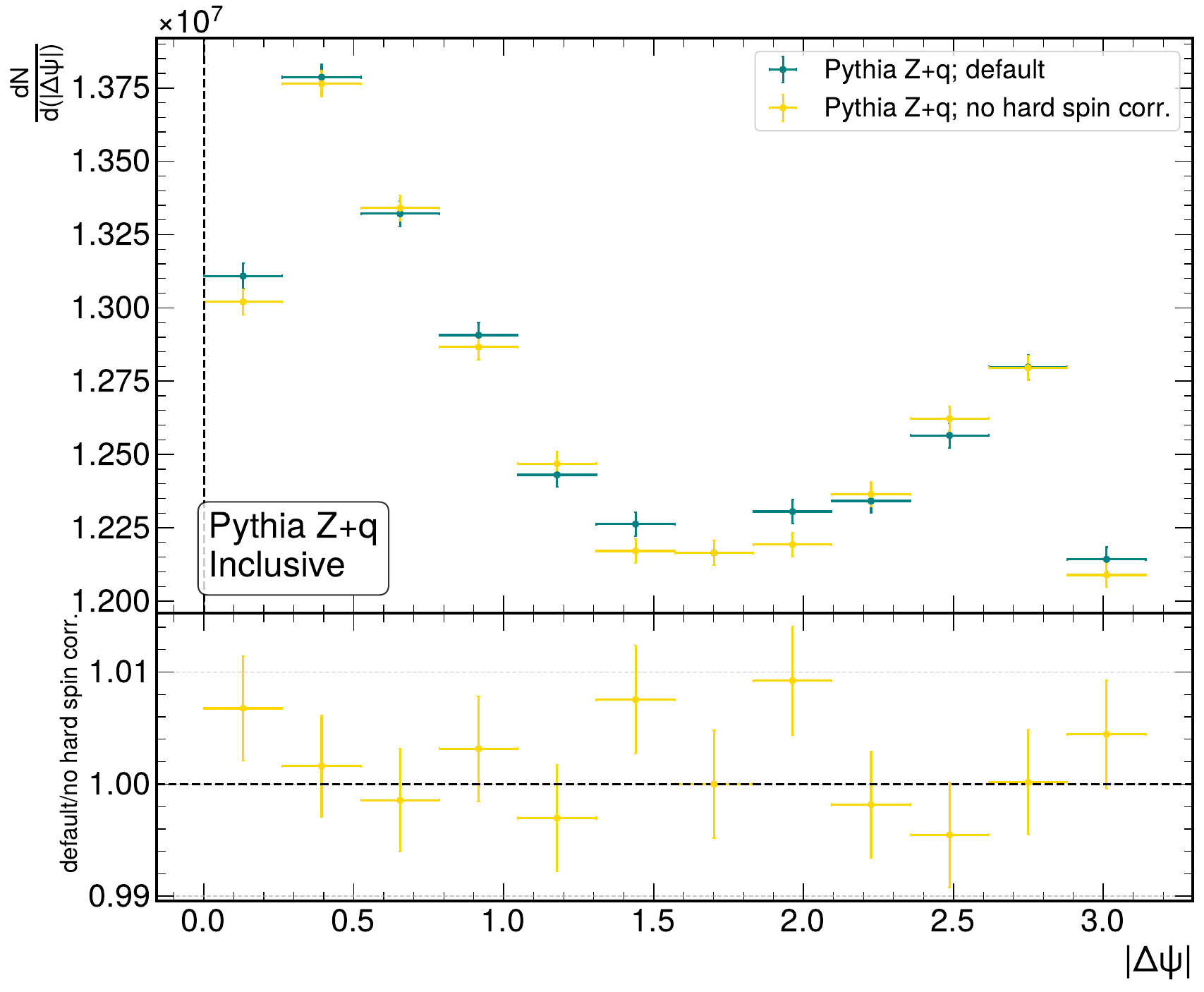}
    \includegraphics[width=0.49\textwidth]{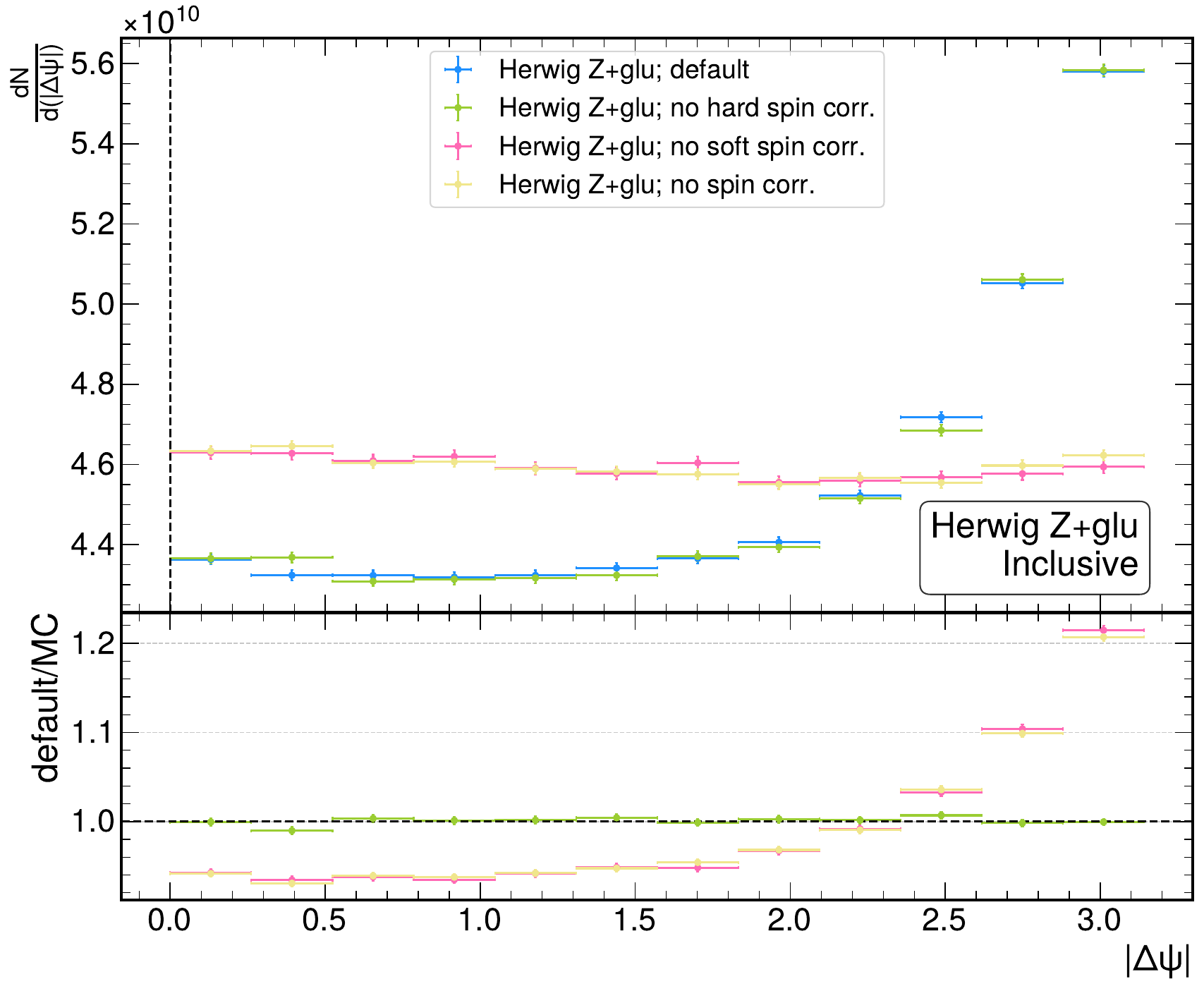}
    \includegraphics[width=0.49\textwidth]{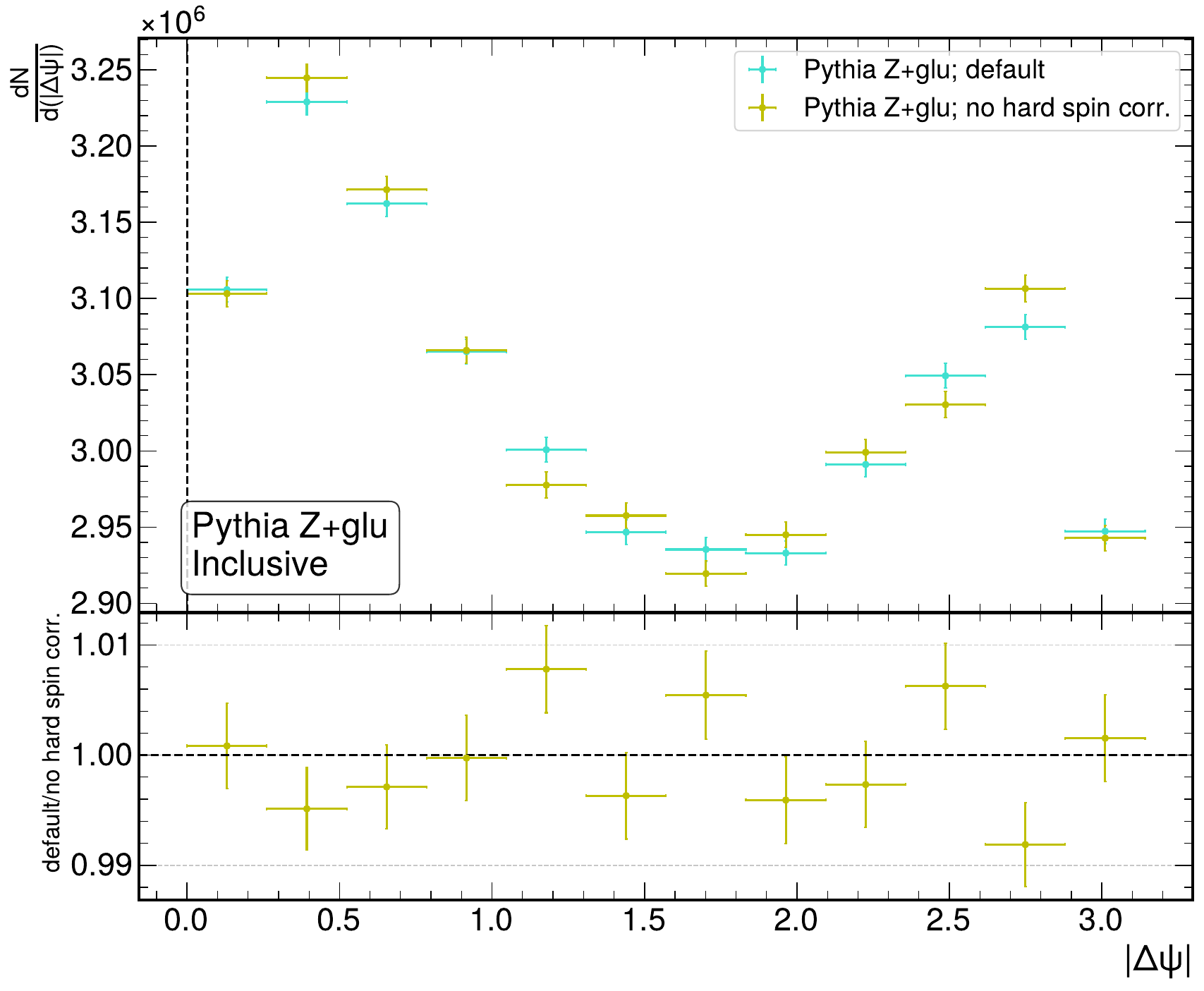}
    \caption{
        The distribution of the angle between the splitting planes of the first two splittings in the parton shower history.
        \textbf{Left:} Herwig; \textbf{Right:} Pythia. \textbf{Top:} $Z+q$ simulation. \textbf{Bottom:} $Z+g$ simulation.
    }
    \label{fig:total_spin_correlations}
\end{figure}

\begin{figure}[h]
    \centering
    \includegraphics[width=0.3\textwidth]{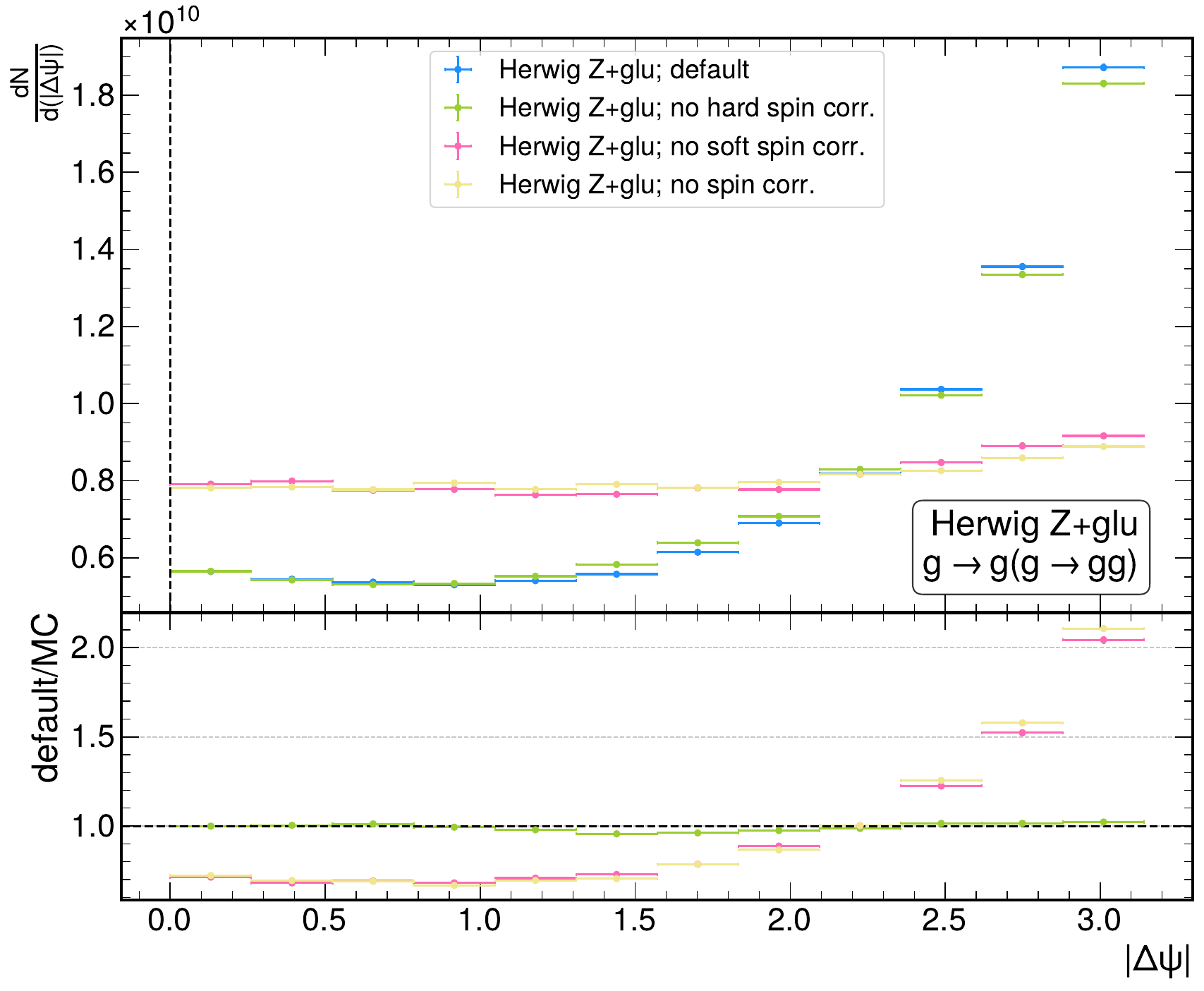}
    \includegraphics[width=0.3\textwidth]{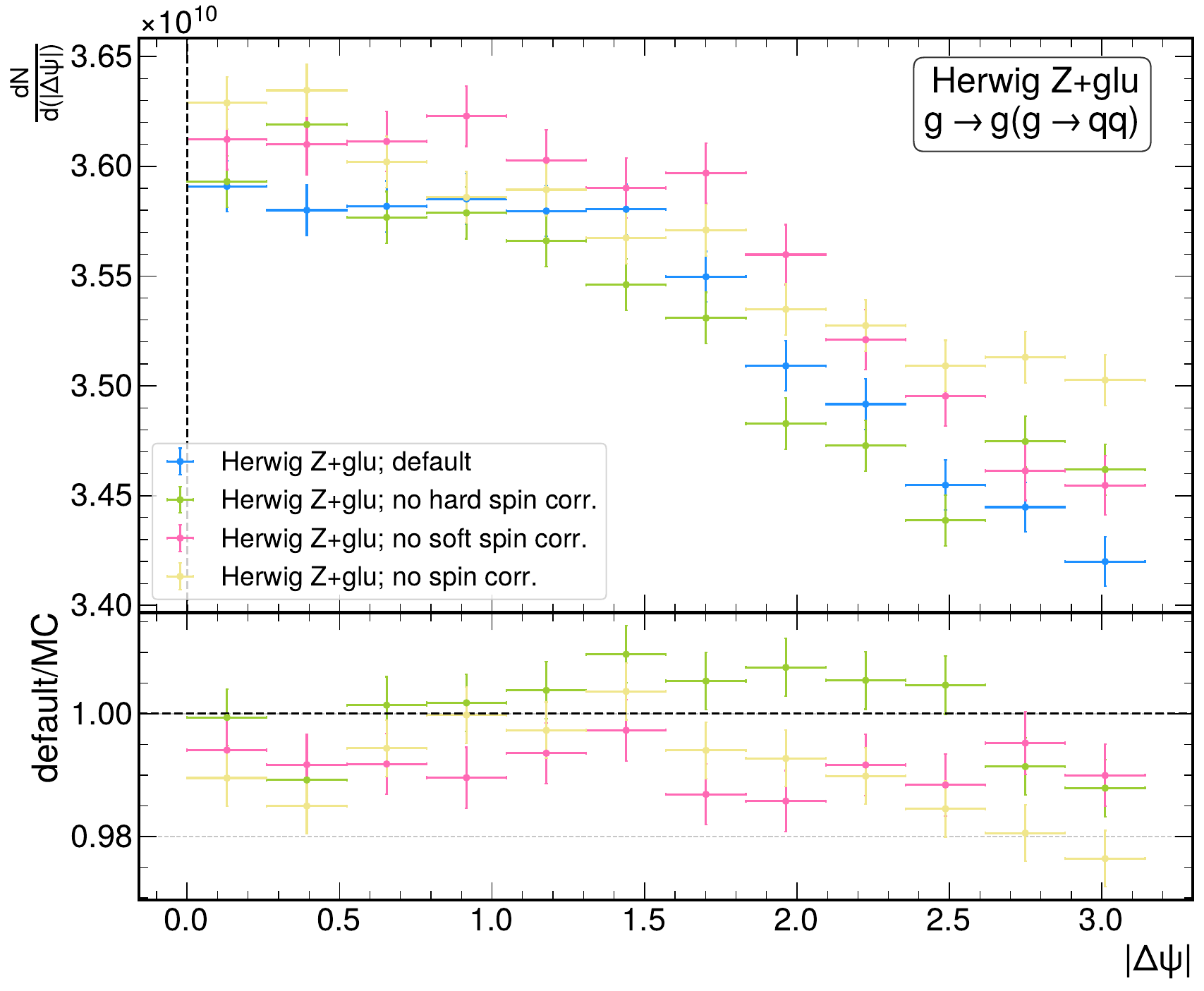}
    \includegraphics[width=0.3\textwidth]{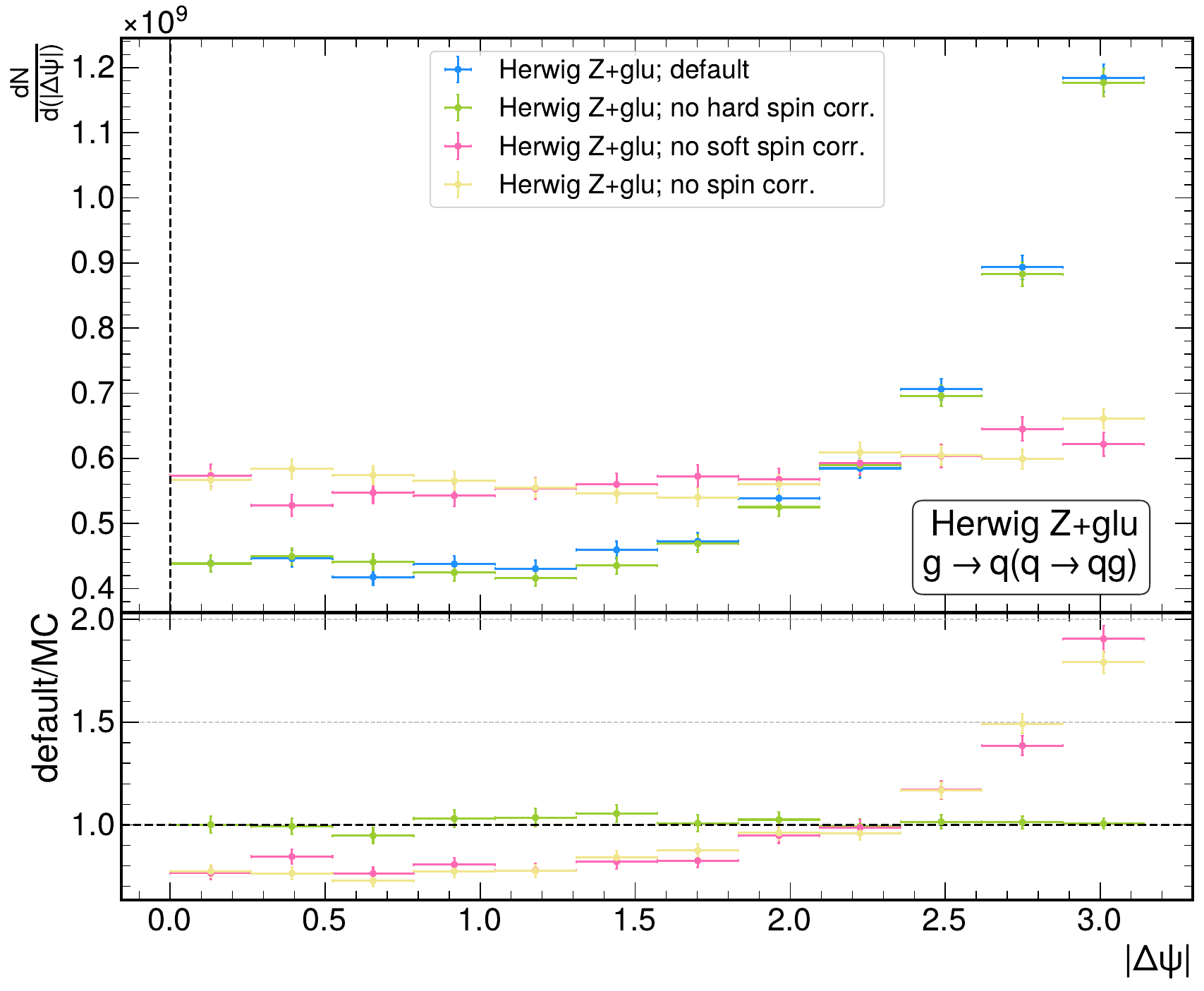}
    \includegraphics[width=0.3\textwidth]{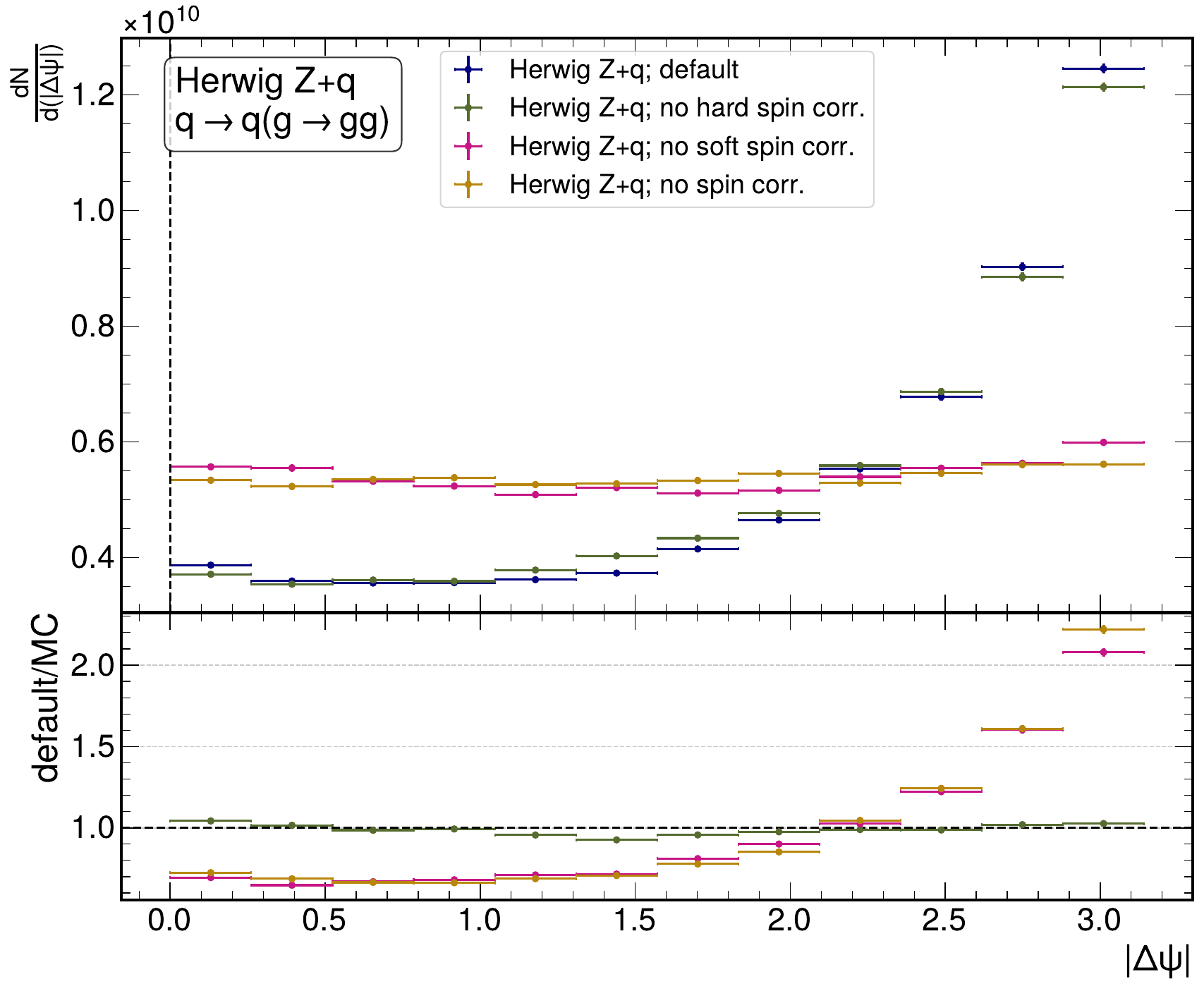}
    \includegraphics[width=0.3\textwidth]{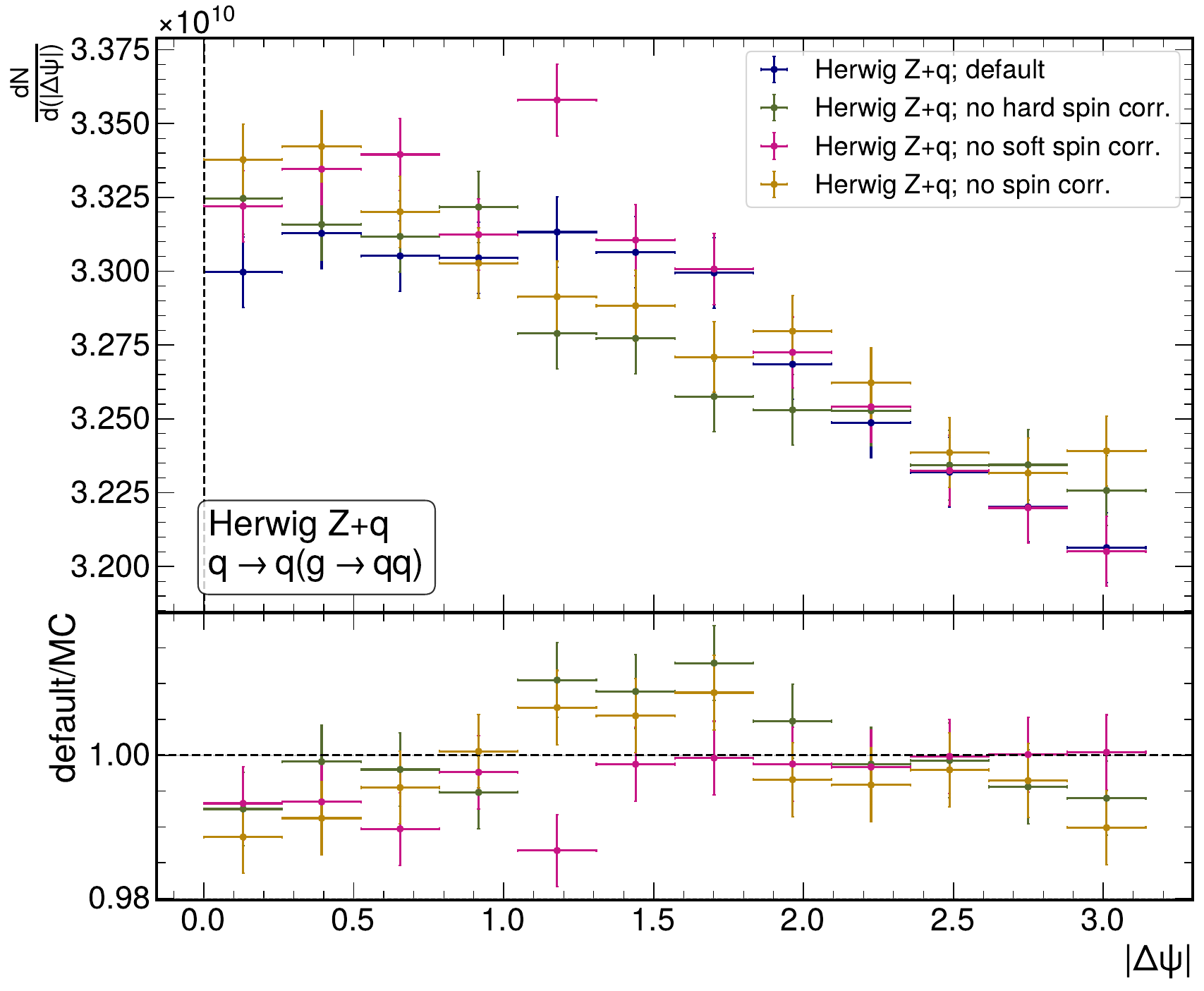}
    \includegraphics[width=0.3\textwidth]{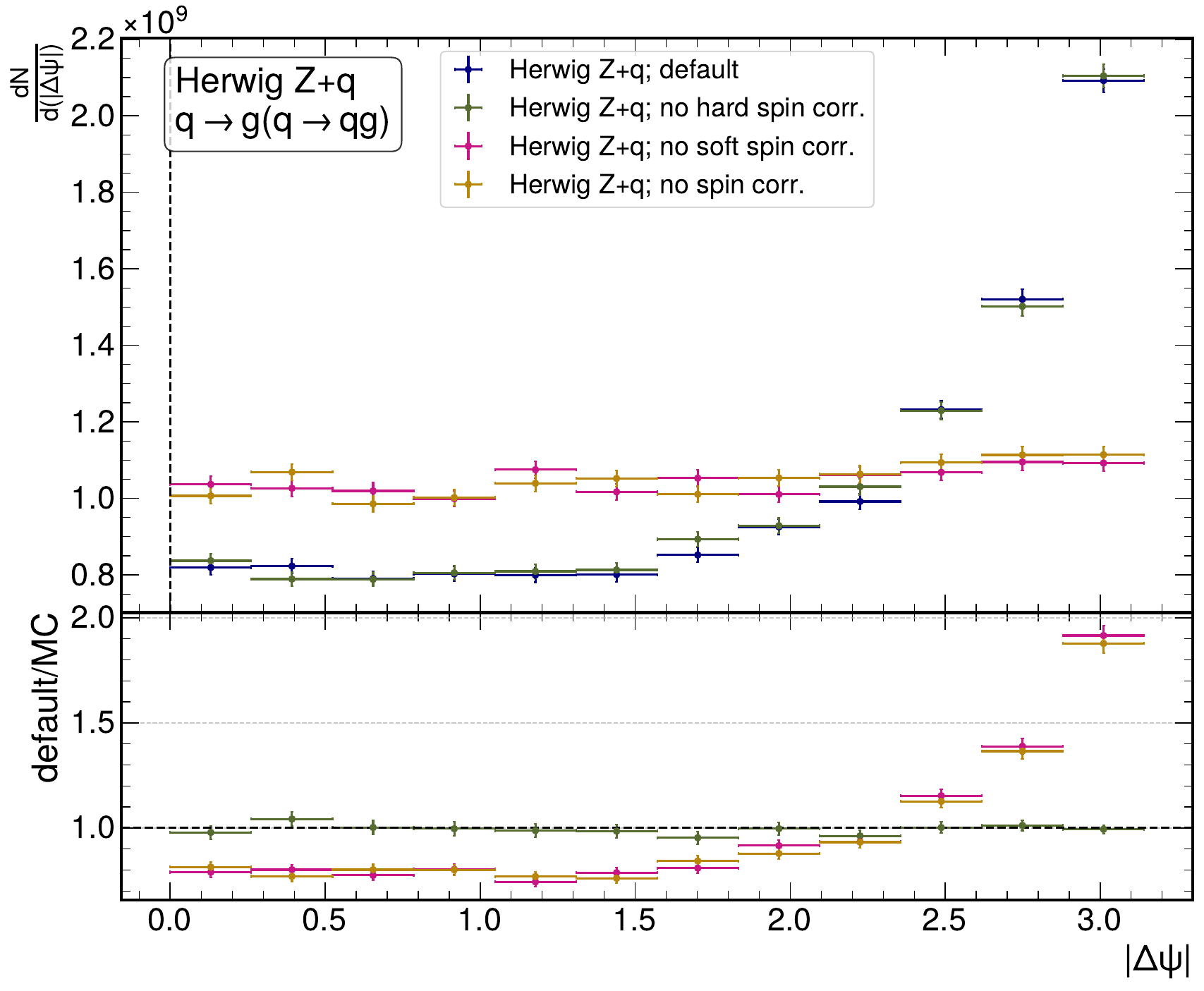}
    \caption{
        The distribution of the angle between the splitting planes of the first two splittings in the Herwig parton shower history, separated by splitting flavor.
        Splittings with intermediate quarks show no hard spin correlations, while splittings with intermediate gluons show the expected $\cos2\Delta\Psi$ modulation, with opposite sign for $g\rightarrow gg$ and $g\rightarrow qq$ splittings.
    }
    \label{fig:herwig_resolved_spin_correlations}
\end{figure}

\begin{figure}[h]
    \centering
    \includegraphics[width=0.3\textwidth]{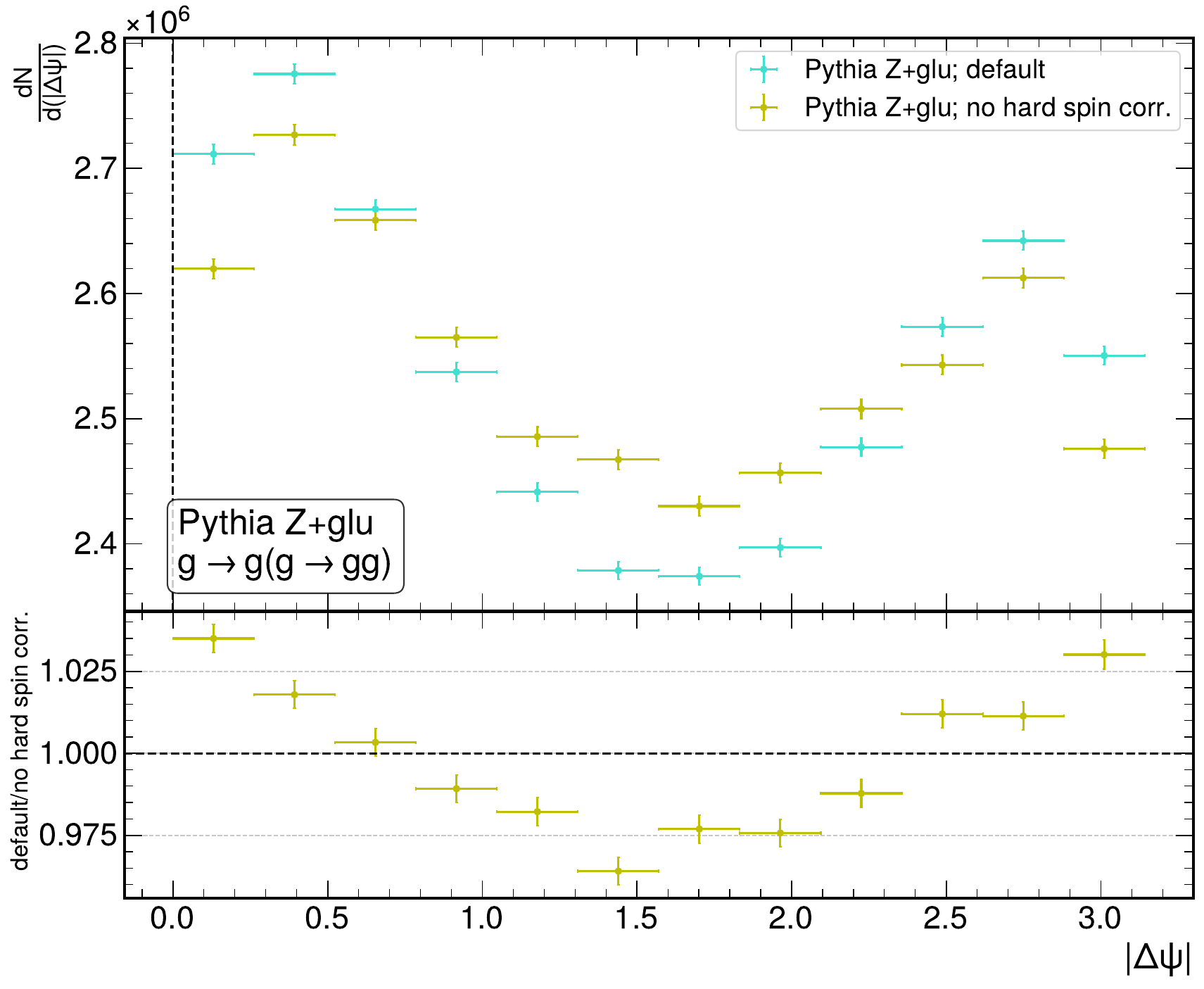}
    \includegraphics[width=0.3\textwidth]{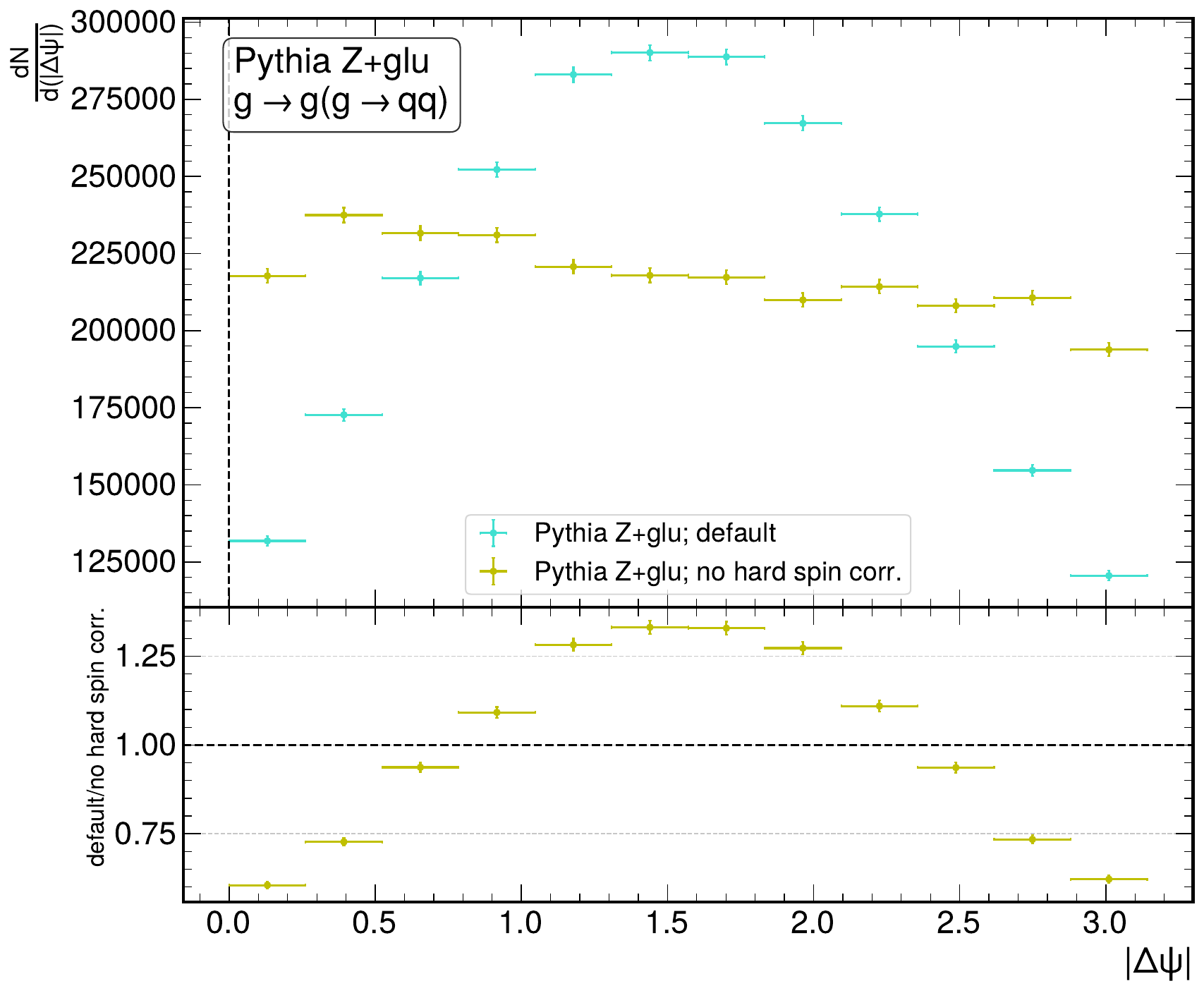}
    \includegraphics[width=0.3\textwidth]{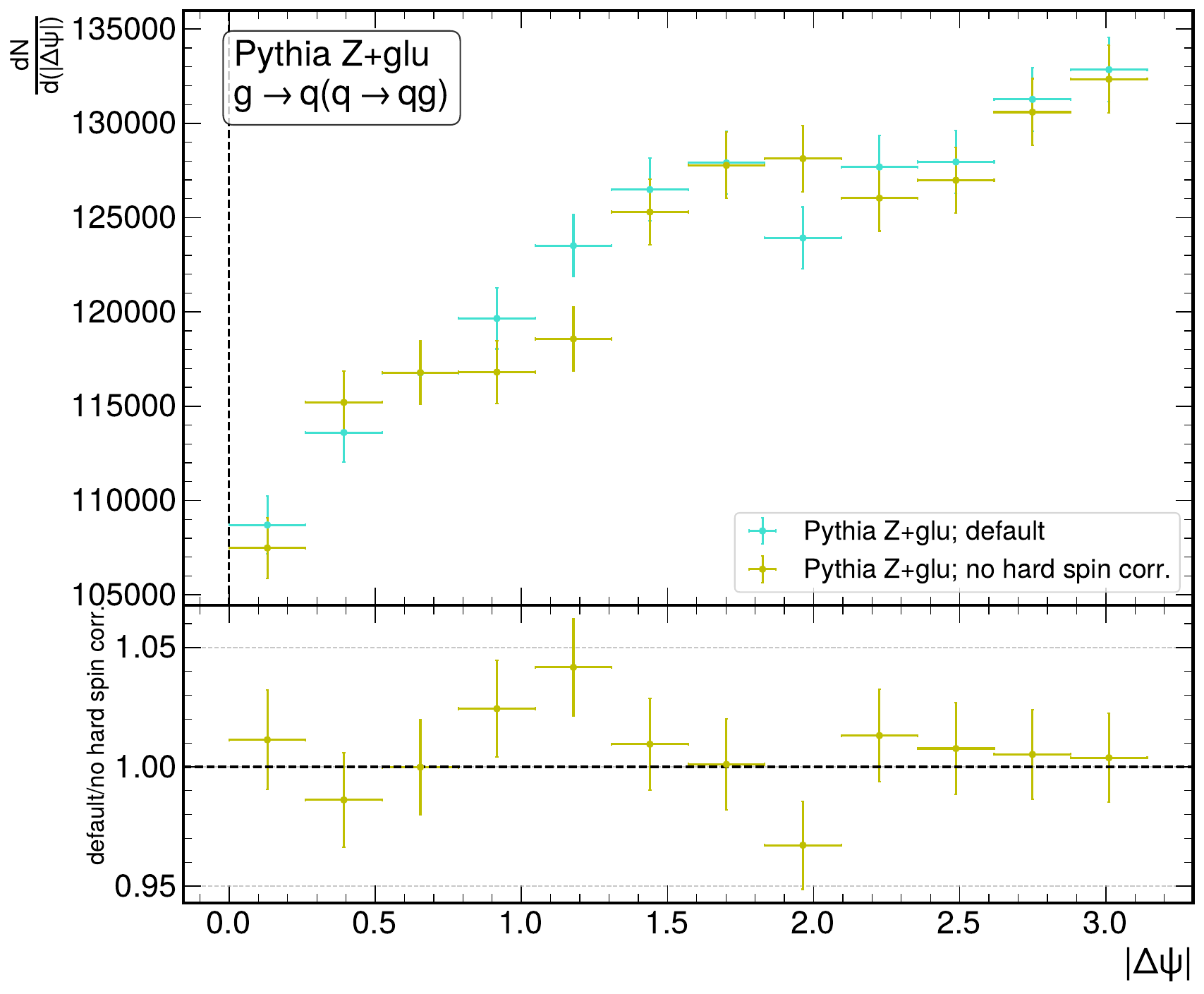}
    \includegraphics[width=0.3\textwidth]{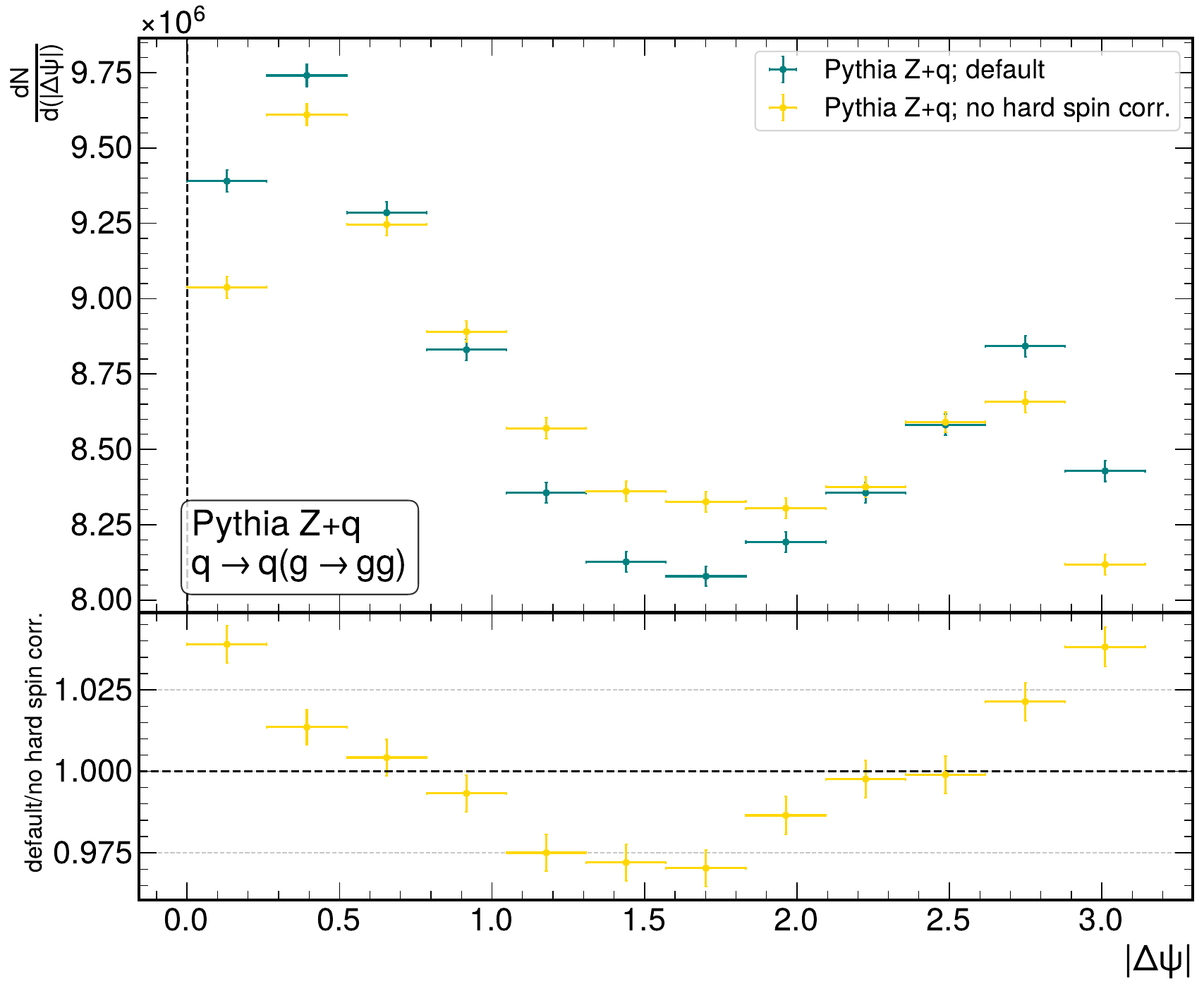}
    \includegraphics[width=0.3\textwidth]{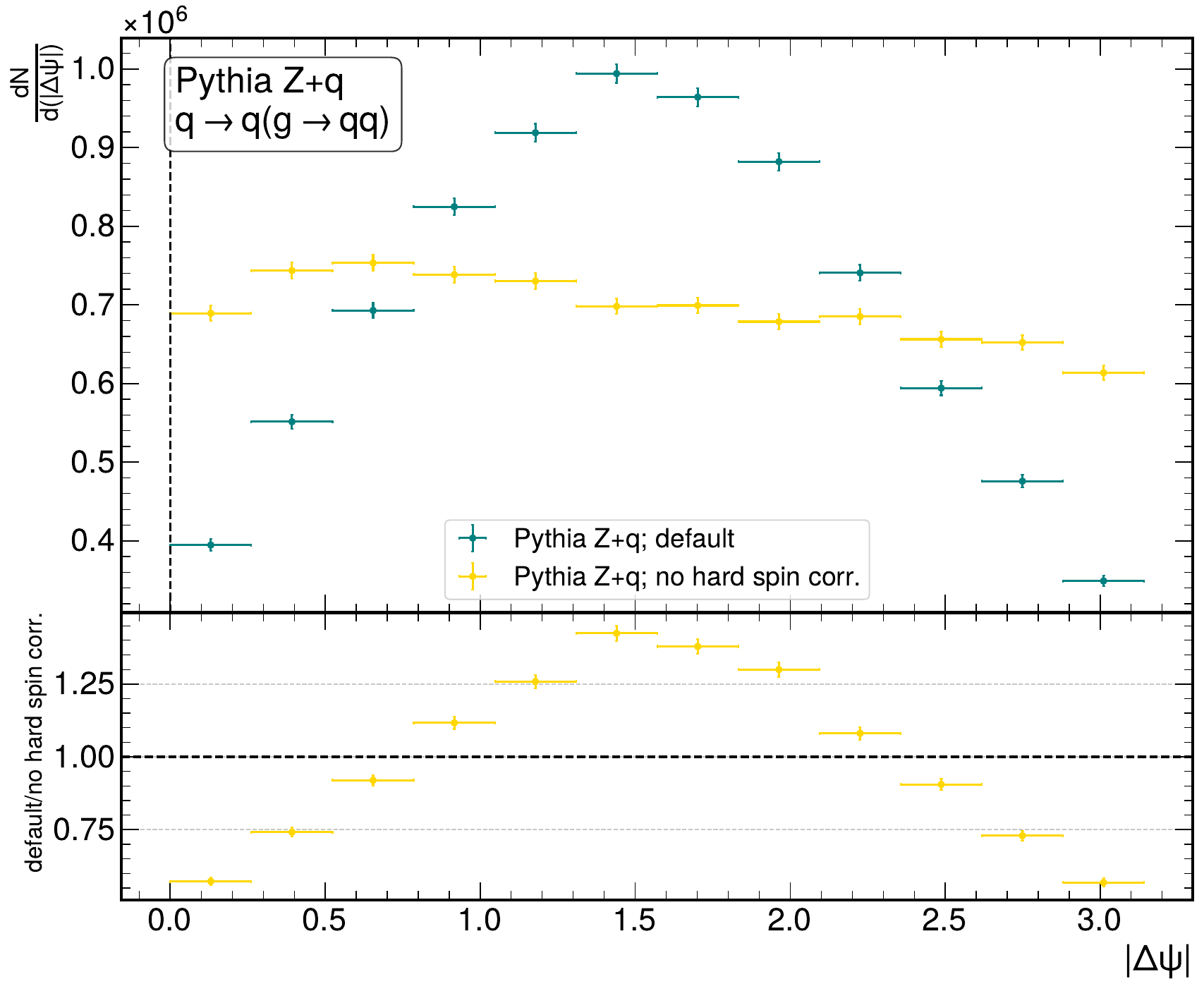}
    \includegraphics[width=0.3\textwidth]{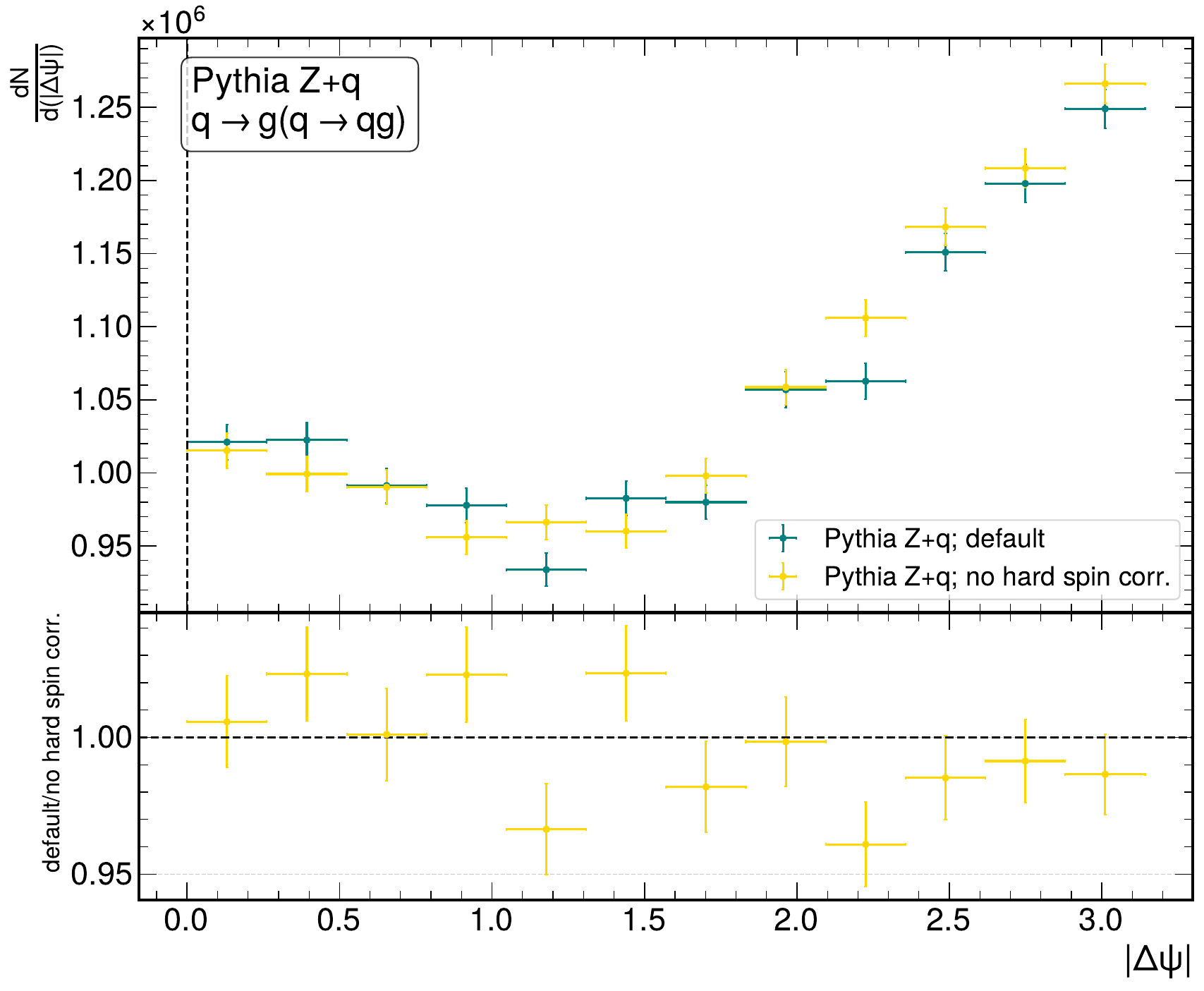}
    \caption{
        The distribution of the angle between the splitting planes of the first two splittings in the Pythia parton shower history, separated by splitting flavor.
        Splittings with intermediate quarks show no hard spin correlations, while splittings with intermediate gluons show the expected $\cos2\Delta\Psi$ modulation, with opposite sign for $g\rightarrow gg$ and $g\rightarrow qq$ splittings.
    }
    \label{fig:pythiaresolved_spin_correlations}
\end{figure}

Figure~\ref{fig:total_spin_correlations} shows the distribution of $\Delta \Psi$ for jets in $Z+q$ simulation, for both Pythia and Herwig, with and without spin correlations.
The soft spin correlations cause a large modulation in the $\Delta \Psi$ distribution in both Pythia and Herwig, and toggling off this effect in the Herwig shower reproduces the expected flat distribution as if the gluon were a scalar. 
On the other hand, the hard spin correlations have a much smaller effect, which is less than the statistical uncertainty. 
This is due to the fact that the hard spin correlations are maximized in the $z \rightarrow 0.5$ limit, while the soft spin correlations are maximized in the $z \rightarrow 0$ limit, and therefore benefit from the infrared singularity in the splitting function.
The residual hard spin correlations effect is further suppressed by the fact that it appears with different sign for $g \rightarrow qq$ and $g \rightarrow gg$ splittings, washing out the effect in the total distribution.
Figures~\ref{fig:herwig_resolved_spin_correlations} and \ref{fig:pythiaresolved_spin_correlations} show the $\Delta \Psi$ distributions for the different splitting flavors in Herwig and Pythia, respectively, revealing the influence of the hard spin correlations effect for splittings with intermediate gluons.

Having validated our understanding of soft and collinear spin correlations in Pythia and Herwig, we now provide additional plots, complimenting \Fig{fig:tee_herwigG_spin_intext} presented in the main text.

\begin{figure}[h]
    \centering
    \includegraphics[width=0.4\textwidth]{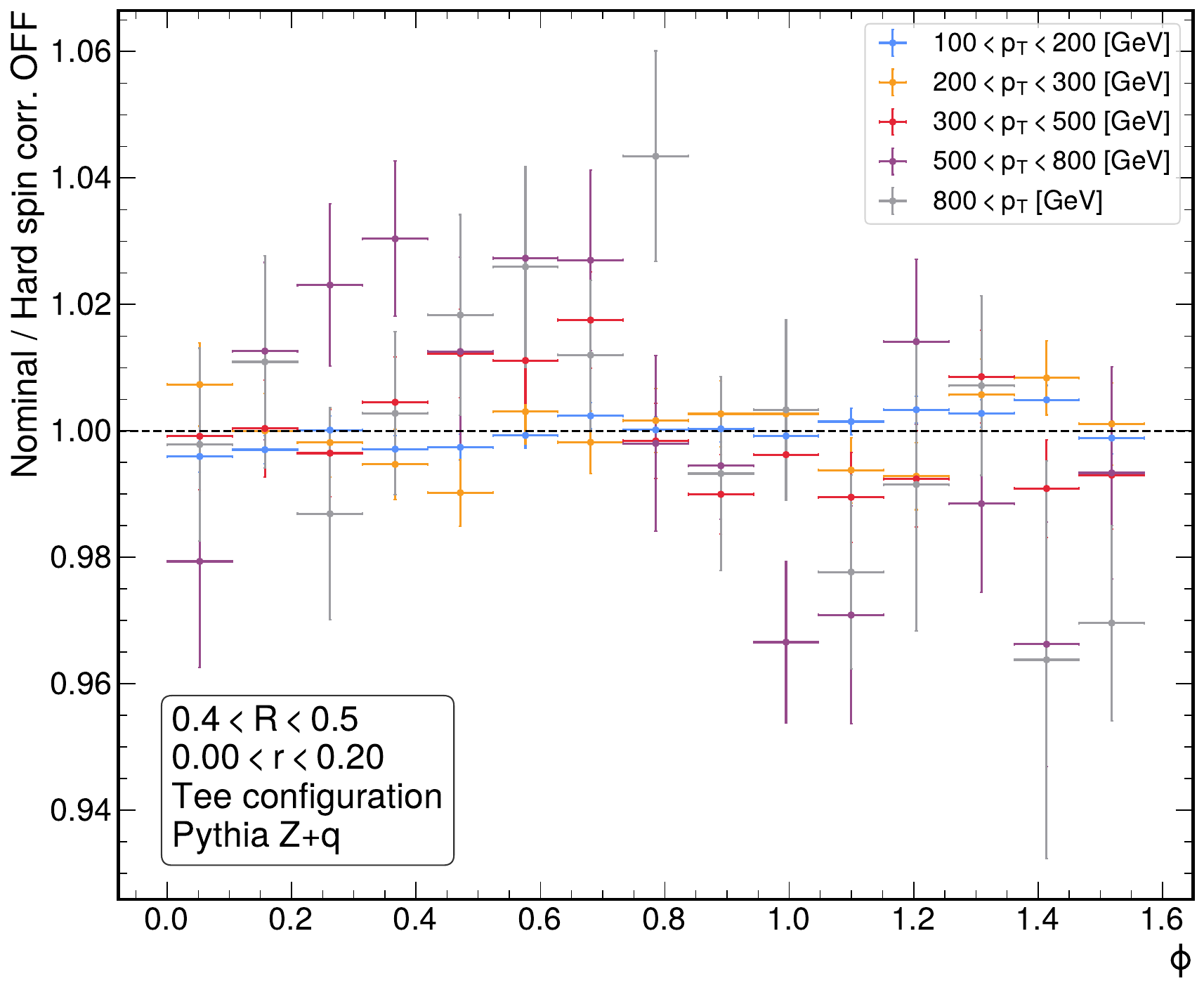} 
    \includegraphics[width=0.4\textwidth]{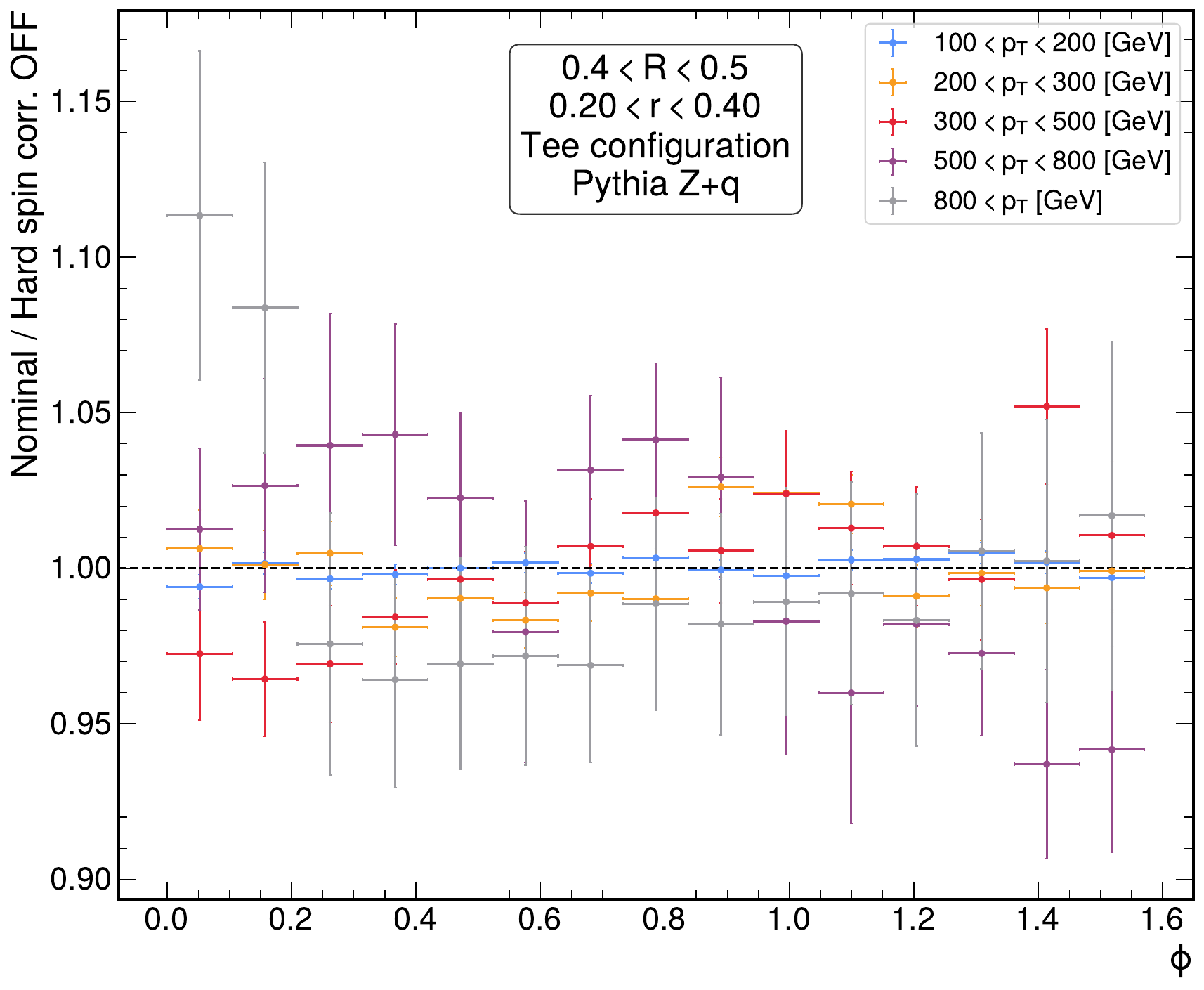} 
    \includegraphics[width=0.4\textwidth]{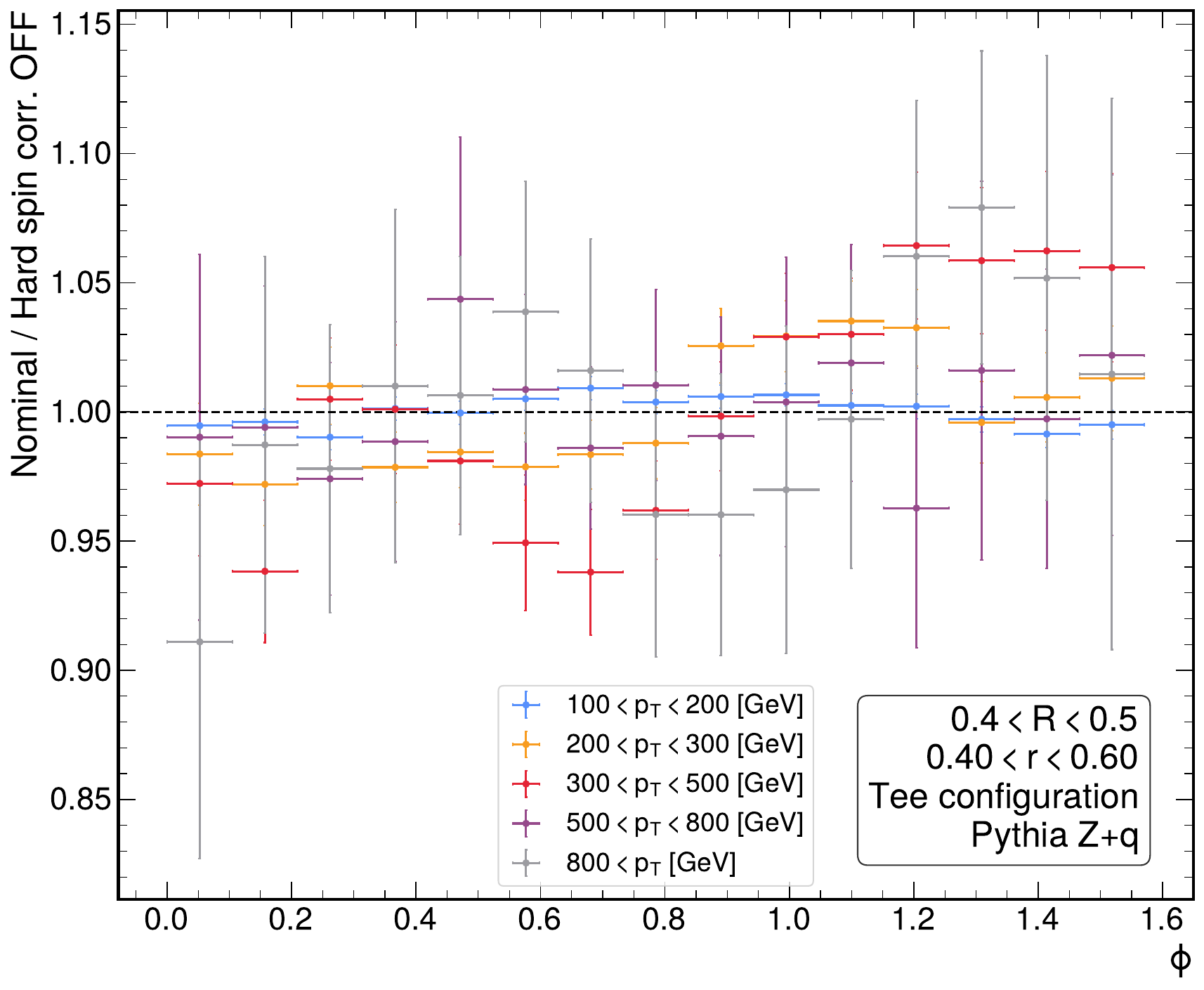} 
    \includegraphics[width=0.4\textwidth]{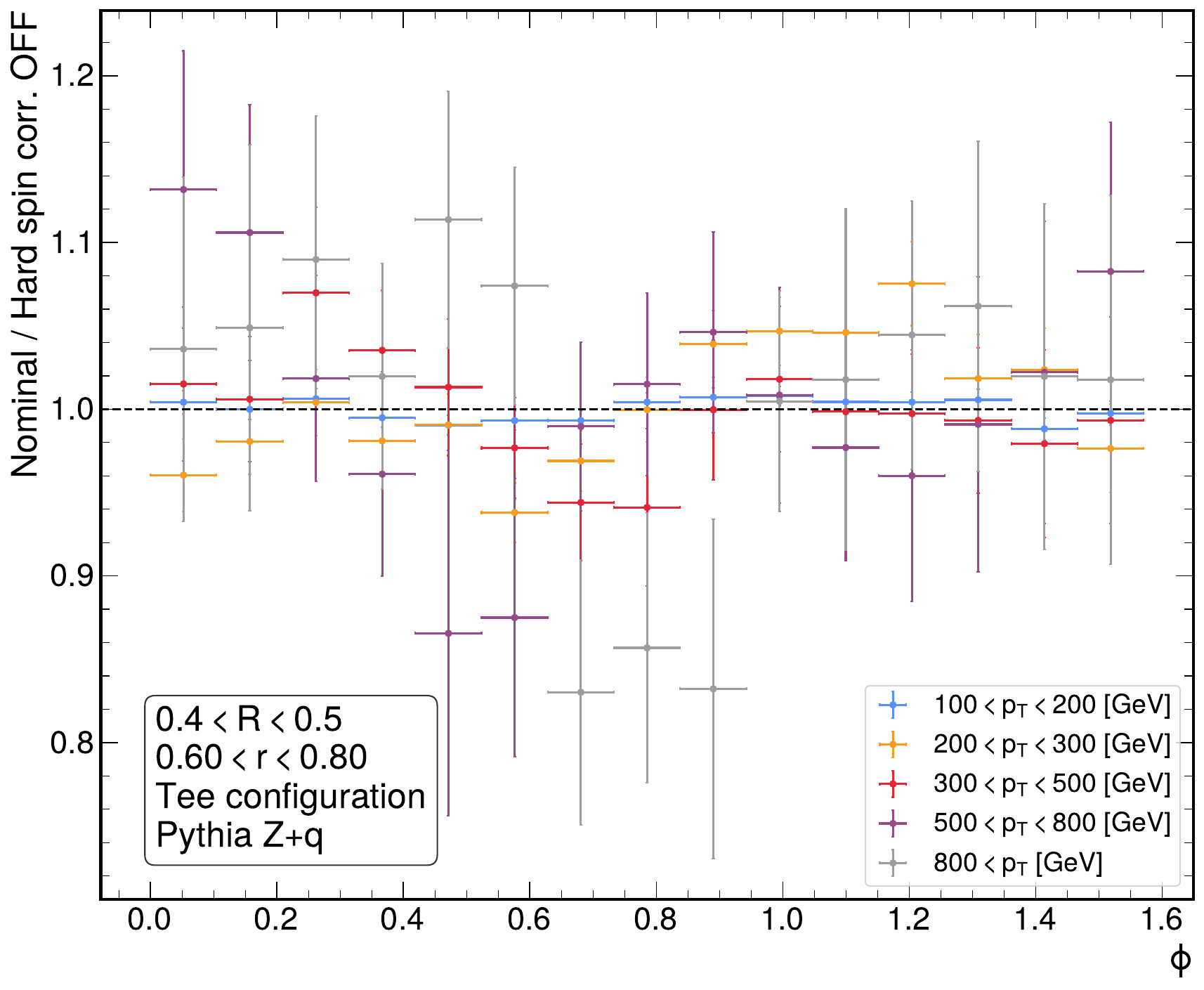} 
    \caption{Ratio between the tee distributions in Pythia $Z+q$ events with and without hard spin correlations. }
    \label{fig:tee_pythiaQ_hardspin}
\end{figure}

\begin{figure}[h]
    \centering
    \includegraphics[width=0.4\textwidth]{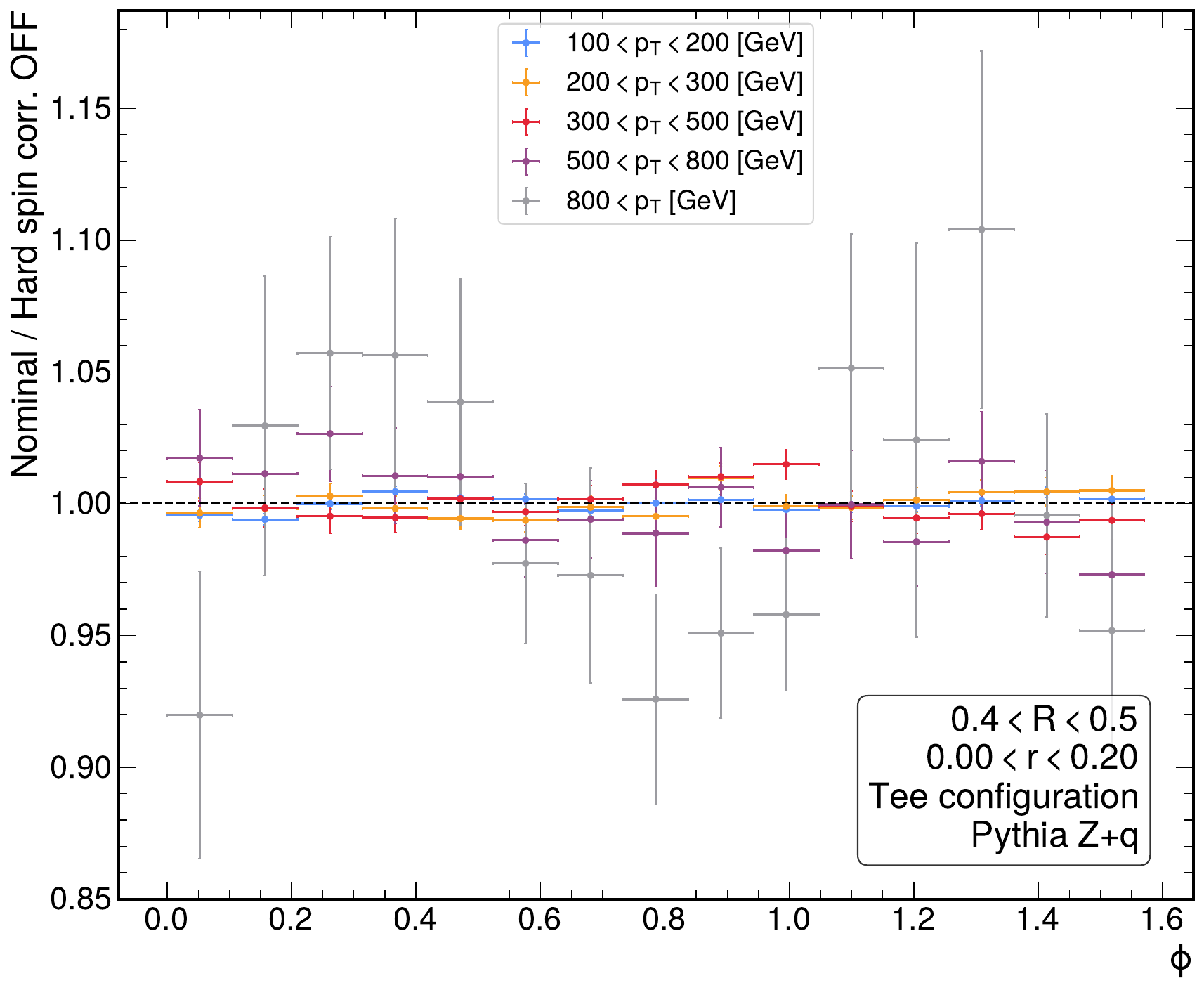} 
    \includegraphics[width=0.4\textwidth]{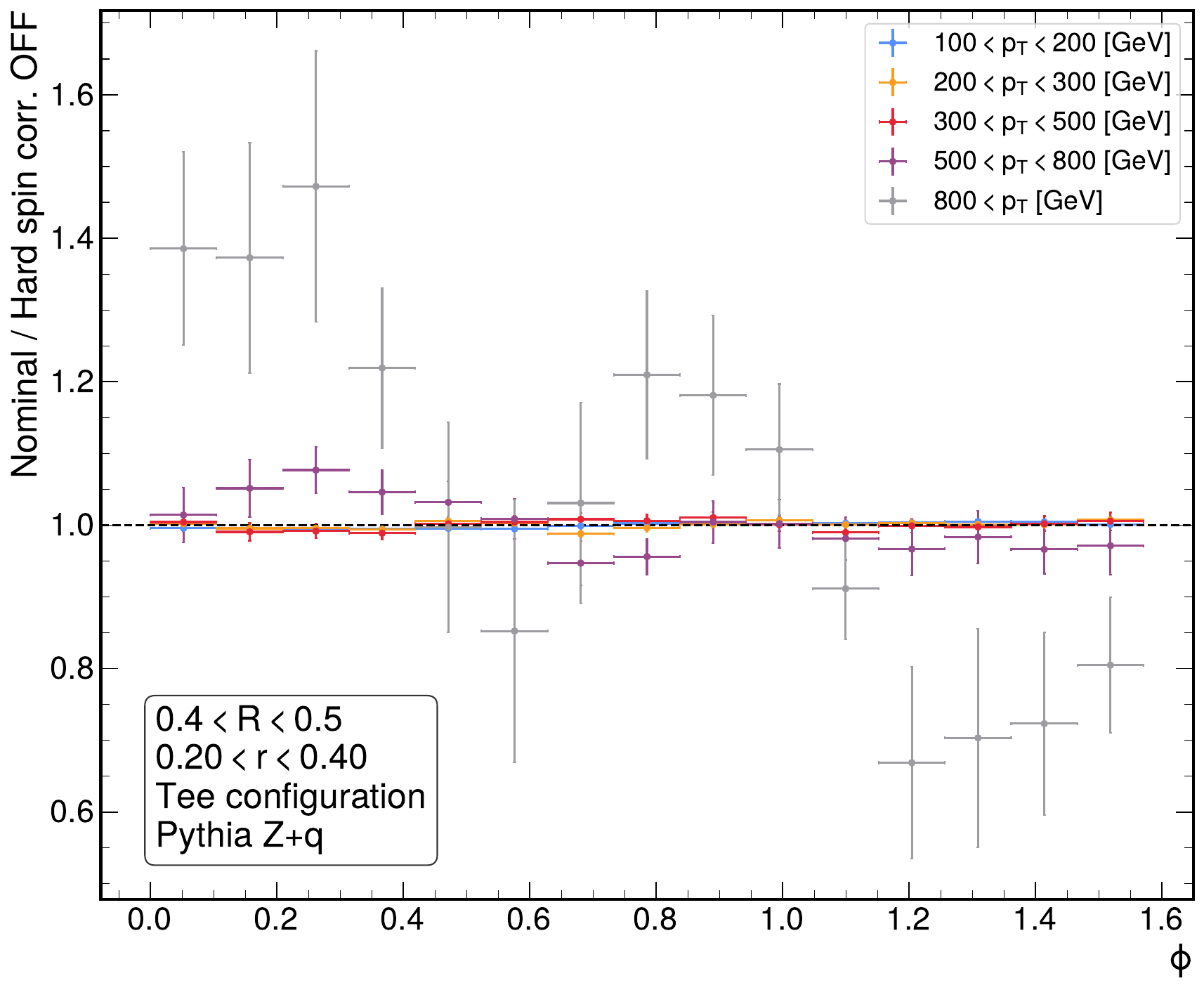} 
    \includegraphics[width=0.4\textwidth]{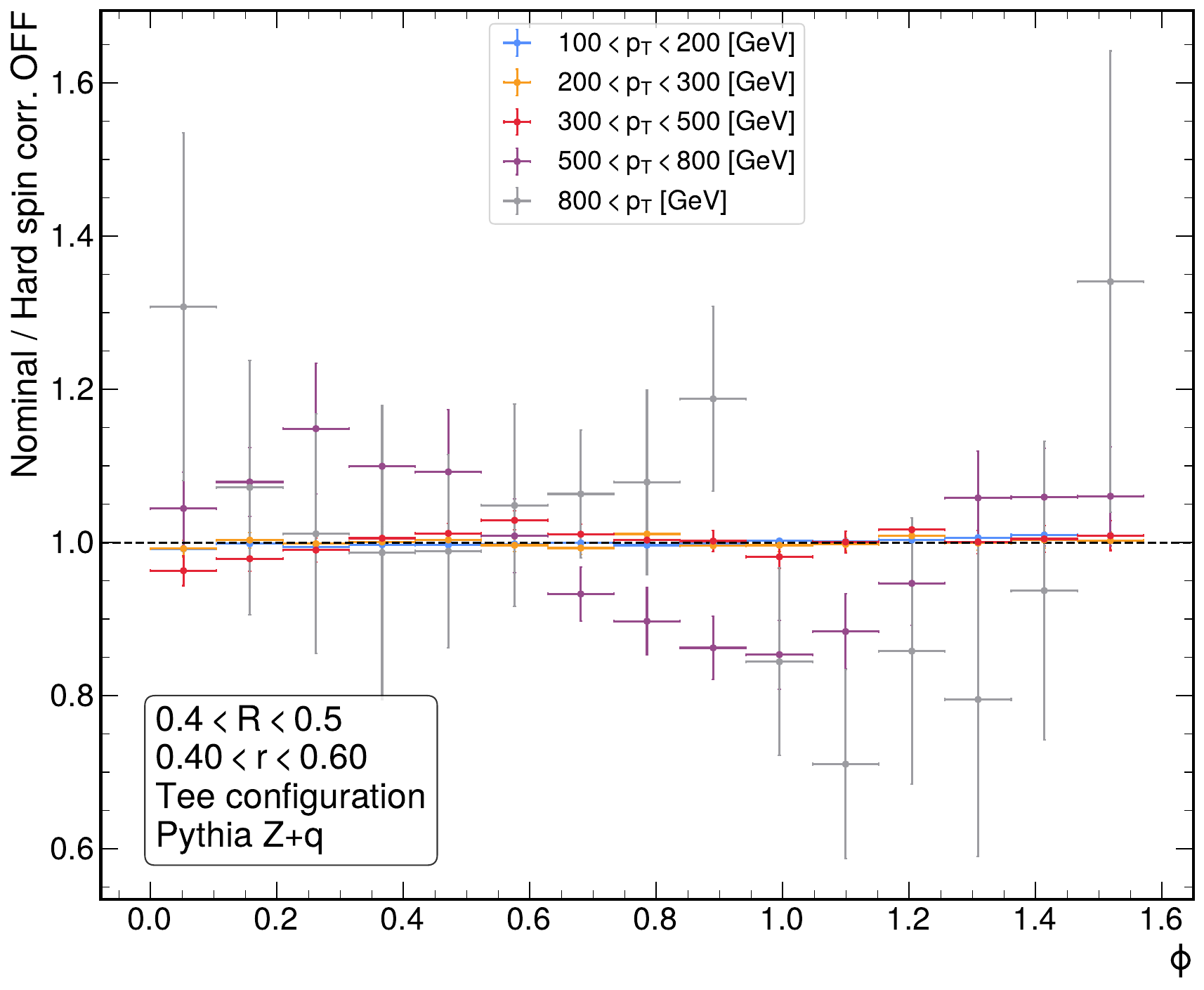} 
    \includegraphics[width=0.4\textwidth]{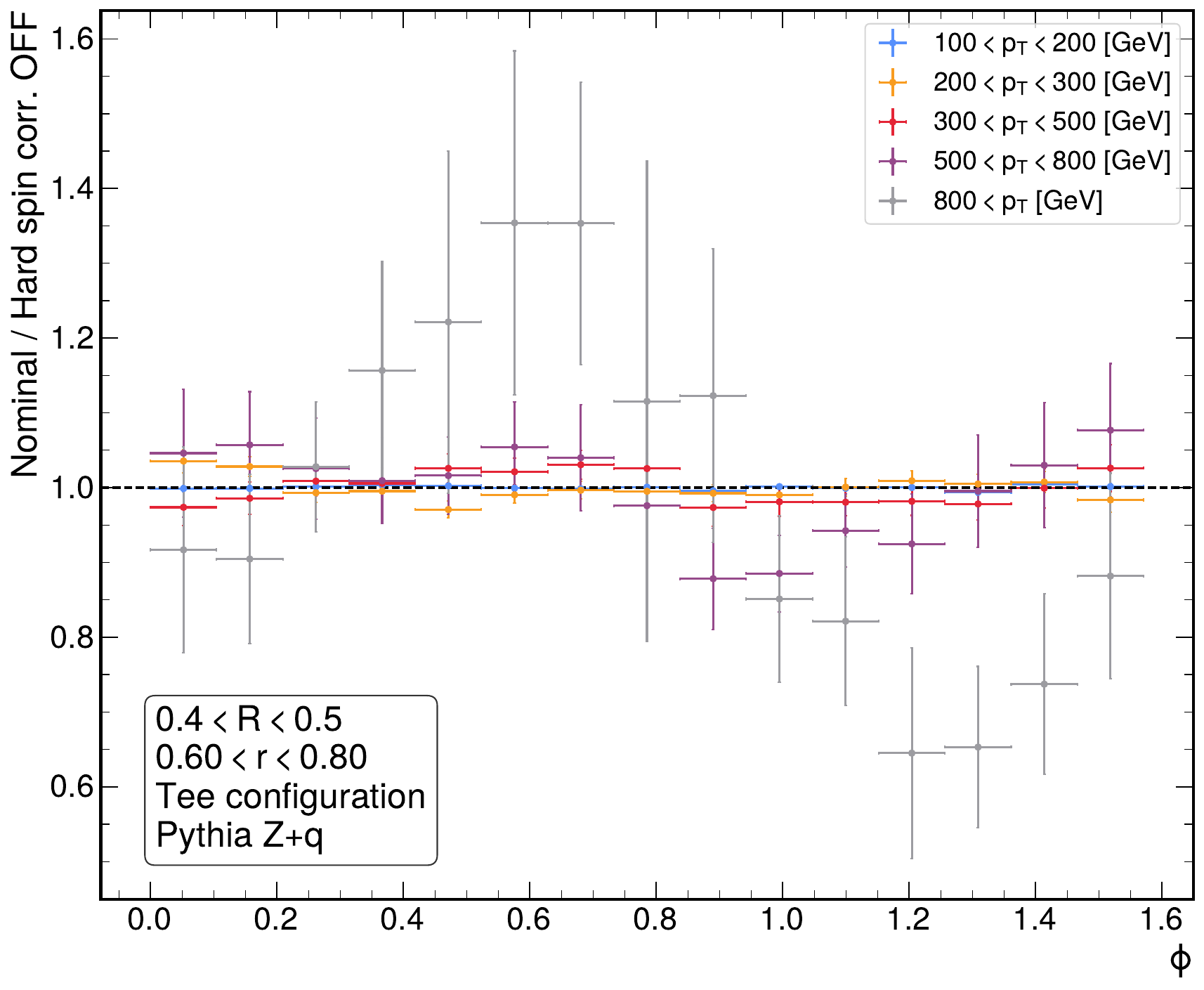} 
    \caption{Ratio between the tee distributions in Pythia $Z+g$ events with and without hard spin correlations.}
    \label{fig:tee_pythiaG_hardspin}
\end{figure}

\begin{figure}[h]
    \centering
    \includegraphics[width=0.4\textwidth]{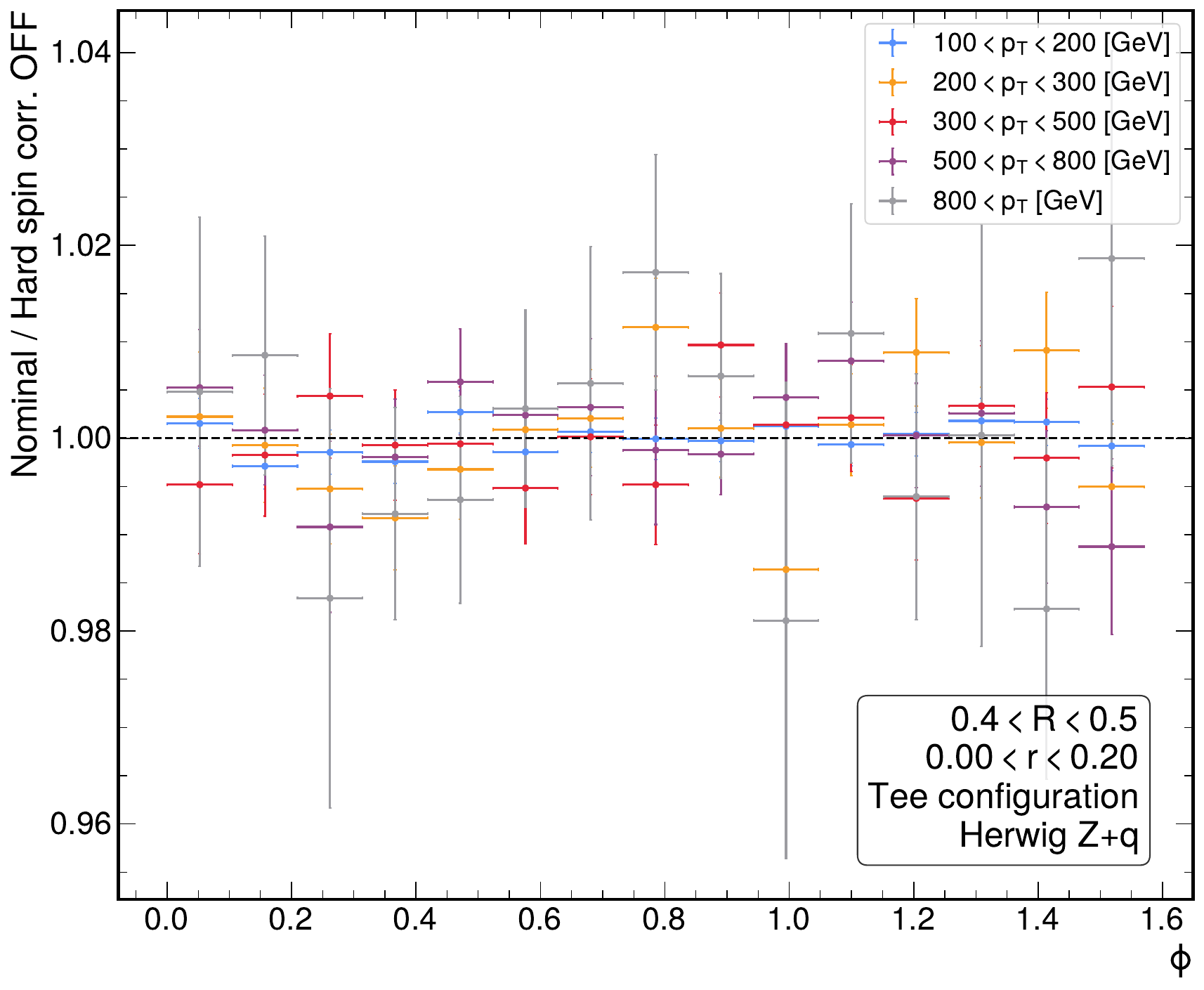} 
    \includegraphics[width=0.4\textwidth]{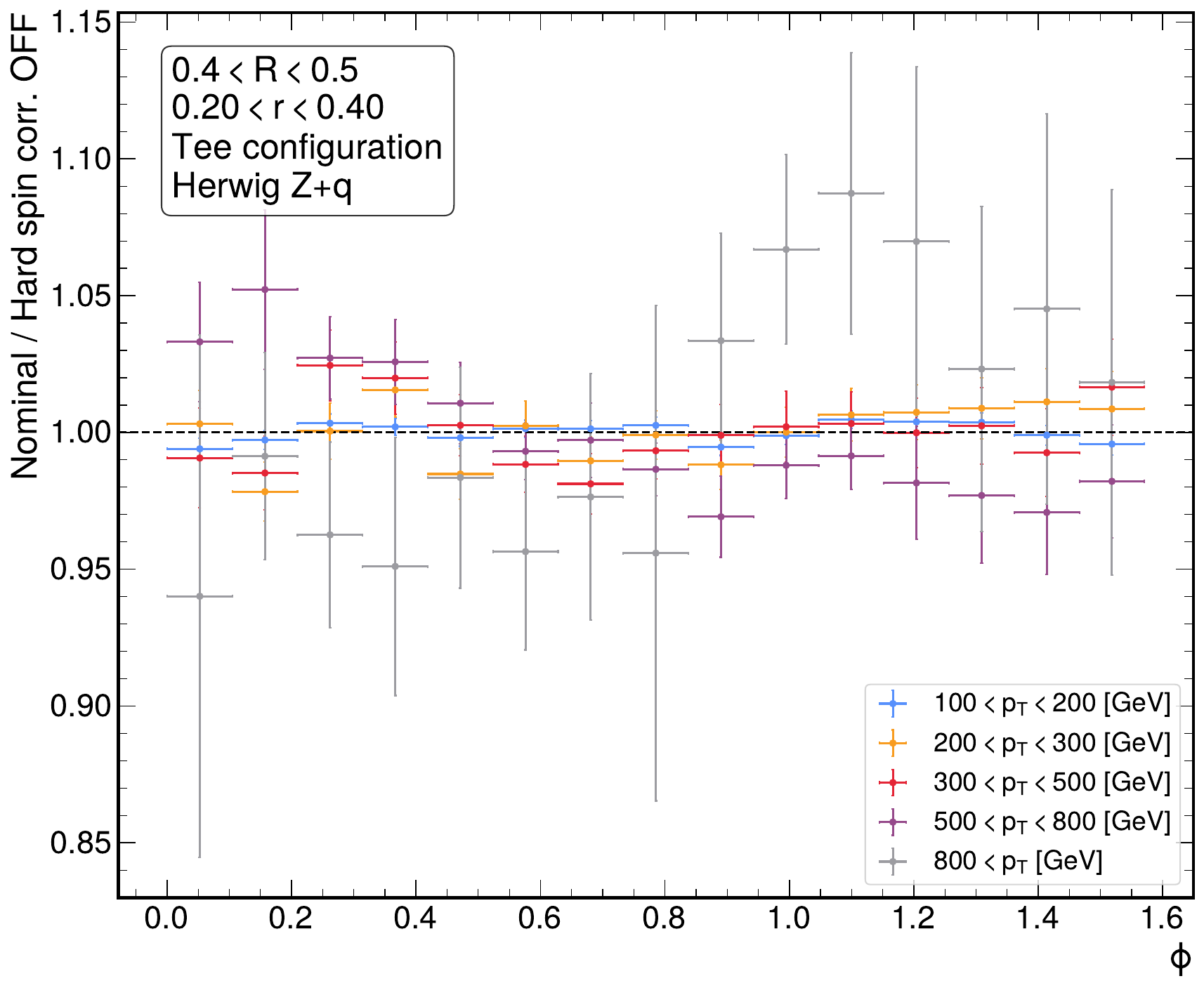} 
    \includegraphics[width=0.4\textwidth]{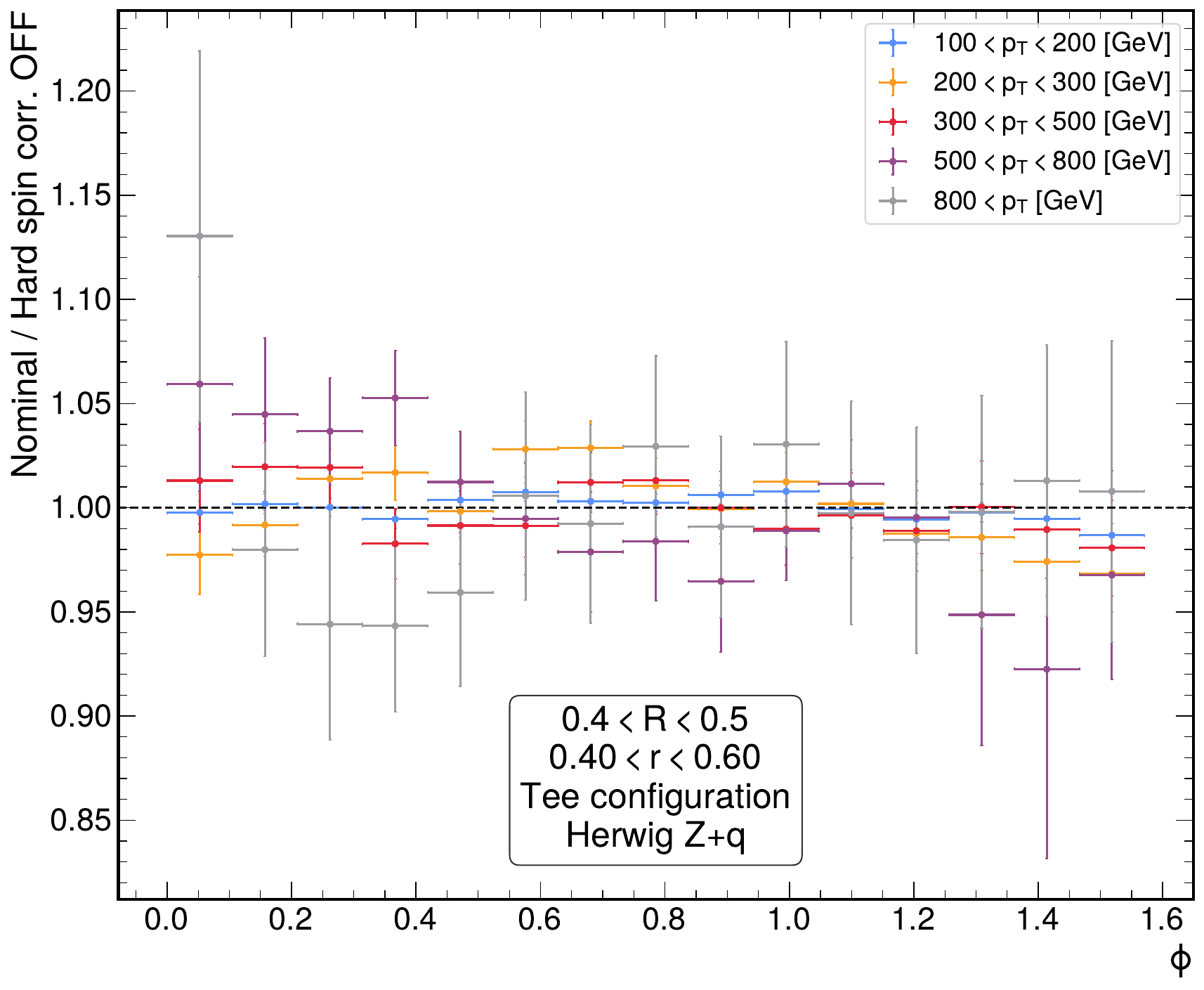} 
    \includegraphics[width=0.4\textwidth]{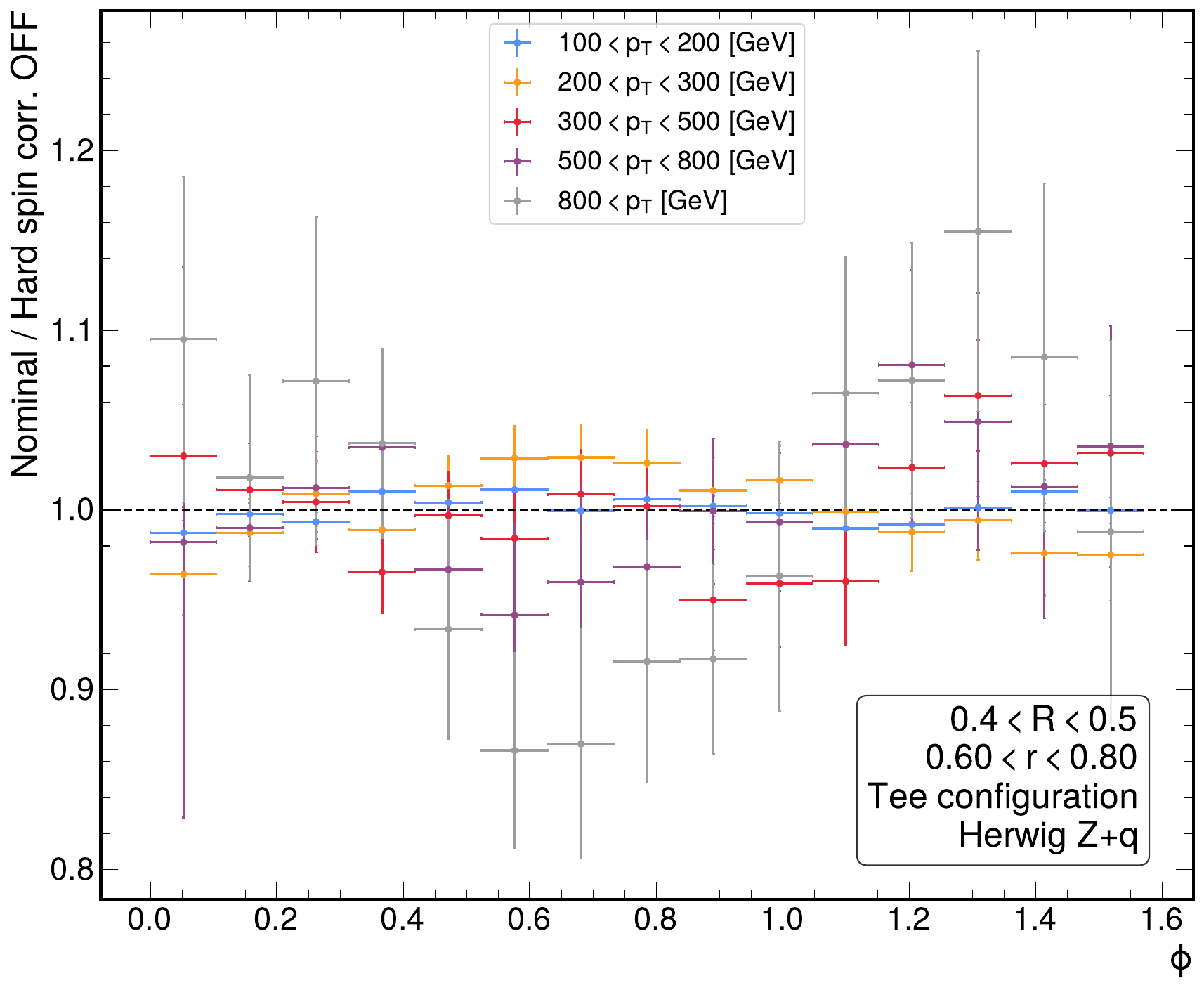} 
    \caption{Ratio between the tee distributions in Herwig $Z+q$ events with and without hard spin correlations. }
    \label{fig:tee_herwigQ_hardspin}
\end{figure}

\begin{figure}[h]
    \centering
    \includegraphics[width=0.4\textwidth]{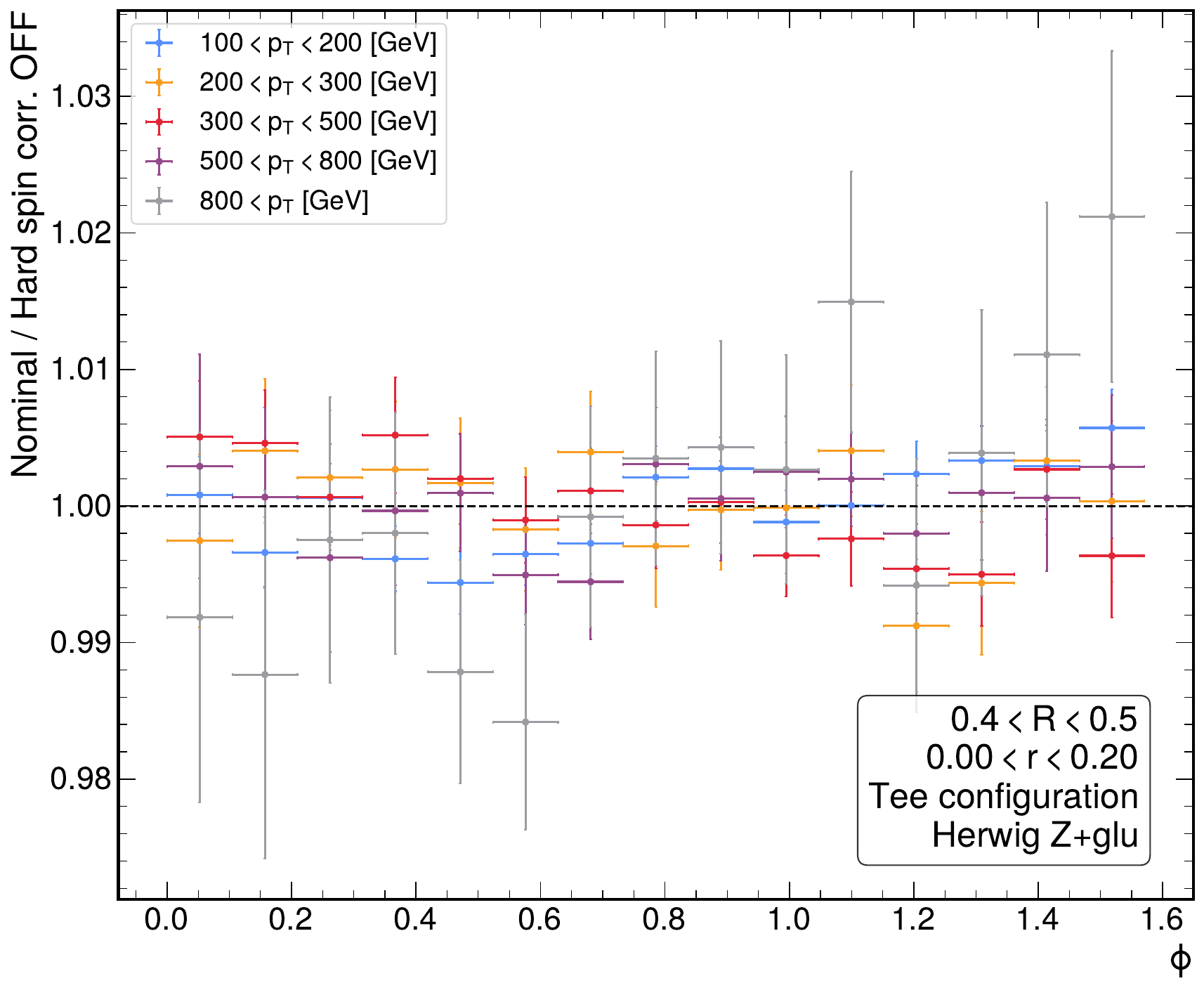} 
    \includegraphics[width=0.4\textwidth]{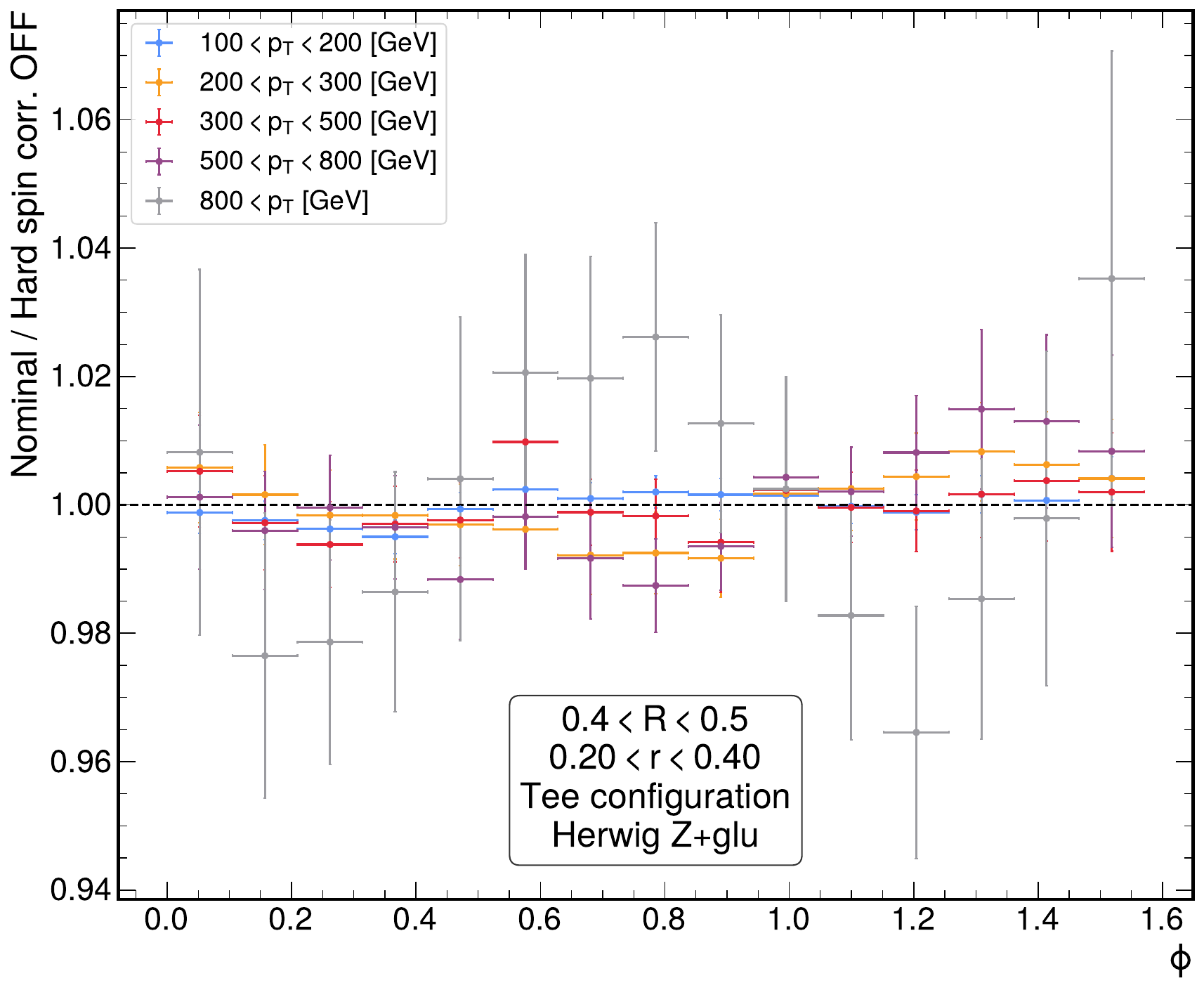} 
    \includegraphics[width=0.4\textwidth]{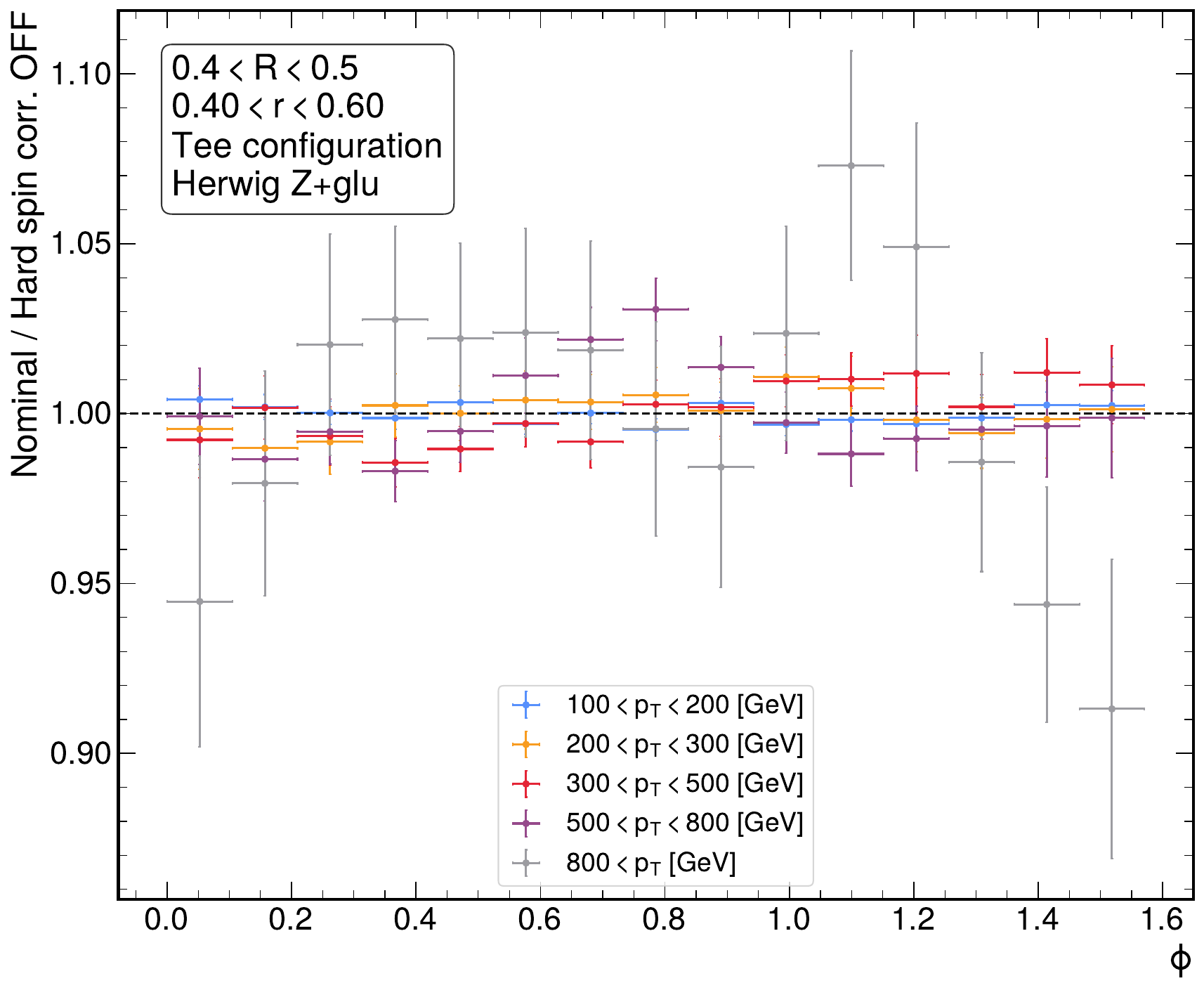} 
    \includegraphics[width=0.4\textwidth]{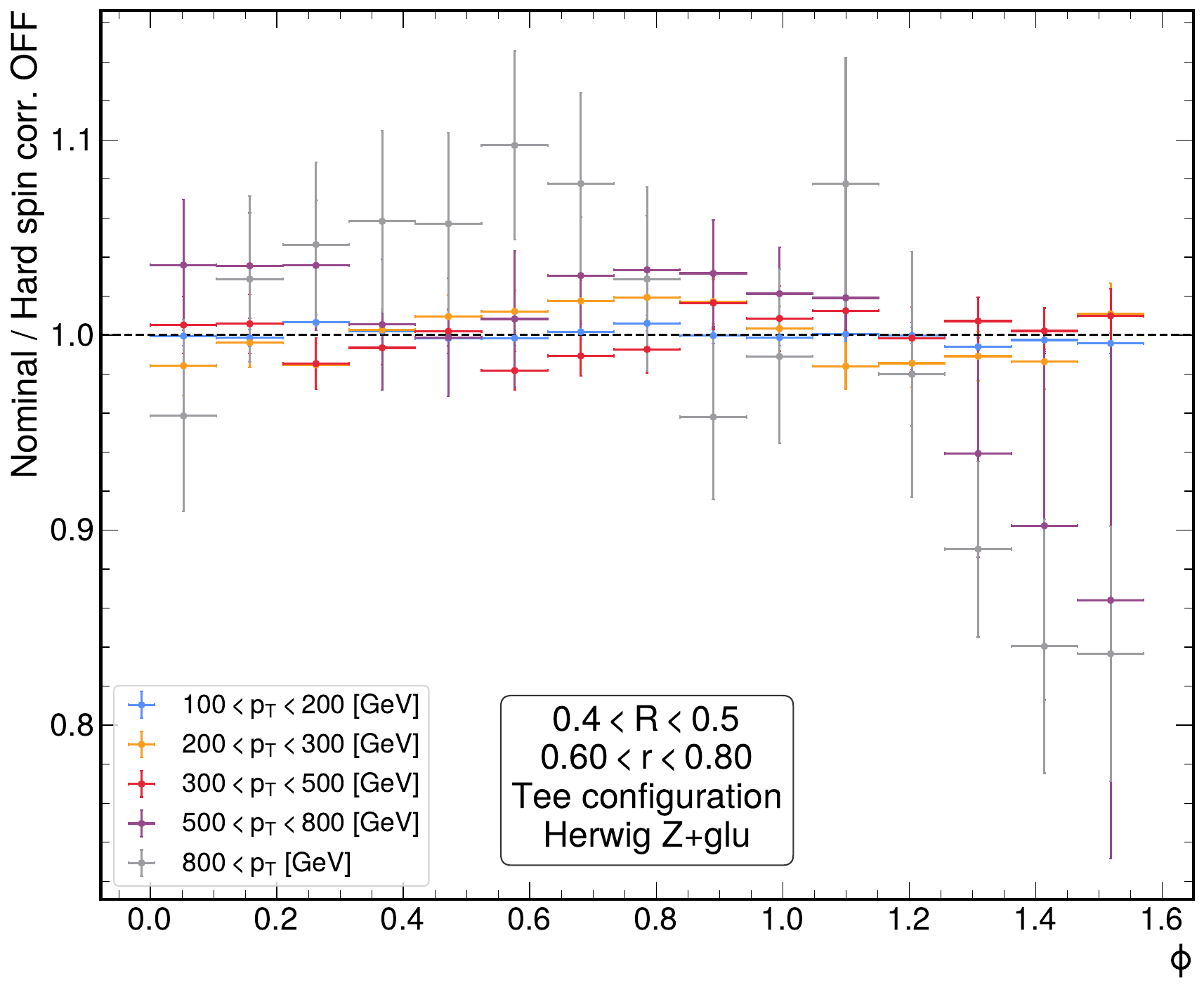} 
    \caption{Ratio between the tee distributions in Herwig $Z+g$ events with and without hard spin correlations.}
    \label{fig:tee_herwigG_hardspin}
\end{figure}

In \Fig{fig:tee_pythiaQ_hardspin} and \Fig{fig:tee_pythiaG_hardspin} we show the ratio of the tee distributions for quark and gluon jets with and without collinear spin correlations as implemented in Pythia. No systematic effect is observed, and no dependence on $r$ is observed. Identical plots generated using Herwig are shown in \Fig{fig:tee_herwigQ_hardspin} and \Fig{fig:tee_herwigG_hardspin}, with consistent conclusions.

\begin{figure}[h]
    \centering
    \includegraphics[width=0.4\textwidth]{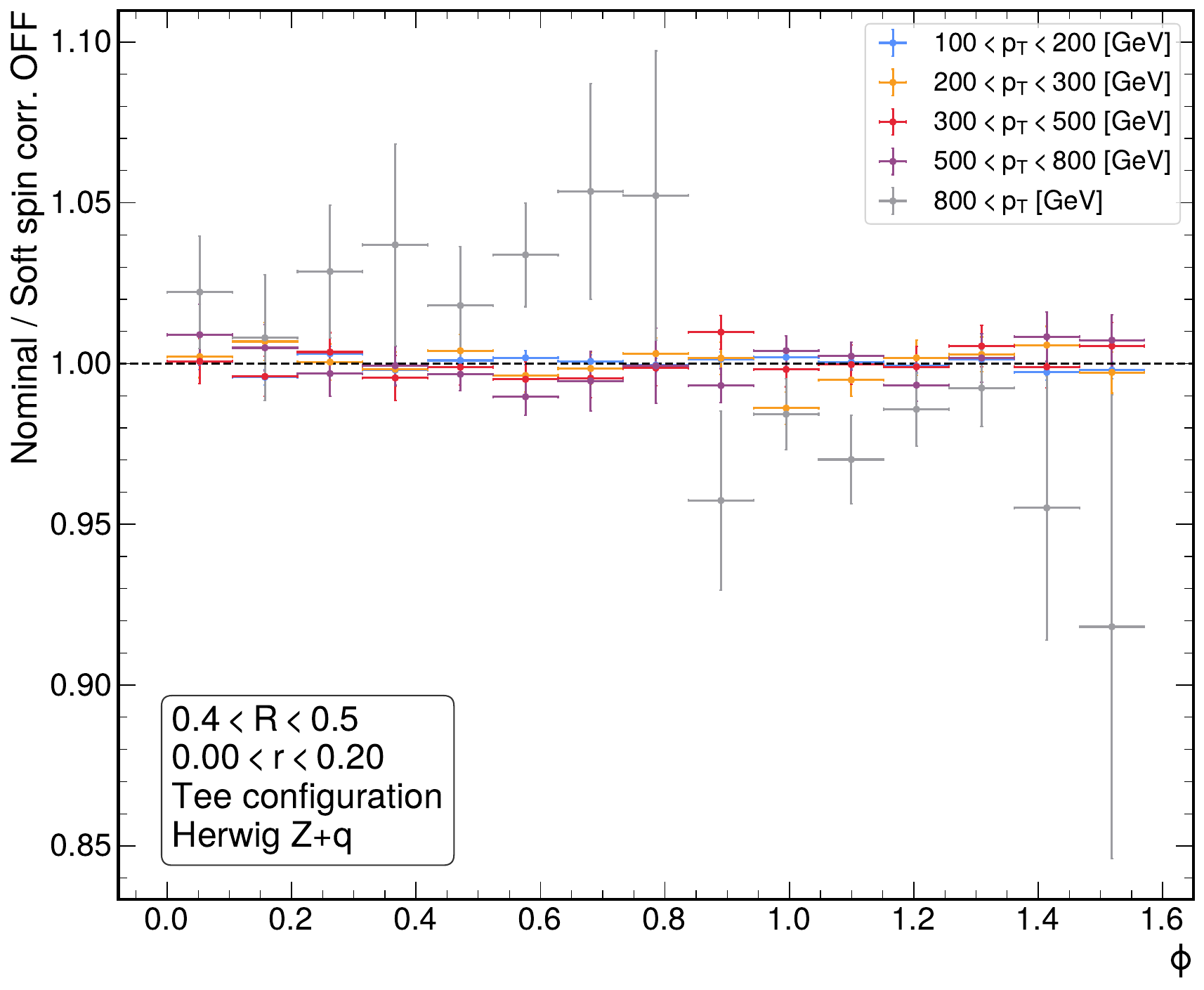} 
    \includegraphics[width=0.4\textwidth]{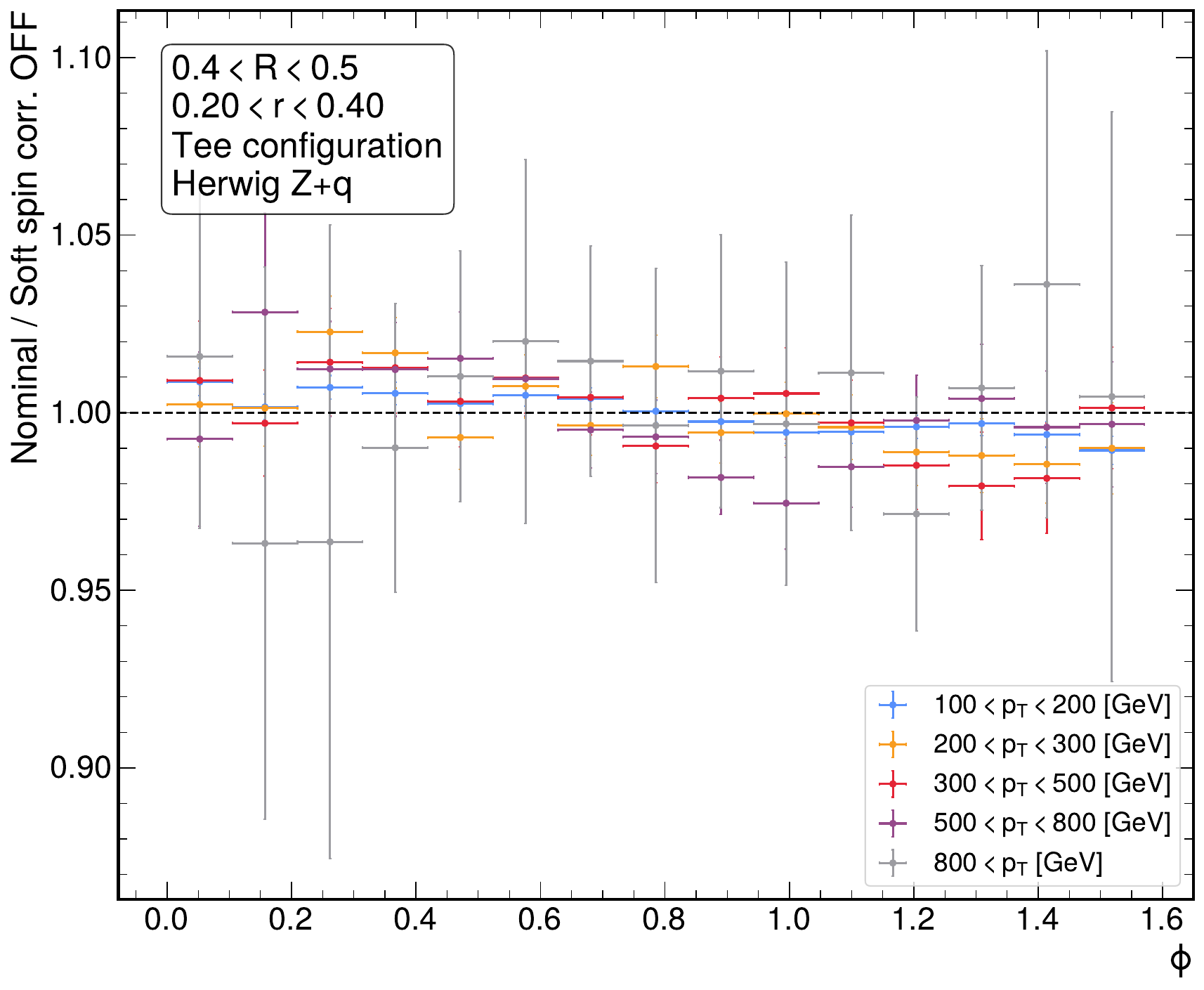} 
    \includegraphics[width=0.4\textwidth]{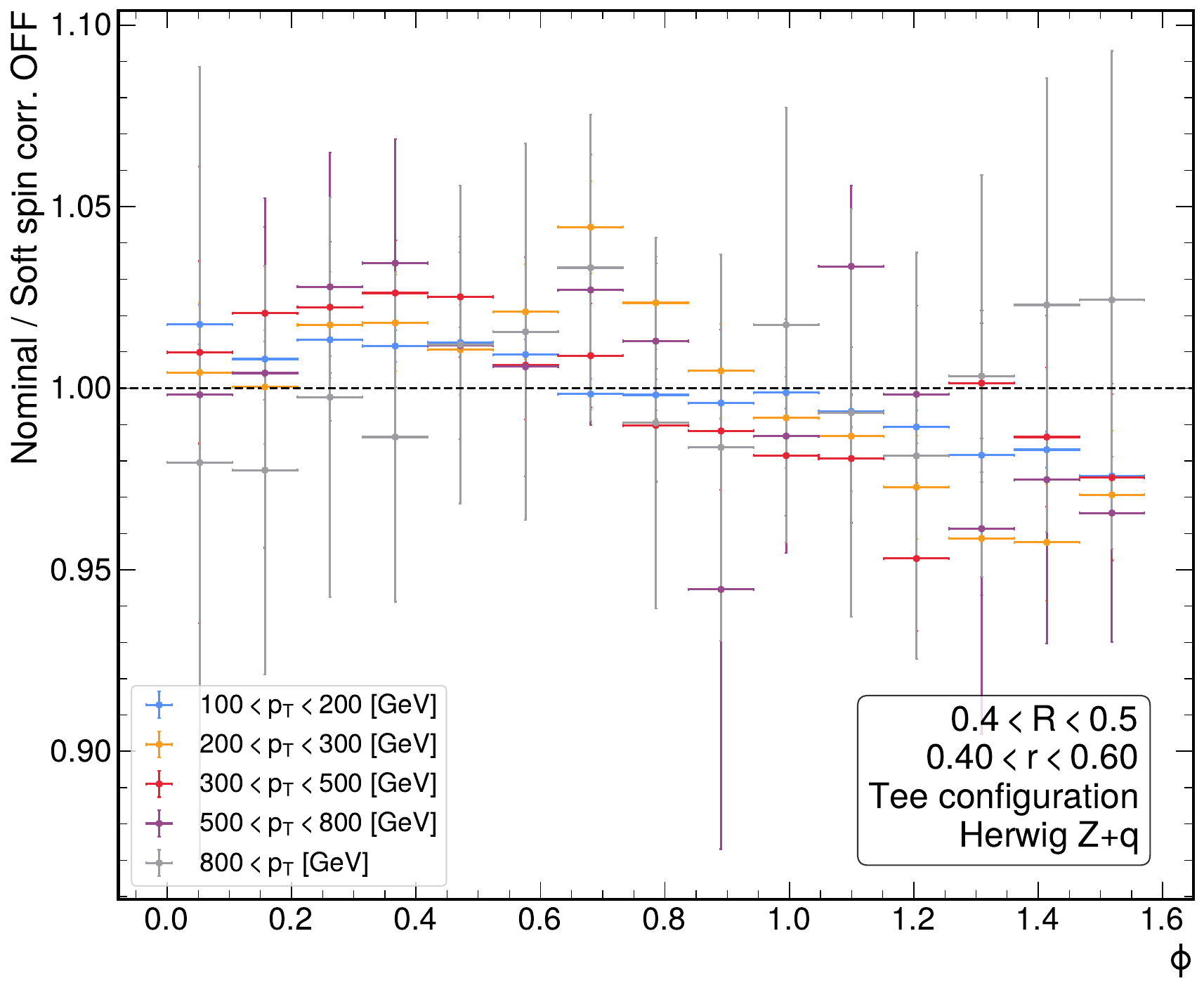} 
    \includegraphics[width=0.4\textwidth]{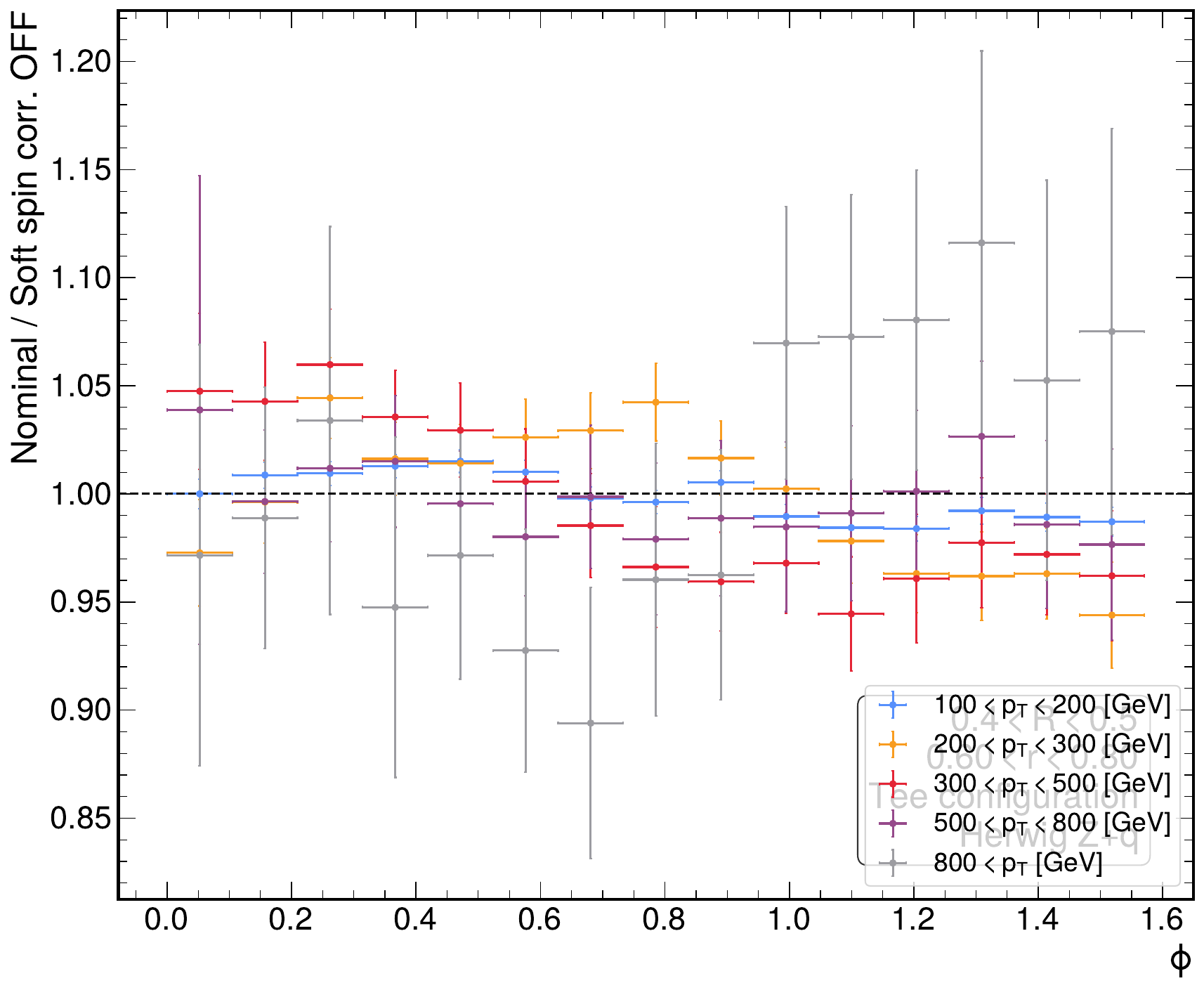} 
    \caption{Ratio between the tee distributions in Herwig $Z+q$ events with and without soft spin correlations. }
    \label{fig:tee_herwigQ_softspin}
\end{figure}

\begin{figure}[h]
    \centering
    \includegraphics[width=0.4\textwidth]{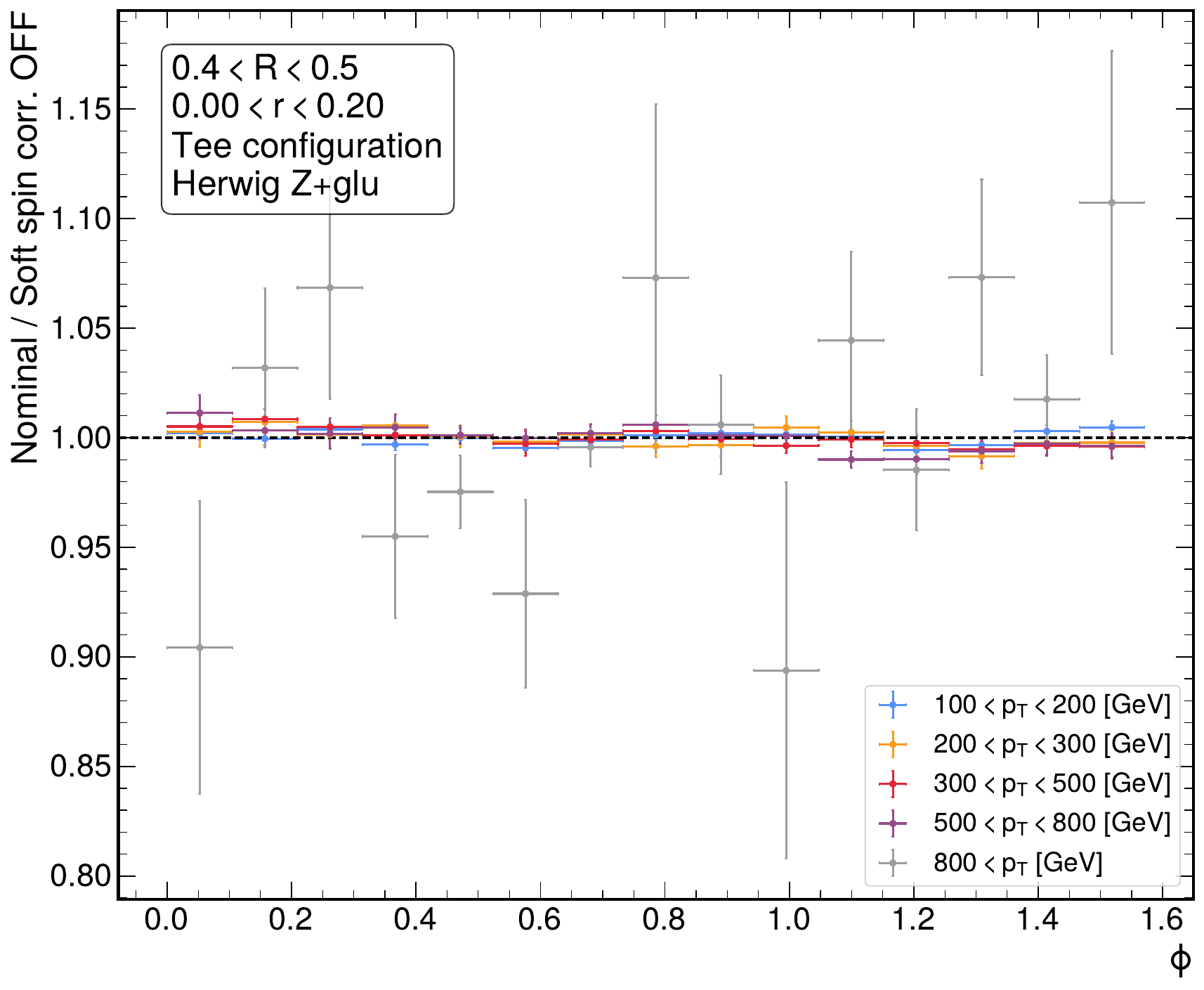} 
    \includegraphics[width=0.4\textwidth]{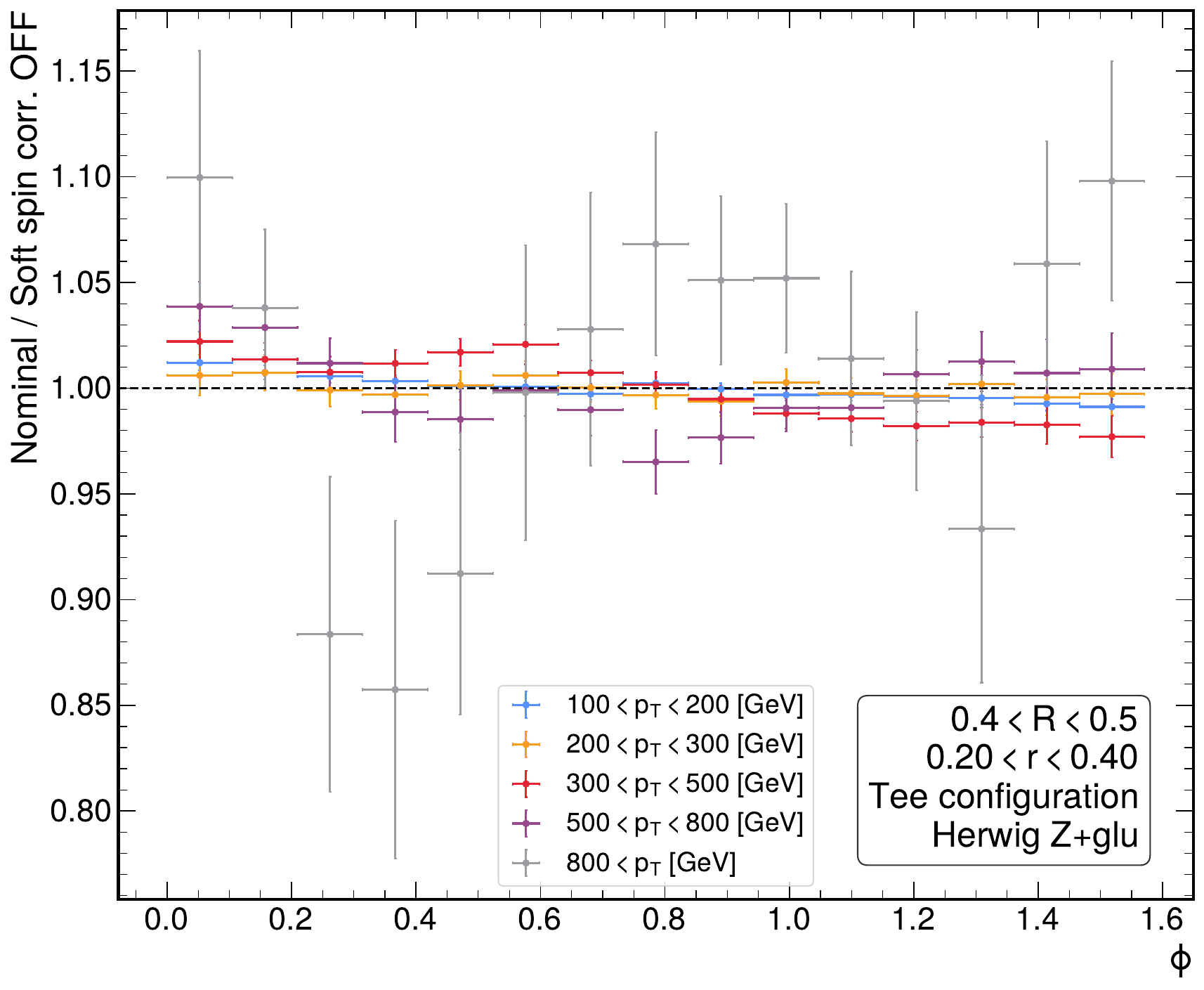} 
    \includegraphics[width=0.4\textwidth]{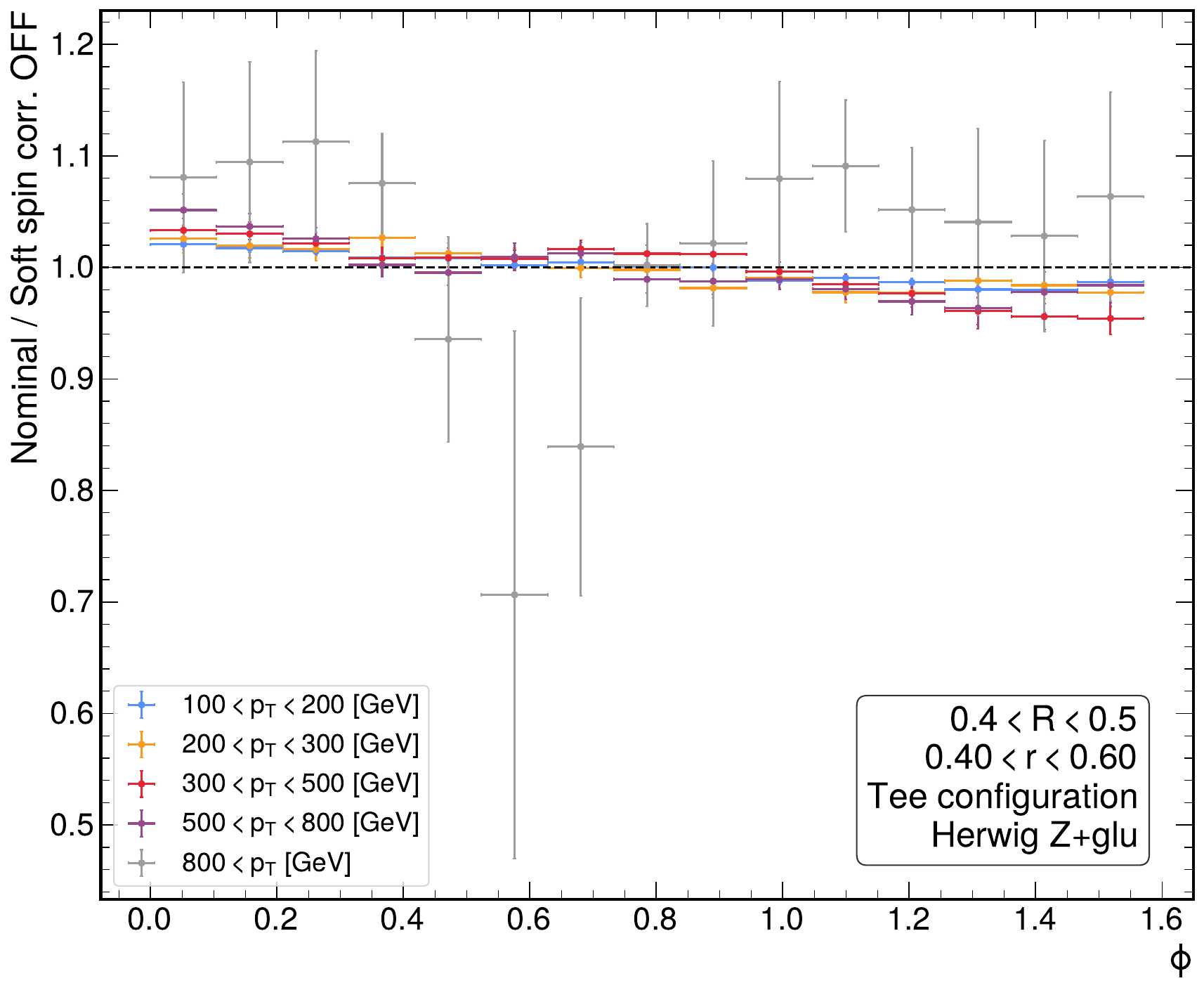} 
    \includegraphics[width=0.4\textwidth]{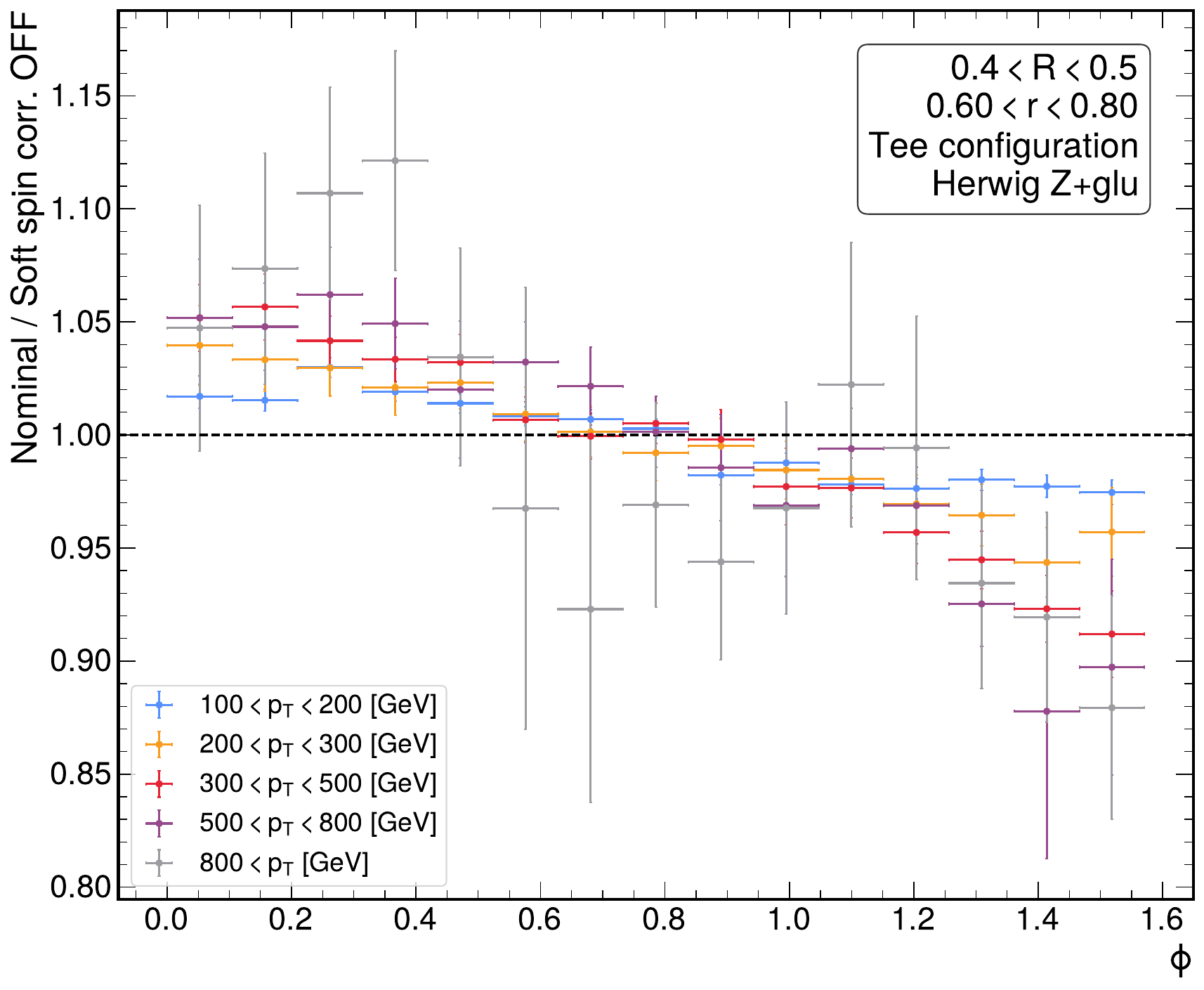} 
    \caption{Ratio between the tee distributions in Herwig $Z+g$ events with and without soft spin correlations.}
    \label{fig:tee_herwigG_softspin}
\end{figure}

Herwig also allows the user to disable soft spin correlations. In \Fig{fig:tee_herwigQ_softspin} and \Fig{fig:tee_herwigG_softspin}, we show identical plots with soft spin correlations turned on and off. While it is hard to draw conclusions due to the statistical uncertainties, the only systematic hint of an effect is for gluons jets in the largest radius bin in \Fig{fig:tee_herwigG_softspin}. This effect is of the order of $5\%$. Again, we emphasize that it is highly subdominant to the $\sim 50\%$ effect we saw for azimuthal correlations from kinematic effects. 

\begin{figure}[h]
    \centering
    \includegraphics[width=0.4\textwidth]{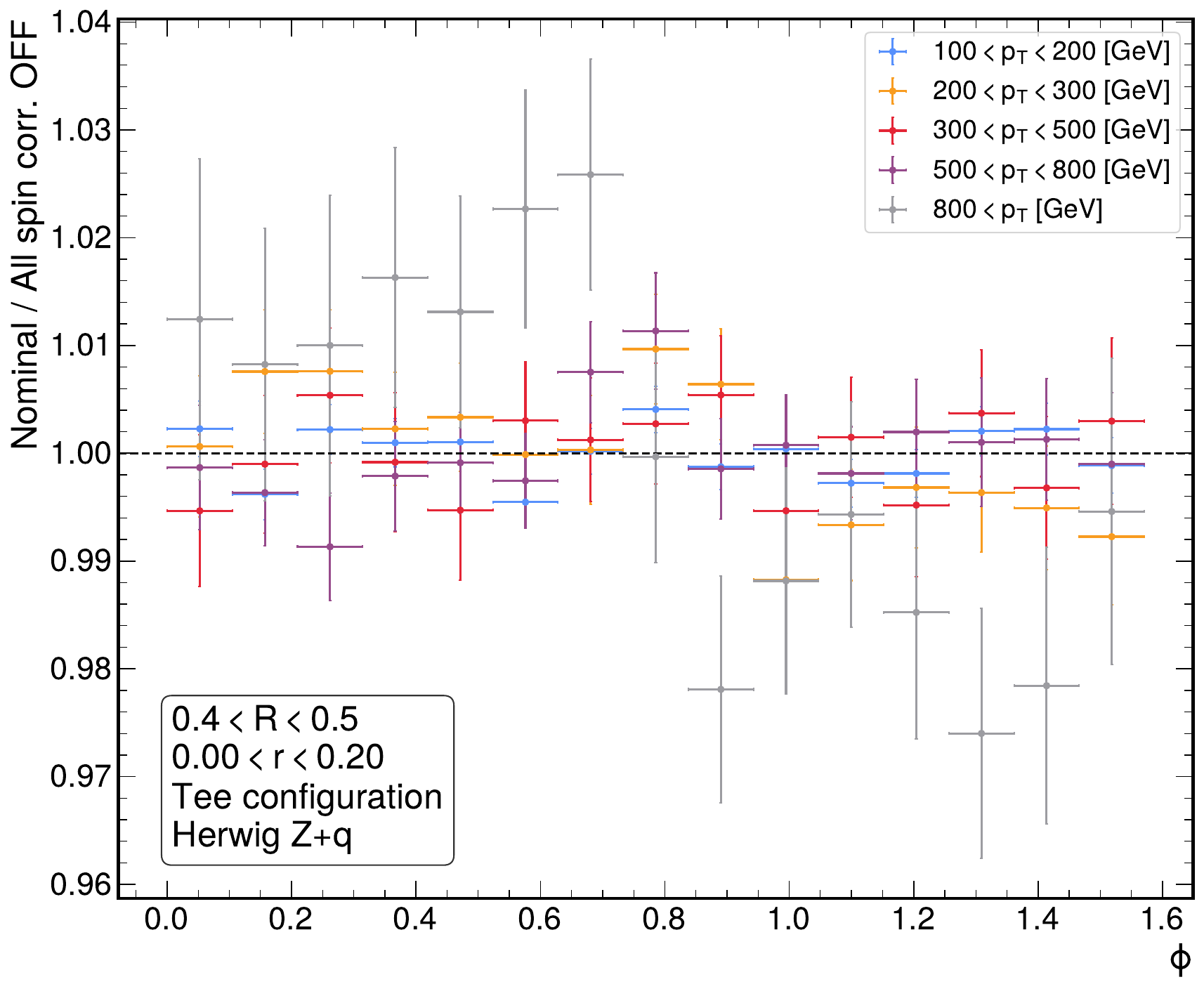} 
    \includegraphics[width=0.4\textwidth]{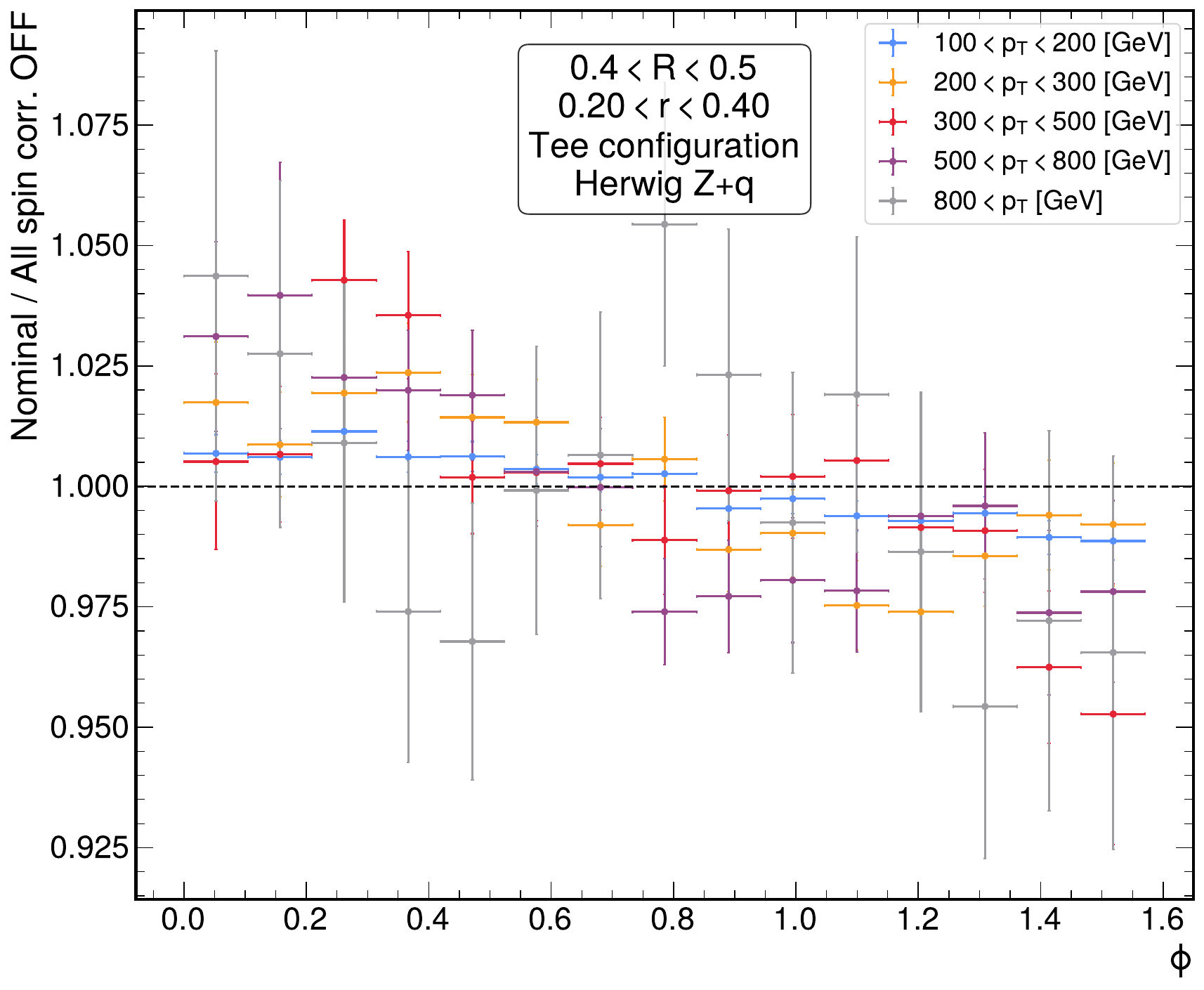} 
    \includegraphics[width=0.4\textwidth]{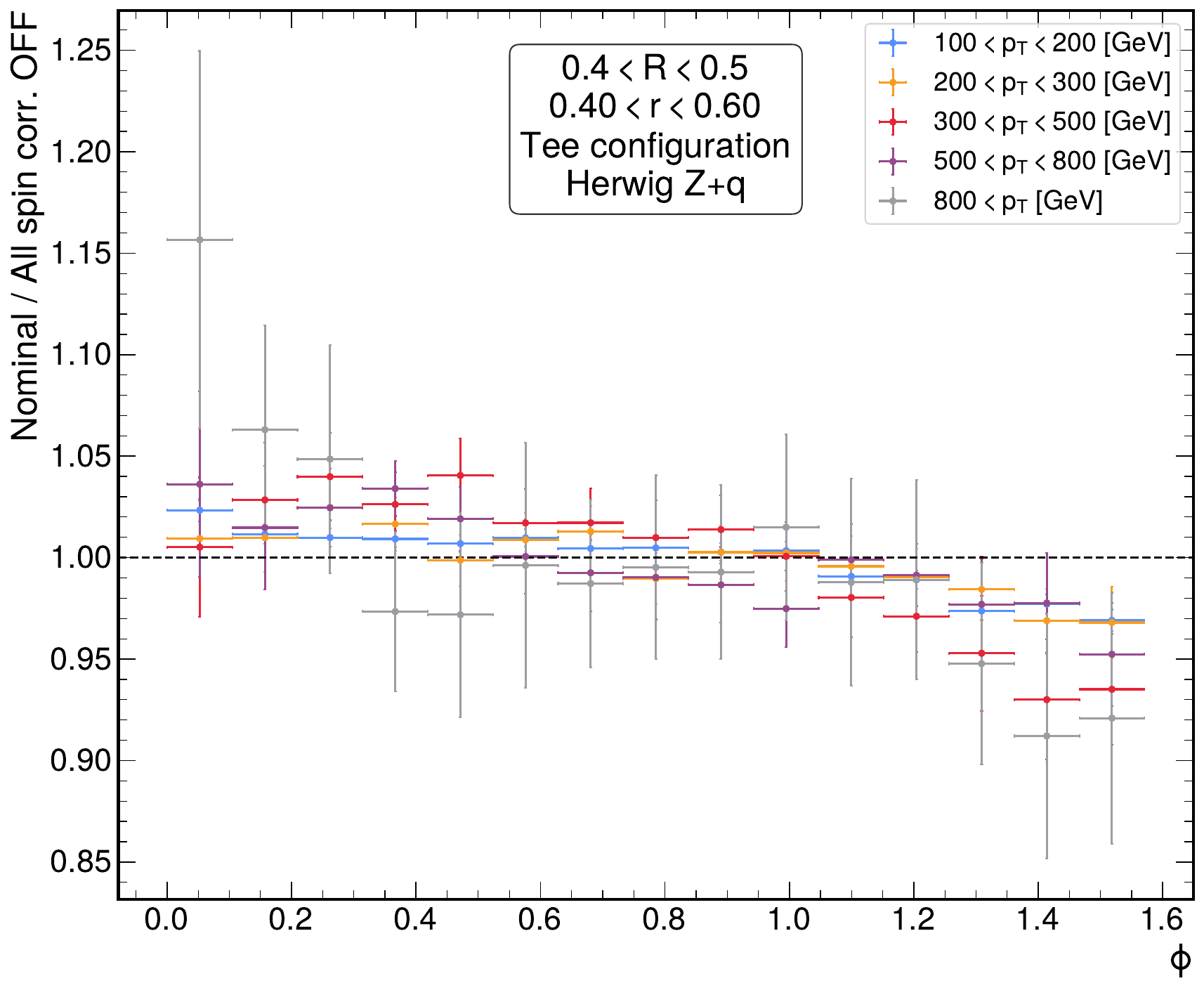} 
    \includegraphics[width=0.4\textwidth]{figures/MCstudy/tee/herwig_q_nominal_vs_nospin/comparison_shapes_comparison_1_cprof_cprofile_R1_r12.pdf} 
    \caption{Ratio between the tee distributions in Herwig $Z+q$ events with and without both hard and soft spin correlations (i.e. comparing gluons behaving as spin-1 vs spin-0).}
    \label{fig:tee_herwigQ_spin}
\end{figure}

\begin{figure}[h]
    \centering
    \includegraphics[width=0.4\textwidth]{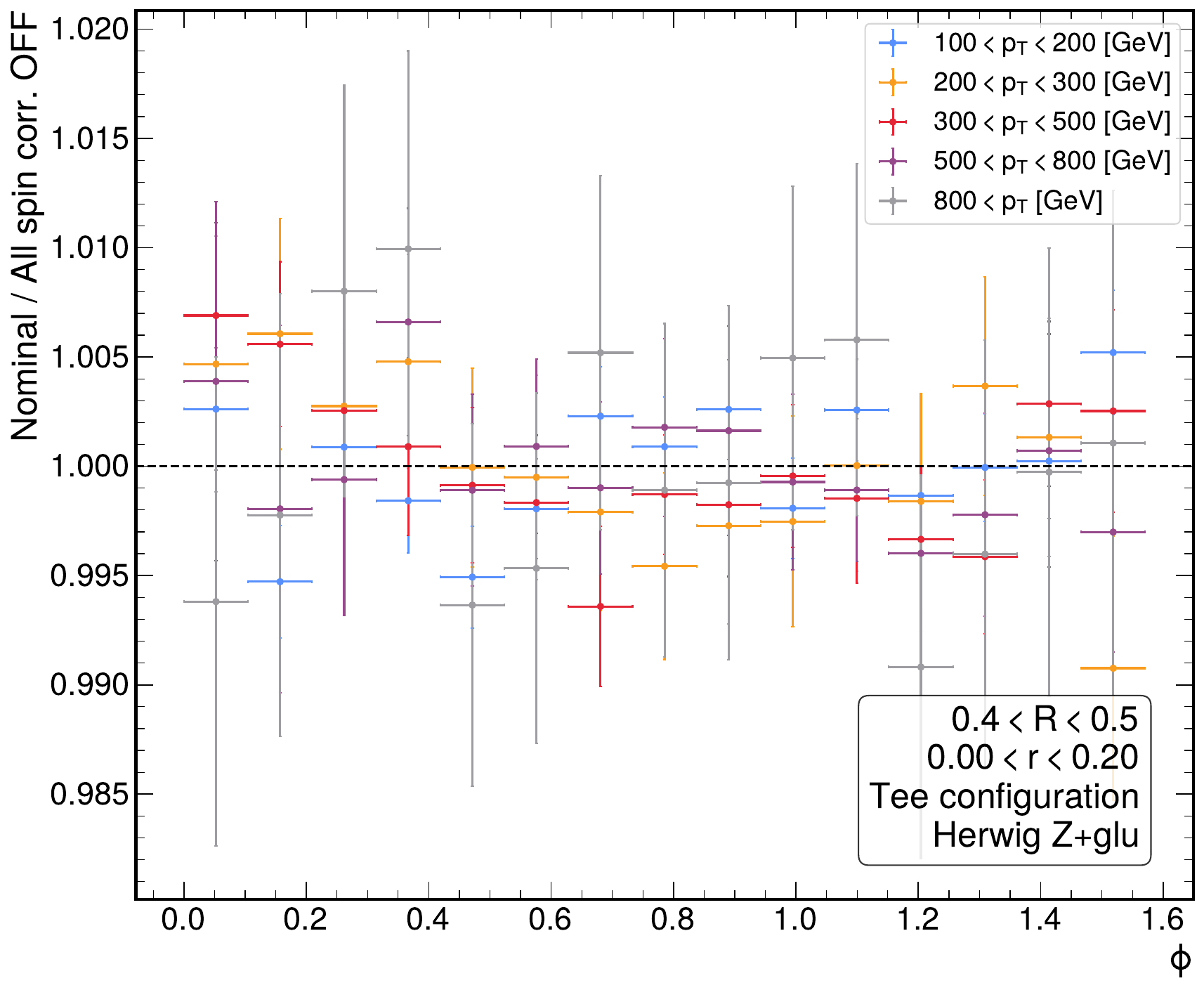} 
    \includegraphics[width=0.4\textwidth]{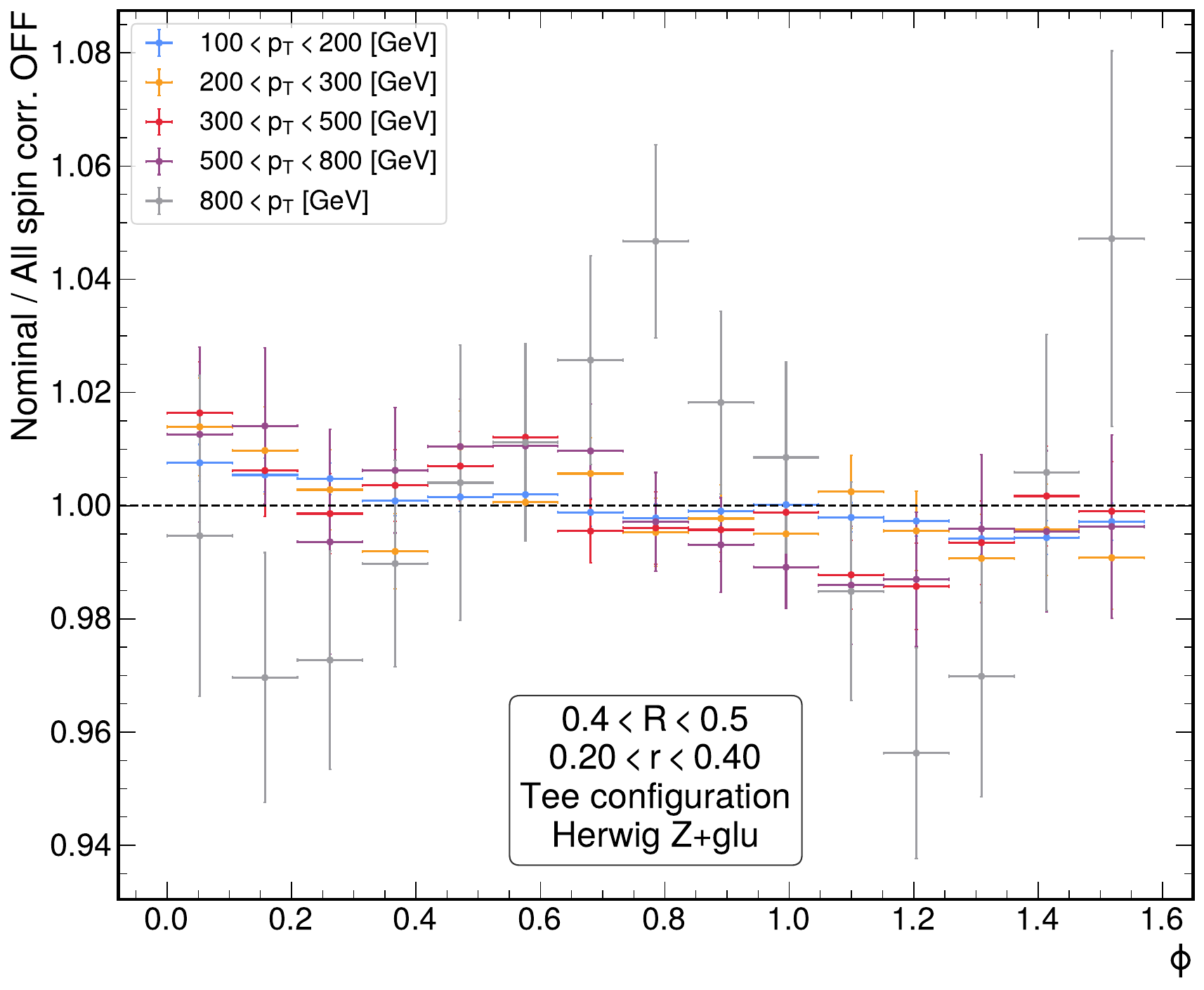} 
    \includegraphics[width=0.4\textwidth]{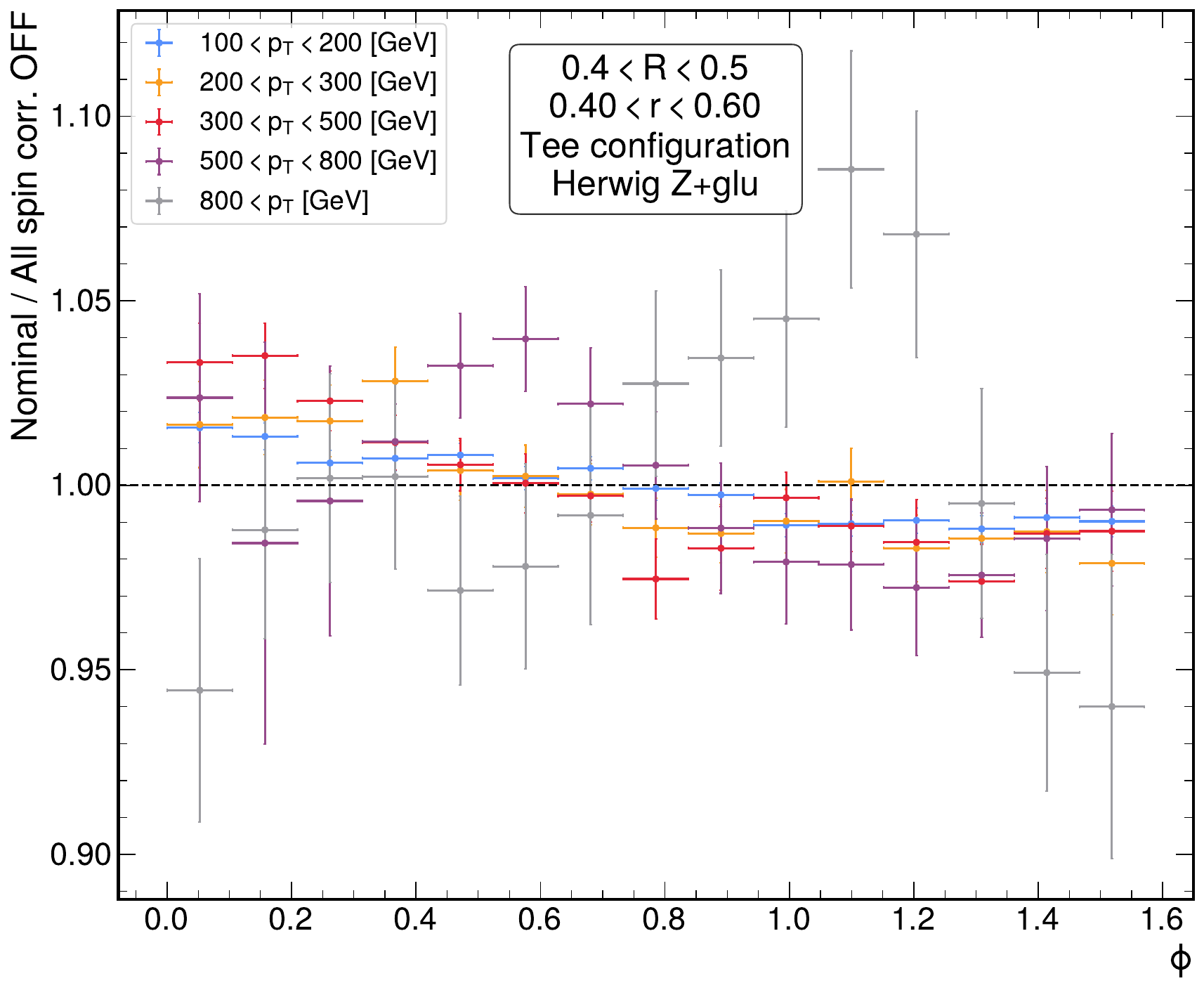} 
    \includegraphics[width=0.4\textwidth]{figures/MCstudy/tee/herwig_glu_nominal_vs_nospin/comparison_shapes_comparison_1_cprof_cprofile_R1_r12.pdf} 
    \caption{Ratio between the tee distributions in Herwig $Z+g$ events with and without both hard and soft spin correlations (i.e. comparing gluons behaving as spin-1 vs spin-0).}
    \label{fig:tee_herwigG_spin}
\end{figure}

Finally, in \Fig{fig:tee_herwigQ_spin} and \Fig{fig:tee_herwigG_spin} we show results in Herwig with both hard and soft spin correlations turned off. Again, a hint of a mild, $\sim 5\%$ effect is observed for gluon jets at large $r$.

\bibliography{four_point_pheno.bib}

@article{Chang:2013rca,
    author = "Chang, Hsi-Ming and Procura, Massimiliano and Thaler, Jesse and Waalewijn, Wouter J.",
    title = "{Calculating Track-Based Observables for the LHC}",
    eprint = "1303.6637",
    archivePrefix = "arXiv",
    primaryClass = "hep-ph",
    reportNumber = "MIT-CTP-4449, MIT--CTP-4449",
    doi = "10.1103/PhysRevLett.111.102002",
    journal = "Phys. Rev. Lett.",
    volume = "111",
    pages = "102002",
    year = "2013"
}

@article{Chang:2013iba,
    author = "Chang, Hsi-Ming and Procura, Massimiliano and Thaler, Jesse and Waalewijn, Wouter J.",
    title = "{Calculating Track Thrust with Track Functions}",
    eprint = "1306.6630",
    archivePrefix = "arXiv",
    primaryClass = "hep-ph",
    reportNumber = "MIT--CTP-4476",
    doi = "10.1103/PhysRevD.88.034030",
    journal = "Phys. Rev. D",
    volume = "88",
    pages = "034030",
    year = "2013"
}

@article{Chen:2022muj,
    author = "Chen, Hao and Jaarsma, Max and Li, Yibei and Moult, Ian and Waalewijn, Wouter J. and Zhu, Hua Xing",
    title = "{Collinear parton dynamics beyond Dokshitzer-Gribov-Lipatov-Altarelli-Parisi framework}",
    eprint = "2210.10061",
    archivePrefix = "arXiv",
    primaryClass = "hep-ph",
    doi = "10.1103/PhysRevD.111.076021",
    journal = "Phys. Rev. D",
    volume = "111",
    number = "7",
    pages = "076021",
    year = "2025"
}

@article{Jaarsma:2023ell,
    author = "Jaarsma, Max and Li, Yibei and Moult, Ian and Waalewijn, Wouter J. and Zhu, Hua Xing",
    title = "{Energy correlators on tracks: resummation and non-perturbative effects}",
    eprint = "2307.15739",
    archivePrefix = "arXiv",
    primaryClass = "hep-ph",
    doi = "10.1007/JHEP12(2023)087",
    journal = "JHEP",
    volume = "12",
    pages = "087",
    year = "2023"
}

@article{Chen:2022pdu,
    author = "Chen, Hao and Jaarsma, Max and Li, Yibei and Moult, Ian and Waalewijn, Wouter J. and Zhu, Hua Xing",
    title = "{Multi-collinear splitting kernels for track function evolution}",
    eprint = "2210.10058",
    archivePrefix = "arXiv",
    primaryClass = "hep-ph",
    doi = "10.1007/JHEP07(2023)185",
    journal = "JHEP",
    volume = "07",
    pages = "185",
    year = "2023"
}

@article{Li:2021zcf,
    author = "Li, Yibei and Moult, Ian and van Velzen, Solange Schrijnder and Waalewijn, Wouter J. and Zhu, Hua Xing",
    title = "{Extending Precision Perturbative QCD with Track Functions}",
    eprint = "2108.01674",
    archivePrefix = "arXiv",
    primaryClass = "hep-ph",
    doi = "10.1103/PhysRevLett.128.182001",
    journal = "Phys. Rev. Lett.",
    volume = "128",
    number = "18",
    pages = "182001",
    year = "2022"
}

@article{Jaarsma:2022kdd,
    author = "Jaarsma, Max and Li, Yibei and Moult, Ian and Waalewijn, Wouter and Zhu, Hua Xing",
    title = "{Renormalization group flows for track function moments}",
    eprint = "2201.05166",
    archivePrefix = "arXiv",
    primaryClass = "hep-ph",
    doi = "10.1007/JHEP06(2022)139",
    journal = "JHEP",
    volume = "06",
    pages = "139",
    year = "2022"
}

@article{Jaarsma:2025tck,
    author = "Jaarsma, Max and Li, Yibei and Moult, Ian and Waalewijn, Wouter J. and Zhu, Hua Xing",
    title = "{From DGLAP to Sudakov: Precision Predictions for Energy-Energy Correlators}",
    eprint = "2512.11950",
    archivePrefix = "arXiv",
    primaryClass = "hep-ph",
    reportNumber = "MITP-24-091",
    month = "12",
    year = "2025"
}

@article{Larkoski:2017jix,
    author = "Larkoski, Andrew J. and Moult, Ian and Nachman, Benjamin",
    title = "{Jet Substructure at the Large Hadron Collider: A Review of Recent Advances in Theory and Machine Learning}",
    eprint = "1709.04464",
    archivePrefix = "arXiv",
    primaryClass = "hep-ph",
    doi = "10.1016/j.physrep.2019.11.001",
    journal = "Phys. Rept.",
    volume = "841",
    pages = "1--63",
    year = "2020"
}

@article{Asquith:2018igt,
    author = "Kogler, Roman and others",
    title = "{Jet Substructure at the Large Hadron Collider: Experimental Review}",
    eprint = "1803.06991",
    archivePrefix = "arXiv",
    primaryClass = "hep-ex",
    reportNumber = "FERMILAB-PUB-18-123-PPD",
    doi = "10.1103/RevModPhys.91.045003",
    journal = "Rev. Mod. Phys.",
    volume = "91",
    number = "4",
    pages = "045003",
    year = "2019"
}

@article{Sveshnikov:1995vi,
    author = "Sveshnikov, N. A. and Tkachov, F. V.",
    editor = "Levchenko, B. B. and Savrin, V. I.",
    title = "{Jets and quantum field theory}",
    eprint = "hep-ph/9512370",
    archivePrefix = "arXiv",
    doi = "10.1016/0370-2693(96)00558-8",
    journal = "Phys. Lett. B",
    volume = "382",
    pages = "403--408",
    year = "1996"
}

@article{Tkachov:1995kk,
    author = "Tkachov, Fyodor V.",
    title = "{Measuring multi - jet structure of hadronic energy flow or What is a jet?}",
    eprint = "hep-ph/9601308",
    archivePrefix = "arXiv",
    reportNumber = "FERMILAB-PUB-95-191-T-REV, FERMILAB-PUB-95-191-T",
    doi = "10.1142/S0217751X97002899",
    journal = "Int. J. Mod. Phys. A",
    volume = "12",
    pages = "5411--5529",
    year = "1997"
}

@article{Korchemsky:1999kt,
    author = "Korchemsky, Gregory P. and Sterman, George F.",
    title = "{Power corrections to event shapes and factorization}",
    eprint = "hep-ph/9902341",
    archivePrefix = "arXiv",
    reportNumber = "ITP-SB-98-73, LPT-ORSAY-98-80",
    doi = "10.1016/S0550-3213(99)00308-9",
    journal = "Nucl. Phys. B",
    volume = "555",
    pages = "335--351",
    year = "1999"
}

@article{Bauer:2008dt,
    author = "Bauer, Christian W. and Fleming, Sean P. and Lee, Christopher and Sterman, George F.",
    title = "{Factorization of e+e- Event Shape Distributions with Hadronic Final States in Soft Collinear Effective Theory}",
    eprint = "0801.4569",
    archivePrefix = "arXiv",
    primaryClass = "hep-ph",
    reportNumber = "UCB-PTH-08-02, YITP-SB-08-02",
    doi = "10.1103/PhysRevD.78.034027",
    journal = "Phys. Rev. D",
    volume = "78",
    pages = "034027",
    year = "2008"
}

@article{Hofman:2008ar,
    author = "Hofman, Diego M. and Maldacena, Juan",
    title = "{Conformal collider physics: Energy and charge correlations}",
    eprint = "0803.1467",
    archivePrefix = "arXiv",
    primaryClass = "hep-th",
    doi = "10.1088/1126-6708/2008/05/012",
    journal = "JHEP",
    volume = "05",
    pages = "012",
    year = "2008"
}

@article{Belitsky:2013xxa,
    author = "Belitsky, A. V. and Hohenegger, S. and Korchemsky, G. P. and Sokatchev, E. and Zhiboedov, A.",
    title = "{From correlation functions to event shapes}",
    eprint = "1309.0769",
    archivePrefix = "arXiv",
    primaryClass = "hep-th",
    reportNumber = "CERN-PH-TH-2013-211, IPHT-T13-210, LAPTH-047-13",
    doi = "10.1016/j.nuclphysb.2014.04.020",
    journal = "Nucl. Phys. B",
    volume = "884",
    pages = "305--343",
    year = "2014"
}

@article{Belitsky:2013bja,
    author = "Belitsky, A. V. and Hohenegger, S. and Korchemsky, G. P. and Sokatchev, E. and Zhiboedov, A.",
    title = "{Event shapes in $\mathcal{N} = 4$ super-Yang-Mills theory}",
    eprint = "1309.1424",
    archivePrefix = "arXiv",
    primaryClass = "hep-th",
    reportNumber = "CERN-PH-TH-2013-212",
    doi = "10.1016/j.nuclphysb.2014.04.019",
    journal = "Nucl. Phys. B",
    volume = "884",
    pages = "206--256",
    year = "2014"
}

@article{Kravchuk:2018htv,
    author = "Kravchuk, Petr and Simmons-Duffin, David",
    title = "{Light-ray operators in conformal field theory}",
    eprint = "1805.00098",
    archivePrefix = "arXiv",
    primaryClass = "hep-th",
    reportNumber = "CALT-TH 2018-018",
    doi = "10.1007/JHEP11(2018)102",
    journal = "JHEP",
    volume = "11",
    pages = "102",
    year = "2018"
}

@article{Basham:1979gh,
    author = "Basham, C. Louis and Brown, Lowell S. and Ellis, Stephen D. and Love, Sherwin T.",
    title = "{Energy Correlations in Perturbative Quantum Chromodynamics: A Conjecture for All Orders}",
    reportNumber = "RLO-1388-786",
    doi = "10.1016/0370-2693(79)90601-4",
    journal = "Phys. Lett. B",
    volume = "85",
    pages = "297--299",
    year = "1979"
}

@article{Basham:1978zq,
    author = "Basham, C. L. and Brown, L. S. and Ellis, S. D. and Love, S. T.",
    title = "{Energy Correlations in electron-Positron Annihilation in Quantum Chromodynamics: Asymptotically Free Perturbation Theory}",
    reportNumber = "RLO-1388-761",
    doi = "10.1103/PhysRevD.19.2018",
    journal = "Phys. Rev. D",
    volume = "19",
    pages = "2018",
    year = "1979"
}

@article{Basham:1978bw,
    author = "Basham, C. Louis and Brown, Lowell S. and Ellis, Stephen D. and Love, Sherwin T.",
    title = "{Energy Correlations in electron - Positron Annihilation: Testing QCD}",
    reportNumber = "RLO-1388-759",
    doi = "10.1103/PhysRevLett.41.1585",
    journal = "Phys. Rev. Lett.",
    volume = "41",
    pages = "1585",
    year = "1978"
}

@article{Basham:1977iq,
    author = "Basham, C. Louis and Brown, Lowell S. and Ellis, S. D. and Love, S. T.",
    title = "{Electron - Positron Annihilation Energy Pattern in Quantum Chromodynamics: Asymptotically Free Perturbation Theory}",
    reportNumber = "RLO-1388-746",
    doi = "10.1103/PhysRevD.17.2298",
    journal = "Phys. Rev. D",
    volume = "17",
    pages = "2298",
    year = "1978"
}

@article{SLD:1994idb,
    author = "Abe, K. and others",
    collaboration = "SLD",
    title = "{Measurement of alpha-s (M(Z)**2) from hadronic event observables at the Z0 resonance}",
    eprint = "hep-ex/9501003",
    archivePrefix = "arXiv",
    reportNumber = "SLAC-PUB-6641",
    doi = "10.1103/PhysRevD.51.962",
    journal = "Phys. Rev. D",
    volume = "51",
    pages = "962--984",
    year = "1995"
}

@article{L3:1992btq,
    author = "Adrian, O. and others",
    collaboration = "L3",
    title = "{Determination of alpha-s from hadronic event shapes measured on the Z0 resonance}",
    reportNumber = "CERN-PPE-92-058, CERN-PPE-92-58",
    doi = "10.1016/0370-2693(92)90463-E",
    journal = "Phys. Lett. B",
    volume = "284",
    pages = "471--481",
    year = "1992"
}

@article{OPAL:1991uui,
    author = "Acton, P. D. and others",
    collaboration = "OPAL",
    title = "{An Improved measurement of alpha-s (M (Z0)) using energy correlations with the OPAL detector at LEP}",
    reportNumber = "CERN-PPE-91-214",
    doi = "10.1016/0370-2693(92)91681-X",
    journal = "Phys. Lett. B",
    volume = "276",
    pages = "547--564",
    year = "1992"
}

@article{TOPAZ:1989yod,
    author = "Adachi, I. and others",
    collaboration = "TOPAZ",
    title = "{Measurements of $\alpha^- s$ in $e^+ e^-$ Annihilation at $\sqrt{s}=53$.3-{GeV} and 59.5-{GeV}}",
    reportNumber = "KEK-Preprint-89-37",
    doi = "10.1016/0370-2693(89)90969-6",
    journal = "Phys. Lett. B",
    volume = "227",
    pages = "495--500",
    year = "1989"
}

@article{TASSO:1987mcs,
    author = "Braunschweig, W. and others",
    collaboration = "TASSO",
    title = "{A Study of Energy-energy Correlations Between 12-{GeV} and 46.8-{GeV} {CM} Energies}",
    reportNumber = "DESY-87-081, FTUAM-EP-87-04, SI-87-12",
    doi = "10.1007/BF01573928",
    journal = "Z. Phys. C",
    volume = "36",
    pages = "349--361",
    year = "1987"
}

@article{JADE:1984taa,
    author = "Bartel, W. and others",
    collaboration = "JADE",
    title = "{Measurements of Energy Correlations in $e^+ e^- \to$ Hadrons}",
    reportNumber = "DESY-84-050",
    doi = "10.1007/BF01547922",
    journal = "Z. Phys. C",
    volume = "25",
    pages = "231",
    year = "1984"
}

@article{Fernandez:1984db,
    author = "Fernandez, E. and others",
    title = "{A Measurement of Energy-energy Correlations in $e^+ e^- \to$ Hadrons at $\sqrt{s}=29$-{GeV}}",
    reportNumber = "SLAC-PUB-3385",
    doi = "10.1103/PhysRevD.31.2724",
    journal = "Phys. Rev. D",
    volume = "31",
    pages = "2724",
    year = "1985"
}

@article{Wood:1987uf,
    author = "Wood, D. R. and others",
    title = "{Determination of $\alpha^- s$ From Energy-energy Correlations in $e^+ e^-$ Annihilation at 29-{GeV}}",
    reportNumber = "SLAC-PUB-4374, LBL-23812",
    doi = "10.1103/PhysRevD.37.3091",
    journal = "Phys. Rev. D",
    volume = "37",
    pages = "3091",
    year = "1988"
}

@article{CELLO:1982rca,
    author = "Behrend, H. J. and others",
    collaboration = "CELLO",
    title = "{Analysis of the Energy Weighted Angular Correlations in Hadronic $e^+ e^-$ Annihilations at 22-{GeV} and 34-{GeV}}",
    reportNumber = "Print-82-0345 (DESY), DESY-82-022",
    doi = "10.1007/BF01495029",
    journal = "Z. Phys. C",
    volume = "14",
    pages = "95",
    year = "1982"
}

@article{PLUTO:1985yzc,
    author = "Berger, Christoph and others",
    collaboration = "PLUTO",
    title = "{A Study of Energy-energy Correlations in $e^+ e^-$ Annihilations at $\sqrt{s}=34$.6-{GeV}}",
    reportNumber = "DESY-85-039",
    doi = "10.1007/BF01413599",
    journal = "Z. Phys. C",
    volume = "28",
    pages = "365",
    year = "1985"
}

@article{Dixon:2019uzg,
    author = "Dixon, Lance J. and Moult, Ian and Zhu, Hua Xing",
    title = "{Collinear limit of the energy-energy correlator}",
    eprint = "1905.01310",
    archivePrefix = "arXiv",
    primaryClass = "hep-ph",
    reportNumber = "SLAC-PUB-17427, SLAC--PUB--17427",
    doi = "10.1103/PhysRevD.100.014009",
    journal = "Phys. Rev. D",
    volume = "100",
    number = "1",
    pages = "014009",
    year = "2019"
}

@article{Chen:2020vvp,
    author = "Chen, Hao and Moult, Ian and Zhang, XiaoYuan and Zhu, Hua Xing",
    title = "{Rethinking jets with energy correlators: Tracks, resummation, and analytic continuation}",
    eprint = "2004.11381",
    archivePrefix = "arXiv",
    primaryClass = "hep-ph",
    doi = "10.1103/PhysRevD.102.054012",
    journal = "Phys. Rev. D",
    volume = "102",
    number = "5",
    pages = "054012",
    year = "2020"
}

@article{Komiske:2022enw,
    author = "Komiske, Patrick T. and Moult, Ian and Thaler, Jesse and Zhu, Hua Xing",
    title = "{Analyzing N-Point Energy Correlators inside Jets with CMS Open Data}",
    eprint = "2201.07800",
    archivePrefix = "arXiv",
    primaryClass = "hep-ph",
    reportNumber = "MIT-CTP 5389",
    doi = "10.1103/PhysRevLett.130.051901",
    journal = "Phys. Rev. Lett.",
    volume = "130",
    number = "5",
    pages = "051901",
    year = "2023"
}

@article{CMS:2024mlf,
    author = "Hayrapetyan, Aram and others",
    collaboration = "CMS",
    title = "{Measurement of Energy Correlators inside Jets and Determination of the Strong Coupling {\ensuremath{\alpha}}S(mZ)}",
    eprint = "2402.13864",
    archivePrefix = "arXiv",
    primaryClass = "hep-ex",
    reportNumber = "CMS-SMP-22-015, CERN-EP-2024-010",
    doi = "10.1103/PhysRevLett.133.071903",
    journal = "Phys. Rev. Lett.",
    volume = "133",
    number = "7",
    pages = "071903",
    year = "2024"
}

@article{CMS:2024ovv,
    collaboration = "CMS",
    title = "{Energy-energy correlators from PbPb and pp collisions at 5.02 TeV}",
    reportNumber = "CMS-PAS-HIN-23-004",
    year = "2024"
}

@article{Moult:2025nhu,
    author = "Moult, Ian and Zhu, Hua Xing",
    title = "{Energy Correlators: A Journey From Theory to Experiment}",
    eprint = "2506.09119",
    archivePrefix = "arXiv",
    primaryClass = "hep-ph",
    month = "6",
    year = "2025"
}

@article{Chen:2019bpb,
    author = "Chen, Hao and Luo, Ming-Xing and Moult, Ian and Yang, Tong-Zhi and Zhang, Xiaoyuan and Zhu, Hua Xing",
    title = "{Three point energy correlators in the collinear limit: symmetries, dualities and analytic results}",
    eprint = "1912.11050",
    archivePrefix = "arXiv",
    primaryClass = "hep-ph",
    doi = "10.1007/JHEP08(2020)028",
    journal = "JHEP",
    volume = "08",
    number = "08",
    pages = "028",
    year = "2020"
}

@article{Chen:2022swd,
    author = "Chen, Hao and Moult, Ian and Thaler, Jesse and Zhu, Hua Xing",
    title = "{Non-Gaussianities in collider energy flux}",
    eprint = "2205.02857",
    archivePrefix = "arXiv",
    primaryClass = "hep-ph",
    reportNumber = "MIT-CTP 5430",
    doi = "10.1007/JHEP07(2022)146",
    journal = "JHEP",
    volume = "07",
    pages = "146",
    year = "2022"
}

@article{Chicherin:2024ifn,
    author = "Chicherin, Dmitry and Moult, Ian and Sokatchev, Emery and Yan, Kai and Zhu, Yunyue",
    title = "{Collinear limit of the four-point energy correlator in N=4 supersymmetric Yang-Mills theory}",
    eprint = "2401.06463",
    archivePrefix = "arXiv",
    primaryClass = "hep-th",
    reportNumber = "LAPTH-004/24",
    doi = "10.1103/PhysRevD.110.L091901",
    journal = "Phys. Rev. D",
    volume = "110",
    number = "9",
    pages = "L091901",
    year = "2024"
}

@article{Yan:2022cye,
    author = "Yan, Kai and Zhang, Xiaoyuan",
    title = "{Three-Point Energy Correlator in N=4 Supersymmetric Yang-Mills Theory}",
    eprint = "2203.04349",
    archivePrefix = "arXiv",
    primaryClass = "hep-th",
    doi = "10.1103/PhysRevLett.129.021602",
    journal = "Phys. Rev. Lett.",
    volume = "129",
    number = "2",
    pages = "021602",
    year = "2022"
}

@article{Yang:2022tgm,
    author = "Yang, Tong-Zhi and Zhang, Xiaoyuan",
    title = "{Analytic Computation of three-point energy correlator in QCD}",
    eprint = "2208.01051",
    archivePrefix = "arXiv",
    primaryClass = "hep-ph",
    reportNumber = "MSUHEP-22-025, ZU-TH 38/22",
    doi = "10.1007/JHEP09(2022)006",
    journal = "JHEP",
    volume = "09",
    pages = "006",
    year = "2022"
}

@article{Yang:2024gcn,
    author = "Yang, Tong-Zhi and Zhang, Xiaoyuan",
    title = "{Three-point energy correlators in hadronic Higgs boson decays}",
    eprint = "2402.05174",
    archivePrefix = "arXiv",
    primaryClass = "hep-ph",
    reportNumber = "ZU-TH 08/24",
    doi = "10.1103/PhysRevD.109.114036",
    journal = "Phys. Rev. D",
    volume = "109",
    number = "11",
    pages = "114036",
    year = "2024"
}

@article{He:2025zbz,
    author = "He, Song and Li, Xiang and Lin, Jingwen and Liu, Jiahao and Yan, Kai",
    title = "{Bootstrapping form factor squared in ${\cal N}=4$ super-Yang-Mills}",
    eprint = "2506.07796",
    archivePrefix = "arXiv",
    primaryClass = "hep-th",
    month = "6",
    year = "2025"
}

@article{Ma:2025qtx,
    author = "Ma, Rourou and Gong, Jianyu and Lin, Jingwen and Yan, Kai and Yang, Gang and Zhang, Yang",
    title = "{Differential equations for energy correlators in any angle}",
    eprint = "2506.02061",
    archivePrefix = "arXiv",
    primaryClass = "hep-ph",
    reportNumber = "USTC-ICTS/PCFT-25-18",
    doi = "10.1007/JHEP02(2026)025",
    journal = "JHEP",
    volume = "02",
    pages = "025",
    year = "2026"
}

@article{Volovich:2026pup,
    author = "Volovich, Anastasia and Wu, Di and Yan, Kai",
    title = "{Energy Correlators from Star Integrals via Mellin Space}",
    eprint = "2604.01071",
    archivePrefix = "arXiv",
    primaryClass = "hep-th",
    month = "4",
    year = "2026"
}

@article{He:2024hbb,
    author = "He, Song and Jiang, Xuhang and Yang, Qinglin and Zhang, Yao-Qi",
    title = "{From squared amplitudes to energy correlators}",
    eprint = "2408.04222",
    archivePrefix = "arXiv",
    primaryClass = "hep-th",
    month = "8",
    year = "2024"
}

@article{Knowles:1988hu,
    author = "Knowles, I. G.",
    title = "{A Linear Algorithm for Calculating Spin Correlations in Hadronic Collisions}",
    reportNumber = "NSF-ITP-88-78",
    doi = "10.1016/0010-4655(90)90063-7",
    journal = "Comput. Phys. Commun.",
    volume = "58",
    pages = "271--284",
    year = "1990"
}

@article{Knowles:1987cu,
    author = "Knowles, I. G.",
    title = "{Angular Correlations in {QCD}}",
    reportNumber = "CAVENDISH-HEP-87/5",
    doi = "10.1016/0550-3213(88)90653-0",
    journal = "Nucl. Phys. B",
    volume = "304",
    pages = "767--793",
    year = "1988"
}

@article{Knowles:1988vs,
    author = "Knowles, I. G.",
    title = "{Spin Correlations in Parton - Parton Scattering}",
    reportNumber = "Cavendish-HEP-88/5, NSF-ITP-88-36",
    doi = "10.1016/0550-3213(88)90092-2",
    journal = "Nucl. Phys. B",
    volume = "310",
    pages = "571--588",
    year = "1988"
}

@article{Bahr:2008pv,
    author = "Bahr, M. and others",
    title = "{Herwig++ Physics and Manual}",
    eprint = "0803.0883",
    archivePrefix = "arXiv",
    primaryClass = "hep-ph",
    reportNumber = "CERN-PH-TH-2008-038, CAVENDISH-HEP-08-03, KA-TP-05-2008, DCPT-08-22, IPPP-08-11, CP3-08-05",
    doi = "10.1140/epjc/s10052-008-0798-9",
    journal = "Eur. Phys. J. C",
    volume = "58",
    pages = "639--707",
    year = "2008"
}

@article{Bellm:2015jjp,
    author = "Bellm, Johannes and others",
    title = "{Herwig 7.0/Herwig++ 3.0 release note}",
    eprint = "1512.01178",
    archivePrefix = "arXiv",
    primaryClass = "hep-ph",
    reportNumber = "CERN-PH-TH-2015-289, MAN-HEP-2015-15, IFJPAN-IV-2015-13, KA-TP-18-2015, DCPT-15-142, MCNET-15-28, IPPP-15-71, HERWIG-2015-01",
    doi = "10.1140/epjc/s10052-016-4018-8",
    journal = "Eur. Phys. J. C",
    volume = "76",
    number = "4",
    pages = "196",
    year = "2016"
}

@article{Bellm:2019zci,
    author = "Bellm, Johannes and others",
    title = "{Herwig 7.2 release note}",
    eprint = "1912.06509",
    archivePrefix = "arXiv",
    primaryClass = "hep-ph",
    reportNumber = "MAN/HEP/2019/011, CERN-TH-2019-213, IFJPAN-IV-2019-18, HERWIG-2019-02, UWTHPH-19-36, KA-TP-24-2019, LU-TP 19-57, MCnet-19-28, IPPP/19/91",
    doi = "10.1140/epjc/s10052-020-8011-x",
    journal = "Eur. Phys. J. C",
    volume = "80",
    number = "5",
    pages = "452",
    year = "2020"
}

@article{Richardson:2018pvo,
    author = "Richardson, Peter and Webster, Stephen",
    title = "{Spin Correlations in Parton Shower Simulations}",
    eprint = "1807.01955",
    archivePrefix = "arXiv",
    primaryClass = "hep-ph",
    reportNumber = "IPPP/18/55, CERN-TH-2018-154, IPPP-18-55, MCNET-18-12",
    doi = "10.1140/epjc/s10052-019-7429-5",
    journal = "Eur. Phys. J. C",
    volume = "80",
    number = "2",
    pages = "83",
    year = "2020"
}

@article{Karlberg:2021kwr,
    author = "Karlberg, Alexander and Salam, Gavin P. and Scyboz, Ludovic and Verheyen, Rob",
    title = "{Spin correlations in final-state parton showers and jet observables}",
    eprint = "2103.16526",
    archivePrefix = "arXiv",
    primaryClass = "hep-ph",
    doi = "10.1140/epjc/s10052-021-09378-0",
    journal = "Eur. Phys. J. C",
    volume = "81",
    number = "8",
    pages = "681",
    year = "2021"
}

@article{Hamilton:2021dyz,
    author = "Hamilton, Keith and Karlberg, Alexander and Salam, Gavin P. and Scyboz, Ludovic and Verheyen, Rob",
    title = "{Soft spin correlations in final-state parton showers}",
    eprint = "2111.01161",
    archivePrefix = "arXiv",
    primaryClass = "hep-ph",
    doi = "10.1007/JHEP03(2022)193",
    journal = "JHEP",
    volume = "03",
    pages = "193",
    year = "2022"
}

@article{Lee:2023xzv,
    author = "Lee, Kyle and Moult, Ian and Ringer, Felix and Waalewijn, Wouter J.",
    title = "{A formalism for extracting track functions from jet measurements}",
    eprint = "2308.00028",
    archivePrefix = "arXiv",
    primaryClass = "hep-ph",
    reportNumber = "MIT-CTP-5589, JLAB-THY-23-3891",
    doi = "10.1007/JHEP01(2024)194",
    journal = "JHEP",
    volume = "01",
    pages = "194",
    year = "2024"
}

@article{Andres:2023ymw,
    author = "Andres, Carlota and Dominguez, Fabio and Holguin, Jack and Marquet, Cyrille and Moult, Ian",
    title = "{Seeing beauty in the quark-gluon plasma with energy correlators}",
    eprint = "2307.15110",
    archivePrefix = "arXiv",
    primaryClass = "hep-ph",
    doi = "10.1103/PhysRevD.110.L031503",
    journal = "Phys. Rev. D",
    volume = "110",
    number = "3",
    pages = "L031503",
    year = "2024"
}

@article{Craft:2022kdo,
    author = "Craft, Evan and Lee, Kyle and Me{\c{c}}aj, Bianka and Moult, Ian",
    title = "{Beautiful and Charming Energy Correlators}",
    eprint = "2210.09311",
    archivePrefix = "arXiv",
    primaryClass = "hep-ph",
    reportNumber = "MIT-CTP 5474",
    month = "10",
    year = "2022"
}

@article{Gao:2026xuq,
    author = "Gao, Anjie and Lee, Kyle and Zhang, Xiaoyuan",
    title = "{Precision Jet Substructure of Boosted Boson Decays with Energy Correlators}",
    eprint = "2601.20933",
    archivePrefix = "arXiv",
    primaryClass = "hep-ph",
    reportNumber = "MIT-CTP 5996, SLAC-PUB-260120",
    month = "1",
    year = "2026"
}

@article{Barata:2025uxp,
    author = "Barata, Jo{\~a}o and Brewer, Jasmine and Lee, Kyle and Silva, Jo{\~a}o M.",
    title = "{Heavy Quark Pair Energy Correlators: From Profiling Partonic Splittings to Probing Heavy-Flavor Fragmentation}",
    eprint = "2508.19404",
    archivePrefix = "arXiv",
    primaryClass = "hep-ph",
    reportNumber = "CERN-TH-2025-172, MIT-CTP 5888",
    month = "8",
    year = "2025"
}

@article{Lee:2023npz,
    author = "Lee, Kyle and Moult, Ian",
    title = "{Energy Correlators Taking Charge}",
    eprint = "2308.00746",
    archivePrefix = "arXiv",
    primaryClass = "hep-ph",
    reportNumber = "MIT-CTP-5590",
    month = "8",
    year = "2023"
}

@article{Lee:2023tkr,
    author = "Lee, Kyle and Moult, Ian",
    title = "{Joint Track Functions: Expanding the Space of Calculable Correlations at Colliders}",
    eprint = "2308.01332",
    archivePrefix = "arXiv",
    primaryClass = "hep-ph",
    reportNumber = "MIT-CTP-5591",
    month = "8",
    year = "2023"
}

@article{Hoche:2017iem,
    author = {H{\"o}che, Stefan and Prestel, Stefan},
    title = "{Triple collinear emissions in parton showers}",
    eprint = "1705.00742",
    archivePrefix = "arXiv",
    primaryClass = "hep-ph",
    reportNumber = "SLAC-PUB-16963, FERMILAB-PUB-17-122-T, MCNET-17-05",
    doi = "10.1103/PhysRevD.96.074017",
    journal = "Phys. Rev. D",
    volume = "96",
    number = "7",
    pages = "074017",
    year = "2017"
}

@article{Dasgupta:2020fwr,
    author = "Dasgupta, Mrinal and Dreyer, Fr{\'e}d{\'e}ric A. and Hamilton, Keith and Monni, Pier Francesco and Salam, Gavin P. and Soyez, Gregory",
    title = "{Parton showers beyond leading logarithmic accuracy}",
    eprint = "2002.11114",
    archivePrefix = "arXiv",
    primaryClass = "hep-ph",
    reportNumber = "CERN-TH-2020-026",
    doi = "10.1103/PhysRevLett.125.052002",
    journal = "Phys. Rev. Lett.",
    volume = "125",
    number = "5",
    pages = "052002",
    year = "2020"
}

@article{Hamilton:2020rcu,
    author = "Hamilton, Keith and Medves, Rok and Salam, Gavin P. and Scyboz, Ludovic and Soyez, Gregory",
    title = "{Colour and logarithmic accuracy in final-state parton showers}",
    eprint = "2011.10054",
    archivePrefix = "arXiv",
    primaryClass = "hep-ph",
    doi = "10.1007/JHEP03(2021)041",
    journal = "JHEP",
    volume = "03",
    number = "041",
    pages = "041",
    year = "2021"
}

@article{Lee:2022uwt,
    author = "Lee, Kyle and Me{\c{c}}aj, Bianka and Moult, Ian",
    title = "{Conformal collider physics meets LHC data}",
    eprint = "2205.03414",
    archivePrefix = "arXiv",
    primaryClass = "hep-ph",
    doi = "10.1103/PhysRevD.111.L011502",
    journal = "Phys. Rev. D",
    volume = "111",
    number = "1",
    pages = "L011502",
    year = "2025"
}

@article{Lee:2024icn,
    author = "Lee, Kyle and Moult, Ian and Zhang, Xiaoyuan",
    title = "{Revisiting single inclusive jet production: timelike factorization and reciprocity}",
    eprint = "2409.19045",
    archivePrefix = "arXiv",
    primaryClass = "hep-ph",
    reportNumber = "MIT-CTP 5766",
    doi = "10.1007/JHEP05(2025)129",
    journal = "JHEP",
    volume = "05",
    pages = "129",
    year = "2025"
}

@article{Lee:2026zyl,
    author = "Lee, Kyle and Li, Yibei and Xu, Zhen and Zhang, Xiaoyuan",
    title = "{Projected Energy Correlators: Two-Loop Jet Functions and NNLL Resummation}",
    eprint = "2606.02714",
    archivePrefix = "arXiv",
    primaryClass = "hep-ph",
    reportNumber = "MITP-26-021, DESY-26-069, MIT-CTP 6028",
    month = "6",
    year = "2026"
}

@article{Lee:2024esz,
    author = "Lee, Kyle and Pathak, Aditya and Stewart, Iain W. and Sun, Zhiquan",
    title = "{Nonperturbative Effects in Energy Correlators: From Characterizing Confinement Transition to Improving {\ensuremath{\alpha}}s Extraction}",
    eprint = "2405.19396",
    archivePrefix = "arXiv",
    primaryClass = "hep-ph",
    reportNumber = "MIT-CTP 5711, DESY-24-064",
    doi = "10.1103/PhysRevLett.133.231902",
    journal = "Phys. Rev. Lett.",
    volume = "133",
    number = "23",
    pages = "231902",
    year = "2024"
}

@article{Chen:2023zlx,
    author = "Chen, Wen and Gao, Jun and Li, Yibei and Xu, Zhen and Zhang, Xiaoyuan and Zhu, Hua Xing",
    title = "{NNLL resummation for projected three-point energy correlator}",
    eprint = "2307.07510",
    archivePrefix = "arXiv",
    primaryClass = "hep-ph",
    doi = "10.1007/JHEP05(2024)043",
    journal = "JHEP",
    volume = "05",
    pages = "043",
    year = "2024"
}

@article{ALICE:2025igw,
    author = "Acharya, Shreyasi and others",
    collaboration = "ALICE",
    title = "{Energy-energy correlators in charm-tagged jets in proton-proton collisions at $\mathbf{\sqrt{s} = 13}$ TeV}",
    eprint = "2504.03431",
    archivePrefix = "arXiv",
    primaryClass = "hep-ex",
    reportNumber = "CERN-EP-2025-082",
    month = "4",
    year = "2025"
}

@article{Lee:2026hub,
    author = "Lee, Kyle and Moult, Ian and Waalewijn, Wouter J.",
    title = "{Putting Jet Substructure on Track(s)}",
    eprint = "2607.00087",
    archivePrefix = "arXiv",
    primaryClass = "hep-ph",
    month = "6",
    year = "2026"
}

@article{Lee:2025okn,
    author = "Lee, Kyle and Stewart, Iain W.",
    title = "{Dihadron Fragmentation and the Confinement Transition in Energy Correlators}",
    eprint = "2507.11495",
    archivePrefix = "arXiv",
    primaryClass = "hep-ph",
    reportNumber = "MIT-CTP 5889",
    doi = "10.1103/m18j-xypt",
    journal = "Phys. Rev. Lett.",
    volume = "136",
    number = "8",
    pages = "081902",
    year = "2026"
}

@article{Hoche:2017hno,
    author = {H{\"o}che, Stefan and Krauss, Frank and Prestel, Stefan},
    title = "{Implementing NLO DGLAP evolution in Parton Showers}",
    eprint = "1705.00982",
    archivePrefix = "arXiv",
    primaryClass = "hep-ph",
    reportNumber = "SLAC-PUB-16965, FERMILAB-PUB-17-134-T, IPPP-17-34, DCPT-17-68, MCNET-17-06",
    doi = "10.1007/JHEP10(2017)093",
    journal = "JHEP",
    volume = "10",
    pages = "093",
    year = "2017"
}

@article{Hoche:2024dee,
    author = {H{\"o}che, Stefan and Krauss, Frank and Reichelt, Daniel},
    title = "{alaric parton shower for hadron colliders}",
    eprint = "2404.14360",
    archivePrefix = "arXiv",
    primaryClass = "hep-ph",
    reportNumber = "FERMILAB-PUB-24-0178-T, IPPP/24/20, MCNET-24-07",
    doi = "10.1103/PhysRevD.111.094032",
    journal = "Phys. Rev. D",
    volume = "111",
    number = "9",
    pages = "094032",
    year = "2025"
}

@article{vanBeekveld:2025lpz,
    author = "van Beekveld, Melissa and Ferrario Ravasio, Silvia and Helliwell, Jack and Karlberg, Alexander and Salam, Gavin P. and Scyboz, Ludovic and Soto-Ontoso, Alba and Soyez, Gregory and Zanoli, Silvia",
    title = "{Logarithmically-accurate and positive-definite NLO shower matching}",
    eprint = "2504.05377",
    archivePrefix = "arXiv",
    primaryClass = "hep-ph",
    reportNumber = "CERN-TH-2025-004, OUTP-25-01P, Nikhef 2025-003",
    doi = "10.1007/JHEP10(2025)038",
    journal = "JHEP",
    volume = "10",
    pages = "038",
    year = "2025"
}

@article{vanBeekveld:2024qxs,
    author = "van Beekveld, Melissa and Dasgupta, Mrinal and El-Menoufi, Basem Kamal and Helliwell, Jack and Monni, Pier Francesco and Salam, Gavin P.",
    title = "{A collinear shower algorithm for NSL non-singlet fragmentation}",
    eprint = "2409.08316",
    archivePrefix = "arXiv",
    primaryClass = "hep-ph",
    reportNumber = "CERN-TH-2024-125, Nikhef 2024-013, OUTP-24-05P",
    doi = "10.1007/JHEP03(2025)209",
    journal = "JHEP",
    volume = "03",
    pages = "209",
    year = "2025"
}

@article{FerrarioRavasio:2023kyg,
    author = "Ferrario Ravasio, Silvia and Hamilton, Keith and Karlberg, Alexander and Salam, Gavin P. and Scyboz, Ludovic and Soyez, Gregory",
    title = "{Parton Showering with Higher Logarithmic Accuracy for Soft Emissions}",
    eprint = "2307.11142",
    archivePrefix = "arXiv",
    primaryClass = "hep-ph",
    reportNumber = "CERN-TH-2023-127, OUTP-23-07P",
    doi = "10.1103/PhysRevLett.131.161906",
    journal = "Phys. Rev. Lett.",
    volume = "131",
    number = "16",
    pages = "161906",
    year = "2023"
}

@article{Hamilton:2023dwb,
    author = "Hamilton, Keith and Karlberg, Alexander and Salam, Gavin P. and Scyboz, Ludovic and Verheyen, Rob",
    title = "{Matching and event-shape NNDL accuracy in parton showers}",
    eprint = "2301.09645",
    archivePrefix = "arXiv",
    primaryClass = "hep-ph",
    reportNumber = "CERN-TH-2023-004",
    doi = "10.1007/JHEP03(2023)224",
    journal = "JHEP",
    volume = "03",
    pages = "224",
    year = "2023",
    note = "[Erratum: JHEP 11, 060 (2023)]"
}

@article{vanBeekveld:2022ukn,
    author = "van Beekveld, Melissa and Ferrario Ravasio, Silvia and Hamilton, Keith and Salam, Gavin P. and Soto-Ontoso, Alba and Soyez, Gregory and Verheyen, Rob",
    title = "{PanScales showers for hadron collisions: all-order validation}",
    eprint = "2207.09467",
    archivePrefix = "arXiv",
    primaryClass = "hep-ph",
    reportNumber = "OUTP-22-10P",
    doi = "10.1007/JHEP11(2022)020",
    journal = "JHEP",
    volume = "11",
    pages = "020",
    year = "2022"
}

@article{vanBeekveld:2022zhl,
    author = "van Beekveld, Melissa and Ferrario Ravasio, Silvia and Salam, Gavin P. and Soto-Ontoso, Alba and Soyez, Gregory and Verheyen, Rob",
    title = "{PanScales parton showers for hadron collisions: formulation and fixed-order studies}",
    eprint = "2205.02237",
    archivePrefix = "arXiv",
    primaryClass = "hep-ph",
    reportNumber = "OUTP-22-06P",
    doi = "10.1007/JHEP11(2022)019",
    journal = "JHEP",
    volume = "11",
    pages = "019",
    year = "2022"
}

@article{Chen:2021gdk,
    author = "Chen, Hao and Moult, Ian and Zhu, Hua Xing",
    title = "{Spinning gluons from the QCD light-ray OPE}",
    eprint = "2104.00009",
    archivePrefix = "arXiv",
    primaryClass = "hep-ph",
    doi = "10.1007/JHEP08(2022)233",
    journal = "JHEP",
    volume = "08",
    pages = "233",
    year = "2022"
}

@article{Chen:2020adz,
    author = "Chen, Hao and Moult, Ian and Zhu, Hua Xing",
    title = "{Quantum Interference in Jet Substructure from Spinning Gluons}",
    eprint = "2011.02492",
    archivePrefix = "arXiv",
    primaryClass = "hep-ph",
    doi = "10.1103/PhysRevLett.126.112003",
    journal = "Phys. Rev. Lett.",
    volume = "126",
    number = "11",
    pages = "112003",
    year = "2021"
}

@article{Catani:1998nv,
    author = "Catani, Stefano and Grazzini, Massimiliano",
    title = "{Collinear factorization and splitting functions for next-to-next-to-leading order QCD calculations}",
    eprint = "hep-ph/9810389",
    archivePrefix = "arXiv",
    reportNumber = "CERN-TH-98-325, ETH-TH-98-27",
    doi = "10.1016/S0370-2693(98)01513-5",
    journal = "Phys. Lett. B",
    volume = "446",
    pages = "143--152",
    year = "1999"
}

@article{Campbell:1997hg,
    author = "Campbell, John M. and Glover, E. W. Nigel",
    title = "{Double unresolved approximations to multiparton scattering amplitudes}",
    eprint = "hep-ph/9710255",
    archivePrefix = "arXiv",
    reportNumber = "DTP-97-82",
    doi = "10.1016/S0550-3213(98)00295-8",
    journal = "Nucl. Phys. B",
    volume = "527",
    pages = "264--288",
    year = "1998"
}

@article{Braun-White:2022rtg,
    author = "Braun-White, Oscar and Glover, Nigel",
    title = "{Decomposition of triple collinear splitting functions}",
    eprint = "2204.10755",
    archivePrefix = "arXiv",
    primaryClass = "hep-ph",
    reportNumber = "IPPP/22/21",
    doi = "10.1007/JHEP09(2022)059",
    journal = "JHEP",
    volume = "09",
    pages = "059",
    year = "2022"
}

@article{Craft:2023aew,
    author = "Craft, Evan and Gonzalez, Mark and Lee, Kyle and Mecaj, Bianka and Moult, Ian",
    title = "{The 1 {\textrightarrow} 3 massive splitting functions from QCD factorization and SCET}",
    eprint = "2310.06736",
    archivePrefix = "arXiv",
    primaryClass = "hep-ph",
    reportNumber = "MIT-CTP-5605",
    doi = "10.1007/JHEP07(2024)080",
    journal = "JHEP",
    volume = "07",
    pages = "080",
    year = "2024"
}

@article{Hoche:2025vto,
    author = {H{\"o}che, Stefan and LeBlanc, Matt and Roloff, Jennifer and Whitman, Grant},
    title = "{Massive tree-level splitting functions beyond kinematical limits}",
    eprint = "2512.07025",
    archivePrefix = "arXiv",
    primaryClass = "hep-ph",
    reportNumber = "FERMILAB-PUB-25-0815-T, MCNET-25-28",
    doi = "10.1103/7jj6-qfqh",
    journal = "Phys. Rev. D",
    volume = "113",
    number = "5",
    pages = "054009",
    year = "2026"
}

@article{Dhani:2023uxu,
    author = "Dhani, Prasanna K. and Rodrigo, Germ{\'a}n and Sborlini, German F. R.",
    title = "{Triple-collinear splittings with massive particles}",
    eprint = "2310.05803",
    archivePrefix = "arXiv",
    primaryClass = "hep-ph",
    reportNumber = "IFIC/23-46",
    doi = "10.1007/JHEP12(2023)188",
    journal = "JHEP",
    volume = "12",
    pages = "188",
    year = "2023"
}

@article{Campbell:2025lrs,
    author = {Campbell, John M. and H{\"o}che, Stefan and Knobbe, Max and Preuss, Christian T. and Reichelt, Daniel},
    title = "{QCD splitting functions beyond kinematical limits}",
    eprint = "2505.10408",
    archivePrefix = "arXiv",
    primaryClass = "hep-ph",
    reportNumber = "FERMILAB-PUB-25-0276-T, CERN-TH-2025-091, MCNET-25-09",
    doi = "10.1103/f5c5-hsdd",
    journal = "Phys. Rev. D",
    volume = "113",
    number = "5",
    pages = "054031",
    year = "2026"
}

@article{DelDuca:2020vst,
    author = "Del Duca, Vittorio and Duhr, Claude and Haindl, Rayan and Lazopoulos, Achilleas and Michel, Martin",
    title = "{Tree-level splitting amplitudes for a gluon into four collinear partons}",
    eprint = "2007.05345",
    archivePrefix = "arXiv",
    primaryClass = "hep-ph",
    doi = "10.1007/JHEP10(2020)093",
    journal = "JHEP",
    volume = "10",
    pages = "093",
    year = "2020"
}

@article{DelDuca:2019ggv,
    author = "Del Duca, Vittorio and Duhr, Claude and Haindl, Rayan and Lazopoulos, Achilleas and Michel, Martin",
    title = "{Tree-level splitting amplitudes for a quark into four collinear partons}",
    eprint = "1912.06425",
    archivePrefix = "arXiv",
    primaryClass = "hep-ph",
    reportNumber = "CERN-TH-2019-221, CP3-19-57",
    doi = "10.1007/JHEP02(2020)189",
    journal = "JHEP",
    volume = "02",
    pages = "189",
    year = "2020"
}

@article{Bossi:2024qho,
    author = "Bossi, Hannah and Kudinoor, Arjun Srinivasan and Moult, Ian and Pablos, Daniel and Rai, Ananya and Rajagopal, Krishna",
    title = "{Imaging the wakes of jets with energy-energy-energy correlators}",
    eprint = "2407.13818",
    archivePrefix = "arXiv",
    primaryClass = "hep-ph",
    reportNumber = "MIT-CTP-5739",
    doi = "10.1007/JHEP12(2024)073",
    journal = "JHEP",
    volume = "12",
    pages = "073",
    year = "2024"
}

@article{Barata:2025fzd,
    author = "Barata, Jo{\~a}o and Moult, Ian and Sadofyev, Andrey V. and Silva, Jo{\~a}o M.",
    title = "{Dissecting Jet Modification in the QGP with Multi-Point Energy Correlators}",
    eprint = "2503.13603",
    archivePrefix = "arXiv",
    primaryClass = "hep-ph",
    reportNumber = "CERN-TH-2025-029",
    month = "3",
    year = "2025"
}

@article{Holguin:2023bjf,
    author = {Holguin, Jack and Moult, Ian and Pathak, Aditya and Procura, Massimiliano and Sch{\"o}fbeck, Robert and Schwarz, Dennis},
    title = "{Using the W Boson as a Standard Candle to Reach the Top: Calibrating Energy-Correlator-Based Top Mass Measurements}",
    eprint = "2311.02157",
    archivePrefix = "arXiv",
    primaryClass = "hep-ph",
    reportNumber = "UWThPh 2023-26; DESY-23-176",
    doi = "10.1103/j4sp-fcmd",
    journal = "Phys. Rev. Lett.",
    volume = "134",
    number = "23",
    pages = "231903",
    year = "2025"
}

@article{Holguin:2022epo,
    author = "Holguin, Jack and Moult, Ian and Pathak, Aditya and Procura, Massimiliano",
    title = "{New paradigm for precision top physics: Weighing the top with energy correlators}",
    eprint = "2201.08393",
    archivePrefix = "arXiv",
    primaryClass = "hep-ph",
    doi = "10.1103/PhysRevD.107.114002",
    journal = "Phys. Rev. D",
    volume = "107",
    number = "11",
    pages = "114002",
    year = "2023"
}

@article{Holguin:2024tkz,
    author = {Holguin, Jack and Moult, Ian and Pathak, Aditya and Procura, Massimiliano and Sch{\"o}fbeck, Robert and Schwarz, Dennis},
    title = "{Top quark mass extractions from energy correlators: a feasibility study}",
    eprint = "2407.12900",
    archivePrefix = "arXiv",
    primaryClass = "hep-ph",
    reportNumber = "DESY-24-107;UWThPh 2024-14, DESY-24-107, UWThPh 2024-14",
    doi = "10.1007/JHEP04(2025)072",
    journal = "JHEP",
    volume = "04",
    pages = "072",
    year = "2025"
}

@article{Alipour-fard:2024szj,
    author = "Alipour-fard, Samuel and Budhraja, Ankita and Thaler, Jesse and Waalewijn, Wouter J.",
    title = "{New Angles on Energy Correlators}",
    eprint = "2410.16368",
    archivePrefix = "arXiv",
    primaryClass = "hep-ph",
    reportNumber = "MIT-CTP/5794",
    doi = "10.1103/l6nj-2gsh",
    journal = "Phys. Rev. Lett.",
    volume = "134",
    number = "23",
    pages = "231902",
    year = "2025"
}

@article{Arkani-Hamed:2015bza,
    author = "Arkani-Hamed, Nima and Maldacena, Juan",
    title = "{Cosmological Collider Physics}",
    eprint = "1503.08043",
    archivePrefix = "arXiv",
    primaryClass = "hep-th",
    month = "3",
    year = "2015"
}

@article{Arkani-Hamed:2018kmz,
    author = "Arkani-Hamed, Nima and Baumann, Daniel and Lee, Hayden and Pimentel, Guilherme L.",
    title = "{The Cosmological Bootstrap: Inflationary Correlators from Symmetries and Singularities}",
    eprint = "1811.00024",
    archivePrefix = "arXiv",
    primaryClass = "hep-th",
    doi = "10.1007/JHEP04(2020)105",
    journal = "JHEP",
    volume = "04",
    pages = "105",
    year = "2020"
}

@article{Chen:2022jhb,
    author = "Chen, Hao and Moult, Ian and Sandor, Joshua and Zhu, Hua Xing",
    title = "{Celestial blocks and transverse spin in the three-point energy correlator}",
    eprint = "2202.04085",
    archivePrefix = "arXiv",
    primaryClass = "hep-ph",
    doi = "10.1007/JHEP09(2022)199",
    journal = "JHEP",
    volume = "09",
    pages = "199",
    year = "2022"
}

@article{Alwall:2011uj,
    author = "Alwall, Johan and Herquet, Michel and Maltoni, Fabio and Mattelaer, Olivier and Stelzer, Tim",
    title = "{MadGraph 5 : Going Beyond}",
    eprint = "1106.0522",
    archivePrefix = "arXiv",
    primaryClass = "hep-ph",
    reportNumber = "FERMILAB-PUB-11-448-T",
    doi = "10.1007/JHEP06(2011)128",
    journal = "JHEP",
    volume = "06",
    pages = "128",
    year = "2011"
}

@article{Sjostrand:2014zea,
    author = {Sj{\"o}strand, Torbj{\"o}rn and Ask, Stefan and Christiansen, Jesper R. and Corke, Richard and Desai, Nishita and Ilten, Philip and Mrenna, Stephen and Prestel, Stefan and Rasmussen, Christine O. and Skands, Peter Z.},
    title = "{An introduction to PYTHIA 8.2}",
    eprint = "1410.3012",
    archivePrefix = "arXiv",
    primaryClass = "hep-ph",
    reportNumber = "LU-TP-14-36, MCNET-14-22, CERN-PH-TH-2014-190, FERMILAB-PUB-14-316-CD, DESY-14-178, SLAC-PUB-16122",
    doi = "10.1016/j.cpc.2015.01.024",
    journal = "Comput. Phys. Commun.",
    volume = "191",
    pages = "159--177",
    year = "2015"
}

@article{CMS:2022awf,
    author = "Tumasyan, Armen and others",
    collaboration = "CMS",
    title = "{CMS pythia ~8 colour reconnection tunes based on underlying-event data}",
    eprint = "2205.02905",
    archivePrefix = "arXiv",
    primaryClass = "hep-ex",
    reportNumber = "CMS-GEN-17-002, CERN-EP-2022-005",
    doi = "10.1140/epjc/s10052-023-11630-8",
    journal = "Eur. Phys. J. C",
    volume = "83",
    number = "7",
    pages = "587",
    year = "2023"
}

@article{Gieseke:2012ft,
    author = "Gieseke, Stefan and Rohr, Christian and Siodmok, Andrzej",
    title = "{Colour reconnections in Herwig++}",
    eprint = "1206.0041",
    archivePrefix = "arXiv",
    primaryClass = "hep-ph",
    reportNumber = "MCNET-12-06, KA-TP-17-2012, MAN-HEP-2012-03",
    doi = "10.1140/epjc/s10052-012-2225-5",
    journal = "Eur. Phys. J. C",
    volume = "72",
    pages = "2225",
    year = "2012"
}

@article{CMS:2026jnt,
    author = "Hayrapetyan, Aram and others",
    collaboration = "CMS",
    title = "{Measurement of angular correlations inside jets induced by gluon polarization in proton-proton collisions at $\sqrt{s}$ = 13.6 TeV}",
    eprint = "2603.03689",
    archivePrefix = "arXiv",
    primaryClass = "hep-ex",
    reportNumber = "CMS-SMP-25-006, CERN-EP-2026-022",
    month = "3",
    year = "2026"
}

@article{Fischer:2017htu,
    author = "Fischer, Nadine and Lifson, Andrew and Skands, Peter",
    title = "{Helicity Antenna Showers for Hadron Colliders}",
    eprint = "1708.01736",
    archivePrefix = "arXiv",
    primaryClass = "hep-ph",
    reportNumber = "COEPP-MN-17-5, MCNET-17-13, CoEPP-MN-17-5",
    doi = "10.1140/epjc/s10052-017-5306-7",
    journal = "Eur. Phys. J. C",
    volume = "77",
    number = "10",
    pages = "719",
    year = "2017"
}

@article{Andersson:1983ia,
    author = "Andersson, Bo and Gustafson, G. and Ingelman, G. and Sjostrand, T.",
    title = "{Parton Fragmentation and String Dynamics}",
    reportNumber = "LU-TP-83-10",
    doi = "10.1016/0370-1573(83)90080-7",
    journal = "Phys. Rept.",
    volume = "97",
    pages = "31--145",
    year = "1983"
}

@article{Sjostrand:1984ic,
    author = "Sjostrand, Torbjorn",
    title = "{Jet Fragmentation of Nearby Partons}",
    reportNumber = "DESY-T-84-01",
    doi = "10.1016/0550-3213(84)90607-2",
    journal = "Nucl. Phys. B",
    volume = "248",
    pages = "469--502",
    year = "1984"
}

@article{Webber:1983if,
    author = "Webber, B. R.",
    title = "{A QCD Model for Jet Fragmentation Including Soft Gluon Interference}",
    reportNumber = "CERN-TH-3713",
    doi = "10.1016/0550-3213(84)90333-X",
    journal = "Nucl. Phys. B",
    volume = "238",
    pages = "492--528",
    year = "1984"
}

@article{ParticleDataGroup:2024cfk,
    author = "Navas, S. and others",
    collaboration = "Particle Data Group",
    title = "{Review of particle physics}",
    doi = "10.1103/PhysRevD.110.030001",
    journal = "Phys. Rev. D",
    volume = "110",
    number = "3",
    pages = "030001",
    year = "2024"
}

@article{Cacciari:2008gp,
    author = "Cacciari, Matteo and Salam, Gavin P. and Soyez, Gregory",
    title = "{The anti-$k_t$ jet clustering algorithm}",
    eprint = "0802.1189",
    archivePrefix = "arXiv",
    primaryClass = "hep-ph",
    reportNumber = "LPTHE-07-03",
    doi = "10.1088/1126-6708/2008/04/063",
    journal = "JHEP",
    volume = "04",
    pages = "063",
    year = "2008"
}

@article{Catani:1999ss,
    author = "Catani, Stefano and Grazzini, Massimiliano",
    title = "{Infrared factorization of tree level QCD amplitudes at the next-to-next-to-leading order and beyond}",
    eprint = "hep-ph/9908523",
    archivePrefix = "arXiv",
    reportNumber = "CERN-TH-99-263, ETH-TH-99-22",
    doi = "10.1016/S0550-3213(99)00778-6",
    journal = "Nucl. Phys. B",
    volume = "570",
    pages = "287--325",
    year = "2000"
}

@article{CMS:2020dqt,
    author = "Sirunyan, Albert M and others",
    collaboration = "CMS",
    title = "{Development and validation of HERWIG 7 tunes from CMS underlying-event measurements}",
    eprint = "2011.03422",
    archivePrefix = "arXiv",
    primaryClass = "hep-ex",
    reportNumber = "CMS-GEN-19-001, CERN-EP-2020-182",
    doi = "10.1140/epjc/s10052-021-08949-5",
    journal = "Eur. Phys. J. C",
    volume = "81",
    number = "4",
    pages = "312",
    year = "2021"
}
\bibliographystyle{JHEP}

\end{document}